\documentclass{article}

\PassOptionsToPackage{numbers, compress}{natbib}

\usepackage[final]{neurips_2025}

\usepackage[utf8]{inputenc} 
\usepackage[T1]{fontenc}    
\usepackage{hyperref}       
\usepackage{url}            
\usepackage{booktabs}       
\usepackage{amsfonts}       
\usepackage{nicefrac}       
\usepackage{microtype}      
\usepackage{xcolor}         

\title{Neural Atlas Graphs for Dynamic Scene Decomposition and Editing}

\author{%
    \bgroup
    
    \begin{tabular}{c c c}
         \textbf{Jan Philipp Schneider}$^{1,2}$ &  \textbf{Pratik Singh Bisht}$^{1}$ & \textbf{Ilya Chugunov}$^{2}$ \\
         \textbf{Andreas Kolb}$^{1}$ & \textbf{Michael Moeller}$^{1, 3}$ & \textbf{Felix Heide}$^{2, 4}$ \\
    \end{tabular}
    \egroup
    \\
  \newline
  \bgroup
  
  \begin{tabular}{c c c c}
         $^{1}$University of Siegen & $^{2}$Princeton University  & $^{3}$ Lamarr Institute &  $^{4}$ Torc Robotics \\
  \end{tabular}
  \egroup
  }

\usepackage{keyval}
\usepackage{graphicx}
\usepackage{amsmath}
\usepackage{usebib}
\usepackage{blkarray}
\usepackage{multirow}
\usepackage{nicematrix}
\usepackage[nolist]{acronym}
\usepackage{amssymb}
\usepackage{comment}
\usepackage{caption}
\usepackage{subcaption}
\usepackage{tcolorbox}
\usepackage{adjustbox}
\usepackage{calc}
\usepackage{tabularray}
\usepackage{rotating}
\usepackage{booktabs}
\usepackage{siunitx}
\usepackage{placeins}
\bibinput{main}
\usepackage{wrapfig}
\usepackage{amsmath}
\usepackage{mathtools}
\usepackage{hyperref}
\usepackage[toc,page,header]{appendix}
\usepackage{minitoc}
\makeatletter
\newcommand\footnoteref[1]{\protected@xdef\@thefnmark{\ref{#1}}\@footnotemark}
\makeatother

\newcommand{\todo}[1]{{\color{red}#1}}

\newcommand{\figref}[1]{Fig. \ref{#1}}
\newcommand{\secref}[1]{Sec. \ref{#1}}
\newcommand{\tabref}[1]{Tab. \ref{#1}}

\begin{acronym}
    \acro{psf}[PSF]{Panoptic Sprite Fields}
    \acro{nag}[NAG]{Neural Atlas Graphs}
    \acro{gs}[3DGS]{3D Gaussian Splatting}
    \acro{nerf}[NeRF]{Neural Radiance Fields}
    \acro{vslam}[vSLAM]{visual Simultaneous Localization and Mapping}
    \acro{lna}[LNA]{Layered Neural Atlases}
    \acro{orf}[ORF]{OmnimatteRF}
    \acro{ore}[ORe]{OmniRe}
    \acro{erf}[ERF]{EmerNeRF}
    \acro{fvd}[FVD]{Fr\'echet Video Distance}
    \acro{fid}[FID]{Fr\'echet Inception Distance}
    \acro{pvg}[PVG]{Periodic Vibration Gaussian}
    \acro{nsg}[NSG]{Neural Scene Graphs}
    \acro{deformgs}[DGS]{Deformable 3D Gaussians}
    \acro{streetgs}[SGS]{Street Gaussians}
\end{acronym}

\newcommand\Alpha{\mathrm{A}}
\newcommand{\scene}{\mathcal{I}}
\newcommand{\timestamp}{t}
\newcommand{\objectIndex}{i}
\newcommand{\translation}[1][]{T_{#1}(\timestamp)}
\newcommand{\baseTranslation}[1][\timestamp]{\tilde{T}_{#1}}

\newcommand{\rotation}[1][]{R_{#1}(\timestamp)}
\newcommand{\baseRotation}[1][\timestamp]{\tilde{R}_{#1}}

\newcommand{\positionVar}{\mathcal{P}}

\newcommand{\pRotationCam}[1][\text{cam}]{\positionVar_{#1}^{\text{R}}}
\newcommand{\pRotationObj}{\positionVar_{\objectIndex}^{\text{R}}}

\newcommand{\pTranslationCam}[1][\text{cam}]{\positionVar_{#1}^{\text{T}}}
\newcommand{\pTranslationObj}{\positionVar_{\objectIndex}^{\text{T}}}

\newcommand{\rotationWeight}{\eta_{\text{R}}}
\newcommand{\translationWeight}{\eta_{\text{T}}}
\newcommand{\colorWeight}[1][]{\eta_{#1c}}
\newcommand{\alphaWeight}[1][]{\eta_{#1\alpha}}
\newcommand{\flowWeight}[1][]{\eta_{#1f}}
\newcommand{\viewWeight}[1][]{\eta_{#1\phi}}

\newcommand{\nFrames}{F}
\newcommand{\nControlPoints}{P}
\newcommand{\nFlowControlPoints}{P_f}
\newcommand{\background}{B}
\newcommand{\spline}{S}
\newcommand{\splineFnc}[1]{\spline(#1)}
\newcommand{\R}{\mathbb{R}}
\newcommand{\Quaternion}{\mathbb{H}}
\newcommand{\rotvecToQuat}{q}
\newcommand{\rotvecToQuatFnc}[1]{\rotvecToQuat(#1)}

\newcommand{\masks}{\mathcal{M}}

\newcommand{\nObjects}{N}

\newcommand{\node}{\mathcal{N}}

\newcommand{\globalPositionBackground}{g_{\background}}
\newcommand{\neuralField}{\mathcal{F}}
\newcommand{\encoding}{\mathcal{H}}
\newcommand{\genericField}[1][\objectIndex]{\neuralField_{#1}}
\newcommand{\network}[1][\objectIndex]{\mathcal{N}_{#1}}
\newcommand{\colorField}{\neuralField_{\objectIndex,c}}
\newcommand{\alphaField}{\neuralField_{\objectIndex,\alpha}}
\newcommand{\flowField}{\neuralField_{\objectIndex,f}}
\newcommand{\viewField}[1][\objectIndex, \phi]{\neuralField_{#1}}

\newcommand{\flowEncoding}{\encoding_{\objectIndex,f}}
\newcommand{\viewEncoding}{\encoding_{\objectIndex,\phi}}
\newcommand{\sparsityMarker}{\tau}
\newcommand{\sparsityFunction}[1]{\text{sparse}(#1, \sparsityMarker)}

\newcommand{\queryPoint}{x}
\newcommand{\colorObj}[1][\objectIndex]{c_{#1}}
\newcommand{\alphaObj}[1][\objectIndex]{\alpha_{#1}}
\newcommand{\flowObj}[1][\objectIndex]{f_{#1}}
\newcommand{\positionObj}[1][\objectIndex]{g_{#1}}
\newcommand{\sizeObj}[1][\objectIndex]{s_{#1}}

\newcommand{\colorFunction}[1][\objectIndex]{\colorObj[#1](\queryPoint)}
\newcommand{\alphaFunction}[1][\objectIndex]{\alphaObj[#1](\queryPoint)}
\newcommand{\flowFunction}[1][\objectIndex]{f_{#1}}
\newcommand{\colorBase}{\tilde{C}_{\objectIndex}}
\newcommand{\alphaBase}{\tilde{A}_{\objectIndex}}

\newcommand{\viewAngle}{\phi}

\newcommand{\modelOutput}{\hat{y}}
\newcommand{\gtValues}{y}
\newcommand{\atlasLoss}{\mathcal{L}_{\text{atlas}}}
\newcommand{\gtMask}{m}
\newcommand{\outputMask}{\hat{a}}
\newcommand{\textureMap}{\mathcal{T}}
\newcommand{\textureMapColor}[1][\textureMap]{\hat{c}}
\newcommand{\textureMapAlpha}[1][\textureMap]{\hat{\alpha}}

\newcommand{\rayDirection}{d}
\newcommand{\rayOrigin}{o}
\newcommand{\focalLength}{f}

\makeatletter
\newcommand{\pushright}[1]{\ifmeasuring@#1\else\omit\hfill$\displaystyle#1$\fi\ignorespaces}
\newcommand{\pushleft}[1]{\ifmeasuring@#1\else\omit$\displaystyle#1$\hfill\fi\ignorespaces}
\newcommand{\specialcell}[1]{\ifmeasuring@#1\else\omit$\displaystyle#1$\ignorespaces\fi}
\makeatother

\begin{document}

\iftrue
\title{Neural Atlas Graphs for Dynamic Scene Decomposition and Editing}
\maketitle
\vspace{-0.7cm}
\begin{abstract}
Learning editable high-resolution scene representations for dynamic scenes is an open problem with applications across the domains from autonomous driving to creative editing -- the most successful approaches today make a trade-off between editability and supporting scene complexity: neural atlases represent dynamic scenes as two deforming image layers, foreground and background, which are editable in 2D, but break down when multiple objects occlude and interact. In contrast, scene graph models make use of annotated data such as masks and bounding boxes from autonomous-driving datasets to capture complex 3D spatial relationships, but their implicit volumetric node representations are challenging to edit view-consistently. We propose Neural Atlas Graphs (NAGs), a hybrid high-resolution scene representation, where every graph node is a view-dependent neural atlas, facilitating both 2D appearance editing and 3D ordering and positioning of scene elements. Fit at test-time, NAGs achieve state-of-the-art quantitative results on the Waymo Open Dataset -- by 5 dB PSNR increase compared to existing methods -- and make environmental editing possible in high resolution and visual quality -- creating counterfactual driving scenarios with new backgrounds and edited vehicle appearance. We find that the method also generalizes beyond driving scenes and compares favorably - by more than 7 dB in PSNR - to recent matting and video editing baselines on the DAVIS video dataset with a diverse set of human and animal-centric scenes. Project Page: \url{https://princeton-computational-imaging.github.io/nag/}
\end{abstract}    
\doparttoc 
\faketableofcontents 
\section{Introduction}
\label{sec:intro}

There has been growing demand in graphics and vision for high-fidelity static 3D~\cite{mildenhall_nerf_2021,kerbl_3d_2023} and dynamic 4D~\cite{stearns2024dynamic, tretschk2021non} reconstruction models, and in particular for \textit{editable} representations which decompose the scene into semantically meaningful components~\cite{yang2024emernerf}. For autonomous driving, where large collections of labeled video data are required to train driving behavior~\cite{hwang2024emma}, editable scene representations offer a direct approach to simulate counterfactual driving scenarios -- removing, re-timing, or repositioning vehicles and pedestrians to generate new trajectories, or editing the visual elements of the scene to reflect new environmental conditions. This enables systems to expand a limited collected real-world dataset into a richer and more diverse training set while preserving photo-realism and semantic consistency.

In this setting, neural scene graphs~\cite{ost2021neural} have emerged as a versatile hierarchical scene representation \cite{kundu2022panoptic, turki2023suds, yang2023unisim, liu2023real}; providing structured, object-centric models that enable repositioning and re-rendering elements with high visual quality. These methods model the scene as set of connected nodes -- e.g., vehicles, pedestrians, backgrounds -- represented as individual radiance fields~\cite{fischer2024multi} or collections of Gaussians~\cite{chen2025omnire}. The nodes are composited to render views of the scene, optimized during test time to fit an input driving sequence. Beyond visual data, these neural scene graph models can also ingest LiDAR and bounding box information, readily available in autonomous driving datasets~\cite{sun2020scalability,yu2020bdd100k}, to better localize objects in 3D space over time. While nodes in the scene graph can be removed, rotated, or translated while preserving 3D view consistency, directly modifying their appearance is significantly more challenging. This requires altering the underlying \ac{nerf} or \ac{gs} models, which requires a method to propagate 2D edits into 3D space~\cite{kuang2023palettenerf} to preserve view-consistency.

Neural atlas representations~\cite{kasten2021layered}, a parallel line of work primarily focused on video editing, offer an alternative approach for this kind of appearance manipulation. These methods learn to map a time-varying 3D environment to a set of static lower-dimensional 2D ''atlases'', analogous to traditional UV unwrapping~\cite{heckbert1986survey} of an object surface. During the process of fitting the scene, these models can disentangle a foreground object and its visual effects -- e.g., the shadows it casts  -- from its background. These atlases are then \emph{edited like a regular raster image}, with changes to object textures propagating correctly through the video~\cite{lee2025generative}. Unlike neural scene graphs, however, these models do not represent scene elements explicitly in 3D space, and do not have an ordering between them -- i.e., they cannot distinguish if one object is in front or behind another. When there are multiple overlapping objects in motion, neural atlas approaches resolve this by learning multiple non-overlapping alpha masks, with ''ordering'' achieved by the foreground mask cutting a hole out the background mask~\cite{Lu_2021_CVPR}. While this does not pose a problem for editing videos with a single primary subject,  for settings such as driving scenes, this makes it impossible to remove or reposition overlapping vehicles without introducing visual artifacts.

In this work, we introduce \textit{Neural Atlas Graphs} (NAG) as a hybrid high-resolution representation without these limitations. Given input 3D bounding boxes and segmentation masks from autonomous driving stacks, a NAG represents each scene element as a 2D plane with a 3D trajectory through space and time. Each of these planes acts as an independent neural atlas, capturing object motion, parallax, and lighting effects in a view-dependent neural field model. By explicitly modeling object depths and ordering, a NAG allows for flexible appearance editing at high resolution -- directly propagating changes from the 2D atlases to the reconstructed video -- without introducing distortions between occluded layers. Designed as an inverse problem, the object trajectories and appearance fields are learned jointly, using ray-casting and efficient ray-plane intersections to accumulate colors and opacities along each ray. By fitting to recorded high-resolution images, an accurate scene decomposition evolves naturally based on the varying motion patterns and provided masks. This enables our approach to perform visually consistent removals, additions, rearrangements and texture editing of scene elements in complex, multi-object environments. \\

We validate the proposed method on automotive scenes~\cite{sun2020scalability} and confirm that the method outperforms recent object-specific \ac{gs} baselines in visual quality by almost 5 dB PSNR on overall scene quality, and up to 11.2 dB PSNR for dynamic objects. This confirms that the method is able to learn accurate scene representations even under fast object motion, diverse reflections or non-rigid motion patterns, while keeping positional and textual editability. For diverse outdoor scenes, with significantly less geometric prior available, covering various (non-)rigid actors, the method performs favorably to recent matting and neural atlas approaches with 7.3 dB PSNR margin, confirming the generality of our approach.

\section{Related Work}

\paragraph{Video Layer Decomposition}
Representing videos as a composition of individually deforming layers is a long-studied problem~\cite{wang1994representing} with roots in seamless video editing~\cite{bregler1998videosprites, jojic2001sprites}. While more recent works~\cite{ye2022deformable, Lu_2021_CVPR} achieve this via trained optical flow networks~\cite{teed2020raft} and UNet-based masking~\cite{ronneberger2015u}, the core principles remain the same: estimate alpha masks and motion for dynamic elements to separate them from a more static background for individual editing and re-composition. Atlas-style methods -- e.g., \textit{Unwrap Mosaics}~\cite{ravacha2008unwrap} and later \textit{Layered Neural Atlases}~\cite{kasten2021layered} -- learn 2D-to-2D warps that map scene points onto an unwrapped canvas, similar to a UV map~\cite{heckbert1986survey}. Recent works explore neural field~\cite{kasten2021layered} and neural radiance field~\cite{lin2023omnimatterf} scene representations, in which compact networks are optimized at test time to map continuous coordinates in the input video sequence to view-consistent color. However, existing approaches assume videos that consist of a primary object (and its visual effects, e.g., shadows) set against a relatively static background, limiting their effectiveness in complex, multi-object scenes. When multiple objects overlap, these methods rely on single-layer masks, resulting in visual holes where foreground elements cut into background objects~\cite{lin2023omnimatterf}. While \textit{Generative OmniMatte}~\cite{lee2025generative} presents a method to generate realistic content to fill these occlusions, it reconstructs this content in the background canvas and does not model interactions between overlapping dynamic objects. In this work, we propose \textit{Neural Atlas Graphs} with neural atlas representations that \textit{explicitly model multiple interacting layers}, where a point in the scene can be mapped simultaneously to multiple time-consistent objects, enabling robust editing and re-composition even under complex occlusion scenarios.

\paragraph{3D Dynamic Scene Models}
Implicit representations such as \ac{nerf}s \cite{mildenhall_nerf_2021, mueller2022instant, barron2023zip} and explicit representations such as \ac{gs} \cite{kerbl_3d_2023, guedon2024sugar, yu2024mip} both enable high-fidelity photorealistic reconstructions of static scenes, and can be extended to fit dynamic scenes via learned deformation and flow fields~\cite{pumarola2021d, stearns2024dynamic, yang2024emernerf}. However, in either case, editing the scene --- and propagating those edits in a view-consistent manner -- proves highly non-trivial~\cite{dong2023vica, wang2024view, fischer2024multi} as these representations require changes to be carefully localized to avoid editing the wrong part of the scene, e.g., editing the subject and not their background. Neural Scene Graphs~\cite{ost2021neural, yang2023unisim} address this by factorizing the scene into per-object radiance fields, treated as graph nodes, enabling simple repositioning or removal of individual objects via edits to their corresponding nodes. This is made possible partially by the structured predictions of modern driving stacks~\cite{Man2023_BEVGuide} and annotations offered by autonomous driving datasets such as KITTI~\cite{geiger2012we}, nuScenes~\cite{caesar2020nuscenes}, and Waymo~\cite{sun2020scalability}, which in addition to image data offer LiDAR point clouds, depth maps, 3D bounding boxes, instance, and object or camera trajectories to instantiate graph nodes in 3D space. Recent hybrid methods combine scene graphs with 3DGS~\cite{chen2025omnire, cheng2025graphgs} to offer more efficient rendering times, but fine-grained editing remains challenging, as changes to individual nodes must be propagated in a view-consistent manner, which remains an open problem. We propose a hybrid representation that builds a 3D scene graph from structured bounding‑box, mask, and trajectory data, but models each graph node as a neural atlas~\cite{kasten2021layered}, allowing both \emph{direct object appearance editing} through manipulation of a 2D canvas, and object repositioning in 3D space.

\section{Neural Atlas Graphs}

Our proposed \textit{Neural Atlas Graph} (NAG) illustrated in \figref{fig:method} represents the scene as a graph of moving planes oriented in 3D space, with one plane per moving object -- e.g., pedestrian, vehicle, bicyclist -- plus a background plane. Each plane follows a learned rigid trajectory and carries a surface-aligned optical-flow field together with view-dependent color–alpha maps -- capturing non-rigid motion, parallax, and illumination changes. Rendering is done via depth-ordered ray-casting and alpha compositing across the planes. The subsequent sections detail these components.

\subsection{Representation}
\vspace{-6pt}
\label{sec:setup}
Our \textit{Neural Atlas Graph} (NAG) representation takes as input an arbitrary video scene $\scene \in \R^{\nFrames \times W \times H \times 3}$, comprised of $\nFrames$ 3-channel RGB images of size $W \times H$, and a stack of coarse masks $\masks \in \{0, 1\}^{\nFrames \times W \times H \times \nObjects}$ corresponding to $\nObjects$-many foreground objects -- nodes in our graph. The remaining unmasked region is represented with an additional background node. To establish the initial position, rotation, and ordering of nodes within the scene graph we primarily rely on the orientation of supplied 3D bounding boxes. If 3D bounding boxes are not available, we fall back to homographies determined from monocular depth estimation for plane initialization~\cite{yang2024depth}. Unfortunately, dynamic scene models such as \ac{nag}  inherit the same scene and camera motion ambiguities as found in \ac{vslam} problems. Therefore, we also require an initial estimation of the camera extrinsics and intrinsics, similar to common \ac{nerf} \cite{mildenhall_nerf_2021} and \ac{gs} \cite{kerbl_3d_2023} approaches. We represent each node $\{\node_i\}_{i=0}^{\nObjects}$  as a tuple $\node_{i} = (\colorObj, \alphaObj, \flowObj, \positionObj, \sizeObj)$ containing color $\colorObj:[0, 1]^{2} \to \R^{3}$, opacity $\alphaObj:[0, 1]^{2} \to \R$, optical flow field $\flowObj:[0, 1]^{2} \times [0, 1] \to \R^{2}$, time-dependent position $\positionObj: [0, 1] \to \R^{4 \times 4}$ and plane size $\sizeObj \in \R^{2}$. 

\begin{figure*}[t!]
 \centering
\includegraphics[width=\textwidth]{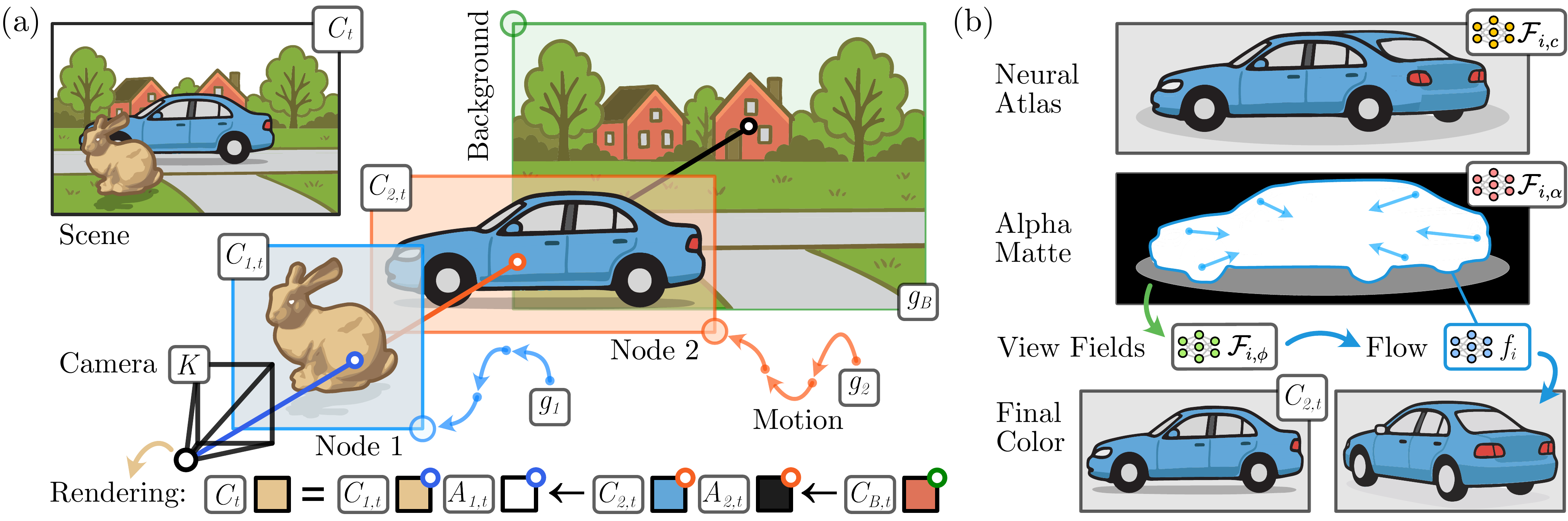} 
  \caption{Neural Atlas Graphs - A NAG represents dynamic scenes (a) as a graph of moving 3D planes (one per object/background). Each plane undergoes rigid transformations and encodes view-dependent appearance/transparency using neural fields $\genericField[]$ (b) along a learned trajectory $\positionObj$. The planar optical flow $\flowFunction$ models non-rigid motion and parallax, while learning the representation and rendering is done via opacity-weighted ray casting of $C_{i,t},\, \Alpha_{i,t}$ using position based z-buffering.}
\label{fig:method}
\end{figure*}

\subsection{Image Formation}
\vspace{-6pt}
\label{sec:image_formation}

\ac{nag} are rendered with a forward ray-projection model. Each pixel intensity at a timestep $\timestamp \in [0, 1]$ in the image $\scene$ is composed by aggregating radiance at typically a handful of plane/ray intersections for rays with direction $\rayDirection \in \R^{3}$ and origin $\rayOrigin \in \R^{3}$ (see supplementary material for details).
Given a planar node at position $\positionObj$, decomposable into a position $p$ and normal vector $n$, and the ray $r(l) = \rayOrigin + l\rayDirection$, the ray-plane intersection $l = (p - \rayOrigin) \cdot n / (d \cdot n)$ may yield an intersection point\footnote{Given non-parallel rays, e.g. $d \cdot n \neq 0$, and assuming $o$ is not within the plane.} $x_{\text{world}} = \rayOrigin + l\rayDirection$. When projected to a finite sized plane $i$, resulting in $x \in [0, 1]^{2}$  (\secref{sec:node_extend}), and applied for each node in the \ac{nag}, the object color $C_{i,t}$ and opacity $\Alpha_{i,t}$  at time $t$ can be determined via a planar optical flow field $\flowObj$:

\begin{equation}
    \label{eq:query_net}
    C_{i,t} = \colorObj(x + \flowObj(x, t)),\; \Alpha_{i,t} = \alphaObj(x + \flowObj(x, t)).
\end{equation}

We alpha-composite \cite{kopanas_2022} each plane intersection along the ray, yielding the final per-pixel color $C$
\begin{equation} 
    \label{eq:alpha_composition}
    C = \sum_{i = 0}^{N}C_{i,t}\Alpha_{i,t}\prod_{j=0}^{i-1}(1 - \Alpha_{j,t}),
\end{equation}
 of each ray. Given all object colors $C_{i,t}$ and opacities $A_{i,t}$ have been ordered by the distance to the camera in ascending order. If a ray does not hit the corresponding plane -- falling off the edge of the finite extent $\sizeObj$ of the plane -- we set $\Alpha_{i,t} = 0$.

\vspace{-6pt}
\subsection{Parametrization of Neural Atlas Nodes}
\vspace{-6pt}
\label{sec:neural_atlas_nodes}

Next, we describe the parametrization of the model components inside nodes  $\node_{i} = (\colorObj, \alphaObj, \flowObj, \positionObj, \sizeObj)$. We opt for modeling our \ac{nag} nodes in a 3D space, where each planar node is assigned a 3D position and orientation with 6 DoF assuming a rigid motion model.

\paragraph{Rigid Plane Pose [$g_i$]}
\label{sec:plane_pose}
The affine rigid transformation matrix $g_i(\timestamp)$ encodes the node’s trajectory over time. We decompose this matrix into translation $\translation[\objectIndex]$ and rotation $\rotation[\objectIndex]$ components, which are learned independently. To ensure temporal coherence, we use smooth Hermite splines \cite{deboor_1987} as 
\begin{align}
\label{eq:spline_object_model}  
    \translation[\objectIndex] &= \baseTranslation[\objectIndex, \timestamp] + \translationWeight \cdot \splineFnc{\timestamp,\,\pTranslationObj} .
\end{align}
Here, we learn an offset relative to an initial position $\baseTranslation[\objectIndex] \in \R^{\nFrames \times 3}$. When 3D bounding boxes are available, $\baseTranslation[\objectIndex, \timestamp]$ corresponds to the box center at time $\timestamp$.

The function $\spline{} : [0, 1] \times \R^{\nControlPoints} \rightarrow \R^{\nFrames}$ denotes the piecewise cubic Hermite spline interpolation of $P$ number of zero-initialized, learnable control points $\pTranslationObj \in \R^{\nControlPoints \times 3}$. The weight $\translationWeight = 0.5$ controls the contribution of the spline. Adjusting the number of control points $\nControlPoints$ determines the smoothness of the trajectory. For rotation $\rotation[\objectIndex]$, we use the orientation of the bounding box with an added offset\footnote{The offset is chosen to align the plane with the box's front, side, or diagonal face—whichever yields the smallest inclination to the camera view. This allows planar representations even for turning objects, like cars, as long as their rotation remains under 180°.}. When bounding boxes are not available, we instead learn the rotation using the following offset model
\begin{align}
    \label{eq:spline_object_rotation_model}
        \rotation[\objectIndex] &= \baseRotation[\objectIndex, \timestamp] \cdot \rotvecToQuatFnc{\rotationWeight  \cdot \splineFnc{\timestamp,\,\pRotationObj}},
\end{align}

given an initial rotation $\baseRotation[\objectIndex] \in \Quaternion^{\nFrames}$\footnote{$\Quaternion$ denotes the set of unit quaternions.}, we apply a rotation offset via the rotation-vector-to-quaternion mapping $\rotvecToQuat : [0, 2\pi)^3 \rightarrow \Quaternion$, using zero-initialized learnable offsets $\pRotationObj \in \R^{\nControlPoints \times 3}$. In the absence of bounding boxes, both $\baseTranslation[\objectIndex]$ and $\baseRotation[\objectIndex]$ are estimated by decomposing per-object image homographies \cite{malis2007deeper} into relative 3D translations and rotations. These are then cumulatively applied to a planar projection of a monocular depth estimate, yielding per-frame pose estimates.

\paragraph{Plane Extent [$s_i$]}
\label{sec:node_extend}
To prevent merging of distinct objects exhibiting similar rigid motion and to improve efficiency, we constrain each plane to a finite extent. A point $\queryPoint_{\text{world}} \in \R^4$ in homogeneous world coordinates lies on the finite plane of object $\objectIndex$ if the coordinates of its planar correspondence point $\queryPoint = (\positionObj(t)^{-1} \queryPoint_{\text{world}} + 0.5) \odot  [s_x, s_y, 0, 0]^{\intercal}$ are within [0, 1] range, where $\odot$ denotes element-wise multiplication.\\
The plane scale $\sizeObj = \{s_x, s_y\}$ is pre-estimated from the largest mask within the stack $\masks$, plus a relative margin to incorporate object-associated effects, e.g., shadows, and mask inaccuracies.\\
\paragraph{Node Color [$C_{i,t}$] and Opacity [$\Alpha_{i,t}$]}
Given a ray–plane intersection point $\queryPoint \in [0, 1]^2$ and its spherical view-angle $\viewAngle \in [0, 1]^2$ in the plane coordinate system, we model the color and opacity of each node as the combination of a base color field $\colorBase: [0,1]^2 \rightarrow \mathbb{R}^3$ and opacity field $\alphaBase: [0,1]^2 \rightarrow \R$, each augmented by two separate learned neural fields. The base appearance, estimated via forward projection onto masked regions, is parameterized on a pixel grid and extended to arbitrary $x \in [0,1]^2$ via bilinear interpolation. We use two types of neural fields: view-agnostic fields $\colorField: [0, 1]^2 \to \R^3$ and $\alphaField: [0, 1]^2 \to \R$, and view-dependent fields $\viewField[\objectIndex, \phi, c]: [0, 1]^4 \to \R^3$ and $\viewField[\objectIndex, \phi, \alpha]: [0, 1]^4 \to \R$, the latter implicitly regularized via a coarse-to-fine scheme \footnote{We detail the coarse-to-fine scheme, along with our \textit{phase-based learning} within the supplementary material.}. This combination balances editability -- requiring time-consistent atlas content -- and adaptivity to scene dynamics, lighting, and view direction, resulting in both temporal stability and high visual fidelity. Our representation is then 
\begin{alignat}{2}
\label{eq:color_model}
    \colorFunction &=\, && \colorBase(\queryPoint)
    + \colorWeight \cdot \colorField(\queryPoint)
     + \viewWeight \cdot \viewField[\objectIndex, \phi, c](\queryPoint, \viewAngle),  \\
    \label{eq:alpha_model}
    \alphaFunction &=\, && -\log\left(\dfrac{1}
    {\alphaBase(\queryPoint)} - 1\right) 
     + \alphaWeight \cdot \alphaField(\queryPoint) 
     + \viewWeight \cdot \viewField[\objectIndex, \phi, \alpha](\queryPoint, \viewAngle),
\end{alignat}
which we use in \eqref{eq:query_net} to represent $C_{i,t}$ and $\Alpha_{i,t}$. While using the base estimates $\colorBase, \alphaBase$ allows to precondition the object to learn, the subsequent MLPs $\colorField, \alphaField$, including positional encodings \cite{mueller2022instant}, refine projection errors and learn a temporally consistent, editable representation. To capture view- or time-dependent appearance changes, we introduce an additional MLP $\viewField$, which predicts color and opacity offsets based on the planar point $\queryPoint$ and its associated spherical view angle $\viewAngle$ at the ray–plane intersection. We weight the contributions of the networks using fixed scalars $\colorWeight= \alphaWeight= \viewWeight = 0.1$, and enforce valid ranges by clamping color values and applying a sigmoid to opacity, ensuring $\colorFunction, \alphaFunction \in [0, 1]$.

\paragraph{Flow Field [$f_i$]} We rely on a temporally changing flow field $f_i: [0,1]^2 \times [0, 1] \to \R^{2}$ attached to each node to model small non-rigid deformations and depth-induced parallax, c.f. \eqref{eq:color_model}. To ensure smoothness, we use a spline-based flow model \cite{chugunov_2024}, which shifts the ray–plane intersection point $\queryPoint$ before it is passed to the subsequent networks. That is
\begin{align}
    \label{eq:flow_control_points}
    \flowFunction(\queryPoint, \timestamp) &= \flowWeight \cdot \spline(\timestamp,\flowField(\queryPoint)).
\end{align}
The flow field $\flowField: [0,1]^2 \to \R^{\nFlowControlPoints \times 2}$ is implemented as an MLP similar to the color and opacity networks. However, instead of predicting a single flow vector, it outputs a set of $\nFlowControlPoints$ control points, which are interpolated using a cubic Hermite spline $\spline(\timestamp,\cdot)$ (cf. \secref{sec:plane_pose}) to produce a smooth, time-varying flow. We employ a coarse-to-fine fitting strategy by progressively masking the encoding dimensions (see supplementary material). Similar to color and alpha, the flow is modeled as an offset and scaled by a fixed weight $\flowWeight = 0.1$.

\subsection{Background Atlas}
\label{sec:background_atlas}
\vspace{-6pt}
In addition to the dynamic foreground objects, we model the background using a dedicated node covering all non-masked regions. The background is represented as a fixed planar atlas with no rigid transformation or opacity variation. Its global pose $\globalPositionBackground \in \R^{4 \times 4}$ is held constant throughout the sequence and placed behind all object bounding boxes. Its orientation and size are chosen to ensure full visibility across the entire camera trajectory, including rotations and translations. The background has the same color and flow model as the foreground nodes (see \eqref{eq:color_model}, \eqref{eq:flow_control_points}), but uses a fixed opacity $\Alpha_B = 1$ and omits rigid motion to maintain a consistent reference frame.


\subsection{Optimization}
\vspace{-6pt}
We fit the model by sampling mini-batches of random pixels and timestamps from the input video as ground truth values $\gtValues \in \scene$  
and render color predictions $\modelOutput$. We jointly optimize the transformation parameters and network parameters of all nodes with the loss
\begin{equation}
    \atlasLoss = ||\modelOutput - \gtValues||_{1} + \beta \cdot ||\outputMask - \gtMask||_{1},
\end{equation}
which combines a photometric $\ell_1$ term with a mask loss that compares the predicted opacity $\outputMask$ to the input mask $\gtMask \in \masks$. The mask term encourages objects to remain opaque in masked regions while allowing flexibility in unmasked areas to represent shadows or object-induced effects. We empirically set $\beta = 0.005$ to balance these objectives.

\subsection{Atlas Texture Editing}
\vspace{-6pt}
To edit the appearance of an object $\objectIndex$ (or the background), we require a user-provided RGBA texture -- with color $\textureMapColor \in [0, 1]^{W \times H \times 3}$ and alpha $\textureMapAlpha \in [0, 1]^{W \times H}$ -- within the image space of a dedicated reference image $\scene_{\text{ref}}$. This texture gets ray-projected onto the planar object $\node_{i}$. On the plane, we can numerically invert the learned flow model and bilinearly interpolating values, to store the user texture $\hat{c}: [0, 1]^2 \rightarrow \R^3$ and $\hat{\alpha}: [0, 1]^2 \rightarrow \R$ on a regular grid in atlas space. This allows us to sample the texture at arbitrary ray-plane intersection points $x$. Further, we can blend the values at $\hat{c}(x)$, $\hat{\alpha}(x)$ with the original color model \eqref{eq:color_model} via alpha matting, 

\begin{equation}
    c^*_i(x) = (1 - \hat{\alpha}(x))\cdot c_i(x) + \hat{\alpha}(x) \cdot \hat{c}(x),
\end{equation}

to our learned object representation. Using $c^*_i$ instead of $c_i$ within \eqref{eq:query_net} allows us to render the scene including the user-provided texture.\\
Storing the texture representation within atlas space is critical: it ensures the edit is independent of the camera motion (given its definition on the node, which undergoes rigid motion) and enables view-consistent editing of subsequent frames as long as these are accurately modeled within our flow representation. Furthermore, by defining textures in image space and projecting them, we allow for better edit comparability since textures are defined independently of the learned flow, in contrast to \cite{kasten2021layered}. This design also enables the seamless use of standard image editing tools and off-the-shelf image generation models.
\section{Experiments}
To evaluate the efficacy and versatility of the proposed method, we conduct three distinct experiments. First, we assess the method with quantitative and qualitative comparisons to state-of-the-art driving scene reconstruction techniques~\cite{chen2025omnire,yang2024emernerf}. Next, we validate and compare the editing capabilities, such as object removal, and texture editing, on challenging autonomous driving scenes. Finally, we assess the generalization of the approach on diverse non-driving outdoor scenes.

\subsection{Driving Scene Reconstruction}
\vspace{-6pt}

\paragraph{Setup}
For evaluating the quality of our method in autonomous driving scenarios, we rely on a subset of the Waymo \cite{sun2020scalability} open dataset. We select scenes with small ego motion but many objects, occlusions, and diverse and large object motions.
In total, we evaluate on 7 scene segments including up to 199 images each, sampled at 10Hz using the forward camera feed. We divide each data segment into sequences of 21 to 89 images, leading to 25 subsequences to be reconstructed. 

We filtered the 3D bounding boxes provided for actually moving objects and used the sparsely provided instance segmentation masks in a semi-automatic process using SAM2 \cite{ravi2025sam} to get reasonable masks for every image. Notably, an exemplary ablation study in the supplementary material indicates that our approach maintains applicability even when using coarser masks or only bounding boxes. Representative ground truth images of these dataset sequences, including our masks and further training details can be found in our supplementary material or code page: \url{https://github.com/jp-schneider/nag}. %

\paragraph{Assessment}
We evaluate our method against the recent OmniRe \cite{chen2025omnire} method, an object-centric \ac{gs} approach with explicit SMPL \cite{loper2023smpl} human modeling and EmerNeRF \cite{yang2024emernerf}, a recent \ac{nerf} method including static and dynamic radiance fields \footnote{For an detailed introduction on our baselines we refer to our supplementary material.}. \tabref{tab:results_waymo_main} reports PSNR, SSIM \cite{ssim_2004} and LPIPS \cite{zhang2018perceptual} on the individual scenes. The method consistently outperforms the baselines, reaching a significant 5 dB PSNR mean increase over OmniRe and 7 dB over EmerNeRF. To assess these differences visually, we show comparisons in \figref{fig:comparison_vis_quality} but recommend viewing the full-resolution images and videos provided in the supplementary. Especially under rapid motion of objects, OmniRe \cite{chen2025omnire} tends to produce artifacts. This can become apparent in missing or smoothed out objects edges, foots of cyclists, partially visible side-mirrors, missing distant objects or reflections which are more reliably modelled with our approach. EmerNeRF \cite{yang2024emernerf} tends to render objects over-smoothed, indicating under-parameterizations due to its scene representation in a static and dynamic radiance field. Our \ac{nag} node formulation is able to precisely model fine high-resolution details, such as spinning wheels, pedestrian motion, reflections or objects in the distance, even if they are part of the background plane.

\begin{table*}[tbh]
    \centering
    \caption{Quantitative Evaluation on Dynamic Driving Sequences of the Waymo~\cite{sun2020scalability} Open Driving Dataset. Best results are in bold. ORe refers to OmniRe \cite{chen2025omnire}, and ERF to EmerNeRF \cite{yang2024emernerf}.
    }
    \label{tab:results_waymo_main}
    \setlength{\tabcolsep}{4pt}
    \begin{tabular}{l S[table-format=2.2] S[table-format=2.2] S[table-format=2.2] S[table-format=1.3] S[table-format=1.3] S[table-format=1.3] S[table-format=1.3] S[table-format=1.3] S[table-format=1.3]}
         \toprule
          \multirow{2}{*}{Seq.} & \multicolumn{3}{c}{PSNR $\uparrow$} & \multicolumn{3}{c}{SSIM $\uparrow$} & \multicolumn{3}{c}{LPIPS $\downarrow$} \\
         & {Ours} & {ORe} & {ERF} & {Ours} & {ORe} & {ERF} & {Ours} & {ORe} & {ERF} \\
         \midrule
      s-975     & \textbf{40.21} & 37.35 & 34.83 & \textbf{0.976} & 0.968 & 0.937 & \textbf{0.058} & 0.080 & 0.143\\
      s-203   & \textbf{43.15} & 36.93 & 36.07 & \textbf{0.978} & 0.966 & 0.936 & \textbf{0.070} & 0.094 & 0.205\\
      s-125     & \textbf{43.32} & 38.74 & 35.20 & \textbf{0.980} & 0.970 & 0.933 & \textbf{0.057} & 0.079 & 0.182 \\
      s-141     & \textbf{42.55} & 36.14 & 34.83 & \textbf{0.978} & 0.964 & 0.924 & \textbf{0.057} & 0.087 & 0.178 \\
      s-952     & \textbf{41.89} & 39.67 & 35.32 & 0.976 & \textbf{0.977} & 0.938 & 0.058 & \textbf{0.050} & 0.120\\
      s-324     & \textbf{40.85} & 32.58 & 33.63 & \textbf{0.977} & 0.953 & 0.926 & \textbf{0.038} & 0.071 & 0.124\\
      s-344     & \textbf{41.84} & 36.67 & 35.24 & \textbf{0.983} & 0.973 & 0.946 & \textbf{0.031} & 0.043 & 0.084\\
      \midrule
      Mean        & \textbf{41.85} & 36.78 & 34.93 & \textbf{0.978} & 0.967 & 0.934 & \textbf{0.051} & 0.070 & 0.142 \\
      \bottomrule
    \end{tabular}
\end{table*}%

We further evaluate the quality of the dynamic objects in isolation, grouped into a rigid ''Vehicle'' class and a non-rigid ''Human'' class, which allows us to differentiate between objects that may benefit differently from our underlying rigid-motion model. \tabref{tab:object_results} validates significant improvements over the best baseline, with PSNR gains of 11.2 dB and 10.7 dB, and SSIM increases of 0.064 and 0.080 for the Vehicle and Human classes, respectively. These findings confirm that the quality of our method does not stem from better background rendering but indeed from the high accuracy of our model in representing even non-rigidly moving actors.
\begin{table*}[tb]
    \centering
    \caption{Quantitative Evaluation of Human and Vehicle Rendering on Waymo~\cite{sun2020scalability} Driving Sequences.
    }
    \label{tab:object_results}
    \setlength{\tabcolsep}{4pt}
    \begin{tabular}{l S[table-format=2.2] S[table-format=2.2] S[table-format=2.2] S[table-format=1.3] S[table-format=1.3] S[table-format=1.3] S[table-format=2.2] S[table-format=2.2] S[table-format=2.2] S[table-format=1.3] S[table-format=1.3] S[table-format=1.3]}
          \toprule
        \multirow{2}{*}{Seq.} & \multicolumn{3}{c}{Vehicle PSNR $\uparrow$} & \multicolumn{3}{c}{Vehicle SSIM $\uparrow$} & \multicolumn{3}{c}{Human PSNR $\uparrow$} & \multicolumn{3}{c}{Human SSIM $\uparrow$} \\
        & {Ours} & {ORe} & {ERF} & {Ours} & {ORe} & {ERF} & {Ours} & {ORe} & {ERF} & {Ours} & {ORe} & {ERF} \\
        \midrule
          s-975 & \textbf{46.79} & 33.09 & 30.21 & \textbf{0.991} & 0.939 & 0.82 & \textbf{45.37} & 32.99 & 28.53 & \textbf{0.989} & 0.927 & 0.777 \\
          s-203 & \textbf{41.90} & 30.45 & 27.10 & \textbf{0.986} & 0.910 & 0.774 & \textbf{45.40} & 34.85 & 33.54 & \textbf{0.986} & 0.950 & 0.901 \\
          s-125 & \textbf{41.00} & 28.72 & 24.55 & \textbf{0.989} & 0.878 & 0.709 & N/A & N/A & N/A & N/A & N/A & N/A \\
          s-141 & \textbf{43.21} & 33.22 & 27.36 & \textbf{0.981} & 0.929 & 0.744 & \textbf{44.22} & 33.31 & 28.86 & \textbf{0.986} & 0.907 & 0.769 \\
          s-952 & \textbf{40.94} & 31.15 & 27.70 & \textbf{0.986} & 0.928 & 0.810 & \textbf{40.45} & 32.32 & 28.10 & \textbf{0.968} & 0.894 & 0.740 \\
          s-324 & \textbf{41.71} & 31.03 & 27.87 & \textbf{0.986} & 0.921 & 0.798 & \textbf{44.12} & 32.09 & 26.40 & \textbf{0.988} & 0.894 & 0.689 \\
          s-344 & \textbf{43.97} & 33.02 & 30.65 & \textbf{0.985} & 0.931 & 0.835 & \textbf{40.99} & 30.20 & 25.94 & \textbf{0.975} & 0.882 & 0.721 \\
          \midrule
          Mean & \textbf{42.88} & 31.69 & 28.09 & \textbf{0.986} & 0.922 & 0.787 & \textbf{42.94} & 32.24 & 27.78 & \textbf{0.981} & 0.901 & 0.744 \\
          \bottomrule
    \end{tabular}
\end{table*}
\begin{figure}[tb]
    \centering
    \newlength{\imageWidth}
    \setlength{\imageWidth}{0.25\textwidth}
    \newlength{\imagerowheight}
    \settoheight{\imagerowheight}{\includegraphics[width=\imageWidth]{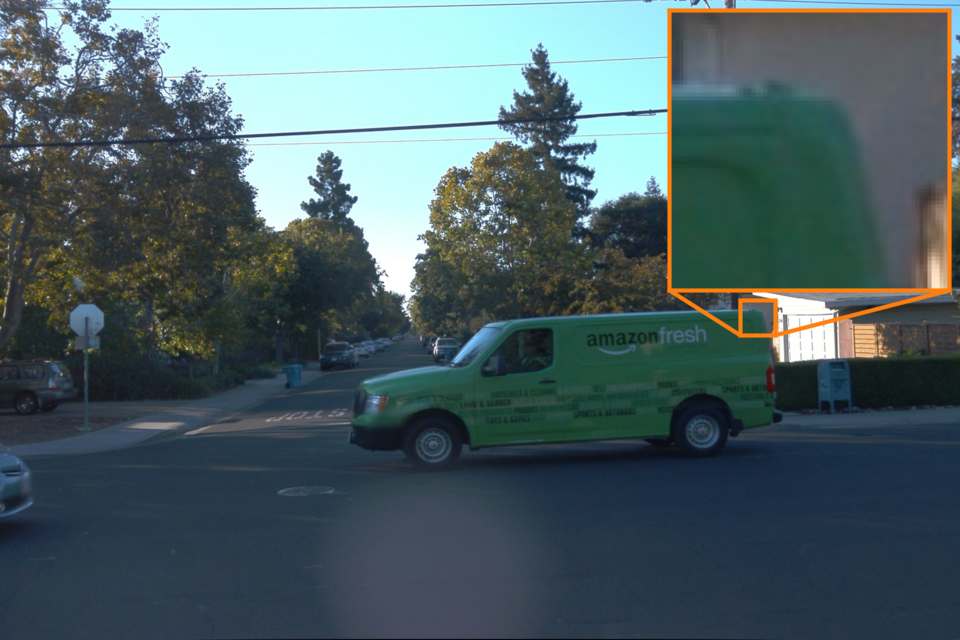}} 
    \newcommand{\gridImage}[1]{%
        \includegraphics[width=\imageWidth]{#1}%
    }
    \begin{tblr}{colspec={c c c c},colsep=0mm, rowsep=-1.2mm, row{1-3}={ht=\imagerowheight}, cells={halign=c, valign=m}}
         \gridImage{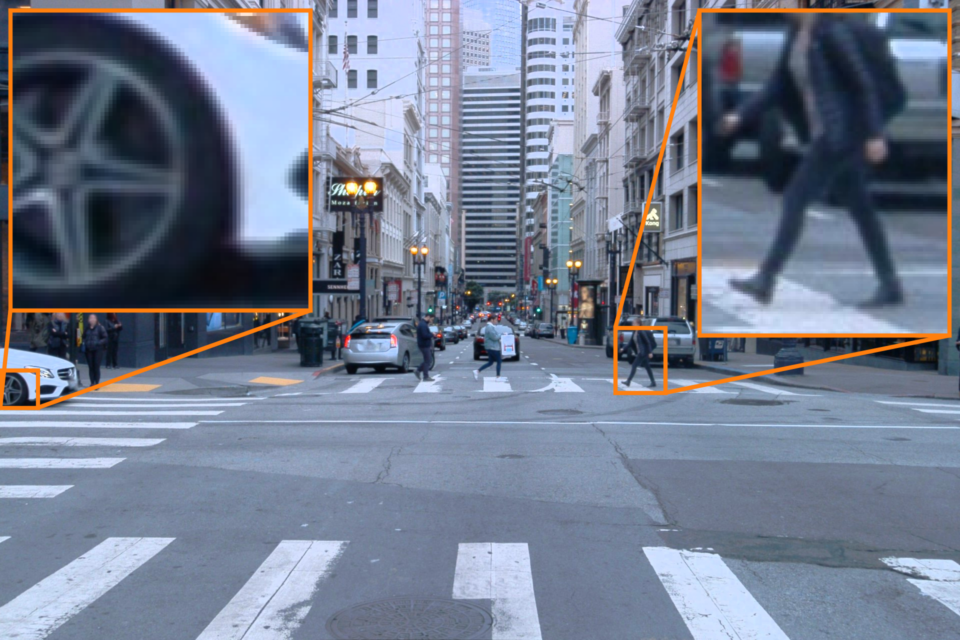}& %
         \gridImage{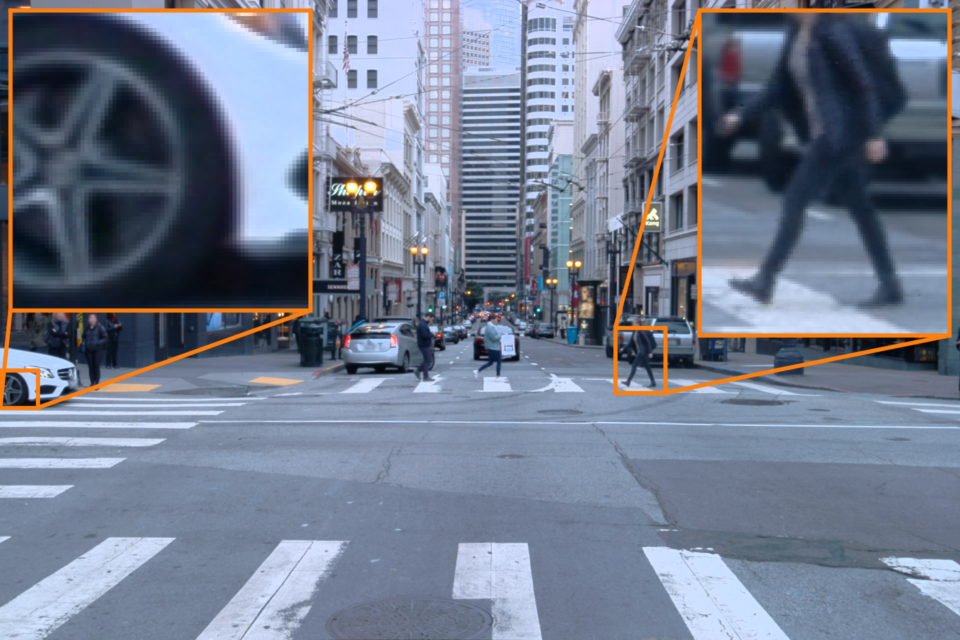}& %
         \gridImage{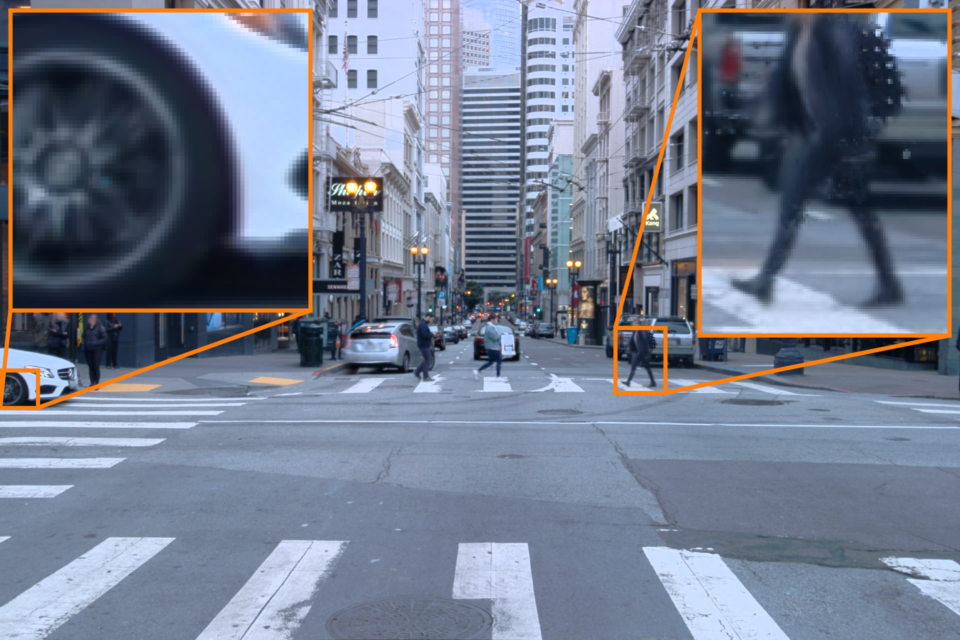}&  %
         \gridImage{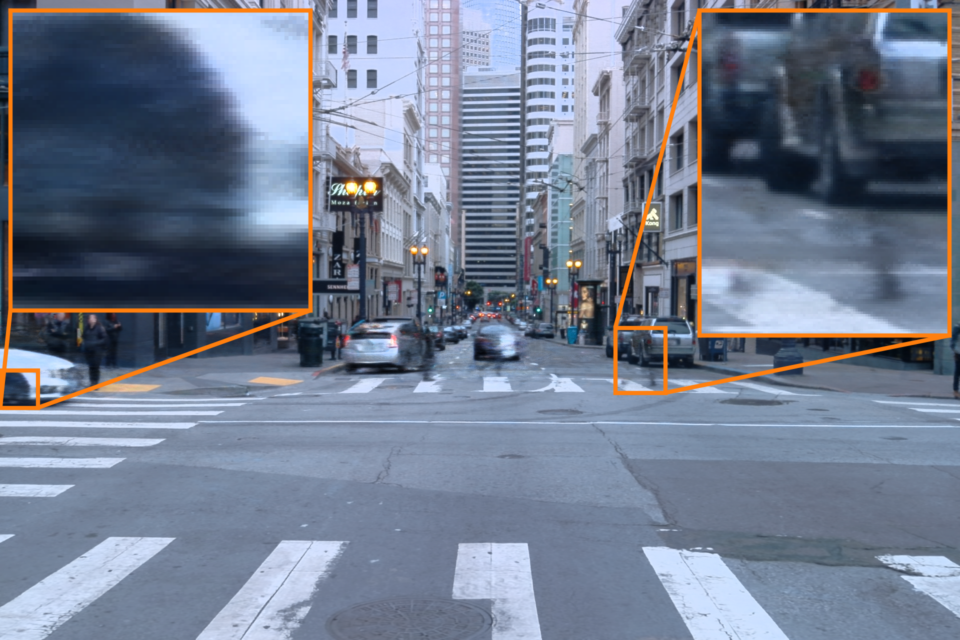}\\
         \gridImage{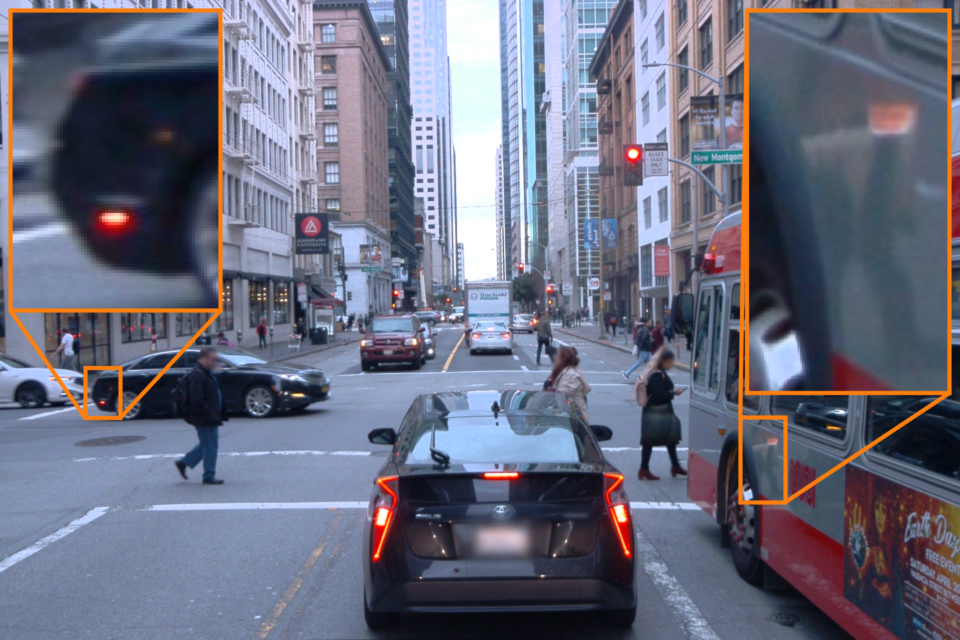} & %
         \gridImage{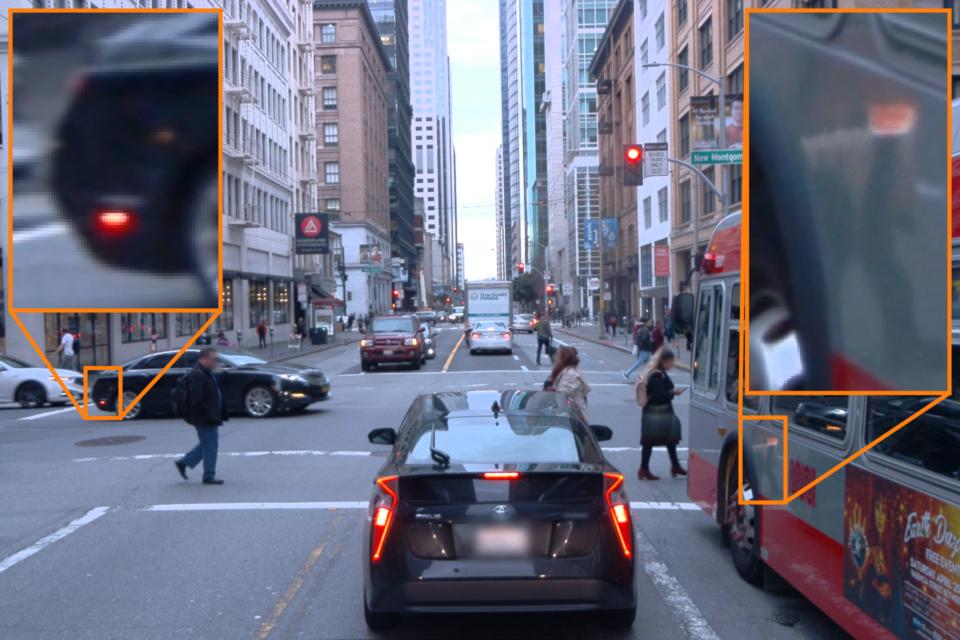}& %
         \gridImage{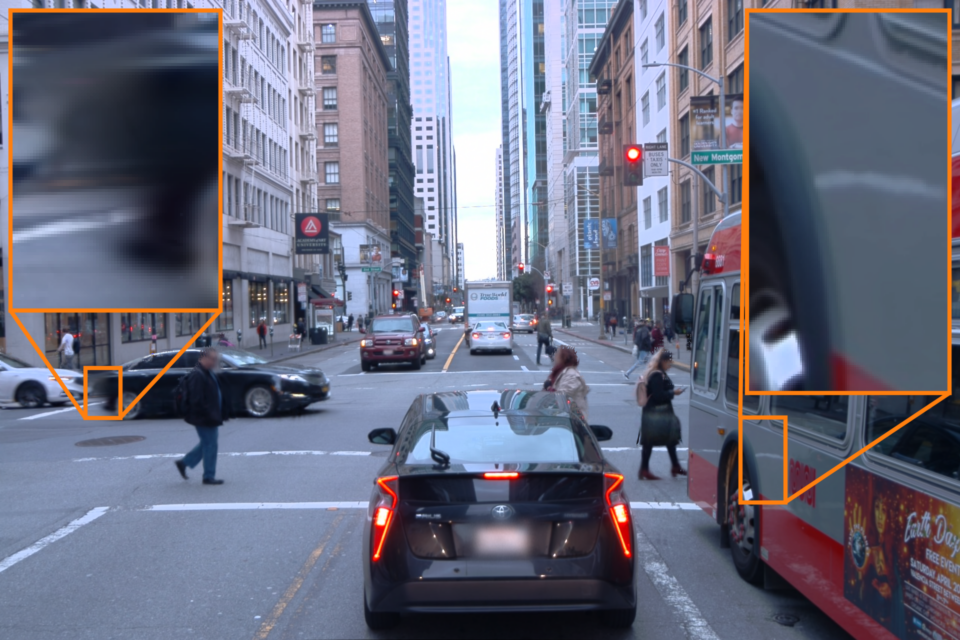} & %
         \gridImage{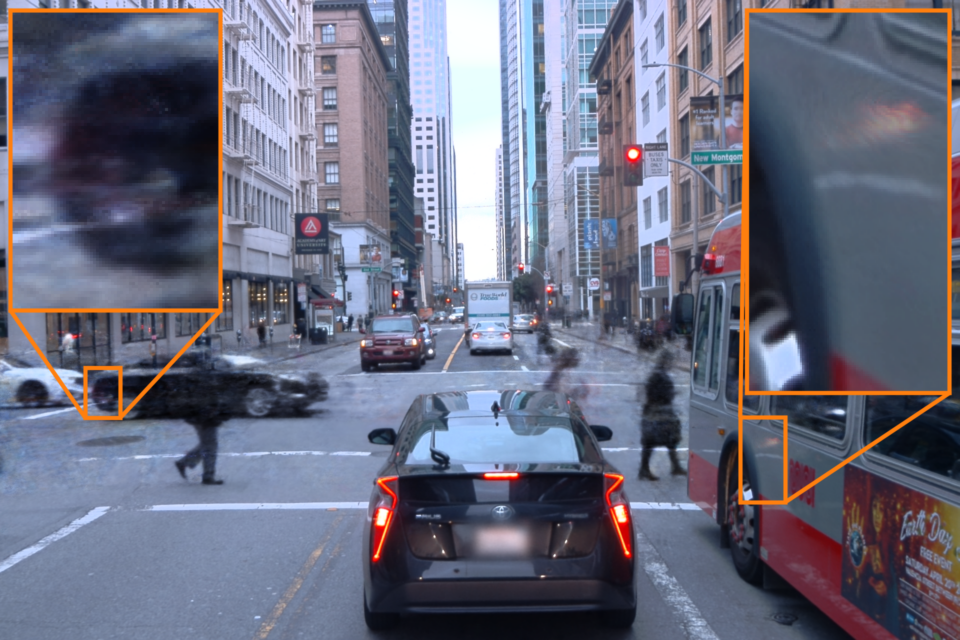}\\
         \gridImage{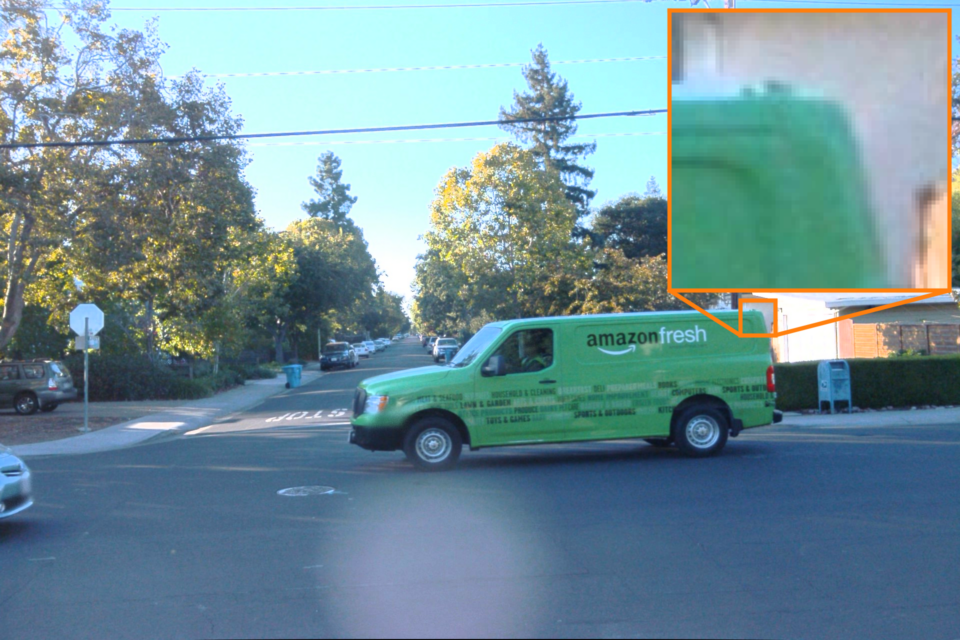} & %
         \gridImage{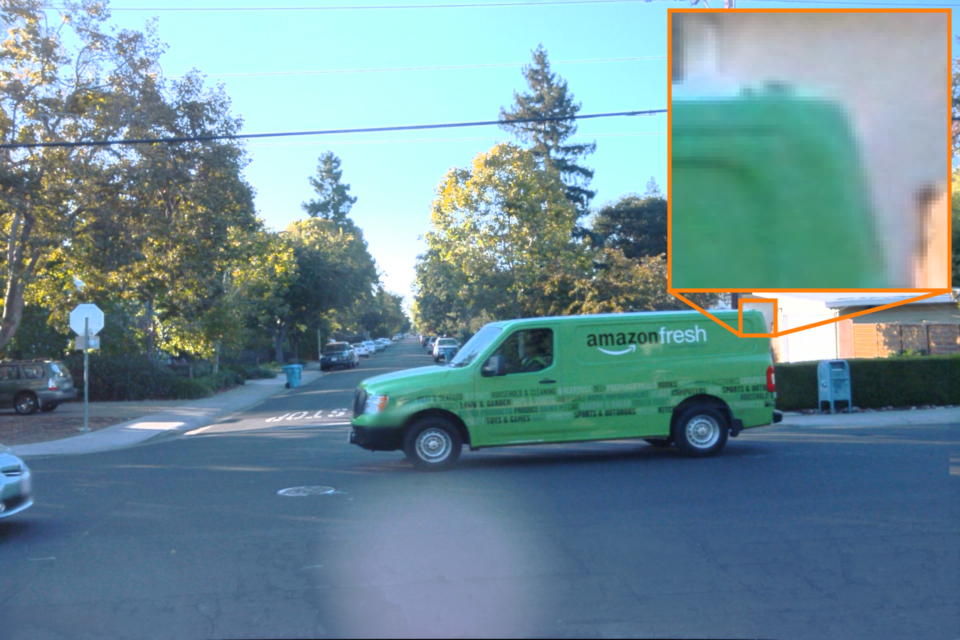} & %
         \gridImage{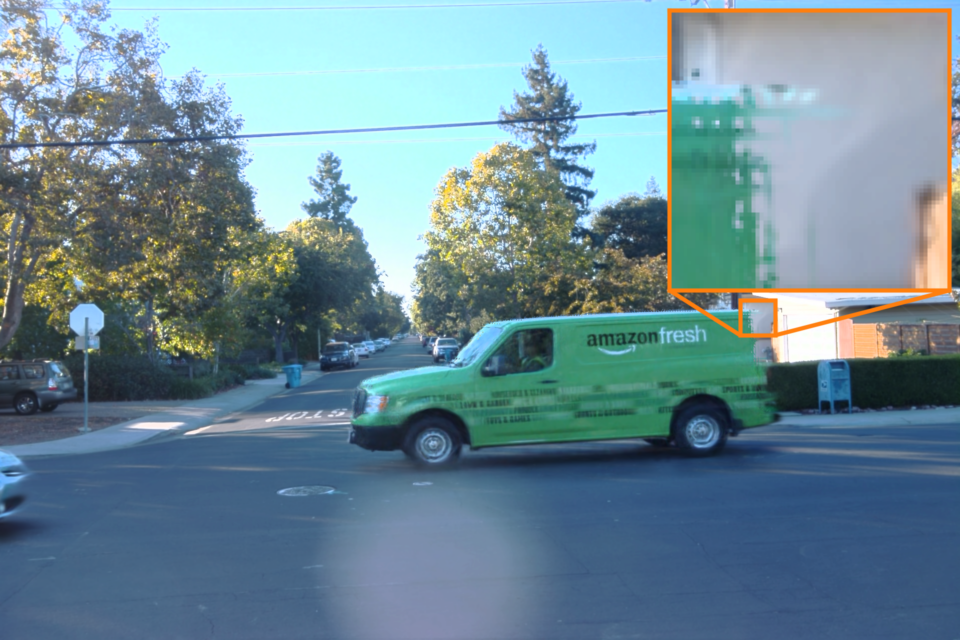} &  %
         \gridImage{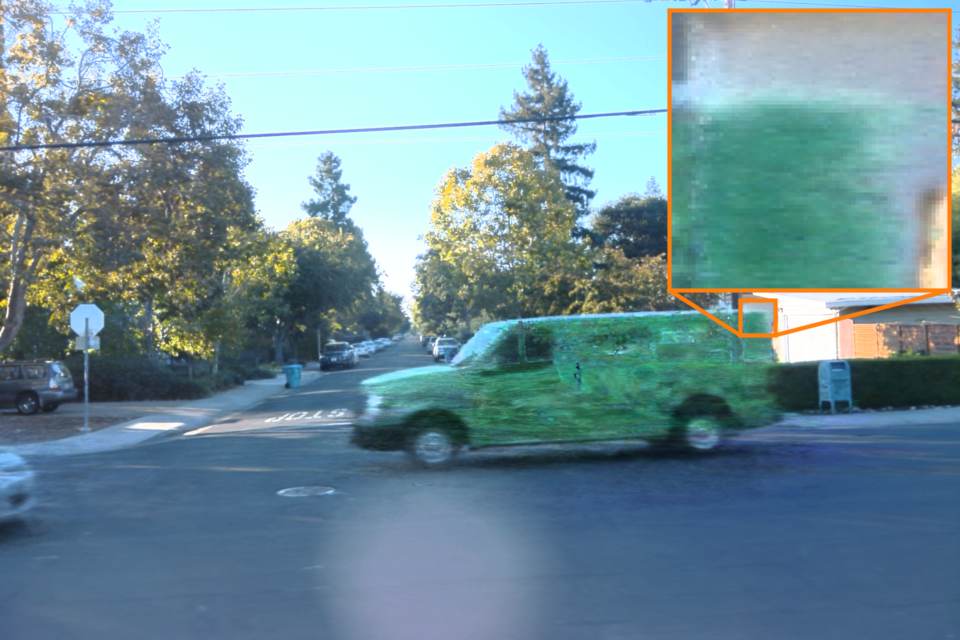}
    \end{tblr}
    \begin{tabular}{cccc}
        \centering \parbox{\dimexpr0.25\textwidth-2\tabcolsep\relax}{\centering Ground Truth} &
        \centering \parbox{\dimexpr0.25\textwidth-2\tabcolsep\relax}{\centering Ours} &
        \centering \parbox{\dimexpr0.25\textwidth-2\tabcolsep\relax}{\centering OmniRe \cite{chen2025omnire}} &
        \centering \parbox{\dimexpr0.25\textwidth-2\tabcolsep\relax}{\centering EmerNeRF \cite{yang2024emernerf}}
    \end{tabular}
    \caption{Visual quality comparisons (sequences s-952, 344, 975) show that our method achieves higher fidelity by producing way fewer artifacts on rapid motion (e.g., spinning wheels, edges), significantly reducing motion blur, and preserving finer details like reflections.}
    \label{fig:comparison_vis_quality}
\end{figure}

\paragraph{Scene Editing}
We showcase a scene decomposition task in \figref{fig:scene_editing}, where all dynamic actors are extracted and separated from the background\footnote{For a clearer illustration, please refer to the supplementary videos.}. Comparing to OmniRe, our method suffers from significantly fewer artifacts around the object peripheries, adding correct shadows and keeping distant information. Given our graph approach, we can further remove, shift, and add additional nodes to the graph and render these into our scene. These functions, along with an evaluation on temporal consistency, are demonstrated in our supplementary material. 
Also, given our planar nodes, we can take arbitrary textures in image space and project them onto one or multiple nodes to change their appearance along time, as demonstrated in \figref{fig:scene_texture_editing}. By combining object removal, and background texture editing, we can generate realistic counterfactual driving scenes. To demonstrate this capability, we execute a realistic editing task within the Waymo s-203 sequence, as shown in \figref{fig:waymo_editing_203}. Here, the car integrates naturally with the edited street, correctly incorporating painted speed signs and a zebra crossing. Crucially, the shadow casts realistically along these edited areas, showcasing our method's ability to maintain natural occlusion behavior. Furthermore, the seamless removal of both the person in the foreground and the car in the background, without recognizable artifacts, highlights the precision required for reliable testing within autonomous driving stacks.

\begin{figure}[tbh]
    \centering
    \setlength{\imageWidth}{0.2\textwidth}
    \settoheight{\imagerowheight}{\includegraphics[width=\imageWidth]{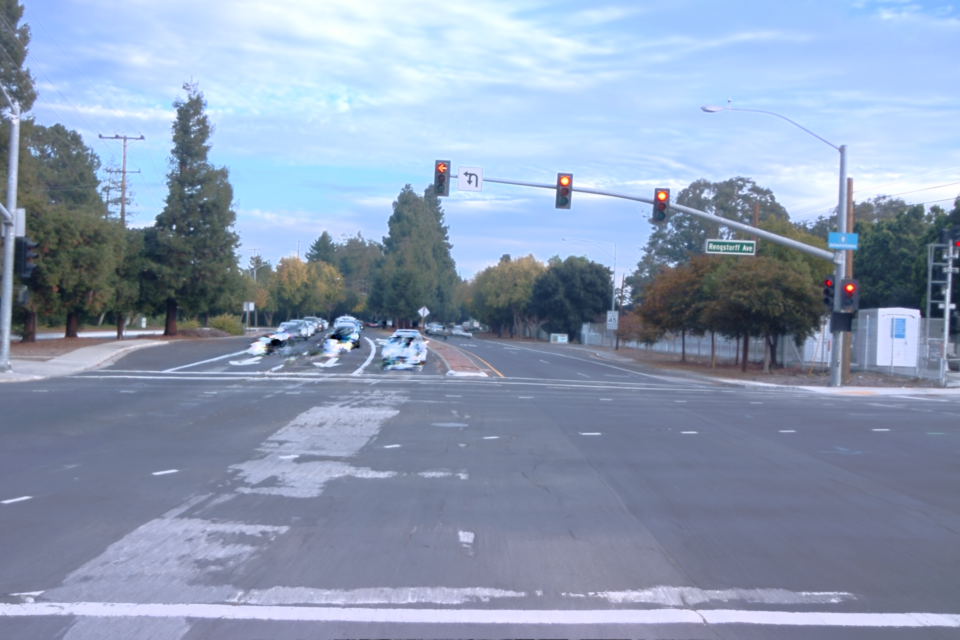}}
    \newcommand{\gridImage}[1]{%
        \includegraphics[width=\imageWidth]{#1}%
    }
    \begin{tblr}{colspec={c c c}, colsep=0mm, rowsep=-1.2mm, row{1-2}={ht=\imagerowheight}, cells={halign=c, valign=m}}
        
        \gridImage{images/edit_results/W125_000/W125_000_015_Ours_BG_g__1_10} &
        \gridImage{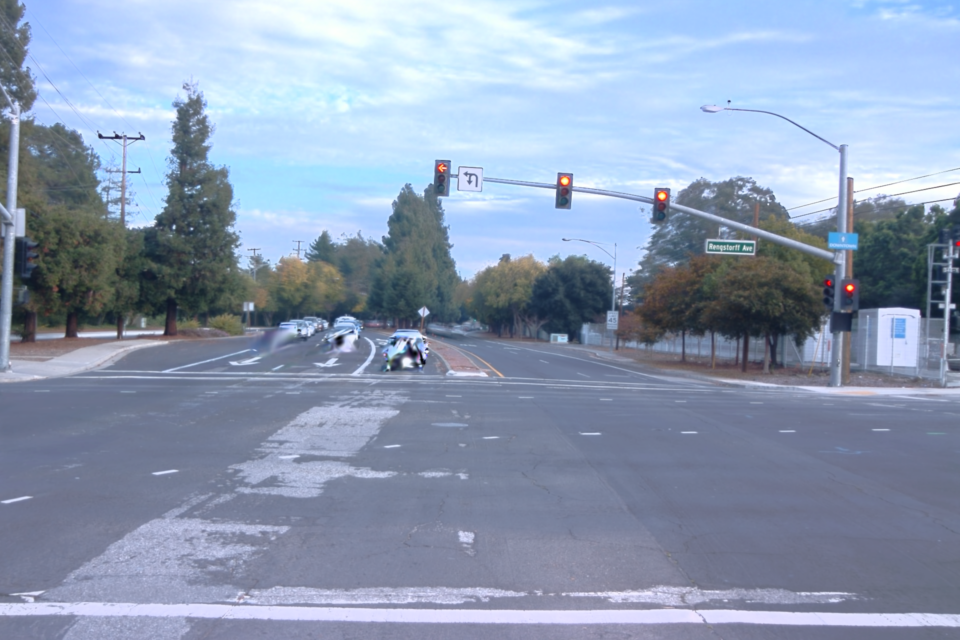} &
        \gridImage{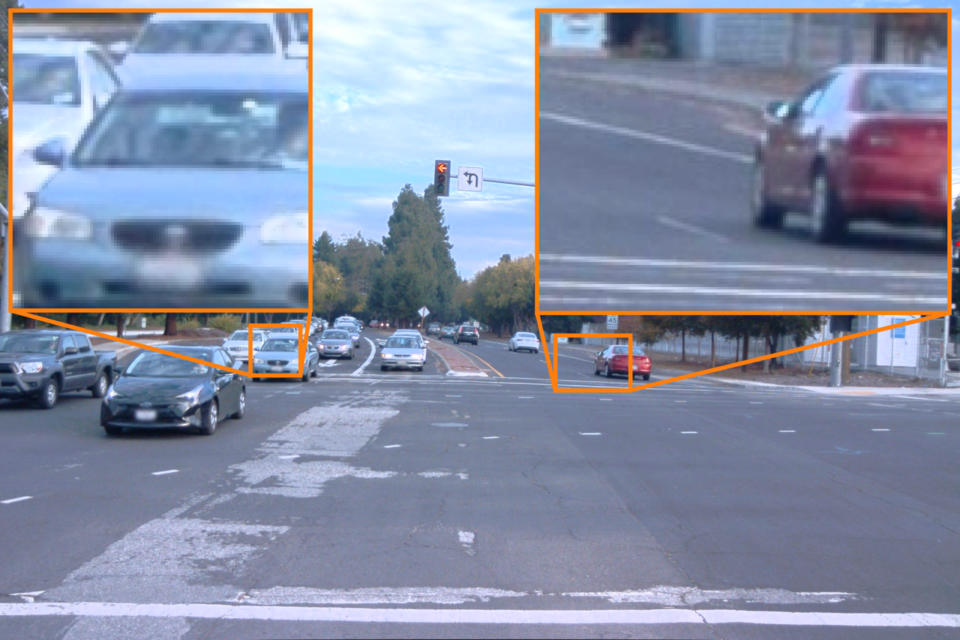} &
        \gridImage{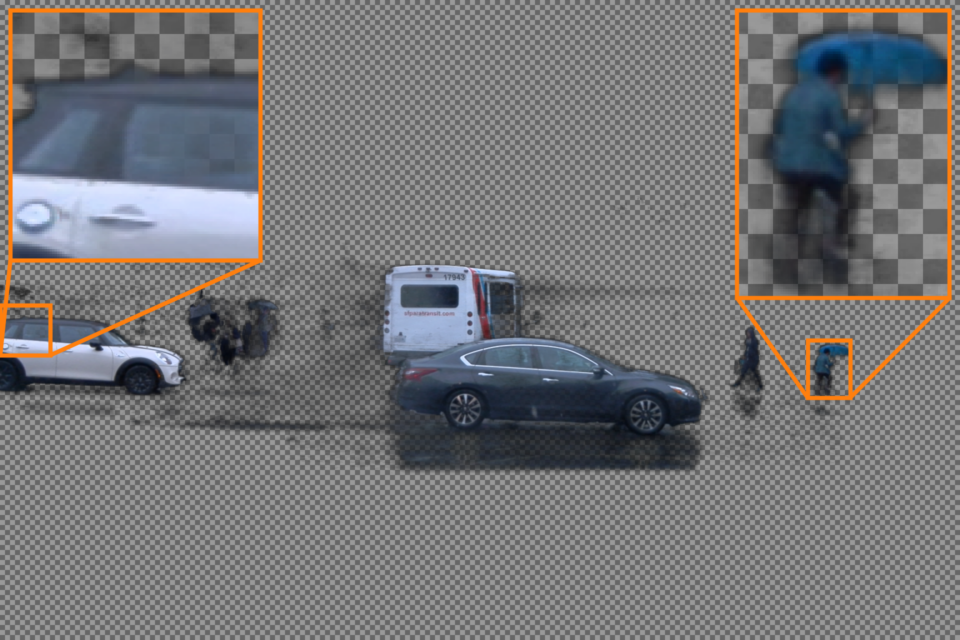} &
        \gridImage{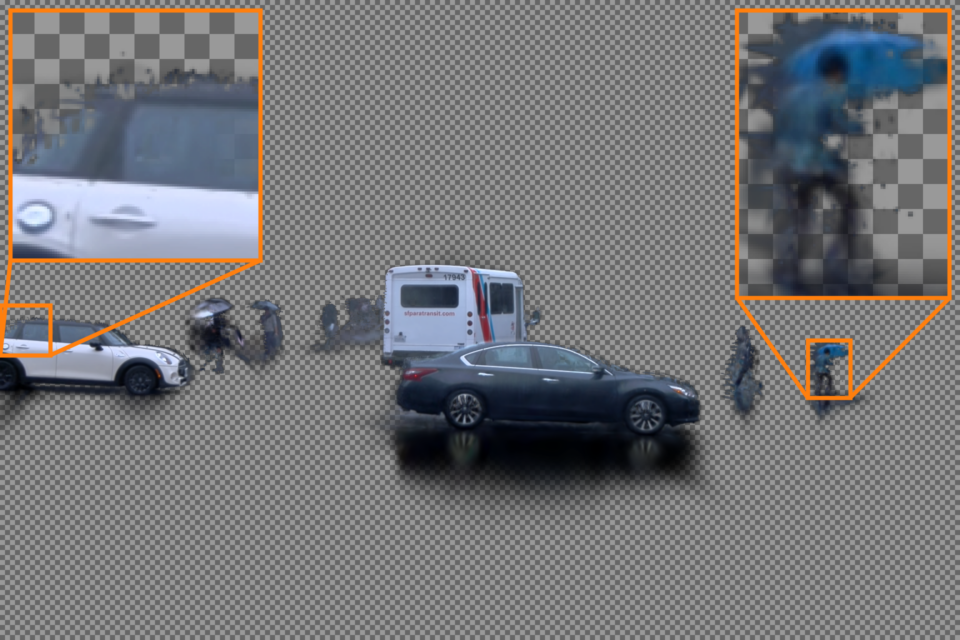} &

        \\
        \gridImage{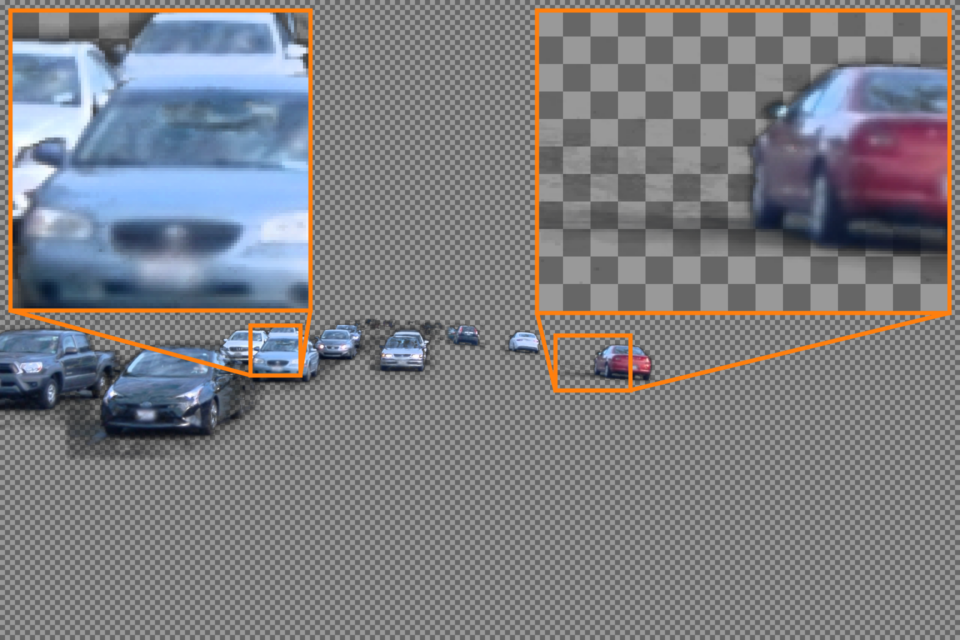} &
        \gridImage{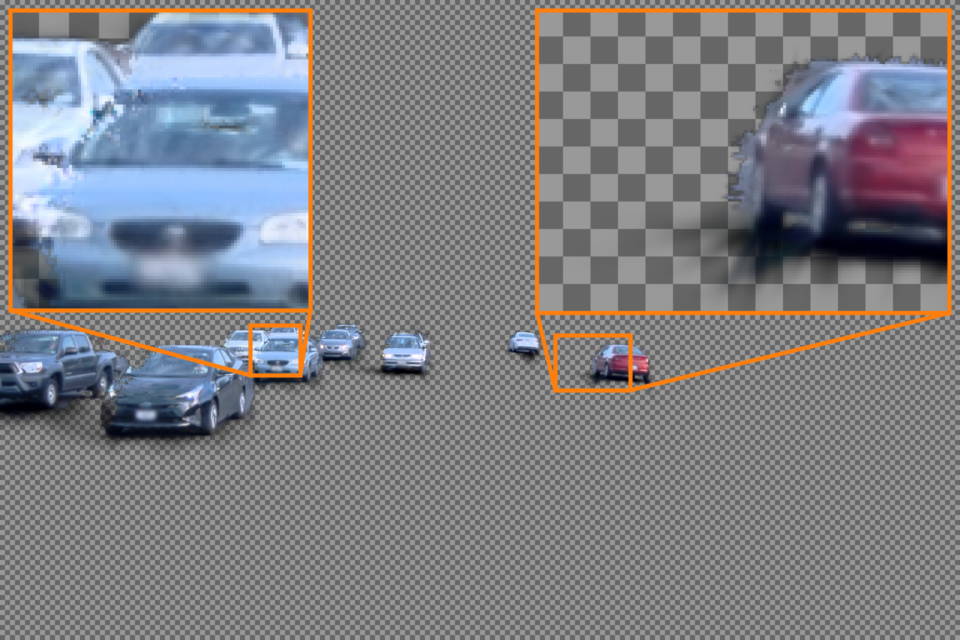} &
        \gridImage{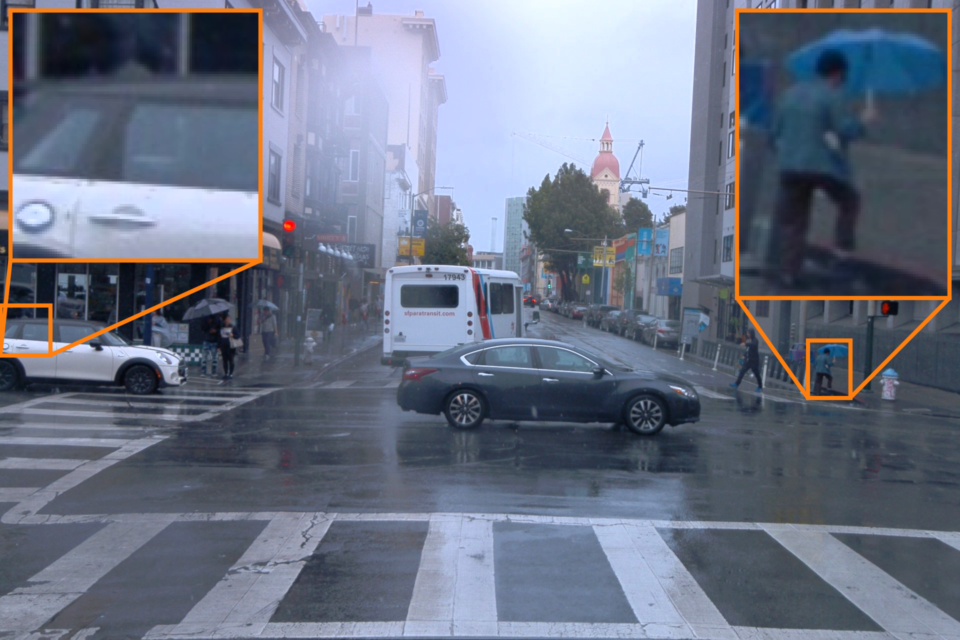} &
        \gridImage{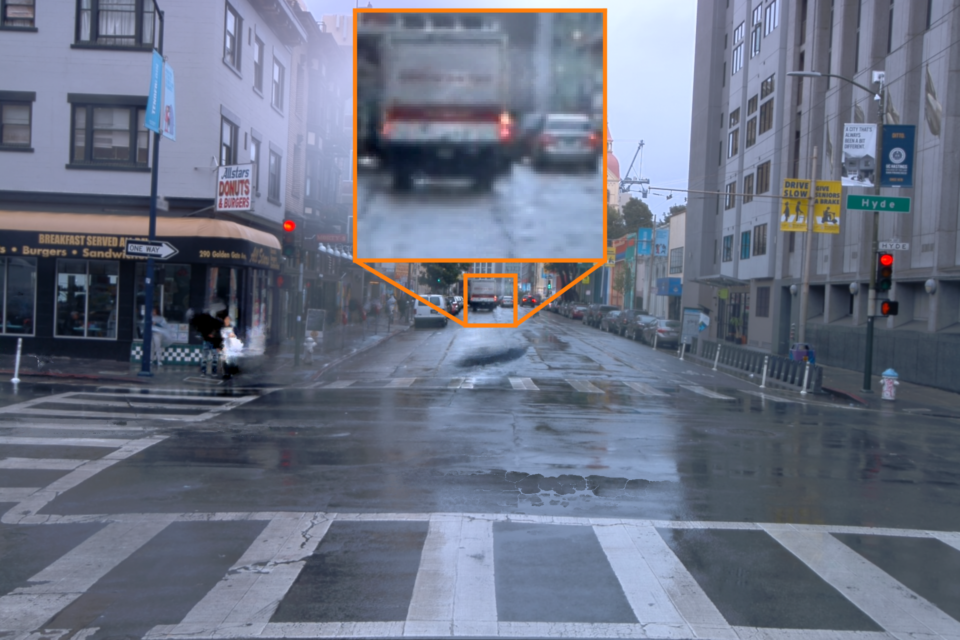} &
        \gridImage{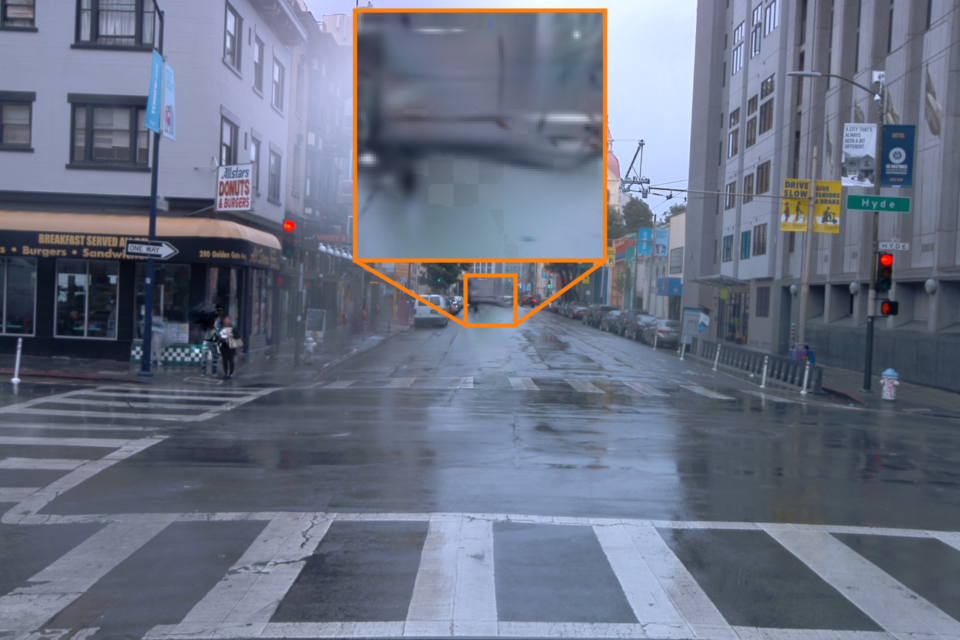} &

        \\
    \end{tblr}
    \begin{tabular}{ccccc}
        \centering \parbox{\dimexpr0.2\textwidth-2\tabcolsep\relax}{\centering Ours} &
        \centering \parbox{\dimexpr0.2\textwidth-2\tabcolsep\relax}{\centering OmniRe\cite{chen2025omnire}} &
        \centering \parbox{\dimexpr0.2\textwidth-2\tabcolsep\relax}{\centering Ground Truth} &
        \centering \parbox{\dimexpr0.2\textwidth-2\tabcolsep\relax}{\centering Ours} &
        \centering \parbox{\dimexpr0.2\textwidth-2\tabcolsep\relax}{\centering OmniRe\cite{chen2025omnire}} 
    \end{tabular}
    \caption{Scene Decomposition on Waymo (sequences s-125 and 141). Visual comparison demonstrates our method's capability for accurate object decomposition, preserving sharp boundaries and shadows for both rigid and non-rigid entities. Furthermore, our approach effectively maintains object information even under occlusion and adverse weather conditions.}
    \label{fig:scene_editing}
\end{figure}
\begin{figure}[tbh]
    \centering
-    \setlength{\imageWidth}{0.2\textwidth}
    \settoheight{\imagerowheight}{\includegraphics[width=\imageWidth]{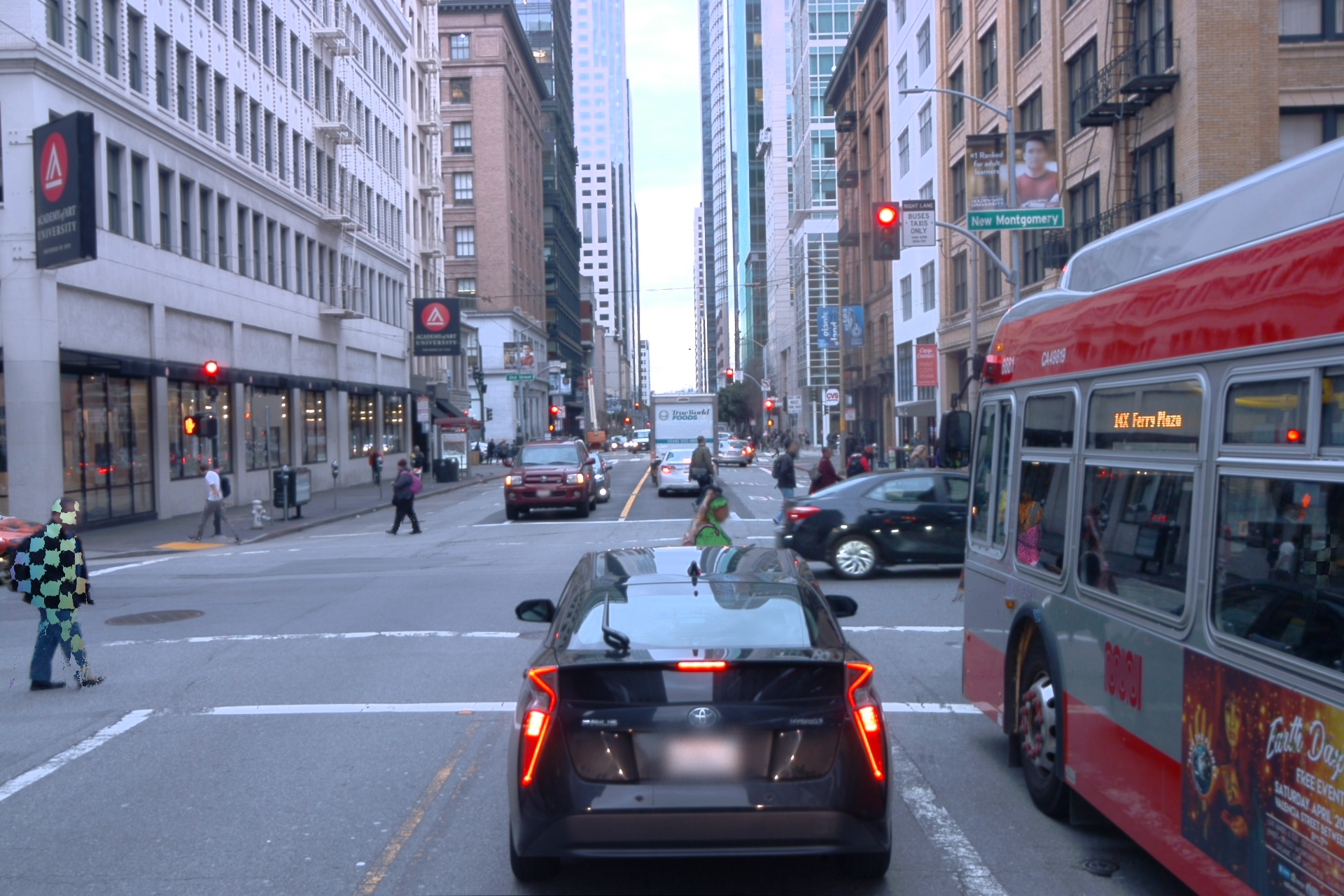}} 
    \newcommand{\gridImage}[1]{%
        \includegraphics[width=\imageWidth]{#1}%
    }
    \begin{tblr}{colspec={c c c c c},colsep=0mm, rowsep=-0.8mm, row{1-2}={ht=\imagerowheight}, cells={halign=c, valign=m}}
        \gridImage{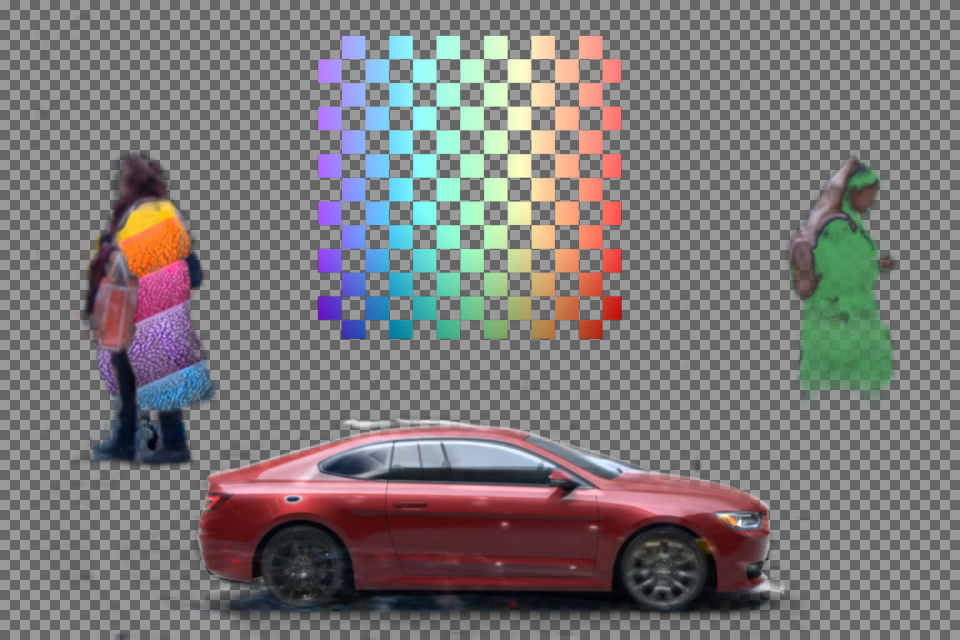} &
        \gridImage{images/editing/W344_000/005_t} &
        \gridImage{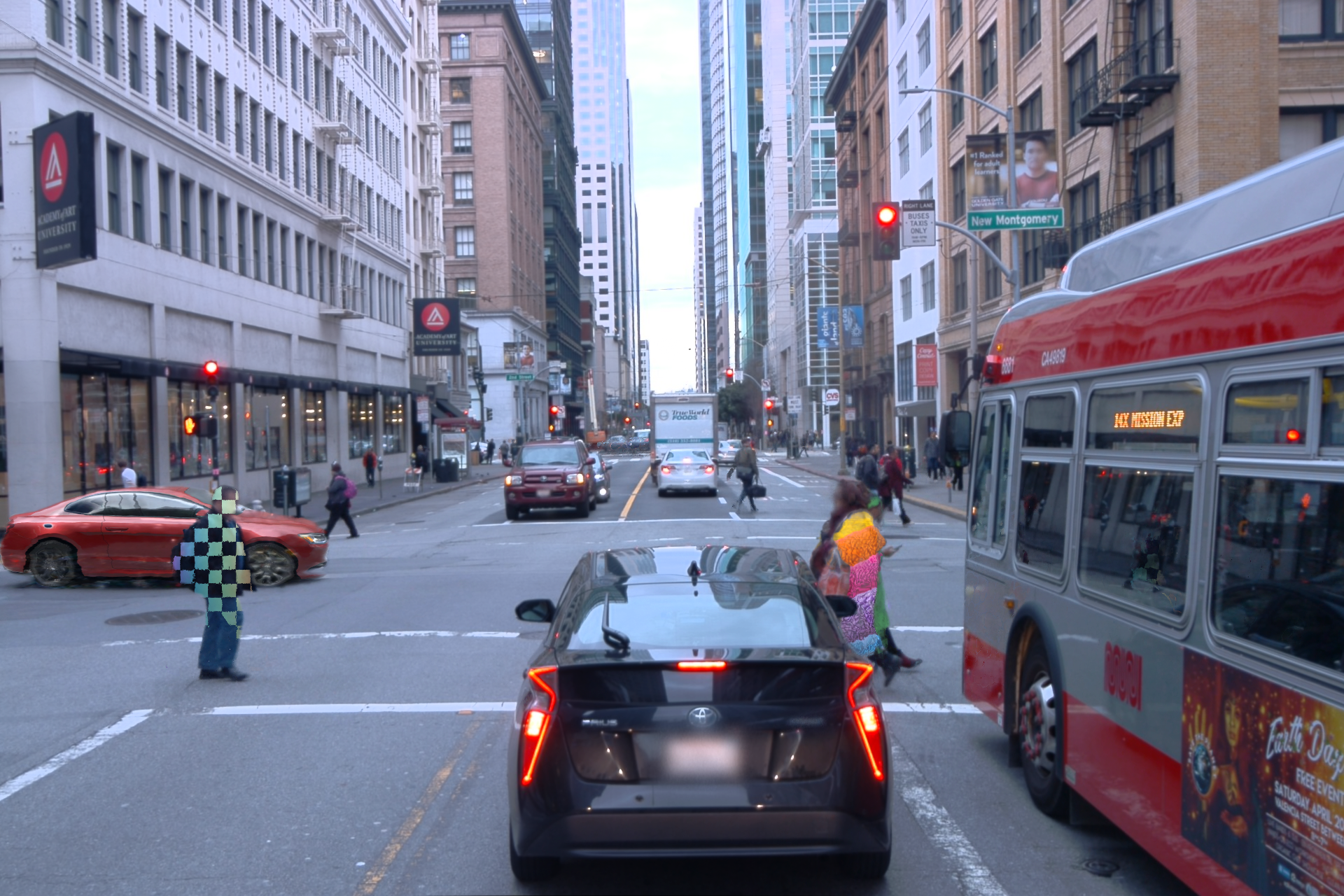} &
        \gridImage{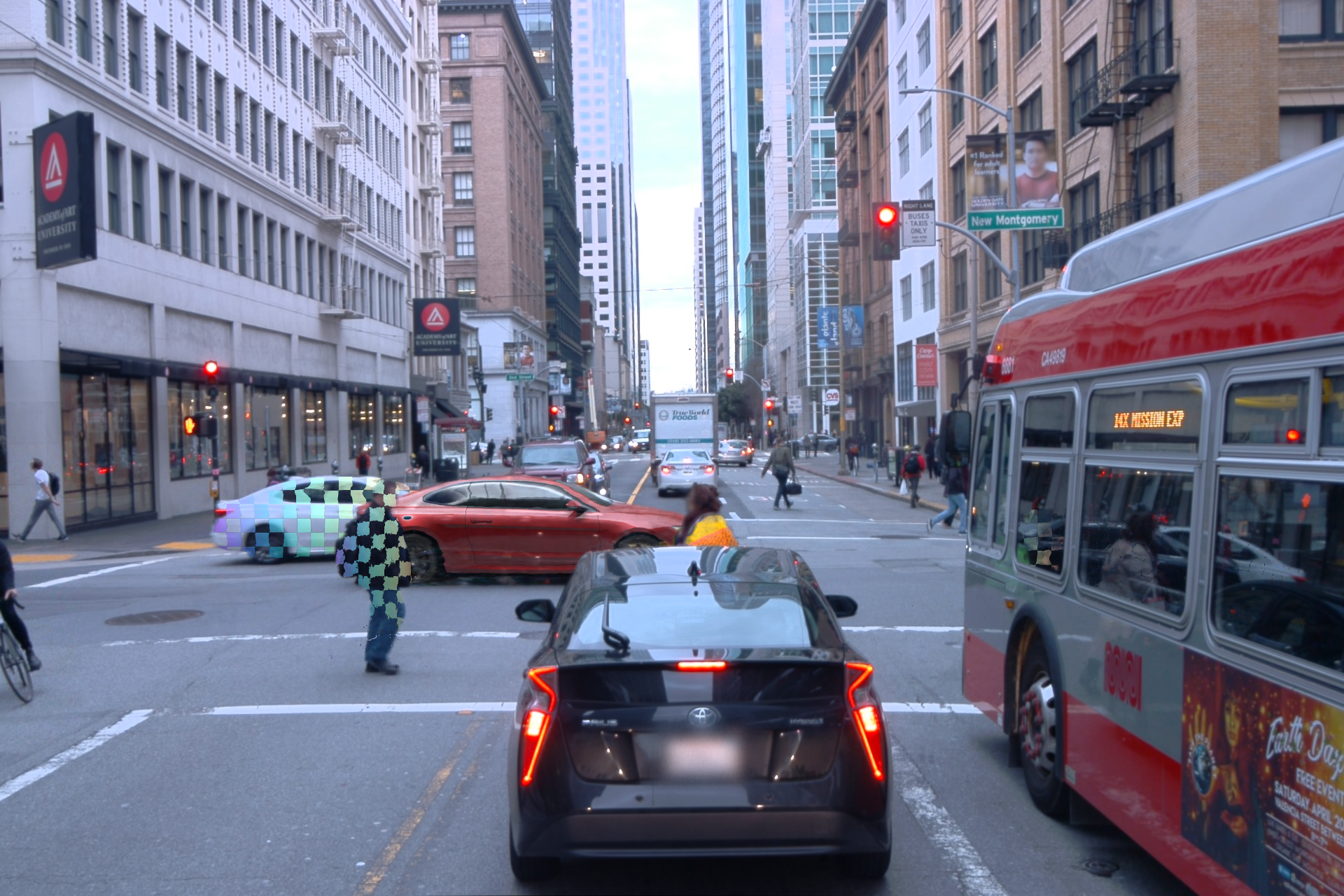} &
        \gridImage{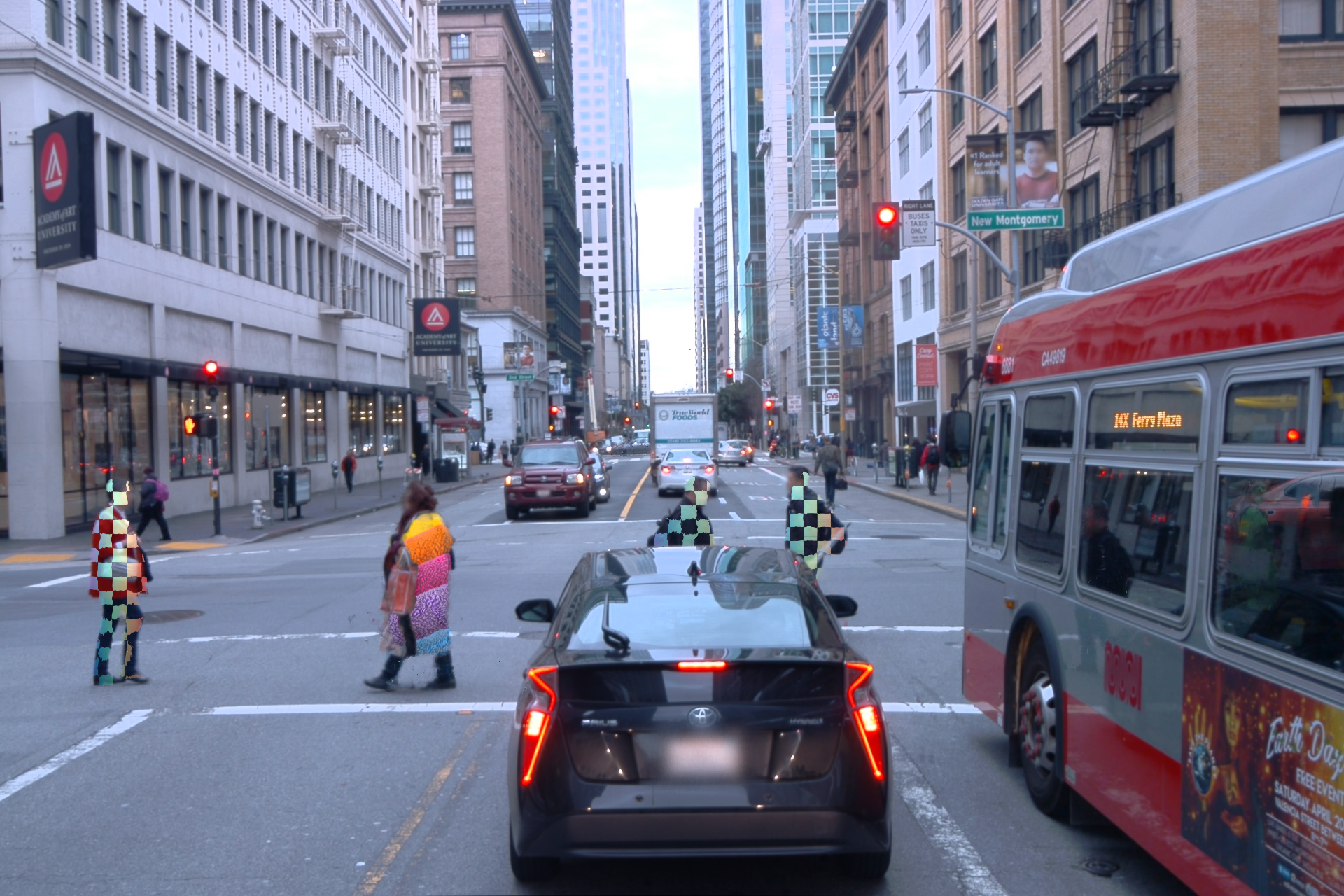} 
        \\
    \end{tblr}
    \caption{Consistent Appearance Editing. We showcase the consistent application of a textures (left) onto dynamically moving objects, changing their appearance and model occlusions accurately.}
    \label{fig:scene_texture_editing}
\end{figure}

\begin{figure}[tbh]
    \setlength{\imageWidth}{0.2\textwidth}
    \settoheight{\imagerowheight}{\includegraphics[width=\imageWidth]{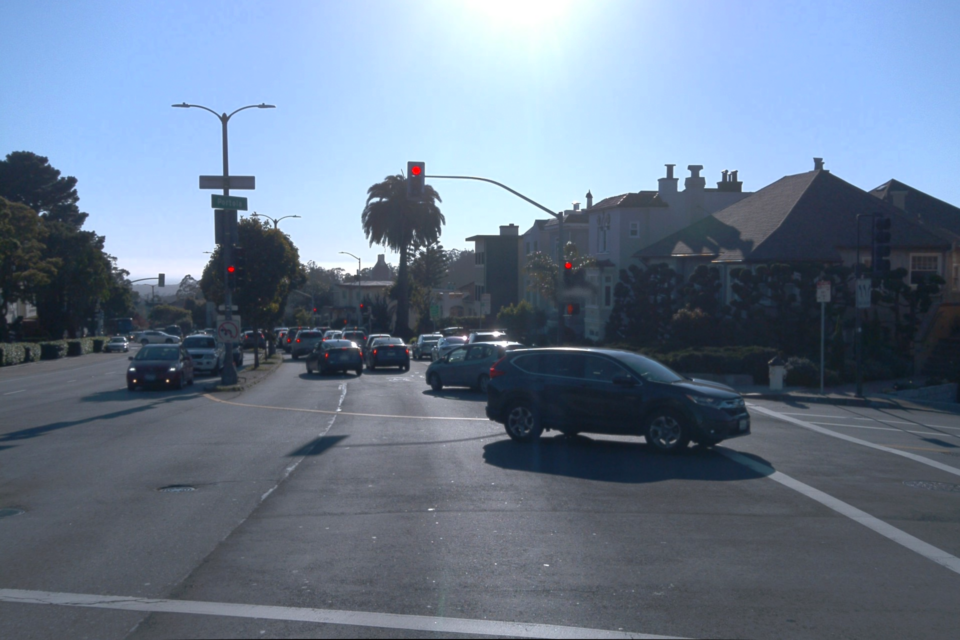}} 
    \newcommand{\gridImage}[1]{%
        \includegraphics[width=\imageWidth]{#1}%
    }
    \begin{tblr}{colspec={c c c c c},colsep=0mm, rowsep=-1.2mm, row{1-2}={ht=\imagerowheight}, cells={halign=c, valign=m}}
         \gridImage{images/edit_results/W203_060/W203_060_002_GT}& %
         \gridImage{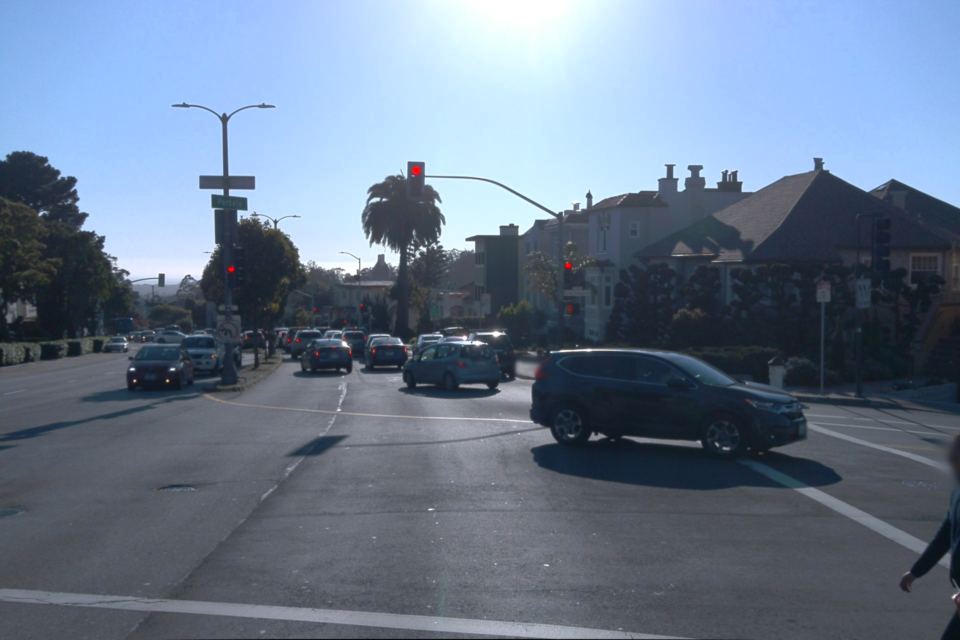}& %
         \gridImage{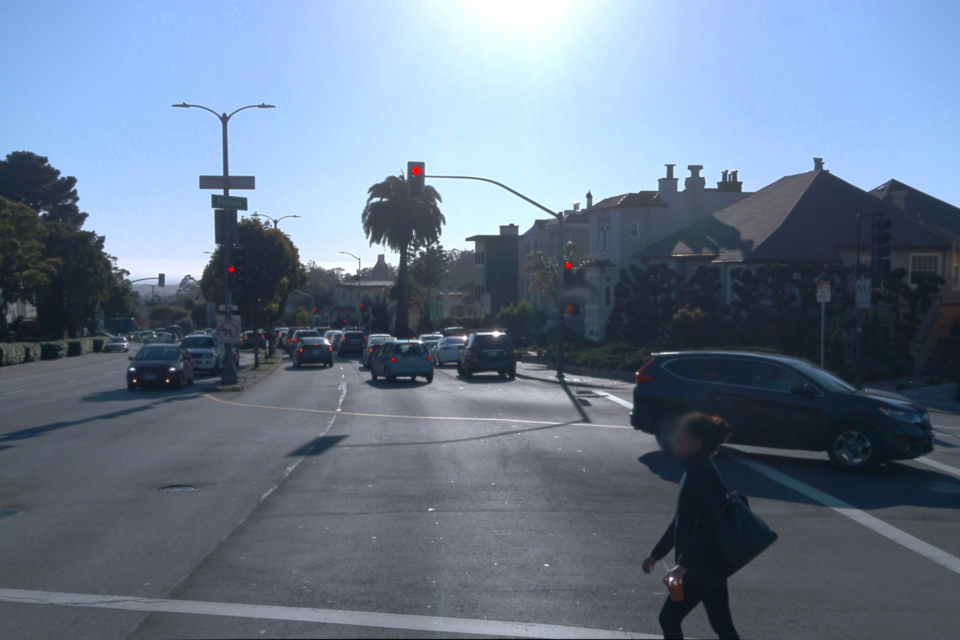}&
         \gridImage{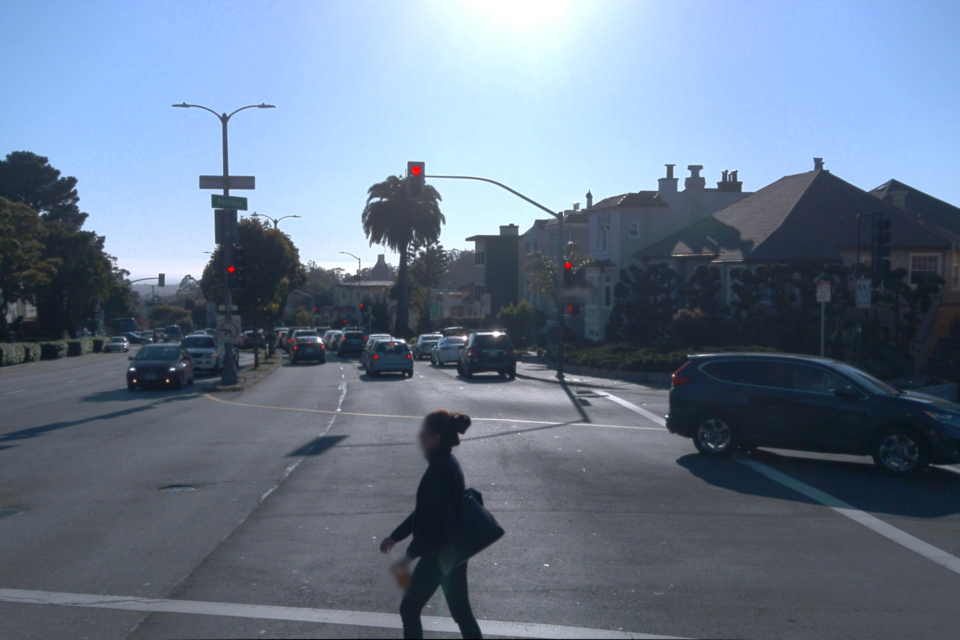}&
         \gridImage{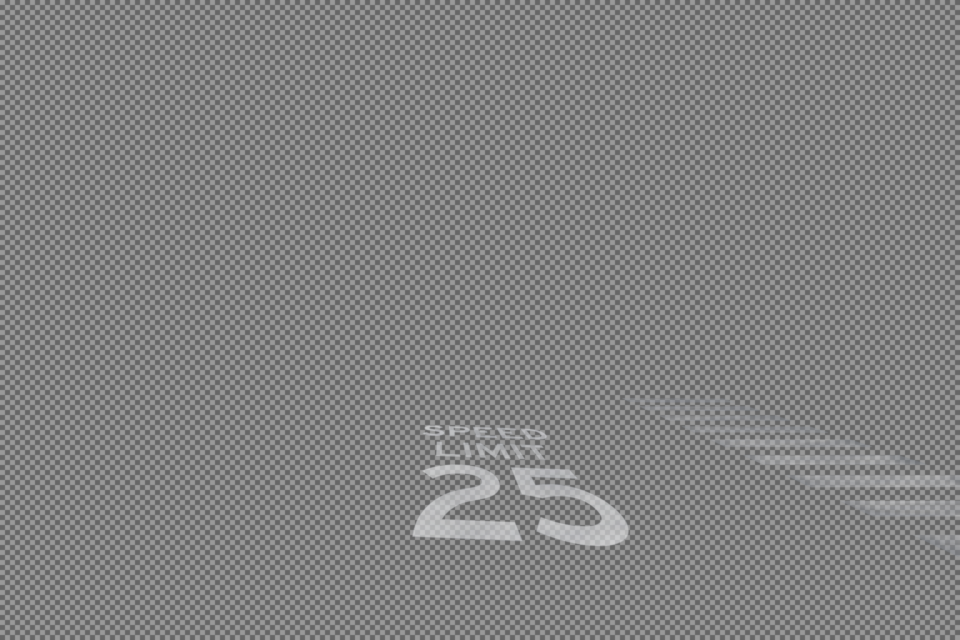}\\
         \gridImage{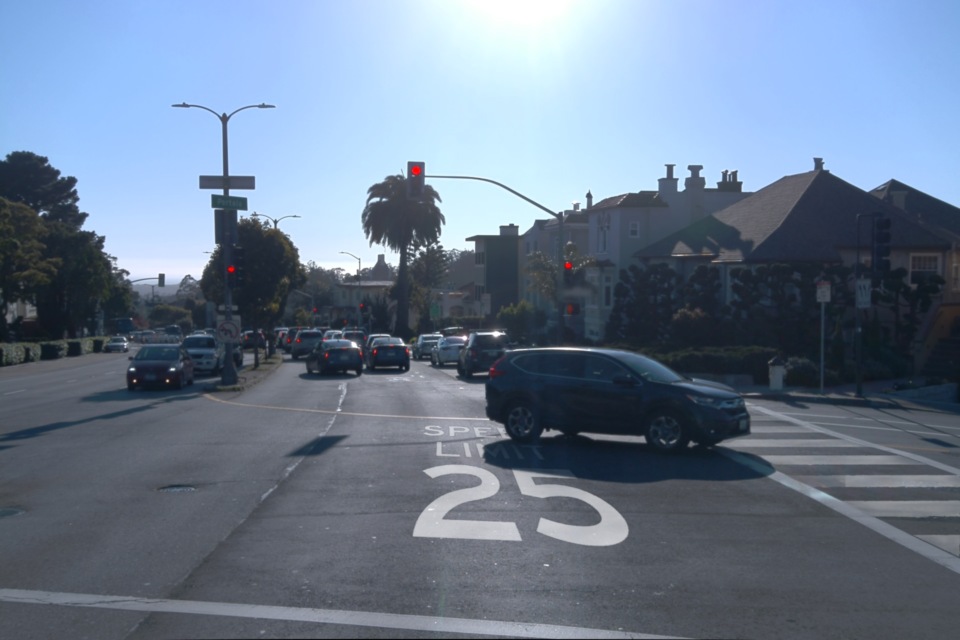}& %
         \gridImage{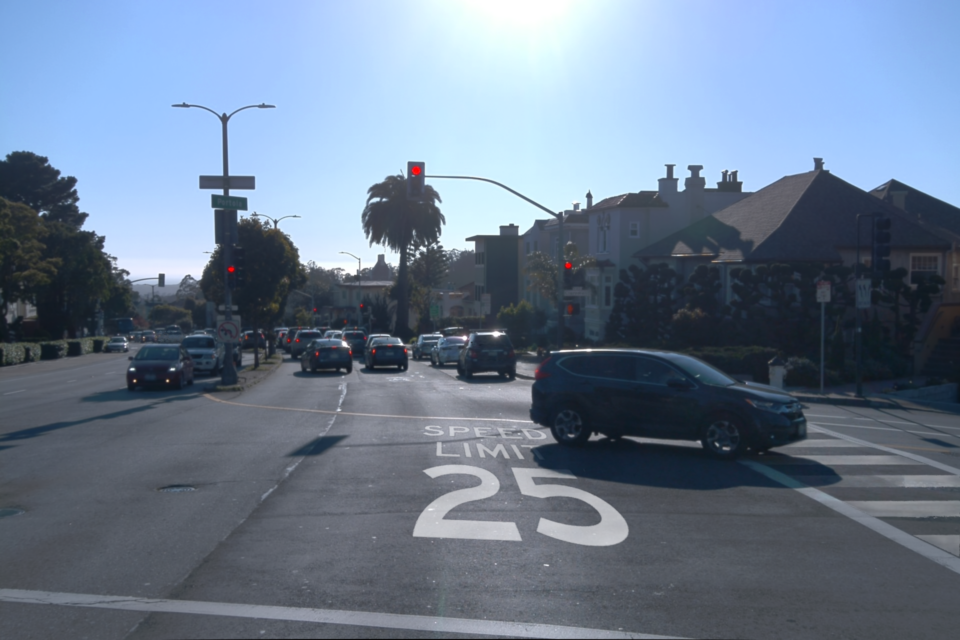}& %
         \gridImage{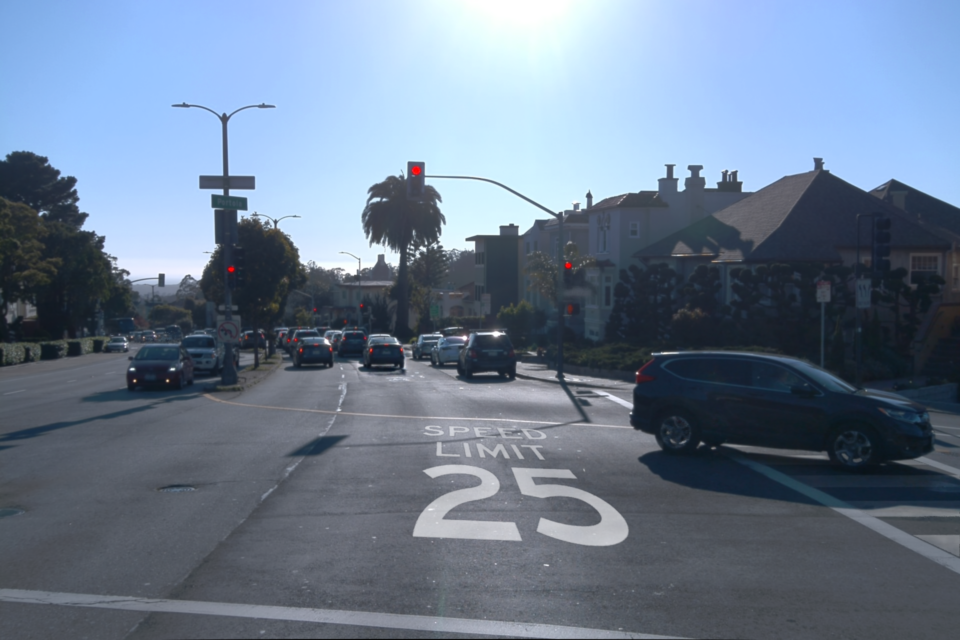}&
         \gridImage{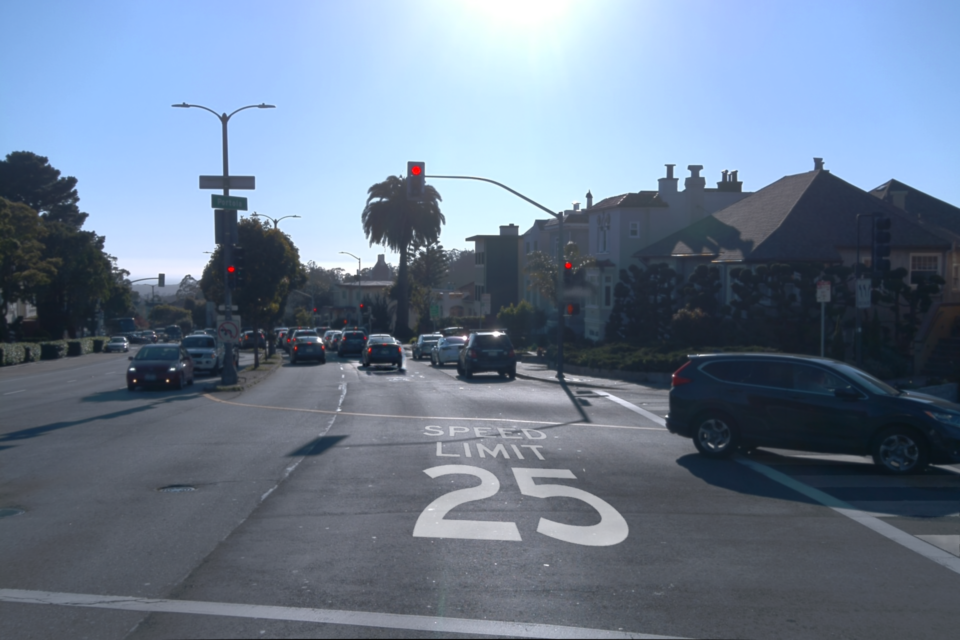}&
         \gridImage{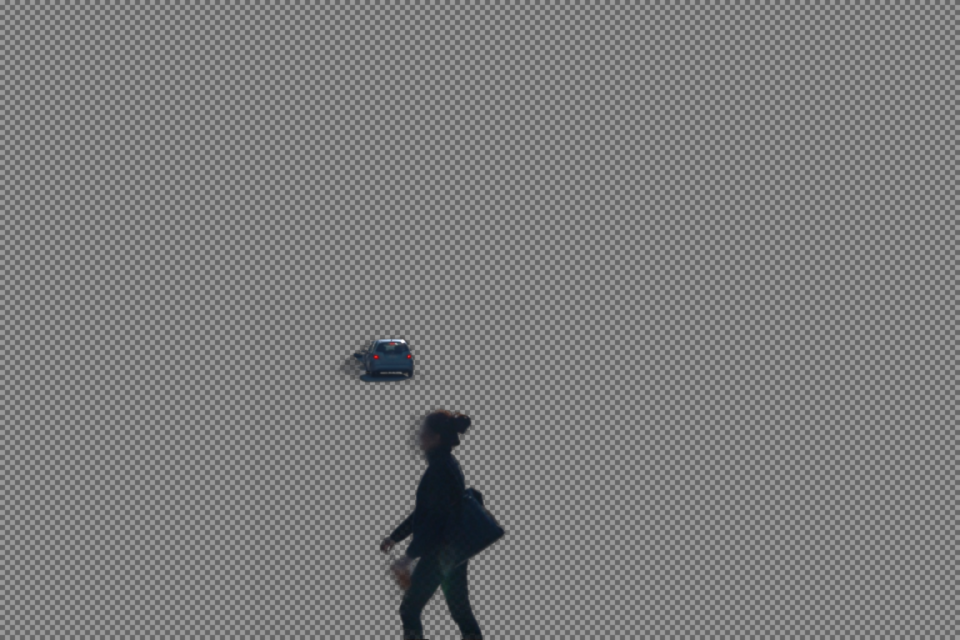}\\
    \end{tblr}
    \caption{Effective scene manipulation in s-203. Four timestamps from the s-203 scene are shown, with ground truth (top) and our edits (bottom). The applied background texture is presented top-right, and the removed objects are shown bottom-right. The natural blending of the car and its shadow with the edits demonstrates our realistic occlusion handling, showcasing our methods applicability in creating counterfactual driving scenarios.}
    \label{fig:waymo_editing_203}
\end{figure}

\subsection{Outdoor Video Scenes}
\vspace{-6pt}
We validate the generalization of our method on diverse outdoor scenes from the DAVIS dataset \cite{Perazzi2016}, which has been regularly used by matting methods \cite{kasten2021layered, Lu_2021_CVPR, lin2023omnimatterf} and provides high-resolution images (up to 1920 $\times$ 1080). We evaluate on 15 specific sequences following \cite{kasten2021layered, lin2023omnimatterf}.
To ensure comparability between our method and the baseline approaches, we match the evaluation setups of all methods, see Supplemental Material for details. 

We report our results in \tabref{tab:davis_results} and find an average 7 dB PSNR improvement w.r.t the best baseline. While the overall performance varies more than in the automotive scenes, due 
complex camera and object motion, the method also outperforms all baseline approaches for all sequences. 
Notably, when objects exhibit rigid-motion behavior (e.g., blackswan, car-shadow or kite-surf), our method achieves its best results. Furthermore, even in scenes involving non-rigid motion (e.g., bear, hike, and elephant), where the challenges are more pronounced, we still outperform the state-of-the-art baselines, validating the generalization to outdoor scenes with various actors. 
We provide additional baseline information, visual examples, and scene decompositions, in our supplementary material.

\begin{table*}[tbh]
\centering
    \caption{Quantitative evaluation results on the Davis Dataset~\cite{Perazzi2016} of diverse outdoor scenes. Our method consistently outperforms the Layered Neural Atlases (LNA) \cite{kasten2021layered} and OmnimatteRF (ORF) \cite{lin2023omnimatterf} baselines, achieving the best metric scores along all sequences. Also scenes with intricate non-rigid motion where alignment is learned are more demanding, our method still yields superior results, with particularly significant gains in sequences well-suited to our motion model, such as motorbike, kite-surf, and car-shadow.}
    \label{tab:davis_results}
    \begin{tabular}{l S[table-format=2.2] S[table-format=2.2] S[table-format=2.2] S[table-format=1.3] S[table-format=1.3] S[table-format=1.3] S[table-format=1.3] S[table-format=1.3] S[table-format=1.3]}
     \toprule
     \multirow{2}{*}{Sequence} & \multicolumn{3}{c}{PSNR $\uparrow$} & \multicolumn{3}{c}{SSIM $\uparrow$} & \multicolumn{3}{c}{LPIPS $\downarrow$} \\
     & {Ours} & {ORF} & {LNA} & {Ours} & {ORF} & {LNA} & {Ours} & {ORF} & {LNA} \\
    \midrule
    bear           & \textbf{33.47} & 24.88 & 26.51 & \textbf{0.934} & 0.658 & 0.771 & \textbf{0.091} & 0.464 & 0.287 \\
    blackswan      & \textbf{36.36} & 26.67 & 29.26 & \textbf{0.938} & 0.739 & 0.815 & \textbf{0.097} & 0.458 & 0.318 \\
    boat           & \textbf{35.83} & 28.63 & 30.15 & \textbf{0.932} & 0.761 & 0.816 & \textbf{0.099} & 0.376 & 0.274 \\
    car-shadow     & \textbf{36.67} & 29.26 & 28.47 & \textbf{0.947} & 0.861 & 0.850 & \textbf{0.084} & 0.313 & 0.269 \\
    elephant       & \textbf{33.91} & 26.94 & 28.34 & \textbf{0.922} & 0.731 & 0.772 & \textbf{0.088} & 0.423 & 0.325 \\
    flamingo       & \textbf{34.96} & 25.74 & 27.10 & \textbf{0.928} & 0.753 & 0.783 & \textbf{0.106} & 0.483 & 0.349 \\
    hike           & \textbf{29.74} & 25.15 & 24.77 & \textbf{0.886} & 0.698 & 0.682 & \textbf{0.108} & 0.388 & 0.343 \\
    horsejump-high & \textbf{34.78} & 28.35 & 27.28 & \textbf{0.932} & 0.846 & 0.830 & \textbf{0.074} & 0.249 & 0.226 \\
    kite-surf      & \textbf{37.96} & 28.04 & 27.88 & \textbf{0.949} & 0.780 & 0.780 & \textbf{0.068} & 0.420 & 0.400 \\
    kite-walk      & \textbf{37.96} & 29.44 & 29.58 & \textbf{0.941} & 0.804 & 0.818 & \textbf{0.070} & 0.367 & 0.334 \\
    libby          & \textbf{38.89} & 29.62 & 29.35 & \textbf{0.949} & 0.819 & 0.828 & \textbf{0.095} & 0.399 & 0.342 \\
    lucia          & \textbf{30.90} & 26.03 & 26.63 & \textbf{0.869} & 0.690 & 0.742 & \textbf{0.178} & 0.407 & 0.329 \\
    motorbike      & \textbf{37.42} & 27.33 & 29.33 & \textbf{0.950} & 0.779 & 0.843 & \textbf{0.082} & 0.376 & 0.241 \\
    swing          & \textbf{35.70} & 26.14 & 27.88 & \textbf{0.926} & 0.722 & 0.808 & \textbf{0.119} & 0.404 & 0.289 \\
    tennis         & \textbf{35.65} & 27.43 & 28.81 & \textbf{0.928} & 0.806 & 0.862 & \textbf{0.120} & 0.328 & 0.209 \\
    \midrule
    Mean           & \textbf{35.35} & 27.31 & 28.09 & \textbf{0.929} & 0.763 & 0.800 & \textbf{0.098} & 0.390 & 0.302 \\
    \bottomrule
    \end{tabular}
\end{table*}

\subsection{Limitations}
\vspace{-6pt}
Our method relies on the assumption that objects and background can be projected onto a plane, imposing limits on camera and object rotation (less than 180 degrees). Correspondingly, Novel View Synthesis and accurate 3D representation is not within the scope of our current manuscript. Accurate initial plane initialization via bounding boxes or geometric priors is recommended, and suboptimal initialization can lead to early failure. Projection errors on non-planar object geometry may accumulate and hinder precise initial position determination.

Representing large camera motion that significantly changes the background is also challenging for the proposed plane assumption, and is further studied in the supplementary material. This is due to the difficulty of capturing large flow vectors, particularly when sampling rays from only a subset of time steps and coordinates as we do. However, we note that the use of view dependency compensates for these errors, although it does come at the cost of reduced texture editability. 

\section{Conclusion}
In this work, we introduce Neural Atlas Graphs (NAGs), an editable hybrid scene representation for high-resolution learning and rendering of dynamic scenes. We find that the hybrid 2.5D representation of NAGs compares favorably in representing driving scenes. Specifically, we validate that the method achieves state-of-the-art results on the Waymo Open Dataset, with 5 dB PSNR improvement overall and 11.2 dB PSNR for rigid and 10.7 dB PSNR for non-rigid actors. We show that NAGs enable high-resolution, view-consistent environmental editing, unlocking the creation of compelling counterfactual driving scenarios and showcasing potential not only in scene arrangement and scene decomposition but also in consistent appearance editing. We also confirm the generalization of NAGs beyond driving scenes for comprehensive scene understanding and manipulation with an evaluation on the outdoor DAVIS video dataset, achieving a 7.3 dB PSNR improvement over existing methods. In the future, the differentiable representation with the low-dimensional neural atlases may allow for task-driven editing, such as learning of counterfactuals specifically to challenge a driving stack, which we exemplary demonstrated in \figref{fig:waymo_editing_203}.

\section{Acknowledgments}

\todo

We extend our sincere gratitude to Julian Ost, Mario Bijelic, Amogh Joshi, and William Koch from Princeton University for their insightful contributions through numerous discussions concerning the autonomous driving context, recent scene representation literature, and methodologies.
This research was supported by the DFG Research Unit 5336 - Learning to Sense (Project No. 459284860) and project funding for "Polymorphic Scene Representation for Enhanced Instant Scene Reconstruction" (Project No. 510825780, Ref. KO 2960/20-1). Ilya Chugunov was supported by NSF GRFP (2039656). Felix Heide was supported by an NSF CAREER Award (2047359), a Packard Foundation Fellowship, a Sloan Research Fellowship, a Disney Research Award, a Sony Young Faculty Award, a Project X Innovation Award, an Amazon Science Research Award, and a Bosch Research Award.
\newpage
{
    \small
    \bibliographystyle{ieeenat_fullname}
    \bibliography{main}
}
\fi

\newpage

\title{Supplementary Material: \\ Neural Atlas Graphs for Dynamic Scene Decomposition and Editing}
\makeatletter
\begin{center}
    \hsize\textwidth
    \linewidth\hsize
    \vskip 0.1in
    \@toptitlebar
    {\LARGE\bf \@title\par}
    \@bottomtitlebar
    \begingroup
        \def\And{%
          \end{tabular}\hfil\linebreak[0]\hfil%
          \begin{tabular}[t]{c}\bf\rule{\z@}{24\p@}\ignorespaces%
        }
        \def\AND{%
          \end{tabular}\hfil\linebreak[4]\hfil%
          \begin{tabular}[t]{c}\bf\rule{\z@}{24\p@}\ignorespaces%
        }
        
        \begin{tabular}[t]{c}\bf\rule{\z@}{24\p@}\@author\end{tabular}%
    \endgroup 
    
    \vskip 0.3in \@minus 0.1in
\end{center}
\makeatother

\appendix
\doparttoc
\part{} 

\acresetall 

This supplementary document provides further method details, including specific aspects of the camera model and our coarse-to-fine training scheme. We also expand on description of our dataset and experimental design, included ablation studies, provide additional quantitative results, visual examples, and edits across both autonomous driving and challenging outdoor video sequences. We recommend reviewing the accompanying \href{https://drive.google.com/file/d/14hPfGsRJuIMDUn2ux1lWMcOc5TiMAlX7/view?usp=drive_link}{video}, which provides a compelling summary of our key visual contributions.

Given the different modalities of our supplementary materials, we provide the high-resolution videos within a \href{https://drive.google.com/drive/folders/1AxR0KoG3knjFDuUZAiDi6Oc_IKCIwrZW?usp=sharing}{google drive folder} and the code within a \href{https://github.com/jp-schneider/nag}{github repository} along this document for further details.

To give an overview of the provided videos within the \href{https://drive.google.com/drive/folders/1AxR0KoG3knjFDuUZAiDi6Oc_IKCIwrZW?usp=sharing}{google drive folder}, we briefly highlight the structure:

\begin{itemize}
    \item \href{https://drive.google.com/file/d/14hPfGsRJuIMDUn2ux1lWMcOc5TiMAlX7/view?usp=sharing}{\path{overview.mp4}} - contains our overview video, showcasing our key visual results and comparisons. For detailed examples see below. 
    \item \href{https://drive.google.com/drive/folders/1ynfwUelfK-wGZWJI3Ql5EshyhFiSE8GX?usp=sharing}{\path{edits/[manuscript|supplement]/figure_[Number]}} - contains the videos matching the given figure number in either our manuscript or this supplementary. 
    \item \href{https://drive.google.com/drive/folders/1NgVLP5Ex8HxLejqZpVwnp_zT6X6YMU3t?usp=drive_link}{\path{visuals/[waymo|davis]/[sequence]}} - contains all reconstruction for Waymo \cite{sun2020scalability} and Davis \cite{Perazzi2016}, and also decompositions for the latter one. We choose the abbreviation ORE for OmniRe \cite{chen2025omnire}, ERF for EmerNeRF \cite{yang2024emernerf}, LNA for Layered Neural Atlases \cite{kasten2021layered}, ORF for OmnimatteRF \cite{lin2023omnimatterf} and GT for the ground truth videos. 
\end{itemize}

To provide a better overview of the remaining supplementary material, we provide a table of contents.

\clearpage

\parttoc 

\section{Additional Method Details}

\subsection{Camera Model} 
\label{sec:camera_model}

For rendering and scene interaction, we require a mapping from image coordinates to a 3D world reference system. We utilize the standard pinhole camera model parameterized by intrinsic $K$ and extrinsic $g(\timestamp)$ matrices. Considering the projection of a single pixel $(u, v)$ at timestamp $t$, the ray origin $\rayOrigin$ and direction $\rayDirection$ in a world reference system can be computed by:

\begin{align}
\label{eq:ray_generation}
\setlength{\tabcolsep}{0pt}
\begin{array}{r l r l}
    \hat{\rayDirection}(u, v) & {= (K^{-1} \odot \left[\begin{array}{c}
         u \\
         v \\
         1 \\
    \end{array}\right] \cdot \focalLength),} & \hat{\rayOrigin}(u, v)\, = & {\hat{\rayDirection}(u, v) - \left[\begin{array}{c} 
         0 \\%
         0 \\%
         \focalLength \\%
    \end{array}\right],} \\%
    \rayOrigin(u, v, t) & = g(\timestamp) \odot \hat{\rayOrigin}(u, v), & \rayDirection(u, v, t)\, = &  \rotation \odot \hat{\rayDirection}(u, v)
\end{array}
\end{align}

for a camera projection plane lying at z=0. For better readability, we avoided stating homogeneous vector conversions. The inverse of the intrinsic matrix $K \in \mathbb{R}^{3 \times 3}$ \eqref{eq:camera_calib} is used to convert a pixel in the camera's local coordinate system. Further, the extrinsic matrix $g(\timestamp) \in \mathbb{R}^{3 \times 4}$, converts from the camera`s local into the world coordinate system. These matrices can be defined as:
\begin{equation}
\label{eq:camera_calib}
g(t) = %
\begin{NiceArray}{[c c c c]}%
1 & -r^z & r^y_i & x_i\\%
r^z_i & 1 & -r^x_i & y_i\\%
-r^y_i & r^x_i & 1 & z_i\\%
\CodeAfter
    \UnderBrace[shorten,yshift=3pt]{1-1}{3-3}{\rotation}
    \UnderBrace[shorten,yshift=3pt]{1-4}{3-4}{\translation}
\end{NiceArray}, K^{-1}=\left[\begin{array}{ c c c }%
1 / fm_x & 0 & -p_x / fm_x \\%
0 & 1 / fm_y & -p_y / fm_y\\%
0 & 0 & 1 %
\end{array}\right]%
\end{equation}
\vspace{.5cm}

The values comprising the intrinsic matrix $K$ are typically provided by the camera manufacturer, whereby $\focalLength \in \mathbb{R}$ defines the focal length, $m_x, m_y$ the image width and height, and $p_x, p_y$ define the principal point. While camera extrinsics are generally provided in autonomous driving datasets, these values are susceptible to inaccuracies due to sensor miscalibration or accumulated odometry drift. Furthermore, when estimated using structure-from-motion or neural methods, such as RoDynRF \cite{liu2023robust} in our outdoor experiments, the resulting poses may also contain noise. To refine these, we utilize the same spline-based offset learning approach \cite{chugunov_2024} as discussed for our nodes to map $t$ to its correspondences control points $\pTranslationCam[\text{cam},i], \pRotationCam[\text{cam},i]$ using interpolation. The learning process will adjust for possible shifts in the camera rotation $\rotation$ and translation $\translation$, recalling our definitions (\ref{eq:spline_object_model}, \ref{eq:spline_object_rotation_model}):%
\begin{equation}
\label{eq:spline_camera_model}
\begin{aligned}
    \translation &= \baseTranslation + \translationWeight \cdot \splineFnc{\timestamp,\,\pTranslationCam}\\
    \rotation &= \baseRotation \cdot \rotvecToQuatFnc{\rotationWeight  \cdot \splineFnc{\timestamp,\,\pRotationCam}}
\end{aligned} 
\end{equation}
\begin{equation}
    \pTranslationCam =  \left\{\left[\begin{array}{c}
         x  \\
         y  \\
         z \\
    \end{array}\right]_{i}\right\}_{i=0}^{P}, \quad
    \pRotationCam =  \left\{\left[\begin{array}{c}
         r^{x}  \\
         r^{y}  \\
         r^{z}  \\
    \end{array}\right]_{i}\right\}_{i=0}^{P}
\end{equation}

whereby $\spline{}:[0, 1] \times \R^{\nControlPoints} \rightarrow \R^{\nFrames}$ denotes the cubic hermite spline interpolation \cite{deboor_1987}, as discussed in \cite{chugunov_2024}, and $\pTranslationCam, \pRotationCam \in \R^{\nControlPoints \times 3}$ being zero-initialized
learnable translation and rotation offsets of the camera. 
We further denote the rotation vector to unit quaternion operation as $\rotvecToQuat: [0, 2\pi)^{3} \rightarrow \Quaternion$ for $\Quaternion$ being the set of unit-quaternions. 

Given such definition, the number of control points $\nControlPoints \in \mathbb{N}$ can be used to encourage smooth motion, e.g. by setting it smaller than the number of frames $\nFrames$ in the video $\scene$ ($\nControlPoints = \nFrames / 2$), or keeping it equal to the number of frames to keep the expressivity. The prior-known positions are stated as $\baseTranslation \in \R^{\nFrames \times 3}\; \text{and} \;\baseRotation \in \Quaternion^{\nFrames}$ describing camera translation and rotation respectively. To control the influence of the learned offsets with introduced temperature weights $\translationWeight = \rotationWeight = 0.5$.

\subsection{Coarse-to-fine Optimization} 
\label{sec:coarse_to_fine_optim}

To limit the expressiveness of the view-dependent fields $\viewField$ to model as few changes as possible, as well as enforcing the planar flow field to firstly learn coarse alignment, we apply a coarse-to-fine learning strategy by masking the hash-encoding using a sparsity function \cite{chugunov_2024} $\sparsityFunction{\cdot}$ based on the training progress $\sparsityMarker \approx \text{clamp}(0.05 + \sin(\text{epoch}\cdot \pi / 1.6 \cdot \text{max\_epoch}))$. For epoch being the current epoch in training and max\_epoch = 80 the total epochs to conduct. Empirically, sparse deactivates several encoding dimensions $E$ from the multi-resolution hash encodings $\encoding_{\objectIndex, \phi}: [0, 1]^4 \to \R^{E}, \encoding_{\objectIndex, f}: [0, 1]^2 \to \R^{E}$ for a node $\objectIndex$, setting their activations to zero and activates them when training progresses. Correspondingly, the view- and flow-neural fields $\viewField, \flowField$ may be rewritten as

\begin{align}
    \viewField(\queryPoint) &= \network[\objectIndex, \phi](\sparsityFunction{\viewEncoding(\queryPoint, \phi)}), \\
    \flowField(\queryPoint) &= \network[\objectIndex, f](\sparsityFunction{\flowEncoding(\queryPoint)}),
\end{align}

for $\network[\objectIndex, \phi]: \R^{E} \to \R^4$ being the view-dependent MLP and  $\network[\objectIndex, f]: \R^{E} \to \R^{\nFlowControlPoints \times 2}$ being the flow MLP, predicting $\nFlowControlPoints$ flow control points, correspondingly. Further, $\queryPoint \in [0, 1]^2$ denotes the intersection point in planar coordinates and $\phi \in [0,1]^2$ its normalized spherical view angle. Additionally, the expressiveness of the model and its learnable parameters are controlled using a \textit{phase-based learning strategy}, which we further detail below.

Note: while we denoted the color and opacity of the view-dependent field $\viewField$ in the main manuscript separately to increase readability, e.g.  $\viewField[\objectIndex, \phi, c], \viewField[\objectIndex, \phi, \alpha]$, they share parameters.  

\subsection{Phase-based Learning} 
\label{sec:phase_based_learning} 
Our training strategy employs a three-phase optimization approach combined with the previously mentioned coarse-to-fine learning strategy to effectively optimize the various components of each node. In the first phase, from epoch 0 to 5, only the positional parameters $\pTranslationObj, \pRotationObj, \pTranslationCam, \pRotationCam$\footnote{Note: $\pRotationObj$ is not optimized in our main automotive experiments.} are optimized to compensate for positional errors of the objects and camera. In the second phase, epoch 5 to 20, the color and opacity fields $\colorField, \alphaField$ are additionally optimized. In the last third phase, starting from epoch 20, all parameters $\pTranslationObj, \pRotationObj, \pTranslationCam, \pRotationCam, \colorField, \alphaField, \flowField, \viewField$ are optimized together.\\

\subsection{Parametrization}
\label{sec:parametrization}
In the following we detail the parametrization of our atlas nodes. Our model fundamentally distinguishes between two types of information for each atlas node: fixed initial conditions (pre-conditioning) and the core learnable neural fields.

\paragraph{Fixed Initial Conditions}
To establish a robust starting point for optimization, we initialize the base color $\colorBase$ and base alpha $\alphaBase$ for each object's appearance. These non-learnable base textures are derived via an initial forward projection (using the camera model in \secref{sec:camera_model}) of the masked image within a single reference frame onto the object's position-initialized plane. We use the image corresponding to the mask with the largest size as reference. The position initialization itself is carried out using our initial translation and rotation parameters $\baseTranslation[\objectIndex, \timestamp], \baseRotation[\objectIndex, \timestamp]$ (\ref{eq:spline_object_model}, \ref{eq:spline_object_rotation_model}), which are extracted from 3D bounding boxes (if available) or image homographies based on the masked region, combined with a monocular depth estimation. These base parameters remain fixed, while the learned components correct for initialization errors.

\paragraph{Learnable Neural Fields}

The core appearance and motion are then captured by our learnable neural fields: the color field $\mathcal{F}_{i, c}$, opacity field $\mathcal{F}_{i, \alpha}$, flow field $\mathcal{F}_{i, f}$, and view-dependent field $\mathcal{F}_{i, \phi}$. These fields are explicitly designed and optimized to serve distinct, disentangled roles:

\begin{itemize}
    \item The color $\mathcal{F}_{i, c}$ and opacity $\mathcal{F}_{i, \alpha}$ fields (\ref{eq:color_model}, \ref{eq:alpha_model}) are primarily responsible for modeling the \textit{view-agnostic, canonical appearance} of the object in its atlas space.
    \item The flow field $\mathcal{F}_{i, f}$ \eqref{eq:flow_control_points} enables the representation of non-rigid motion by warping the canonical appearance across time, facilitating better editability by maintaining consistent base texture across frames.
    \item The view-dependent field $\mathcal{F}_{i, \phi}$ (\ref{eq:color_model}, \ref{eq:alpha_model}) is designed to capture subtle view-dependent effects (e.g., specularities, reflections) that cannot be explained by the canonical appearance or flow alone.
\end{itemize}

The optimization process distinguishes these components through our phase-based (\secref{sec:phase_based_learning}) and coarse-to-fine (\secref{sec:coarse_to_fine_optim}) learning strategy. By initially limiting the expressiveness of the view-dependent field (via sparse encoding) and progressively activating it, we implicitly regularize the model by prioritizing the non-view-dependent fields ($\mathcal{F}_{i, c}, \mathcal{F}_{i, \alpha}, \mathcal{F}_{i, f}$) for primary appearance and motion capture. This forces the components relevant to editing and flow-mapping to learn the majority of the information, ensuring a disentangled representation where $\mathcal{F}_{i, \phi}$ only incorporates subtle, additional changes.

\paragraph{Number of Parameters}

We state the parameterization of the learnable components of each \ac{nag} node in \tabref{tab:parametrization_nag_node}, while we refer to \secref{sec:evaluation_setup} for a description of each field's architecture. The translation and rotation control points $\pTranslationObj \in \R^{\nControlPoints \times 3}, \pRotationObj \in \R^{\nControlPoints \times 3}$ for each object $\objectIndex$ are dependent on the scene length $\nFrames$ and expected smoothness. The camera consists of the same number of translation and rotation parameters. The background will have no learnable position parameters due to its static definition and has no opacity field $\alphaField$ due to its constant opacity of 1. While the number of parameters may be decreased based on the expected size of an object to increase efficiency, we use a single, unified size for all objects for simplicity.

\begin{table}[tbh]
    \centering
    \caption{Number of learnable parameters for a single \ac{nag} node.}
    \label{tab:parametrization_nag_node}
    \begin{tabular}{l l r}
        \toprule
        Component &  &Learnable Parameters \\
        \midrule
        Color Field & $\colorField$ & 3,720,488 \\
        Flow Field & $\flowField$ & 3,720,488 \\
        Opacity Field & $\alphaField$ & 3,720,488 \\
        View-Dependent Field & $\viewField$ & 6,716,320 \\
        Translation (single control point) & $\pTranslationObj$ & 3 \\
        Rotation (single control Point) & $\pRotationObj$ & 3 \\
        \bottomrule
        \end{tabular}
\end{table}

\section{Experimental Details}

\subsection{Baselines}

In the following we briefly describe our comparison baselines within the Automotive and Outdoor domain.

\paragraph{Automotive Baselines}
Within the autonomous driving scenes, we evaluate against \ac{ore} \cite{chen2025omnire}, a recent dynamic \ac{gs} method, which was explicitly designed for autonomous driving scenes including a dedicated SMPL-based human \cite{loper2023smpl} model for pedestrians and showing peak visual performance on the Waymo dataset \cite{sun2020scalability}. Further, given its object-specific architecture, it allows for positional edits, but lacks support for texture editing. We also compare against \ac{erf} \cite{yang2024emernerf}, a state-of-the-art dynamic scene reconstruction method, which leverages learned dynamics models and neural radiance fields to capture complex object motion and interactions, including non-rigid transformations. Although \ac{erf} is not object-specific, it serves as a recent and relevant NeRF baseline. \\
\noindent\textit{Note on \ac{ore} Scene Decomposition}: Although the authors provide visualizations of scene decompositions within their manuscript, we could not find the corresponding implementation in the provided codebase. Therefore, we adapted their evaluation scripts to specify and render individual object IDs for decomposition comparisons, while leaving the core implementation untouched.

\paragraph{Outdoor Baselines}
Our evaluation for outdoor videos, conducted on the DAVIS dataset \cite{Perazzi2016}, compares our approach against both recent texture editing and state-of-the-art video matting methods. 
\ac{lna} \cite{kasten2021layered} serves as our texture editing baseline. \ac{lna} operates by learning a 2D coordinate mapping, at test time, that projects pixels from all video frames onto a single texture atlas. This atlas can then be edited directly, with the changes re-projected back to all frames for scene manipulation. 
\ac{orf} is included as the most recent video matting baseline. These models are designed for robust layer separation, aiming to cover objects and associated effects (such as shadows) by learning a 2D foreground layer per-segmented object in image space, situated on top of a 3D background modeled by a radiance field. Crucially, as the 2D foreground layers are generated on a per-frame basis by a U-Net \cite{ronneberger2015u}, \ac{orf} contains no editable layer representation - all information is encoded within the learned U-Net weights. \\

For all our evaluations, we use the codebase provided by the respective authors unless explicitly stated. We used the recommended settings by the authors and only changed parameters to ensure a fair comparison (e.g., training on a higher resolution). For specific details on parameter changes, we refer to \secref{sec:dataset_details}.

\subsection{Datasets}
\label{sec:dataset_details}

\begin{figure}[t!]
    \centering
    \ifdefined\imageWidth
    \else
        \newlength{\imageWidth}
    \fi
    \setlength{\imageWidth}{0.25\textwidth}
    \ifdefined\imagerowheight
    \else
        \newlength{\imagerowheight}
    \fi
    \settoheight{\imagerowheight}{\includegraphics[width=\imageWidth]{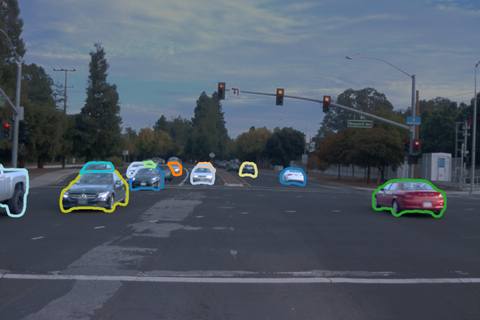}} 
    \newcommand{\gridImage}[1]{%
        \includegraphics[width=\imageWidth]{#1}%
    }
    \begin{tblr}{colspec={c c c c},colsep=0mm, rowsep=-1.2mm, row{1-7}={ht=\imagerowheight}, cells={halign=c, valign=m}}
         \gridImage{images/gt_images_by2/W125_000}& %
         \gridImage{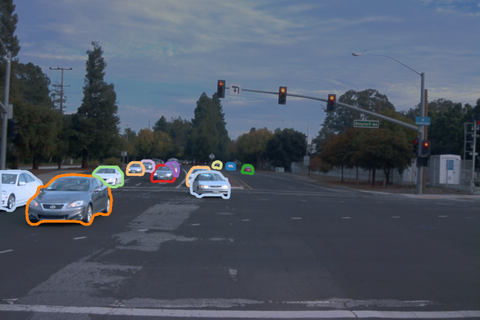}& %
         \gridImage{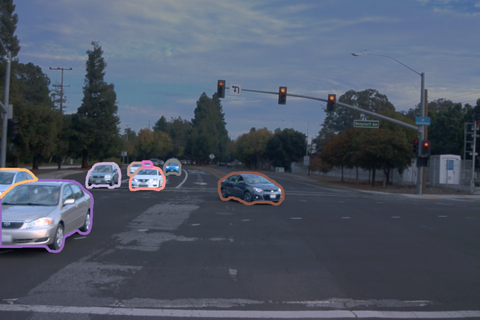}&
         \gridImage{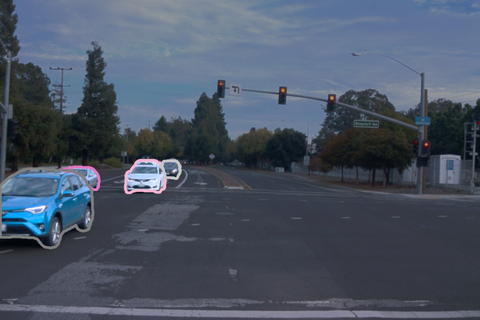}\\
         \gridImage{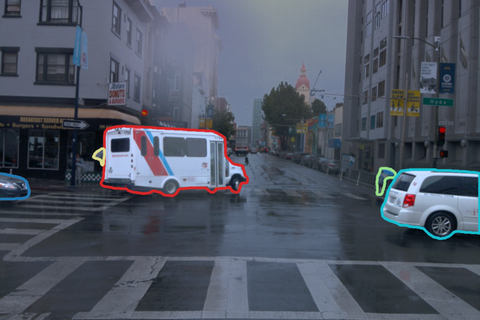}& %
         \gridImage{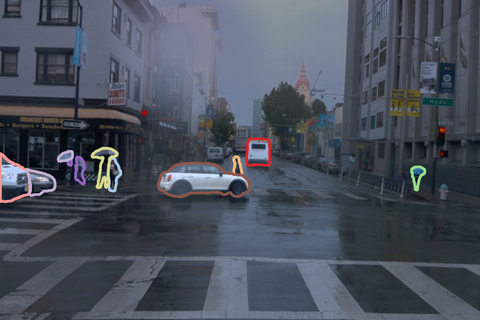}& %
         \gridImage{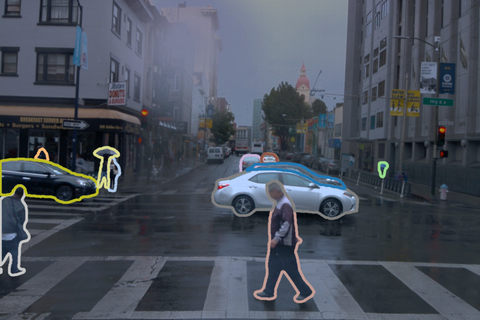}&
         \gridImage{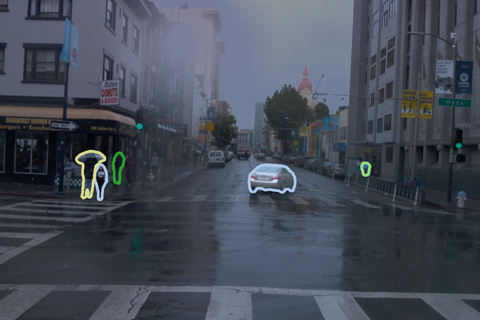}\\
         \gridImage{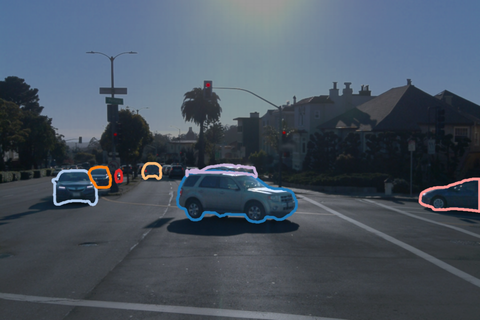}& %
         \gridImage{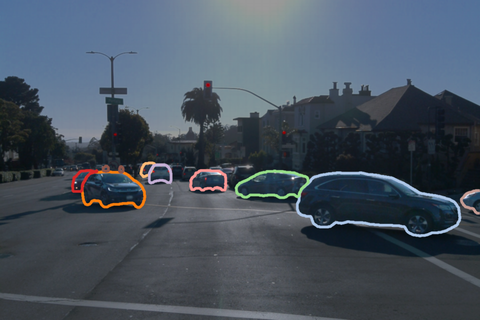}& %
         \gridImage{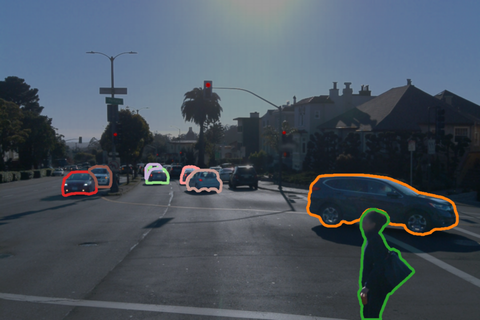}&
         \gridImage{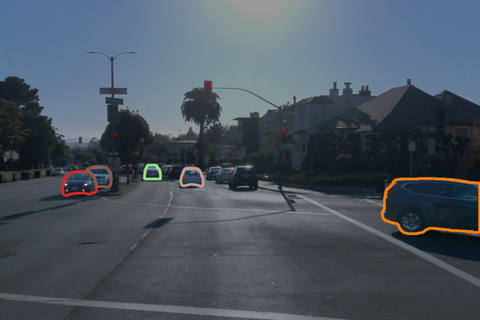}\\
         \gridImage{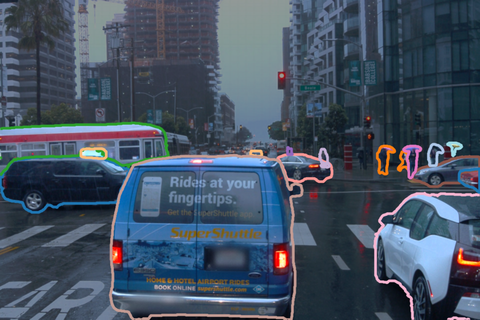}& %
         \gridImage{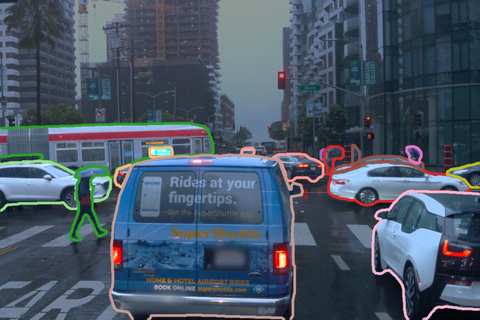}& %
         \gridImage{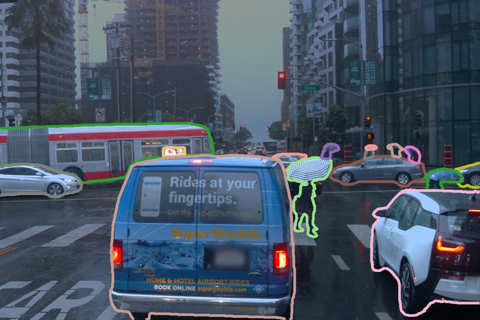}&
         \gridImage{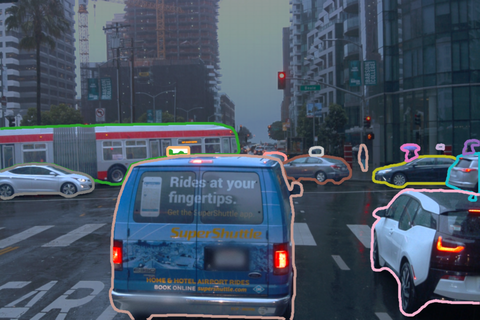}\\
         \gridImage{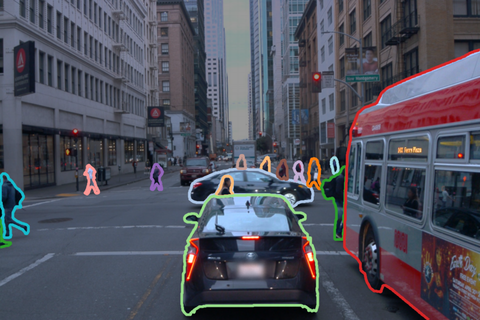}& %
         \gridImage{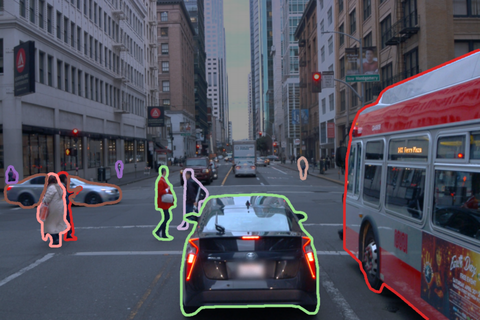}& %
         \gridImage{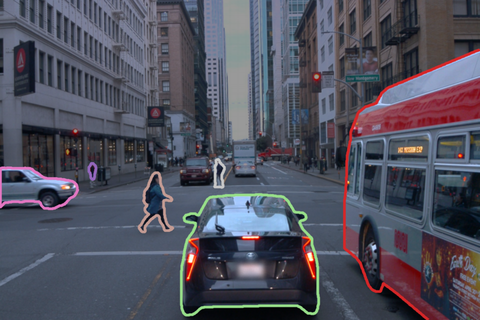}&
         \gridImage{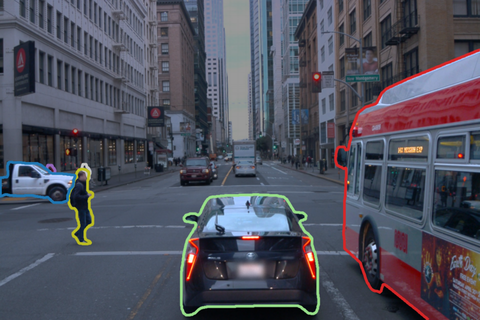}\\
         \gridImage{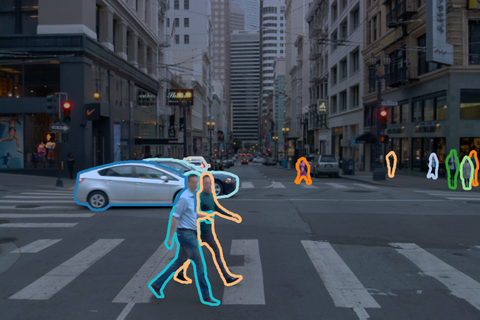}& %
         \gridImage{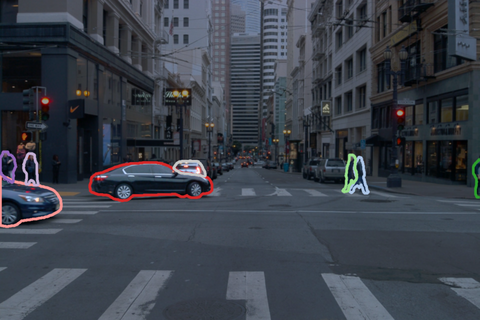}& %
         \gridImage{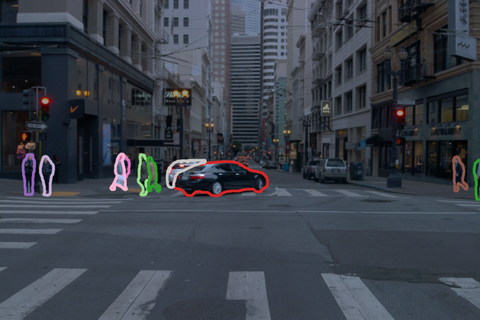}&
         \gridImage{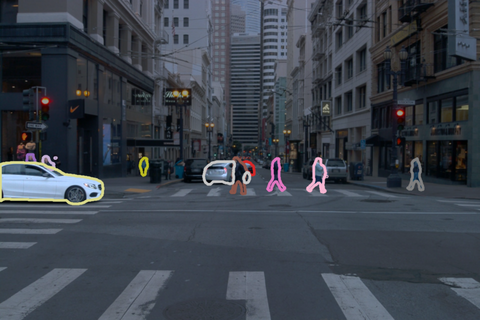}\\
         \gridImage{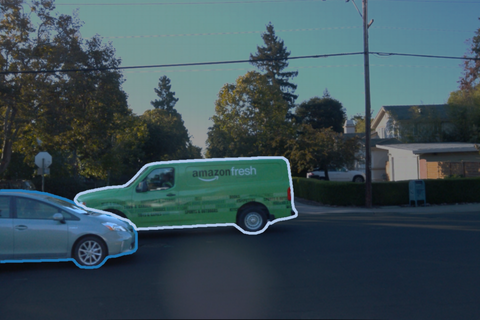}& %
         \gridImage{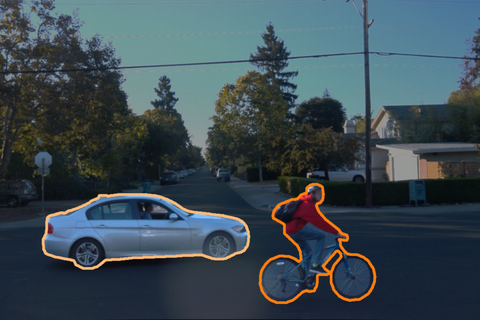}& %
         \gridImage{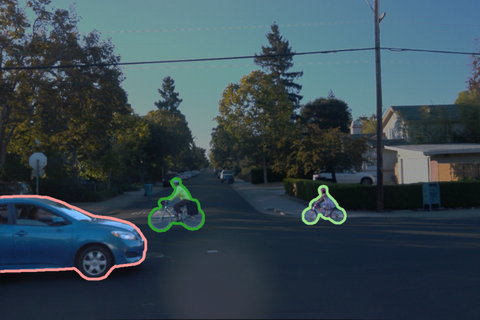}&
         \gridImage{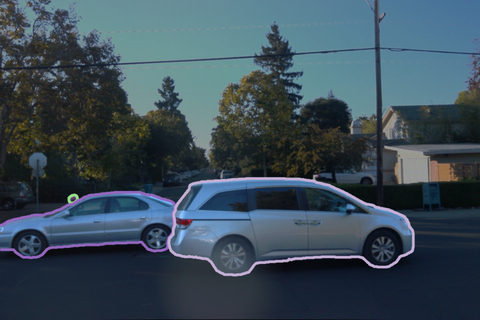}\\

    \end{tblr}
    \caption{Ground truth references and mask examples out of the studied autonomous driving Waymo sequences \cite{sun2020scalability}. Displayed are sequences in order: s-125, s-141, s-203, s-324, s-344, s-952, s-975. The sequences containing various objects and motion patterns. For our \ac{nag} representation, each of the masked instances will be attributed to its own atlas node.}
    \label{fig:dataset_waymo}
\end{figure}

\paragraph{Driving Scenes}
This section provides a detailed description of the specific subset of the Waymo Open Dataset \cite{sun2020scalability} used for evaluating our proposed method. As outlined in the main manuscript, we specifically selected scenes characterized by small ego-vehicle movement but a high density of dynamic objects, frequent occlusions, and significant variations in object motion emphasizing editability. Our evaluation was conducted on a total of 7 distinct scene segments, from which we extracted 25 subsequences, ranging from 21 to 89 frames sampled at 10Hz. During the subsequence creation, we excluded frames containing corrupted bounding box annotations, as well as longer sequences where no significant object intersection occurred. The sequence identifiers and ranges are stated in \tabref{tab:waymo_sequences}. For the remaining images within these subsequences, we segmented all objects\footnote{We provide these masks along our code base.} for which bounding box information was available and that exhibited significant motion or caused substantial occlusions. Representative ground truth images from each of these sequences, showcasing the generated instance segmentation masks, are visualized in \figref{fig:dataset_waymo}. For all methods we used all images of the datasets (as per experiment's subset division) in the native resolution (1920 $\times$ 1280) to train the models, yielding a representation of maximal visual expressiveness. Since \ac{ore} explicitly removes lens distortion during its preprocessing pipeline, we also apply this undistortion step for our method. We note that, due to the ERF model's implementation, no undistortion process is carried out. Consequently, all methods are evaluated against their respective ground truth targets (distorted or undistorted) to ensure a fair scene reconstruction comparison.

\begin{table}[tbh]
    \centering
    \caption{Waymo \cite{sun2020scalability} dataset sequences within our automotive evaluation. We stating the range, as respective inclusive start and end indices, forming our 25 subsequences.}
    \label{tab:waymo_sequences}
    \begin{tabular}{l l r@{ }r@{  }r@{ }r@{  }r@{ }r@{  }r@{ }r}
        \toprule
        Sequence & Segment Specifier & \multicolumn{8}{l}{Range} \\
        \midrule
        s-125 & segment-12511696717465549299 & 0 -& 40,& 40 -& 93,&   93 -& 124,& 124 -& 149 \\
        s-141 & segment-14133920963894906769 & 2 -& 53,& 53 -& 101,& 102 -& 173,& 174 -& 197 \\
        s-203 & segment-2036908808378190283 & 3 -& 58,& 60 -& 107\phantom{,} & & & & \\
        s-324 & segment-3247914894323111613 & 0 -& 42,& 42 -& 96,& 96 -& 161,& 161 -& 197 \\
        s-344 & segment-3441838785578020259 & 0 -& 51,& 52 -& 95,& 95 -& 135,& 135 -& 197 \\
        s-952 & segment-9521653920958139982 & 0 -& 63,& 64 -& 140,& 141 -& 198\phantom{,} & &\\
        s-975 & segment-9758342966297863572 & 0 -& 68,& 69 -& 99,& 99 -& 162,& 175 -& 195 \\
    \bottomrule
    \end{tabular}
\end{table}

\paragraph{Outdoor Scenes}

For evaluating the generalization of our method to diverse outdoor scenarios, we utilized a subset of the high-resolution DAVIS dataset \cite{Perazzi2016}, a common benchmark in video object segmentation and matting. Specifically, we selected the same 15 sequences also used by the baseline methods, \ac{orf} \cite{lin2023omnimatterf} and \ac{lna} \cite{kasten2021layered}, ensuring a direct basis for comparison across varied objects, backgrounds, and camera motion. We employed the dataset provided instance masks, and combined them into a single foreground mask per frame due to \ac{lna}'s implementation. Consistent with \ac{orf}, we used RodynRF \cite{liu2023robust} for initial camera pose estimation. Recognizing that the original evaluation resolutions for \ac{orf} (428 x 270) and \ac{lna} (768 × 432) were significantly lower than our capabilities, we leveraged more computational resources to evaluate our method and \ac{lna} on the full DAVIS resolution (up to 1920 x 1080), with the exception of the \textit{lucia} sequence. Due to \ac{lna}'s high memory demands on this longer scene, we down-sampled \textit{lucia} to 960 x 540 for \ac{lna} only. For \ac{orf}, given its original compute limitations and our focus on high-resolution performance, we down-sampled the input images by a factor of two (e.g., to 960 x 540) across all sequences, followed by bilinear interpolation of its output to the original resolution for accurate comparison.

\subsection{Neural Atlas Graphs Evaluation}
\label{sec:evaluation_setup}

\paragraph{NAG Training}\label{sec:training_setup} Our \ac{nag} is trained for 80 epochs, whereby each epoch consists of $2.8 \times 10^8$ ray-casts into the scene. Each epoch is subdivided into 140 batches, and each batch consists of 100,000 spatial ray-casts which are simultaneously evaluated along 20 random timestamps \footnote{Based on the dynamic architecture of the \ac{nag}, leading to a different total parameter sizes, we decrease in populated scenes the number of ray-casts per batch and increase the batches per epoch to fit the model into the available VRAM.}. We use the Adam \cite{kingma2014adam} optimizer, with an initial learning rate of 0.001 in combination with a ''ReduceLROnPlateau'' scheduler, which will be activated from epoch 20 on.

\paragraph{Atlas Node Architecture}
The neural fields $\colorField, \alphaField, \flowField\, \text{and}\, \viewField$ within every atlas node are parameterized by 5-layer MLPs (64 neurons, ReLU), while the input coordinates are encoded with a 16-level multi-resolution hash encoding \cite{mueller2022instant} (4 features/level, scale 1.61, hashmap size 17, base resolution 4, linear interpolation). We state the actual sizes in a dedicated section \ref{sec:parametrization}. 
For the Waymo \cite{sun2020scalability} dataset, the spline-based motion model of each node utilizes a number of control points $\nControlPoints = \nFrames$ equal to the number of images $\nFrames$ in the sequence. This is necessary to capture the potentially rapid camera motion (10Hz sampling), caused by oscillation on ego vehicle stops, which a lower-resolution spline cannot accurately represent. For the DAVIS \cite{Perazzi2016} dataset, we set the number of control points to $\nControlPoints = \nFrames / 2$, allowing for a smoother representation of the nodes, yielding a slightly more robust approach to compensate for inaccuracies in initialization. The effect of the control points is briefly studied within our ablations \secref{sec:ablation_studies}.
The training runtime ranges from approximately 2 to 6 hours depending on scene complexity and length, using a machine with a NVIDIA L 40 GPU and 64 GB RAM. Our reproducible code and dataset preparation schemes are available at: \url{https://github.com/jp-schneider/nag}.

\section{Results}

This supplementary section provides a comprehensive extension of the results presented in the main manuscript, necessitated by space constraints. We begin by recalling our main quantitative results in \secref{sec:additional_quantitative_results}, now including inter-frame standard variations to rigorously assess significance and temporal consistency. Following this, we dedicate \secref{sec:assessing_editing_quality} to evaluate editing quality using explicit temporal consistency measures. Subsequently, we analyze our model's performance by detailing the partial limitations of the \ac{nag} method under large ego-motion conditions (\secref{sec:quant_eval_large_motion}). We then proceed to ablate our model, conducting extensive ablation studies in \secref{sec:ablation_studies}, where we analyze network sizes, the impact of key components, and sensitivity to input masks. Finally, we provide a rich set of additional visual results in \secref{sec:additional_visual_results}, including automotive scene reconstruction comparisons, editing figures demonstrating positional and time shifts, object insertions, and removals. This is followed by further reconstruction, decomposition, and editing results on DAVIS \cite{Perazzi2016} outdoor scenes. Concluding this additional results section, we benchmark against three additional Gaussian Splatting baselines for autonomous driving in \secref{sec:additional_gaussian_splatting_baselines} and provide a transparent overview of measured training times for representative scenes across all methods of our main manuscript in \secref{sec:training_time}.

\subsection{Additional Quantitative Results}
\label{sec:additional_quantitative_results}

We extend the quantitative analysis from our main manuscript, benchmarking our approach against the recent OmniRe \cite{chen2025omnire} —a 3D \ac{gs} framework — and EmerNeRF \cite{yang2024emernerf} a recent dynamic NeRF model. We now explicitly provide inter-frame standard deviation measures to quantify temporal consistency.
\tabref{tab:std_results_waymo_main} lists the PSNR, SSIM \cite{ssim_2004}, and LPIPS \cite{zhang2018perceptual} scores for each scene, along with their inter-frame standard deviations over all different sub-segments of individual sequences, measuring temporal consistency.

We further isolate and assess the dynamic elements by partitioning them into a rigid “Vehicle” category and a non-rigid “Human” category. This division allows us to specifically evaluate how each class benefits from our underlying rigid-motion model. \tabref{tab:std_object_results} reports per-class PSNR and SSIM \cite{ssim_2004} results, including accompanying inter-object standard deviations over the sub-segments. The results demonstrate substantial improvements over the strongest baseline, confirming that our gains stem not merely from improved background rendering but from the high fidelity of our model in capturing even non-rigid motion.

Beyond the improvements in overall and object-based quality, the competitively low values for the inter-frame standard deviations highlight the high temporal consistency of our method. Our PSNR score of $\pm$ 0.91 compares favorably against OmniRe ($\pm$ 1.39). While this is slightly worse than the $\pm$ 0.78 achieved by EmerNeRF, EmerNeRF achieves this stability at a significantly lower quality (34.93 dB PSNR versus our 41.85 dB PSNR). For SSIM and LPIPS we improve against both baselines. Also, on object-based consistency, we achieve comparable temporal consistency in PSNR, while improving it on SSIM. This demonstrates our method's capability to learn a high-quality and temporally stable scene representation.

To verify generalization, we test our method on diverse outdoor sequences from the DAVIS dataset \cite{Perazzi2016}—a high-resolution (up to 1920 × 1080) benchmark commonly used by matting methods \cite{kasten2021layered, Lu_2021_CVPR, lin2023omnimatterf}. Following the selection of 15 sequences in \cite{kasten2021layered, lin2023omnimatterf} featuring varied objects, complex backgrounds, and dynamic camera moves, we summarize our results in \tabref{tab:std_davis_results}, including inter-frame standard deviations for PSNR, SSIM \cite{ssim_2004}, and LPIPS \cite{zhang2018perceptual}. 
In terms of visual quality, we significantly outperform the baselines in PSNR, SSIM, and LPIPS. While our standard deviations measuring temporal consistency are competitive in SSIM and LPIPS, our PSNR stability is weaker ($\pm$ 1.3 vs. $\pm$ 0.66). Crucially, this is achieved alongside a quality improvement of over 7 dB PSNR. We attribute this weakness to challenging scenes (e.g. tennis) which include highly non-rigid and rapid motion. This type of motion can lead to artifacts due to flow collapse and the difficulty of learning large and rapidly changing flow vectors within our spline-based smooth flow assumption.

\begin{table*}[t]
    \centering
    \caption{Quantitative Evaluation on Dynamic Driving Sequences of the Waymo~\cite{sun2020scalability} Open Driving Dataset. The temporal consistency is measured by the inter-frame standard deviation ($\pm$ STD), which is calculated over sub-segments and mean-aggregated per sequence. Best results are in bold. ORe refers to OmniRe \cite{chen2025omnire}, and ERF to EmerNeRF \cite{yang2024emernerf}. Our method compares very favorably in overall quality, showing higher consistency in SSIM and LPIPS over the baselines and highly competitive consistency in PSNR.
    }
    \label{tab:std_results_waymo_main}
    \setlength{\tabcolsep}{5pt}
    \begin{tabular}{l S[table-format=2.2] S[table-format=2.2] S[table-format=2.2] @{\quad} S[table-format=1.3] S[table-format=1.3] S[table-format=1.3] S[table-format=1.3] S[table-format=1.3] S[table-format=1.3]}
         \toprule
          \multirow{2}{*}{Seq.} & \multicolumn{3}{c}{PSNR $\uparrow$} & \multicolumn{3}{c}{SSIM $\uparrow$} & \multicolumn{3}{c}{LPIPS $\downarrow$} \\
         & {Ours} & {ORe} & {ERF} & {Ours} & {ORe} & {ERF} & {Ours} & {ORe} & {ERF} \\
         \midrule
      \multirow{2}{*}{s-975} & \textbf{40.21} & 37.35 & 34.83 & \textbf{0.976} & 0.968 & 0.937 & \textbf{0.058} & 0.080 & 0.143 \\
      &  \pm1.11 & \pm1.73 & \pm1.65 & \pm0.004 & \pm0.005 & \pm0.013 & \pm0.012 & \pm0.005 & \pm0.017\\
      
      \multirow{2}{*}{s-203} & \textbf{43.15} & 36.93 & 36.07 & \textbf{0.978} & 0.966 & 0.936 & \textbf{0.070} & 0.094 & 0.205\\
      & \pm0.39 & \pm1.14 & \pm0.43 & \pm0.001 & \pm0.002 & \pm0.003 & \pm0.004 & \pm0.003 & \pm0.005\\
      
      \multirow{2}{*}{s-125} & \textbf{43.32} & 38.74 & 35.20 & \textbf{0.980} & 0.970 & 0.933 & \textbf{0.057} & 0.079 & 0.182\\
      & \pm0.49 & \pm0.87 & \pm0.48 & \pm0.003 & \pm0.002 & \pm0.005 & \pm0.007 & \pm0.003 & \pm0.006\\
      
      \multirow{2}{*}{s-141} & \textbf{42.55} & 36.14 & 34.83 & \textbf{0.978} & 0.964 & 0.924 & \textbf{0.057} & 0.087 & 0.178 \\
      & \pm1.60 & \pm1.29 & \pm0.53 & \pm0.003 & \pm0.003 & \pm0.006 & \pm0.005 & \pm0.005 & \pm0.011\\
      
      \multirow{2}{*}{s-952} & \textbf{41.89} & 39.67 & 35.32 & 0.976 & \textbf{0.977} & 0.938 & 0.058 & \textbf{0.050} & 0.120 \\
      & \pm0.59 & \pm0.84 & \pm0.84 & \pm0.003 & \pm0.003 & \pm0.008 & \pm0.006 & \pm0.002 & \pm0.012\\
      
      \multirow{2}{*}{s-324} &  \textbf{40.85} & 32.58 & 33.63 & \textbf{0.977} & 0.953 & 0.926 & \textbf{0.038} & 0.071 & 0.124 \\
      & \pm1.31 & \pm2.21 & \pm0.57 & \pm0.002 & \pm0.010 & \pm0.005 & \pm0.004 & \pm0.009 & \pm0.007\\
      
      \multirow{2}{*}{s-344} & \textbf{41.84} & 36.67 & 35.24 & \textbf{0.983} & 0.973 & 0.946 & \textbf{0.031} & 0.043 & 0.084 \\
      & \pm0.52 & \pm1.40 & \pm0.77 & \pm0.001 & \pm0.003 & \pm0.006 & \pm0.002 & \pm0.003 & \pm0.005 \\
      
      \midrule
      \multirow{2}{*}{Mean} & \textbf{41.85} & 36.78 & 34.93 & \textbf{0.978} & 0.967 & 0.934 & \textbf{0.051} & 0.070 & 0.142 \\
      & \pm0.91 & \pm1.39 & \pm0.78 & \pm0.002 & \pm0.004 & \pm0.007 & \pm0.006 & \pm0.004 & \pm0.010\\
      \bottomrule
    \end{tabular}
\end{table*}

\begin{table*}[tb]
    \centering
    \caption{Quantitative Evaluation of Human and Vehicle Rendering on Waymo~\cite{sun2020scalability} Driving Sequences. The stated standard deviations ($\pm$ STD), are calculated following \tabref{tab:std_results_waymo_main}, and mean-aggregated per object and sub-sequence. 
    }
    \label{tab:std_object_results}
    \setlength{\tabcolsep}{5pt}
    \resizebox{\textwidth}{!}{%
    \begin{tabular}{l S[table-format=2.2] S[table-format=2.2] S[table-format=2.2] S[table-format=1.3] S[table-format=1.3] S[table-format=1.3] S[table-format=2.2] S[table-format=2.2] S[table-format=2.2] S[table-format=1.3] S[table-format=1.3] S[table-format=1.3]}
          \toprule
        \multirow{2}{*}{Seq.} & \multicolumn{3}{c}{Vehicle PSNR $\uparrow$} & \multicolumn{3}{c}{Vehicle SSIM $\uparrow$} & \multicolumn{3}{c}{Human PSNR $\uparrow$} & \multicolumn{3}{c}{Human SSIM $\uparrow$} \\
        & {Ours} & {ORe} & {ERF} & {Ours} & {ORe} & {ERF} & {Ours} & {ORe} & {ERF} & {Ours} & {ORe} & {ERF} \\
        \midrule
          \multirow{2}{*}{s-975} & \textbf{46.79} & 33.09 & 30.21 & \textbf{0.991} & 0.939 & 0.820 & \textbf{45.37} & 32.99 & 28.53 & \textbf{0.989} & 0.927 & 0.777 \\
          & \pm1.21 & \pm3.37 & \pm1.73 & \pm0.001 & \pm0.038 & \pm0.035 & \pm1.58 & \pm2.60 & \pm1.13 & \pm0.002 & \pm0.021 & \pm0.030 \\
          
          \multirow{2}{*}{s-203} & \textbf{41.90} & 30.45 & 27.10 & \textbf{0.986} & 0.910 & 0.774 & \textbf{45.40} & 34.85 & 33.54 & \textbf{0.986} & 0.950 & 0.901 \\
          & \pm1.89 & \pm3.14 & \pm1.89 & \pm0.005 & \pm0.046 & \pm0.053 & \pm1.65 & \pm2.81 & \pm1.29 & \pm0.004 & \pm0.017 & \pm0.016 \\
          
          \multirow{2}{*}{s-125} & \textbf{41.00} & 28.72 & 24.55 & \textbf{0.989} & 0.878 & 0.709 & N/A & N/A & N/A & N/A & N/A & N/A \\
          & \pm1.90 & \pm2.42 & \pm1.24 & \pm0.004 & \pm0.054 & \pm0.049 & N/A & N/A & N/A & N/A & N/A & N/A \\
          
          \multirow{2}{*}{s-141} & \textbf{43.21} & 33.22 & 27.36 & \textbf{0.981} & 0.929 & 0.744 & \textbf{44.22} & 33.31 & 28.86 & \textbf{0.986} & 0.907 & 0.769 \\
          & \pm1.44 & \pm2.05 & \pm1.21 & \pm0.007 & \pm0.028 & \pm0.036 & \pm1.61 & \pm2.55 & \pm1.67 & \pm0.005 & \pm0.044 & \pm0.051 \\
          
          \multirow{2}{*}{s-952} & \textbf{40.94} & 31.15 & 27.70 & \textbf{0.986} & 0.928 & 0.810 & \textbf{40.45} & 32.32 & 28.10 & \textbf{0.968} & 0.894 & 0.740 \\
          & \pm1.46 & \pm2.59 & \pm2.23 & \pm0.004 & \pm0.036 & \pm0.061 & \pm2.82 & \pm2.35 & \pm2.22 & \pm0.021 & \pm0.039 & \pm0.065 \\
          
          \multirow{2}{*}{s-324} & \textbf{41.71} & 31.03 & 27.87 & \textbf{0.986} & 0.921 & 0.798 & \textbf{44.12} & 32.09 & 26.40 & \textbf{0.988} & 0.894 & 0.689 \\
          & \pm1.56 & \pm3.41 & \pm1.96 & \pm0.004 & \pm0.048 & \pm0.053 & \pm1.95 & \pm2.63 & \pm1.78 & \pm0.005 & \pm0.041 & \pm0.065 \\
          
          \multirow{2}{*}{s-344} & \textbf{43.97} & 33.02 & 30.65 & \textbf{0.985} & 0.931 & 0.835 & \textbf{40.99} & 30.20 & 25.94 & \textbf{0.975} & 0.882 & 0.721 \\
          & \pm1.69 & \pm2.29 & \pm1.43 & \pm0.007 & \pm0.019 & \pm0.018 & \pm2.97 & \pm2.57 & \pm1.40 & \pm0.016 & \pm0.045 & \pm0.051 \\
          
          \midrule
          \multirow{2}{*}{Mean} & \textbf{42.88} & 31.69 & 28.09 & \textbf{0.986} & 0.922 & 0.787 & \textbf{42.94} & 32.24 & 27.78 & \textbf{0.981} & 0.901 & 0.744 \\
          & \pm1.56 & \pm2.73 & \pm1.67 & \pm0.005 & \pm0.037 & \pm0.043 & \pm2.21 & \pm2.55 & \pm1.67 & \pm0.010 & \pm0.039 & \pm0.052 \\
          \bottomrule
    \end{tabular}%
    }
\end{table*}

\begin{table*}[tb]
\centering
    \caption{Quantitative evaluation results on the Davis Dataset~\cite{Perazzi2016} of diverse outdoor scenes with inter-frame standard deviation ($\pm$ STD) over all images from each scene. The best results are in bold for all metrics. Our method consistently yields higher quality than its competitors OmnimatteRF (ORF)~\cite{lin2023omnimatterf} and Layered Neural Atlases (LNA)~\cite{kasten2021layered}, while maintaining competitive temporal stability. (Best results are in bold.)}
    \label{tab:std_davis_results}
    \setlength{\tabcolsep}{5pt}
    \begin{tabular}{l S[table-format=2.2] S[table-format=2.2] S[table-format=2.2] S[table-format=1.3] S[table-format=1.3] S[table-format=1.3] S[table-format=1.3] S[table-format=1.3] S[table-format=1.3]}
     \toprule
     \multirow{2}{*}{Sequence} & \multicolumn{3}{c}{PSNR $\uparrow$} & \multicolumn{3}{c}{SSIM $\uparrow$} & \multicolumn{3}{c}{LPIPS $\downarrow$} \\
     & {Ours} & {ORF} & {LNA} & {Ours} & {ORF} & {LNA} & {Ours} & {ORF} & {LNA} \\
    \midrule
    \multirow{2}{*}{bear}           & \textbf{33.47} & 24.88 & 26.51 & \textbf{0.934} & 0.658 & 0.771 & \textbf{0.091} & 0.464 & 0.287 \\
    & \pm1.42 & \pm0.52 & \pm0.72 & \pm0.027 & \pm0.020 & \pm0.018 & \pm0.030 & \pm0.011 & \pm0.015\\
    
    \multirow{2}{*}{blackswan}      & \textbf{36.36} & 26.67 & 29.26 & \textbf{0.938} & 0.739 & 0.815 & \textbf{0.097} & 0.458 & 0.318 \\
    & \pm0.53 & \pm0.98 & \pm0.48 & \pm0.005 & \pm0.031 & \pm0.014 & \pm0.010 & \pm0.032 & \pm0.020\\
    
    \multirow{2}{*}{boat}           & \textbf{35.83} & 28.63 & 30.15 & \textbf{0.932} & 0.761 & 0.816 & \textbf{0.099} & 0.376 & 0.274 \\
    & \pm0.42 & \pm0.31 & \pm0.48 & \pm0.005 & \pm0.012 & \pm0.011 & \pm0.009 & \pm0.013 & \pm0.011\\
    
    \multirow{2}{*}{car-shadow}     & \textbf{36.67} & 29.26 & 28.47 & \textbf{0.947} & 0.861 & 0.850 & \textbf{0.084} & 0.313 & 0.269 \\
    & \pm1.57 & \pm0.38 & \pm0.48 & \pm0.010 & \pm0.014 & \pm0.015 & \pm0.011 & \pm0.014 & \pm0.015\\
    
    \multirow{2}{*}{elephant}       & \textbf{33.91} & 26.94 & 28.34 & \textbf{0.922} & 0.731 & 0.772 & \textbf{0.088} & 0.423 & 0.325 \\
    & \pm2.04 & \pm0.45 & \pm0.55 & \pm0.033 & \pm0.012 & \pm0.013 & \pm0.013 & \pm0.006 & \pm0.010\\
    
    \multirow{2}{*}{flamingo}       & \textbf{34.96} & 25.74 & 27.10 & \textbf{0.928} & 0.753 & 0.783 & \textbf{0.106} & 0.483 & 0.349 \\
    & \pm0.65 & \pm0.73 & \pm1.01 & \pm0.007 & \pm0.018 & \pm0.020 & \pm0.015 & \pm0.008 & \pm0.013\\
    
    \multirow{2}{*}{hike}           & \textbf{29.74} & 25.15 & 24.77 & \textbf{0.886} & 0.698 & 0.682 & \textbf{0.108} & 0.388 & 0.343 \\
    & \pm1.88 & \pm0.25 & \pm0.38 & \pm0.048 & \pm0.019 & \pm0.022 & \pm0.026 & \pm0.019 & \pm0.017\\
    
    \multirow{2}{*}{horsejump-high} & \textbf{34.78} & 28.35 & 27.28 & \textbf{0.932} & 0.846 & 0.830 & \textbf{0.074} & 0.249 & 0.226 \\
    & \pm1.78 & \pm0.41 & \pm0.64 & \pm0.016 & \pm0.019 & \pm0.020 & \pm0.013 & \pm0.023 & \pm0.024\\
    
    \multirow{2}{*}{kite-surf}      & \textbf{37.96} & 28.04 & 27.88 & \textbf{0.949} & 0.780 & 0.780 & \textbf{0.068} & 0.420 & 0.400 \\
    & \pm0.50 & \pm0.70 & \pm0.32 & \pm0.005 & \pm0.026 & \pm0.018 & \pm0.006 & \pm0.031 & \pm0.016\\
    
    \multirow{2}{*}{kite-walk}      & \textbf{37.96} & 29.44 & 29.58 & \textbf{0.941} & 0.804 & 0.818 & \textbf{0.070} & 0.367 & 0.334 \\
    & \pm0.66 & \pm0.38 & \pm0.56 & \pm0.010 & \pm0.009 & \pm0.011 & \pm0.012 & \pm0.007 & \pm0.014\\
    
    \multirow{2}{*}{libby}          & \textbf{38.89} & 29.62 & 29.35 & \textbf{0.949} & 0.819 & 0.828 & \textbf{0.095} & 0.399 & 0.342 \\
    & \pm0.56 & \pm0.94 & \pm0.76 & \pm0.004 & \pm0.028 & \pm0.028 & \pm0.010 & \pm0.031 & \pm0.025\\
    
    \multirow{2}{*}{lucia}          & \textbf{30.90} & 26.03 & 26.63 & \textbf{0.869} & 0.690 & 0.742 & \textbf{0.178} & 0.407 & 0.329 \\
    & \pm1.44 & \pm0.54 & \pm0.65 & \pm0.047 & \pm0.027 & \pm0.036 & \pm0.068 & \pm0.033 & \pm0.040\\
    
    \multirow{2}{*}{motorbike}      & \textbf{37.42} & 27.33 & 29.33 & \textbf{0.950} & 0.779 & 0.843 & \textbf{0.082} & 0.376 & 0.241 \\
    & \pm0.89 & \pm0.93 & \pm1.10 & \pm0.008 & \pm0.023 & \pm0.014 & \pm0.011 & \pm0.020 & \pm0.017\\
    
    \multirow{2}{*}{swing}          & \textbf{35.70} & 26.14 & 27.88 & \textbf{0.926} & 0.722 & 0.808 & \textbf{0.119} & 0.404 & 0.289 \\
    & \pm0.89 & \pm0.54 & \pm0.50 & \pm0.010 & \pm0.021 & \pm0.019 & \pm0.017 & \pm0.027 & \pm0.029\\
    
    \multirow{2}{*}{tennis}         & \textbf{35.65} & 27.43 & 28.81 & \textbf{0.928} & 0.806 & 0.862 & \textbf{0.120} & 0.328 & 0.209 \\
    & \pm4.22 & \pm1.89 & \pm1.32 & \pm0.036 & \pm0.062 & \pm0.044 & \pm0.049 & \pm0.081 & \pm0.054\\
    
    \midrule
    \multirow{2}{*}{Mean}           & \textbf{35.35} & 27.31 & 28.09 & \textbf{0.929} & 0.763 & 0.800 & \textbf{0.098} & 0.390 & 0.302 \\
    & \pm1.30 & \pm0.66 & \pm0.66 & \pm0.018 & \pm0.023 & \pm0.020 & \pm0.020 & \pm0.024 & \pm0.021\\
    \bottomrule
    \end{tabular}
\end{table*}

To further quantify the quality and temporal consistency of our method against the core baselines, we computed the \ac{fvd} \cite{unterthiner2018towards} metric, a dedicated video quality evaluation method originally targeted for generative videos, proposed to better align with human judgment than PSNR or SSIM. Additionally, we evaluated the Temporal (T)-LPIPS, which is applied inter-frame wise to indicate differences in machine perception and can therefore be interpreted as an additional temporal consistency metric. Since inter-frame differences are also induced by changes in the scene or camera, T-LPIPS scores must be interpreted relative to the T-LPIPS of the Ground Truth video.
The \ac{fvd} and T-LPIPS metrics are presented in \tabref{tab:additional_qual_metrics2}. For this evaluation, we use the ablation sequences (s-141, s-975) from the Waymo Open Dataset and the sequences (blackswan, bear, and boat) from DAVIS. 
Our method achieves the best FVD scores across the evaluated sequences, aligning with the perceptual improvements observed in our earlier PSNR, SSIM, and LPIPS evaluations. In terms of temporal consistency (T-LPIPS), our method's score is the closest to that of the Ground Truth (GT) video, indicating a similar degree of fidelity in inter-frame changes. 

\begin{table*}[tb]
\centering
    \caption{Quantitative Comparison of Temporal Consistency (\ac{fvd} and T-LPIPS) against core baselines on a selected subset of the Waymo and DAVIS datasets. Results are reported as mean metric value and standard deviation ($\pm$ STD) across (sub-)sequences. Our method demonstrates favorable \ac{fvd} scores and achieves T-LPIPS closest to the Ground Truth (GT), indicating both high quality and inter-frame stability.
(Best results are shown in bold.)}
    \label{tab:additional_qual_metrics2}
    \setlength{\tabcolsep}{5pt}
    \newlength{\meantostdsep}
    \setlength{\meantostdsep}{2pt}
    \begin{tabular}{l 
    S[table-format=3.0] @{\hspace{\meantostdsep}}r 
    S[table-format=1.3] @{\hspace{\meantostdsep}}r 
    S[table-format=3.0] @{\hspace{\meantostdsep}}r 
    S[table-format=1.3] @{\hspace{\meantostdsep}}r}
     \toprule
     \multirow{2}{*}{Method} & \multicolumn{4}{c}{Waymo} & \multicolumn{4}{c}{DAVIS} \\
     & \multicolumn{2}{c}{FVD $\downarrow$} & \multicolumn{2}{c}{T-LPIPS} & \multicolumn{2}{c}{FVD $\downarrow$} & \multicolumn{2}{c}{T-LPIPS} \\
    \midrule
    Ours & {\textbf{174}} & $\pm$ 238 & 0.063 & $\pm$ 0.031 & {\textbf{108}} & $\pm$\hphantom{0}35 & 0.143 & $\pm$ 0.022 \\
     {ORe}  & 423 & $\pm$ 446 & 0.055 & $\pm$ 0.029 & \multicolumn{2}{c}{N/A} & \multicolumn{2}{c}{N/A} \\
     {ERF}  & 439 & $\pm$ 350 & 0.053 & $\pm$ 0.026 & \multicolumn{2}{c}{N/A} & \multicolumn{2}{c}{N/A} \\
     {ORF}  & \multicolumn{2}{c}{N/A} & \multicolumn{2}{c}{N/A} & 986 & $\pm$153 & 0.104 & $\pm$ 0.023 \\
     {LNA}  & \multicolumn{2}{c}{N/A} & \multicolumn{2}{c}{N/A} & 595 & $\pm$\hphantom{0}38 & 0.116 & $\pm$ 0.022 \\
     \midrule
     GT & \multicolumn{2}{c}{N/A} & 0.079 & $\pm$ 0.029 & \multicolumn{2}{c}{N/A} & 0.155 & $\pm$ 0.020 \\
     \bottomrule
    \end{tabular}
\end{table*}

\subsection{Assessment of Editing Quality}
\label{sec:assessing_editing_quality}
Quantifying video quality for edited scenes, particularly within decomposition-based methods like ours, remains an open problem. Notably, related works such as Layered Neural Atlases (LNA) \cite{kasten2021layered} and OmnimatteRF (ORF) \cite{lin2023omnimatterf} also omit a direct quantitative evaluation of edited video quality. 
We posit that this omission stems from two major hurdles:

Firstly, a direct, quantitative assessment of video quality is complicated by the lack of available ground truth data. This area is currently the subject of intensive research in Blind Video Quality Assessment ($\text{BVQA}$), particularly for generative video models \cite{zheng2024video}. While earlier methods are tied to specific image and video corruptions \cite{hassen2013image, wang2002no, wang2000blind}, \textit{driving} the need for combining multiple scores, newer feature- and learning-based approaches \cite{chen2022no, wang2023exploring, yuan2024ptm} are often model- or domain-dependent and have exhibited questioned robustness \cite{antsiferova2024comparing}.

Secondly, assessing the quality of the decomposition or texture edits itself is highly non-trivial. For scene decomposition, removing an object requires the system to hallucinate the previously occluded geometry and appearance. Since the content of the occluded region (including potential changes) cannot be known without a geometric reference, any arbitrary, visually plausible content may be valid, which greatly complicates traditional quality scoring. Furthermore, for texture edits, separating the quality contribution of the \textit{continuous texture application} (method influence) from the inherent quality of the \textit{user-defined texture} (user influence) presents an additional difficulty, as the latter can heavily skew the rated video quality.

While assessing the perceptual quality of the edits is challenging, an additional evaluation of the temporal consistency of the edited video may provide insights. This approach allows us to quantify the fluctuations introduced during the editing process. To this end, we computed the Temporal (T)-LPIPS score (frame-by-frame perceptual difference) and the \ac{fid} \cite{heusel2017gans} score, applied frame-by-frame wise to the edited video sequence. These scores provide a quantitative measure of temporal stability for the decomposed objects or edited regions, which, combined with a qualitative human assessment, forms the basis of our edit evaluation, as shown in \tabref{tab:temp_consistency}. For T-LPIPS we report the mean and standard deviation over all image-pairs and object regions in case of \figref{fig:scene_editing}, while over the full image when editing, incorporating background edits. For FID we report the standard deviation only for the per-object decomposition evaluation, given that it is computed as a distributional measure. For all stated edits, our consistency results are comparable to our reference method or the respective ground truth (GT).

\begin{table*}[tb]
\centering
    \caption{Temporal Consistency of Editing Figures}
    \label{tab:temp_consistency}
    \setlength{\tabcolsep}{5pt}
    \setlength{\meantostdsep}{2pt}
    \resizebox{\textwidth}{!}{%
    \begin{tabular}{l 
    S[table-format=1.3] @{\hspace{\meantostdsep}}r 
    S[table-format=1.3] @{\hspace{\meantostdsep}}r 
    S[table-format=1.3] @{\hspace{\meantostdsep}}r 
    S[table-format=1.3] @{\hspace{\meantostdsep}}r
    S[table-format=1.3] @{\hspace{\meantostdsep}}r
    S[table-format=1.3] @{\hspace{\meantostdsep}}r
    }
     \toprule
     \multirow{2}{*}{Edit} & \multicolumn{6}{c}{T-LPIPS} & \multicolumn{6}{c}{FID \cite{heusel2017gans}} \\
     & \multicolumn{2}{c}{Ours} & \multicolumn{2}{c}{ORe} & \multicolumn{2}{c}{GT} &  \multicolumn{2}{c}{Ours} & \multicolumn{2}{c}{ORe} & \multicolumn{2}{c}{GT} \\
    \midrule
    \figref{fig:scene_editing}, Seg. 125 & 0,052 & $\pm$ 0,027 &0,054 & $\pm$ 0,032 & 0,081 & $\pm$ 0,033 &	2.244 & $\pm$ 1.857 & 1.972 & $\pm$ 1.326 & 2.228 & $\pm$ 1.578 \\
    \figref{fig:scene_editing}, Seg. 141&0.056 & $\pm$ 0.048 & 0.063 & $\pm$ 0.062&0.103 & $\pm$ 0.072&0.999 & $\pm$ 1.032 & 1.111 & $\pm$ 1.308 & 1.517 & $\pm$ 1.222\\
    \figref{fig:scene_texture_editing}  & 0.037 & $\pm$ 0.006  & \multicolumn{2}{c}{N/A}  & 0.043 & $\pm$ 0.006  & 0.175 & - & \multicolumn{2}{c}{N/A}  & 0.173 & - \\
Supl. \figref{fig:davis_editing_swan}  & 0.249 & $\pm$ 0.028  & \multicolumn{2}{c}{N/A}  & 0.261 & $\pm$ 0.026  & 0.255 & - & \multicolumn{2}{c}{N/A}  & 0.185 & - \\
Supl. \figref{fig:davis_editing_boat}  & 0.058 & $\pm$ 0.012  & \multicolumn{2}{c}{N/A}  & 0.080 & $\pm$ 0.012  & 0.132 & - & \multicolumn{2}{c}{N/A}  & 0.180 & - \\
     \bottomrule
    \end{tabular}}
\end{table*}

\subsection{Evaluation on Large Ego Motion}
\label{sec:quant_eval_large_motion}
\begin{figure}[t!]
    \centering
    \ifdefined\imageWidth
    \else
        \newlength{\imageWidth}
    \fi
    \setlength{\imageWidth}{0.25\textwidth}
    \ifdefined\imagerowheight
    \else
        \newlength{\imagerowheight}
    \fi
    \settoheight{\imagerowheight}{\includegraphics[width=\imageWidth]{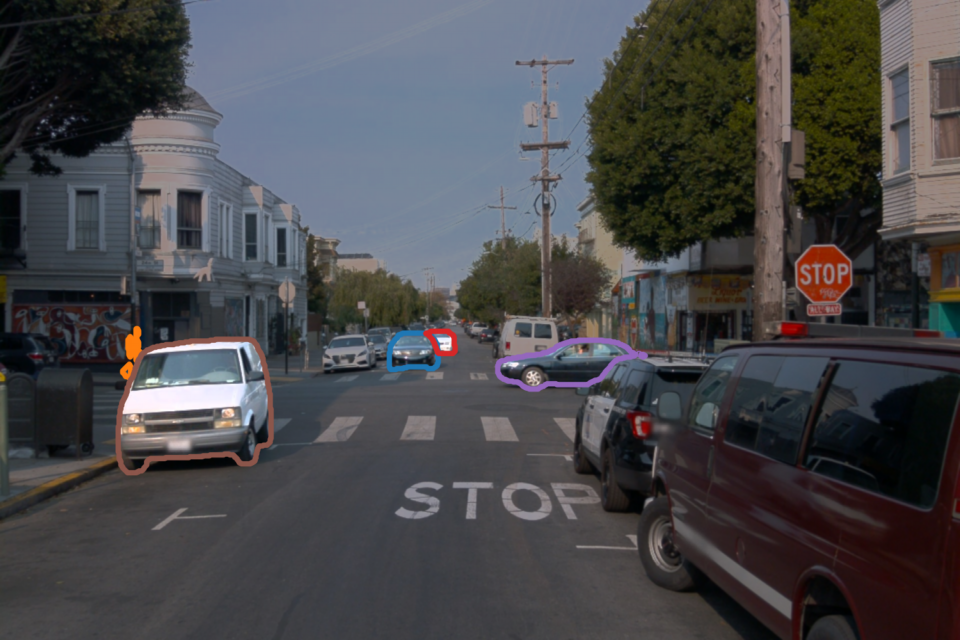}} 
    \newcommand{\gridImage}[1]{%
        \includegraphics[width=\imageWidth]{#1}%
    }
    \begin{tblr}{colspec={c c c c},colsep=0mm, rowsep=-1.2mm, row{1-7}={ht=\imagerowheight}, cells={halign=c, valign=m}}
         \gridImage{images/gt_images_by2/W191_000}& %
         \gridImage{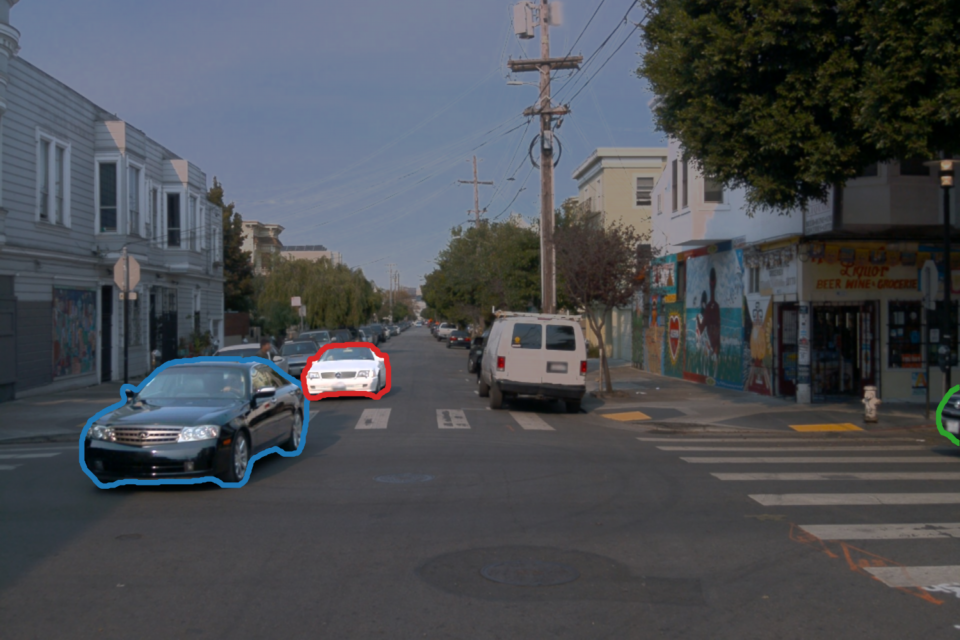}& %
         \gridImage{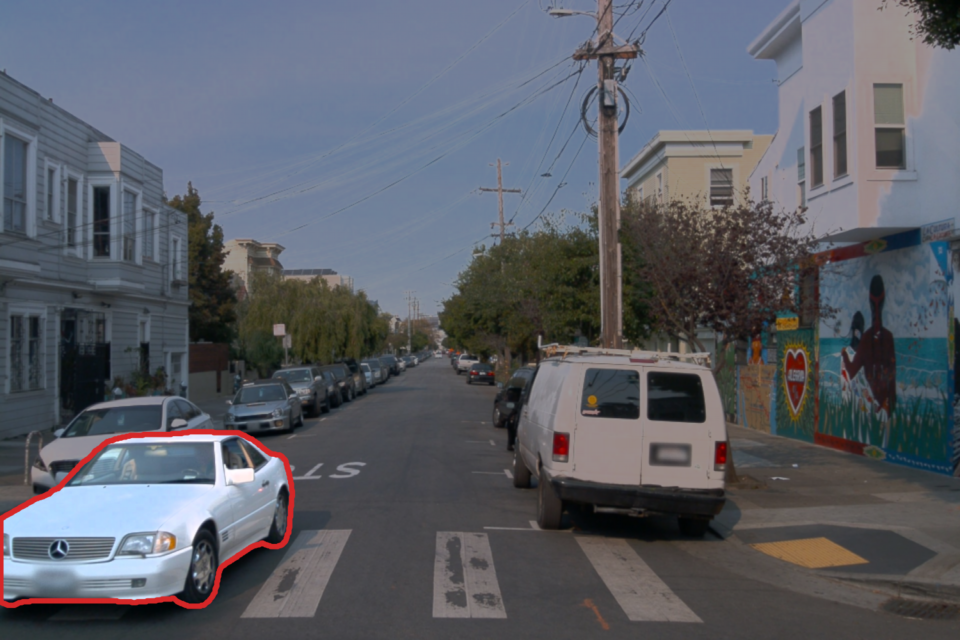}&
         \gridImage{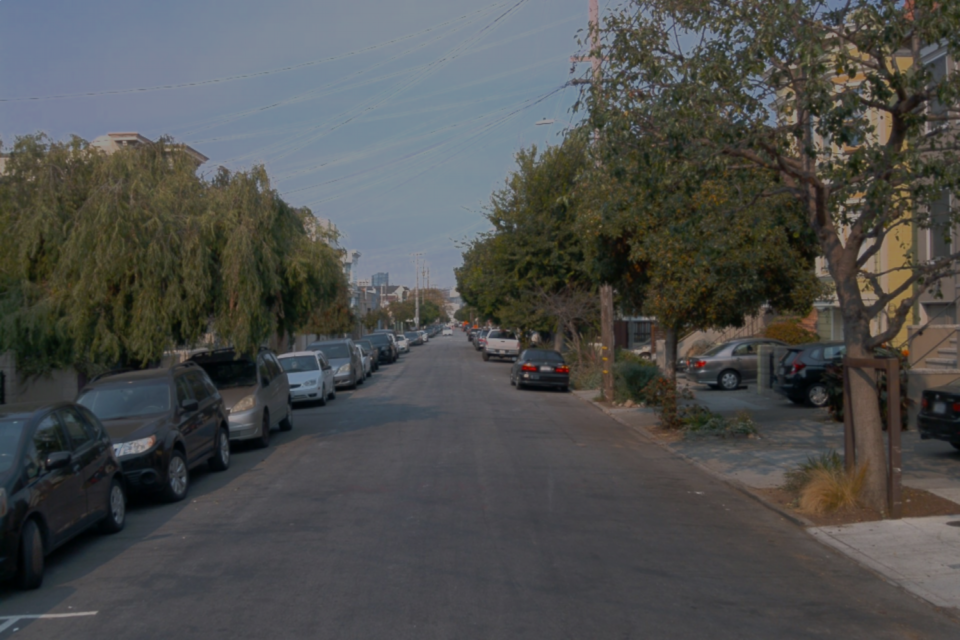}\\
         \gridImage{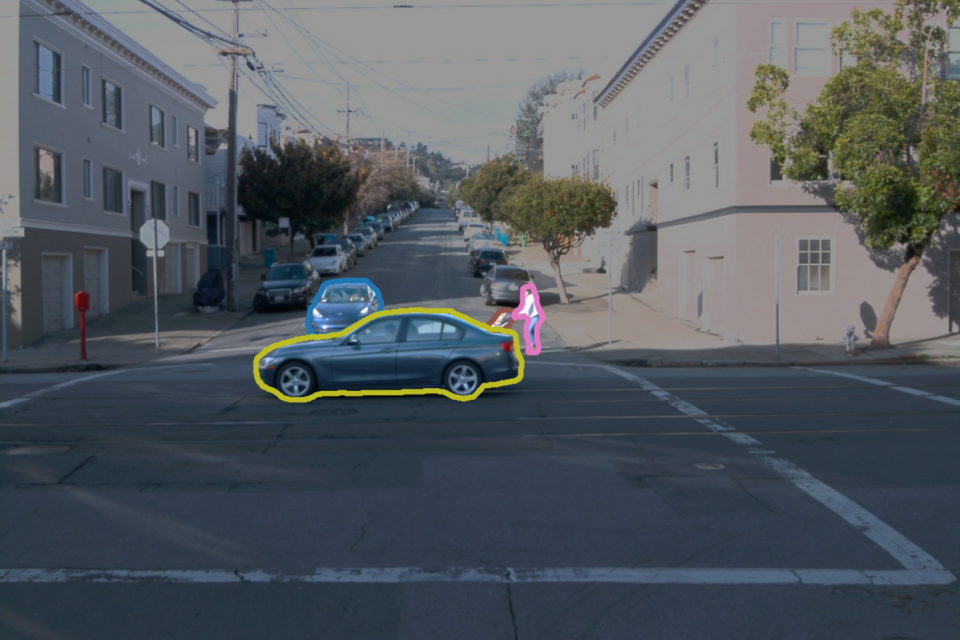}& %
         \gridImage{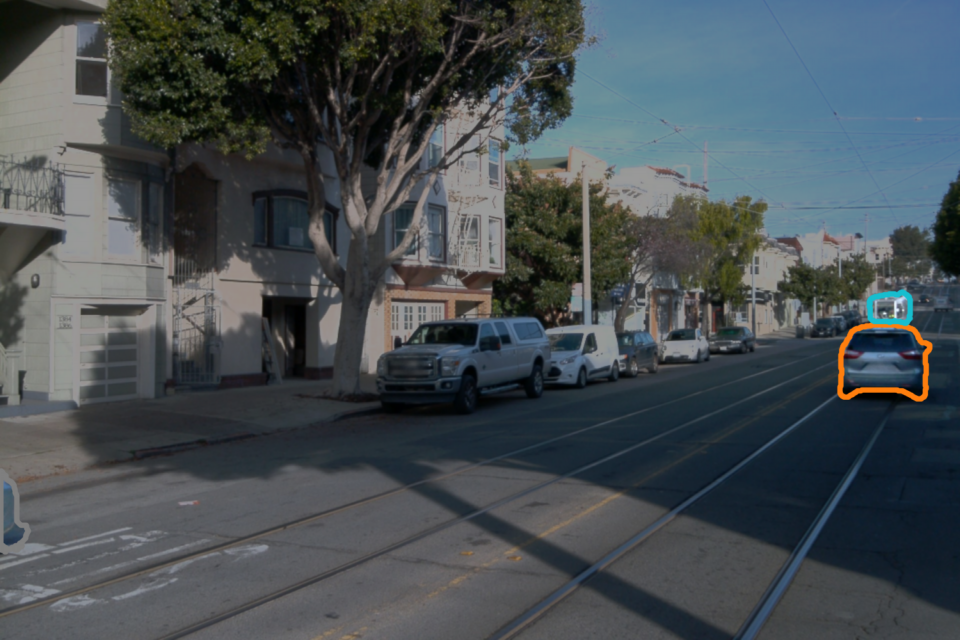}& %
         \gridImage{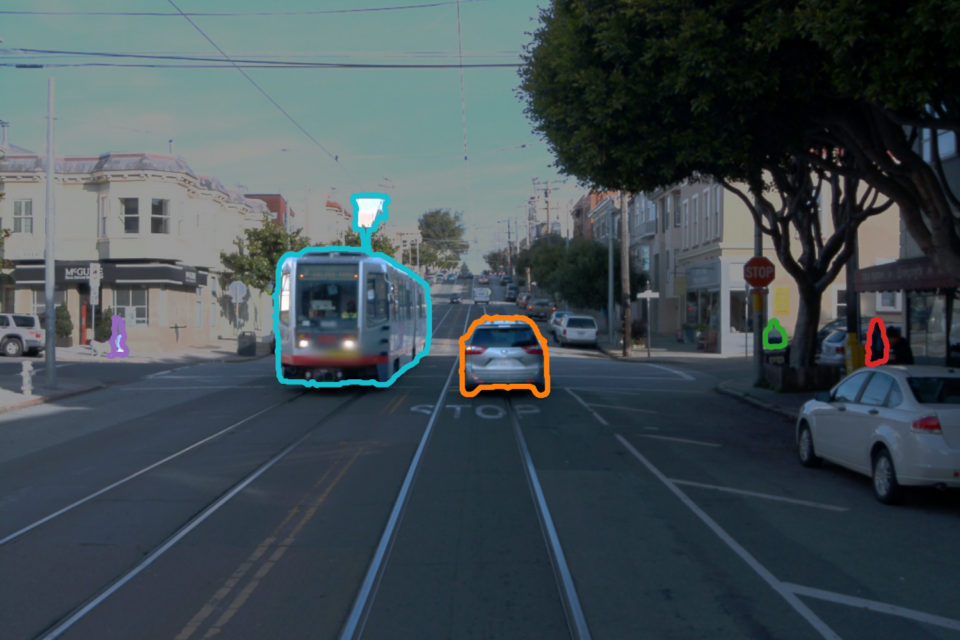}&
         \gridImage{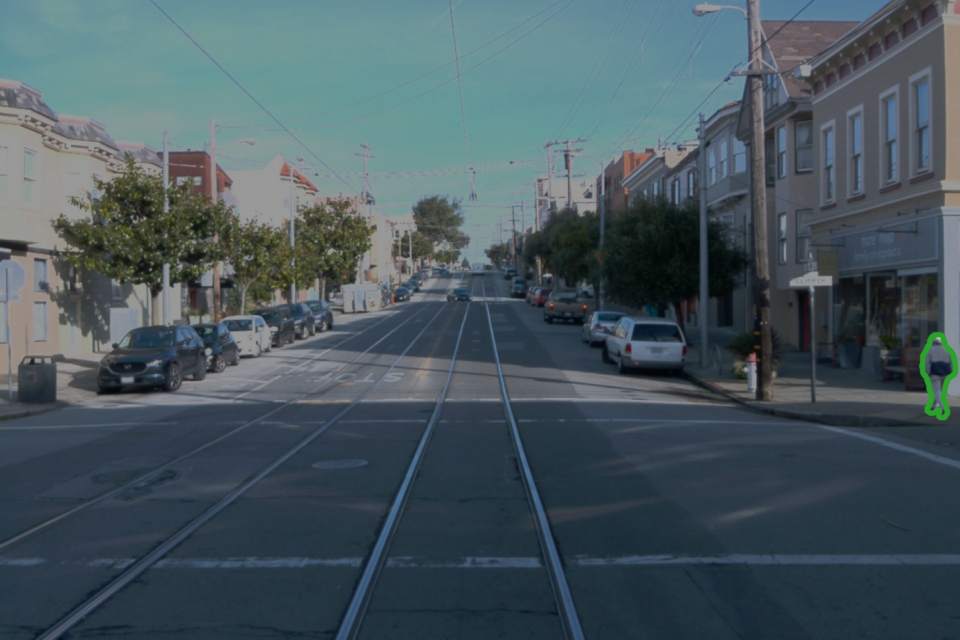}\\
    \end{tblr}
    \caption{Ground truth references and mask examples out of the two tested \textit{large-ego-motion} Waymo sequences \cite{sun2020scalability} (s-191, s-254).}
    \label{fig:dataset_waymo_lem}
\end{figure}

Our method was originally designed with a primary focus on texture editable scenes, often implying lower ego-motion to maintain stable views of objects. However, to rigorously test the robustness of our motion model, we conducted an evaluation on two additional Waymo Open Dataset sequences, s-191 and s-254 \footnote{Referring to segment-1918764220984209654 and segment-2547899409721197155, subdivided in 3 / 4 sub-segments.}, which feature significant camera ego-motion (visualized in \figref{fig:dataset_waymo_lem}).

Following our analysis on other sequences, we evaluated the performance by focusing on foreground objects managed by dedicated nodes (vehicles and humans) and by assessing overall image quality.

For foreground objects (Vehicles and Humans), which are managed by dedicated nodes with a robust rigid motion model, our approach achieves substantial improvements—up to 8 dB PSNR—compared to all baselines (see \tabref{tab:results_waymo_large_ego_vehicle}). This success confirms that our rigid motion model, being independent of the large global camera motion, is suited for handling moving objects in challenging, large ego-motion environments, provided the object's view-angle does not completely change.

The overall image scores (PSNR, SSIM, LPIPS) are presented in \tabref{tab:results_waymo_large_ego}. While these composite scores are comparable to competitors \ac{ore} and \ac{erf}, they are not substantially higher. This trade-off is attributed to the inherent limitations of our $\text{2.5D}$ representation, as discussed in our limitation section. In large ego-motion scenes, our planar background assumption requires the flow network to learn large, complex flow vectors to cover the rapidly changing content. Assuring the fidelity of such long flow vectors purely through photometric loss proves highly challenging. In failure cases, the flow tends to collapse regions onto the plane and unfold them as needed, which unfortunately introduces characteristic \textit{wavy-line artifacts} within the background. 
We showcase these artifacts along with the still highly competitive foreground objects in \figref{fig:comparison_vis_quality_lem}.

This evaluation demonstrates a clear and informative trade-off. We emphasize that our proposed Neural Atlas Graph Model is fundamentally 2.5D (planes + flow), endowed with a large inductive bias that excels at regularizing solutions for low-parallax or sparsely observed scene elements—a strength evident in the robust handling of moving foreground objects and in providing direct texture editability. However, the background artifacts observed within large ego-motion scenes reveal the current limitations of this 2.5D planar representation, which is not as well suited to model complex 3D structures with large amounts of self-occlusion within the background. We believe these findings strongly indicate that extending this architecture towards a hybrid object-centric graph model — combining both 2D and 3D primitives in a single render graph — is a compelling direction for future research.

\begin{table*}[thb]
    \centering
    \caption{Quantitative Evaluation on Large-Ego-Motion Dynamic Driving Sequences of the Waymo~\cite{sun2020scalability} Open Driving Dataset on Vehicle and Human class.
    }
    \label{tab:results_waymo_large_ego_vehicle}
    \setlength{\tabcolsep}{4pt}
    \begin{tabular}{l S[table-format=2.2] S[table-format=2.2] S[table-format=2.2] S[table-format=1.3] S[table-format=1.3] S[table-format=1.3] S[table-format=2.2] S[table-format=2.2] S[table-format=2.2] S[table-format=1.3] S[table-format=1.3] S[table-format=1.3]}
          \toprule
        \multirow{2}{*}{Seq.} & \multicolumn{3}{c}{Vehicle PSNR $\uparrow$} & \multicolumn{3}{c}{Vehicle SSIM $\uparrow$} & \multicolumn{3}{c}{Human PSNR $\uparrow$} & \multicolumn{3}{c}{Human SSIM $\uparrow$} \\
        & {Ours} & {ORe} & {ERF} & {Ours} & {ORe} & {ERF} & {Ours} & {ORe} & {ERF} & {Ours} & {ORe} & {ERF} \\
         \midrule
      s-191 & \textbf{37.60} & 31.20 & 25.78 & \textbf{0.959} & 0.928 & 0.714 & \textbf{34.40} & 27.34 & 22.93 & \textbf{0.908} & 0.808 & 0.589\\
      s-254 & \textbf{35.07} & 30.52 & 26.00 & \textbf{0.926} & 0.928 & 0.735 & \textbf{34.42} & 28.87 & 22.67 & \textbf{0.919} & 0.880 & 0.629\\
      \bottomrule
    \end{tabular}
\end{table*}%

\begin{table*}[tbh]
    \centering
    \caption{Quantitative Evaluation on Large-Ego-Motion Dynamic Driving Sequences of the Waymo~\cite{sun2020scalability} Open Driving Dataset.
    }
    \label{tab:results_waymo_large_ego}
    \setlength{\tabcolsep}{4pt}
    \begin{tabular}{l S[table-format=2.2] S[table-format=2.2] S[table-format=2.2] S[table-format=1.3] S[table-format=1.3] S[table-format=1.3] S[table-format=1.3] S[table-format=1.3] S[table-format=1.3]}
         \toprule
          \multirow{2}{*}{Seq.} & \multicolumn{3}{c}{PSNR $\uparrow$} & \multicolumn{3}{c}{SSIM $\uparrow$} & \multicolumn{3}{c}{LPIPS $\downarrow$} \\
         & {Ours} & {ORe} & {ERF} & {Ours} & {ORe} & {ERF} & {Ours} & {ORe} & {ERF} \\
         \midrule
s-191 & 32.02 & \textbf{33.17} & 29.97 & 0.892 & \textbf{0.952} & 0.864 & 0.209 & \textbf{0.088} & 0.244 \\
s-254 & 31.70 & \textbf{31.91} & 29.32 & 0.911 & \textbf{0.950} & 0.871 & 0.190 & \textbf{0.093} & 0.241 \\
      \bottomrule
    \end{tabular}
\end{table*}%

\begin{figure}[tb]
    \centering
    \ifdefined\imageWidth
    \else
        \newlength{\imageWidth}
    \fi
    \setlength{\imageWidth}{0.25\textwidth}
    \ifdefined\imagerowheight
    \else
        \newlength{\imagerowheight}
    \fi
    \settoheight{\imagerowheight}{\includegraphics[width=\imageWidth]{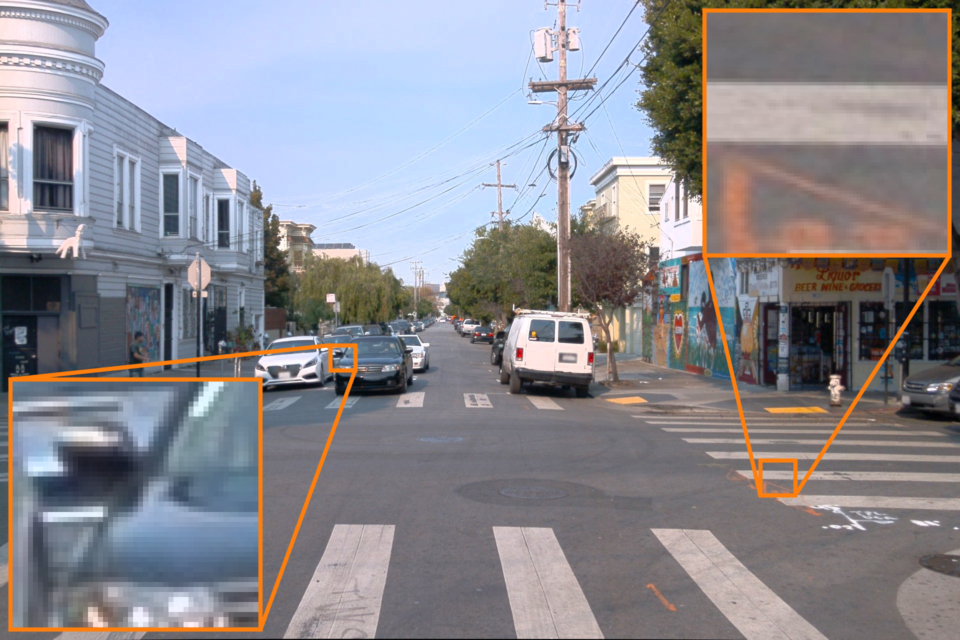}} 
    \ifdefined\gridImage
        \renewcommand{\gridImage}[1]{%
            \includegraphics[width=\imageWidth]{#1}%
        }
    \else
        \newcommand{\gridImage}[1]{%
            \includegraphics[width=\imageWidth]{#1}%
        }
    \fi
    \begin{tblr}{colspec={c c c c},colsep=0mm, rowsep=-1.2mm, row{1-3}={ht=\imagerowheight}, cells={halign=c, valign=m}}
         \gridImage{images/qual_results/W191_000/W191_000_028_GT_g__1_05_z_10__7}& %
         \gridImage{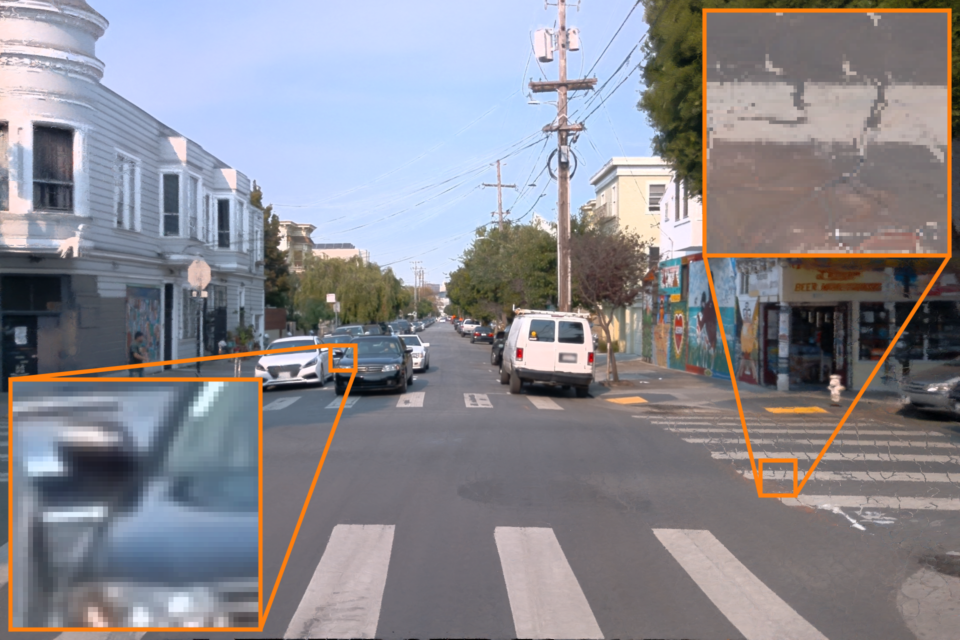}& %
         \gridImage{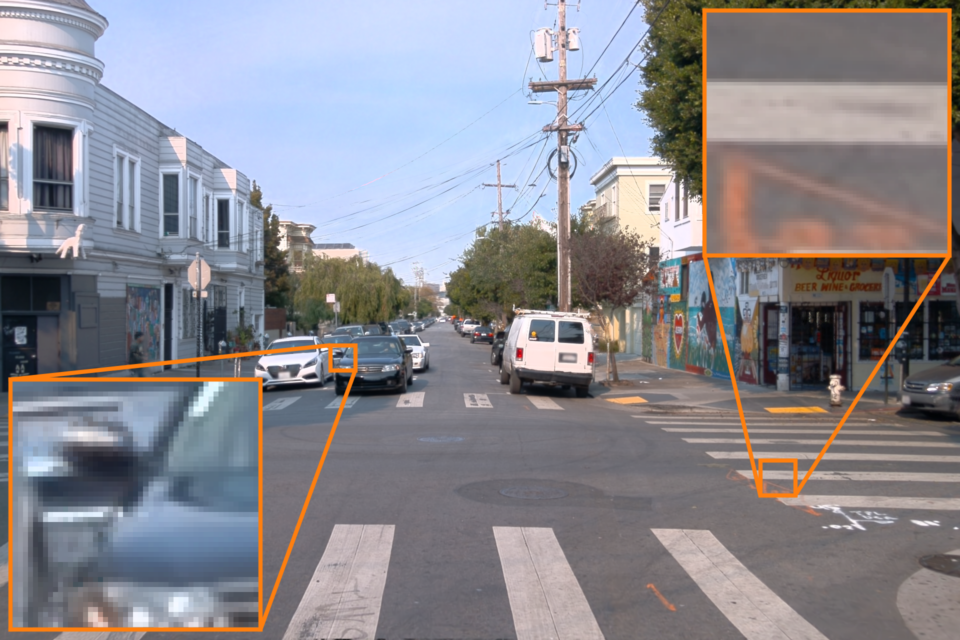}&  %
         \gridImage{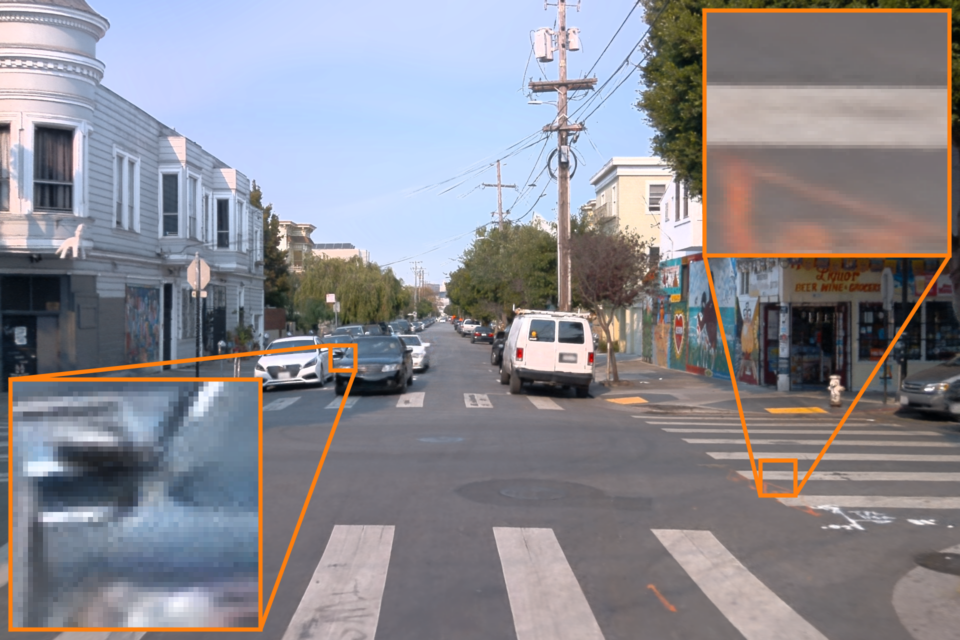}\\
    \end{tblr}
    \begin{tabular}{cccc}
        \centering \parbox{\dimexpr0.25\textwidth-2\tabcolsep\relax}{\centering Ground Truth} &
        \centering \parbox{\dimexpr0.25\textwidth-2\tabcolsep\relax}{\centering Ours} &
        \centering \parbox{\dimexpr0.25\textwidth-2\tabcolsep\relax}{\centering OmniRe \cite{chen2025omnire}} &
        \centering \parbox{\dimexpr0.25\textwidth-2\tabcolsep\relax}{\centering EmerNeRF \cite{yang2024emernerf}}
    \end{tabular}
    \caption{Visual quality comparison on the \textit{large-ego-motion} scene s-191. A clear trade-off is observed. Foreground objects managed by the rigid motion model exhibit increased sharpness and reduced edge artifacts compared to baselines. Conversely, the high flow compensation required by our 2.5D background may cause visual degradation in rapidly changing background regions, manifesting as flow artifacts or blurring.}
    \label{fig:comparison_vis_quality_lem}
\end{figure}

\subsection{Additional Ablation Experiments}
\label{sec:ablation_studies}

\begin{figure}[tb!]
    \centering
    \setlength{\imageWidth}{0.33\textwidth}
    \settoheight{\imagerowheight}{\includegraphics[width=\imageWidth]{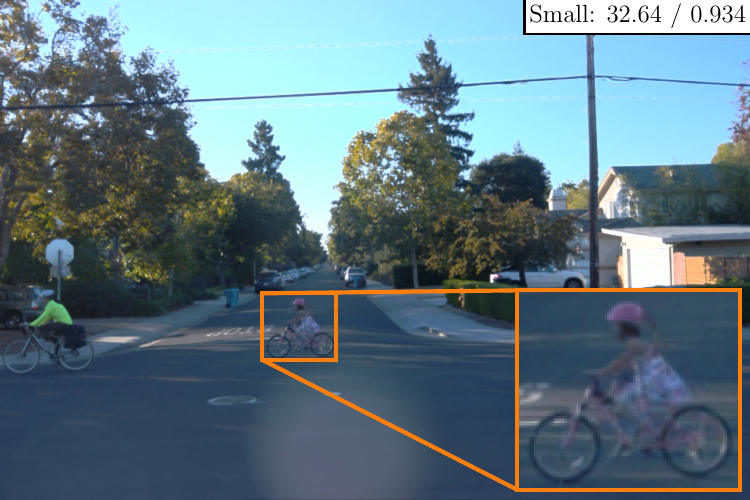}} 
    \newcommand{\gridImage}[1]{%
        \includegraphics[width=\imageWidth]{#1}%
    }
    \begin{tblr}{colspec={c c c},colsep=0mm, rowsep=-1.2mm, row{1-1}={ht=\imagerowheight}, cells={halign=c, valign=m}}
         \gridImage{images/ablation/sizes/W975_099_045_small}& %
         \gridImage{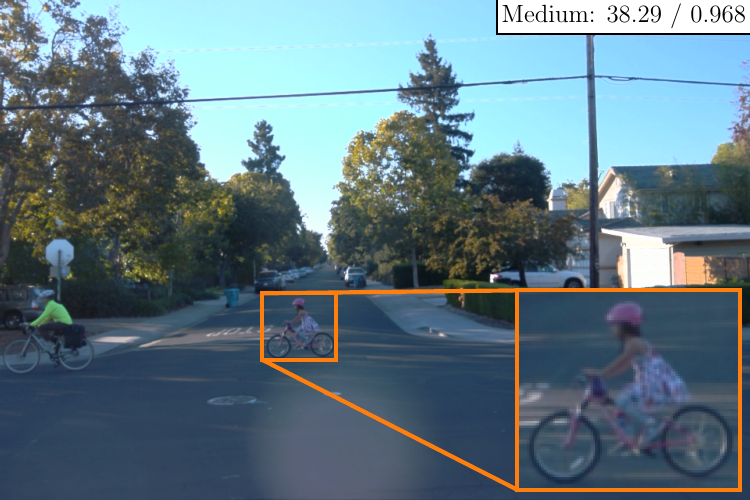}& %
         \gridImage{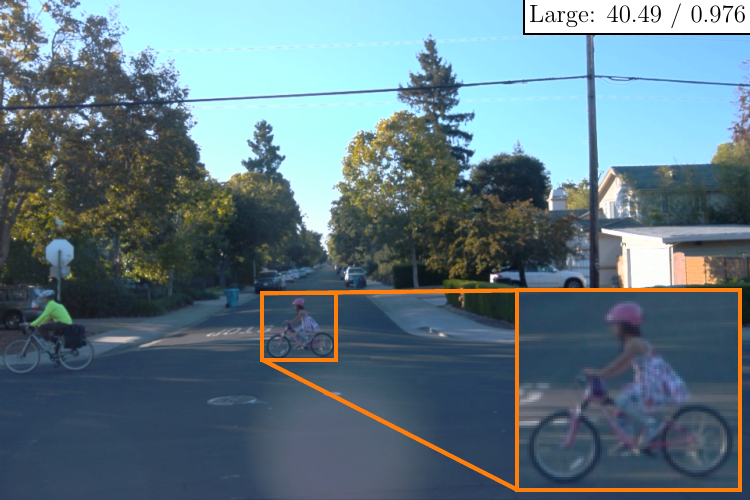}
    \end{tblr}
    \caption{Representative examples of NAG nodes with varying parametrization sizes (Small, Medium, Large), including their PSNR / SSIM scores. Noticeable image quality degradation and flow collapsing artifacts are evident in the small node due to its limited representation, whereas distinguishing visual differences between medium and large nodes is challenging, with only minor lighting variations on the ground.}
    \label{fig:model_sizes}
    
\end{figure}

To assess the contribution of different components of our proposed model, we conducted a comprehensive ablation study on a subset (s-141, s-975) of the Waymo Open Dataset \cite{sun2020scalability}. These sequences were further divided into 8 subsequences, which, given a systematic evaluation of all key model components and hyperparameters, yielded 96 additional experiments. The results of these experiments are detailed in \tabref{tab:ablations}. The top row of the table, labeled "Large", represents the performance of our full reference model as described in the main manuscript. 
\paragraph{Parametrization Sizes}
We evaluated the impact of varying the model size ("Medium" and "Small")\footnote{The sizes "Large", "Medium", "Small" are referring to different parameterizations of our MLP Network and Hash-Grid Configurations. Effectively, they are reducing the number of levels and sizes within the hash-grid encoding, as well as reducing the number of hidden layers within our MLPs. For details we refer to our code base.} and observed a general trend of performance degradation (lower PSNR and SSIM, higher LPIPS) with reduced capacity, highlighting the importance of model scale for achieving optimal reconstruction quality. A representative visual example of these different model sizes and their corresponding PSNR/SSIM scores can be found in \figref{fig:model_sizes}, further illustrating the qualitative differences. 

\paragraph{Initialization \& Coarse-to-fine}
Furthermore, we investigated the significance of several key modules within our architecture by systematically excluding or modifying them. The rows "Coarse Init-Projection" and "Excl. Coarse-to-fine" examine the role of our coarse initialization and the subsequent coarse-to-fine refinement strategy. For the first, we limit the size of our initial estimates for color $\colorBase \in \R^{20 \times 20}$ and opacity $\alphaBase \in \R^{20 \times 20}$ to a much lower spatial extend than the original used, which is based on the mask size. This shall mimic a mean initialization of the objects. The latter, deactivates our coarse-to-fine scheme.
While the performance drop observed may look rather small, the visual changes on decomposition and edits may be very significant, as excluding these components could lead to much more background information in the foreground or vice-versa.

\paragraph{Flow- \& View-Fields}
The rows "Excl. Flow" and "Excl. View-Dependence" quantify the impact of our optical flow estimation and view-dependent modeling components, by disabling them respectively. The substantial decrease in all evaluated metrics upon their removal underscores their critical role in handling motion and viewpoint changes within the driving scenes. Notably, the combined exclusion of both flow and view-dependence ("Excl. Flow + View-Dependence") resulted in the most significant performance decline, emphasizing the synergy between these modules.

\paragraph{Position Learning}
We also assessed the translation learning component ("Excl. Translation Learning"). Excluding this learning component slightly weakens the overall reconstruction quality. While the quantitative impact is minor on the studied Waymo sequence, the degradation would likely be more severe if less precise initializations (e.g., non-3D bounding box initializations) were provided. The model's ability to maintain high object-based scores, despite excluding translation, suggests that the planar flow-, or view-dependent- fields compromises for errors within the rigid motion model.

Additionally, we explored the effect of explicitly learning plane rotations, similar to our DAVIS \cite{Perazzi2016} experiments. The row "Incl. Plane Rotation Learning" shows a slight improvement across all metrics on this specific Waymo subset compared to the "Large" baseline. However, we were unable to consistently verify this improvement across  additional Waymo sequences, suggesting that its benefit might be scene-specific or less pronounced in more diverse scenarios.

\paragraph{Position Granularity}

The two rows ("Num. Position CP $\nControlPoints = \nFrames/2$" and "Num. Position CP $\nControlPoints = \nFrames3/4$") investigate the influence of the number of control points used for our motion model. Employing fewer control points results in a smoother motion trajectory for both the ego-camera and individual objects. Interestingly, the observed improvement in Vehicle PSNR and SSIM with fewer control points is relatively minor and just slightly holds for the Human category, which often exhibits more complex, non-rigid motion. This discrepancy suggests that while a smoother motion constraint might offer a slight benefit for predominantly rigid objects like vehicles, it could be insufficient and potentially detrimental for capturing the intricate deformations and trajectories of non-rigid objects such as pedestrians. Further, over-smoothing the camera motion does negatively impact overall scene alignment, outweighing any minor per-object benefits seen for vehicles, measured in the worse overall scores. \\
Yet, the observed benefits suggest that imposing different smoothness assumptions for rigid objects (like vehicles), non-rigid objects (like pedestrians), as well as the ego-camera, may further improve the overall reconstruction quality, but requires further investigation.

\begin{table}[tb]
    \setlength{\tabcolsep}{4pt}
    \caption{Ablation Experiments. We conducted ablation studies on a subset of our Waymo Datasets \cite{sun2020scalability}, evaluating various components of our model. Best results are \textbf{bold}, second best are \underline{underlined}. The top row (Large) marks the reference model stated in our manuscript. On different model sizes, the scores may degrade significantly. When excluding or changing certain keyparts, we observe degradation of the performance, showing their importance. When also learning the plane rotation (cf. Davis), this slightly benefits performance reported on this subset.}
    \label{tab:ablations}
    \resizebox{\textwidth}{!}{%
    \begin{tabular}{l S[table-format=2.2] S[table-format=1.3] S[table-format=1.3] S[table-format=2.2] S[table-format=1.3] S[table-format=2.2] S[table-format=1.3]}
                \toprule
        \multirow{2}{*}{Abl.} & {\multirow{2}{*}{PSNR $\uparrow$}} & {\multirow{2}{*}{SSIM $\uparrow$}} & {\multirow{2}{*}{LPIPS$\downarrow$}} & \multicolumn{2}{c}{Vehicle} & \multicolumn{2}{c}{Human} \\
        &  & & & PSNR $\uparrow$ & SSIM $\uparrow$ & PSNR $\uparrow$ & SSIM $\uparrow$ \\
        \midrule
        \textbf{Large} & \underline{41.42} & \textbf{0.977} &  \underline{0.057} & 44.94 & \underline{0.986} & 44.65 & \underline{0.987} \\
        Medium & 39.33 & 0.968 & 0.071 & 42.09 & 0.973 & 41.80 & 0.975 \\
        Small & 35.64 & 0.943 & 0.099 & 36.56 & 0.936 & 36.90 & 0.939 \\
        \midrule
        Coarse Init-Projection & 41.27 & \textbf{0.977} & 0.060 & 44.87 & 0.985 & 44.84 & \underline{0.987} \\
        Excl. Coarse-to-fine & 41.37 & \textbf{0.977} & 0.058 & 44.93 & 0.985 & 44.86 & \underline{0.987} \\
        Excl. Flow & 39.44 & 0.974 & 0.063 & 44.47 & 0.981 & 44.26 & 0.985 \\
        Excl. View-Dependence & 38.08 & 0.961 & 0.095 & 34.07 & 0.901 & 34.91 & 0.913 \\
        Excl. Flow \& View-Dependence & 32.29 & 0.936 & 0.110 & 24.71 & 0.775 & 27.92 & 0.808 \\
        Excl. Translation Learning & 39.37 & 0.971 & 0.062 & 45.15 & 0.984 & 44.88 & \underline{0.987} \\
        Incl. Plane Rotation Learning & \textbf{41.46} & \textbf{0.977} & \textbf{0.056} & \textbf{45.35} & \textbf{0.987} & \textbf{46.94} & \textbf{0.992} \\ 
        \midrule
        Num. Position CP $\nControlPoints = \nFrames/2$ & 40.44 & 0.972 & 0.061 & \underline{45.22} & \underline{0.986} & 44.77 & 0.986 \\
        Num. Position CP $\nControlPoints = \nFrames3/4$ & 40.97 & \underline{0.976} & 0.060 & 45.08 & 0.986 & 44.66 & 0.986 \\
        \midrule
        Excl. Mask-Loss & 41.31 & \textbf{0.977} & 0.058 & 44.95 &\underline{0.986} & 44.88 & \underline{0.987} \\
        Morph. Masks & 41.24 & 0.975 & 0.060 & 44.92 & 0.984 & 44.91 & 0.986 \\
        Morph. Masks, Excl. Mask-Loss & 41.29 & 0.975 & 0.060 & 45.00 & 0.985 & \underline{44.94} & 0.986 \\
        Bounding. Masks & 41.15 & 0.974 & 0.061 & 44.82 & 0.982 & 44.87 & 0.986 \\
        Bounding. Masks, Excl. Mask-Loss & 41.20 & 0.975 & 0.060 & 44.80 & 0.982 & 44.88 & 0.986 \\
        \bottomrule
    \end{tabular}%
    }
\end{table}

\paragraph{Mask Quality}

To assess the influence of initial mask quality and the mask loss term on our reconstruction, we conducted an exemplary ablation.
First, we tested the impact of the mask loss itself by setting its weight to 0 on the original data, labeled "Excl. Mask-Loss" in \tabref{tab:ablations}.

\begin{figure}[tbh]
    \centering
    \ifdefined\imageWidth
    \else
        \newlength{\imageWidth}
    \fi
    \setlength{\imageWidth}{0.25\textwidth}
    \ifdefined\imagerowheight
    \else
        \newlength{\imagerowheight}
    \fi
    \settoheight{\imagerowheight}{\includegraphics[width=\imageWidth]{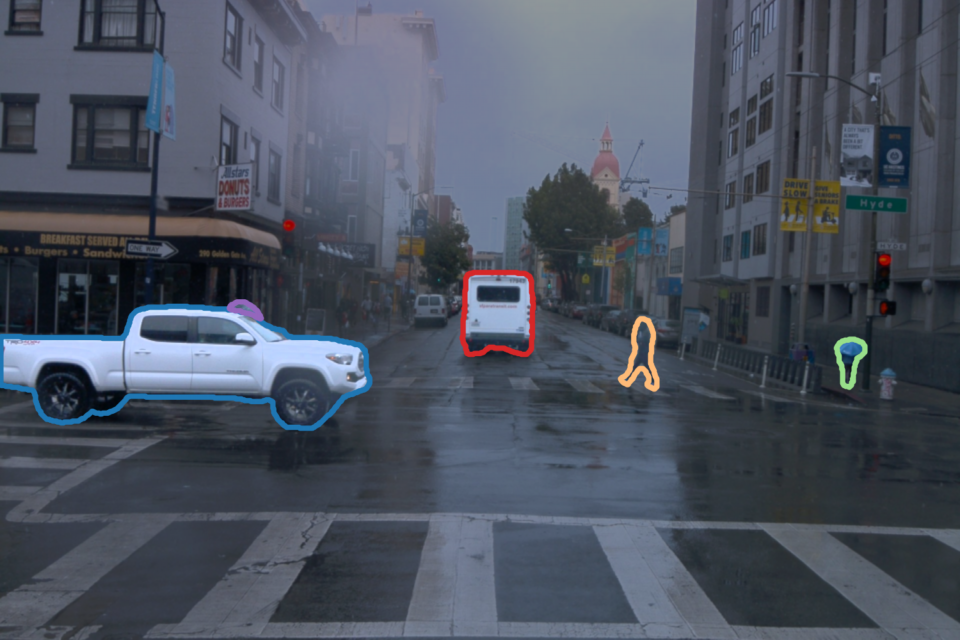}} 
    \ifdefined\gridImage
        \renewcommand{\gridImage}[1]{%
            \includegraphics[width=\imageWidth]{#1}%
        }
    \else
        \newcommand{\gridImage}[1]{%
            \includegraphics[width=\imageWidth]{#1}%
        }
    \fi
    \begin{tblr}{colspec={c c c c},colsep=0mm, rowsep=-1.2mm, row{1-3}={ht=\imagerowheight}, cells={halign=c, valign=m}}
         \gridImage{images/gt_images_adverse_masks/W141_040}& %
         \gridImage{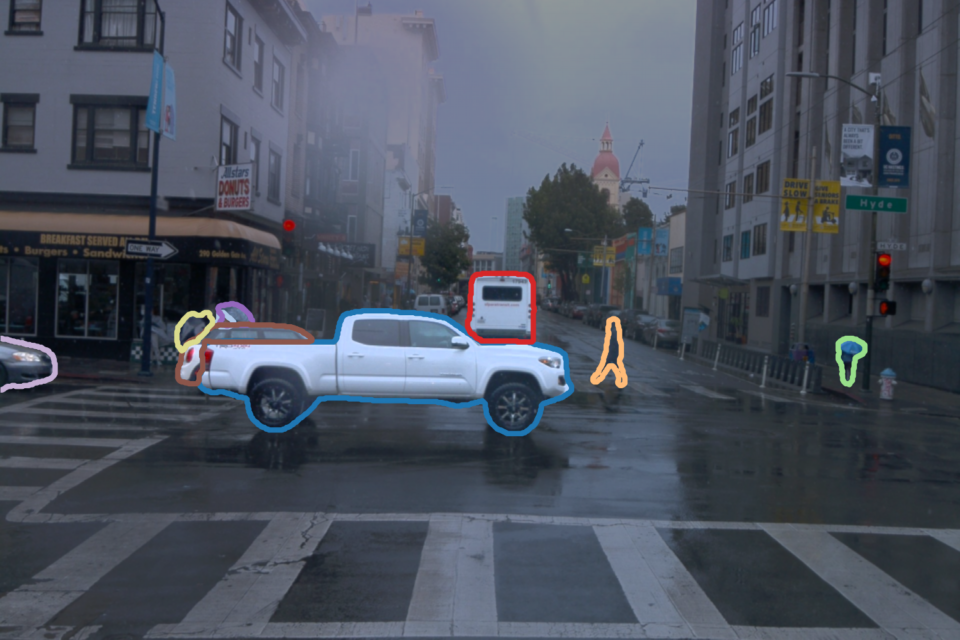}& %
         \gridImage{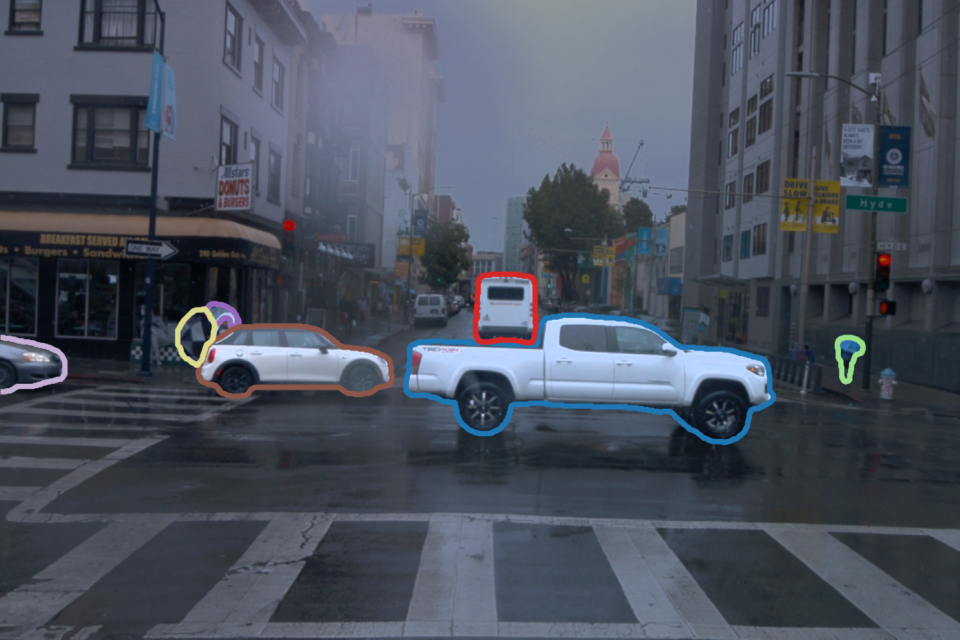}&  %
         \gridImage{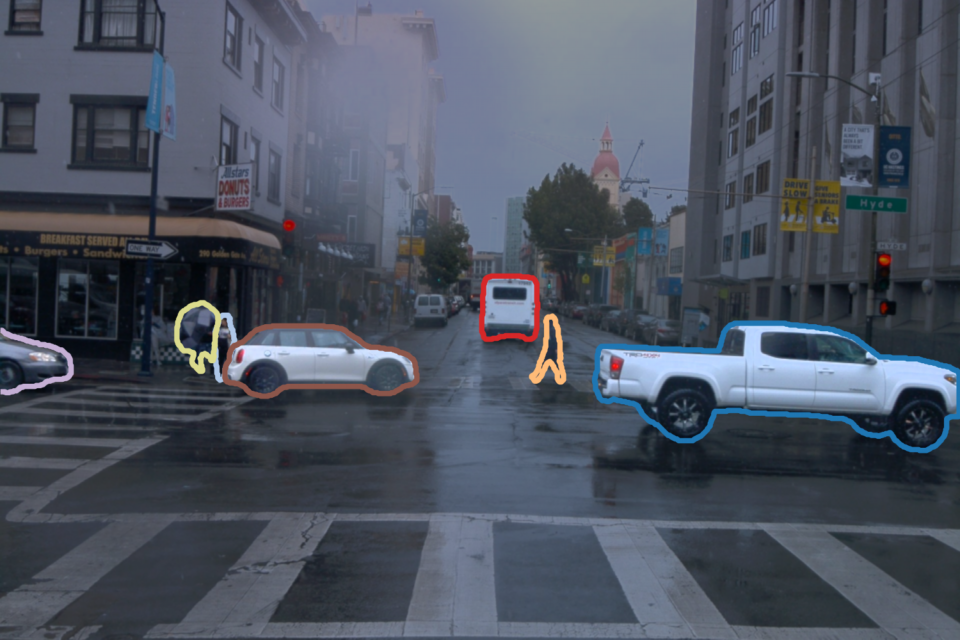}\\
         \gridImage{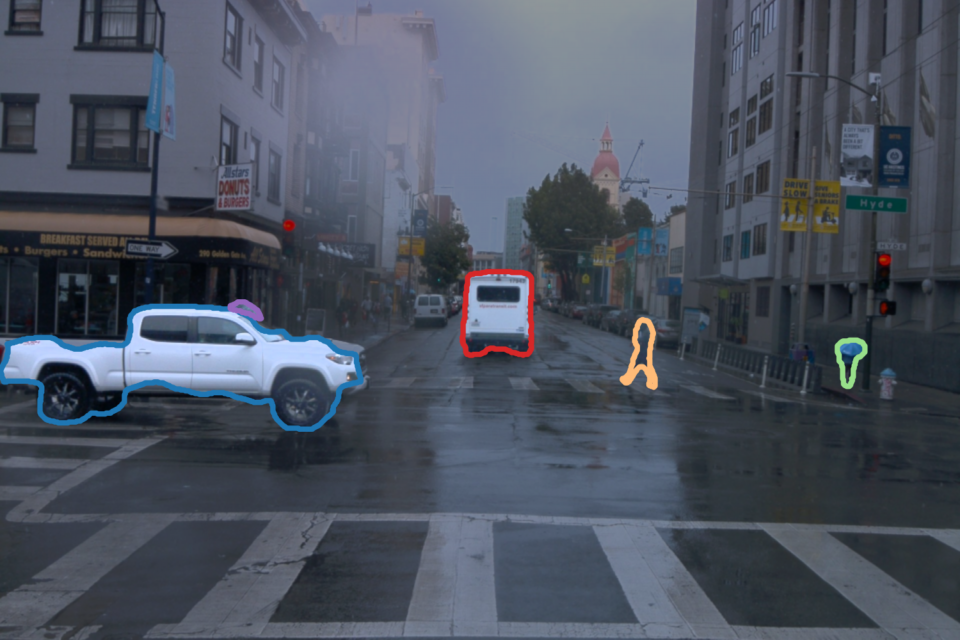}& %
         \gridImage{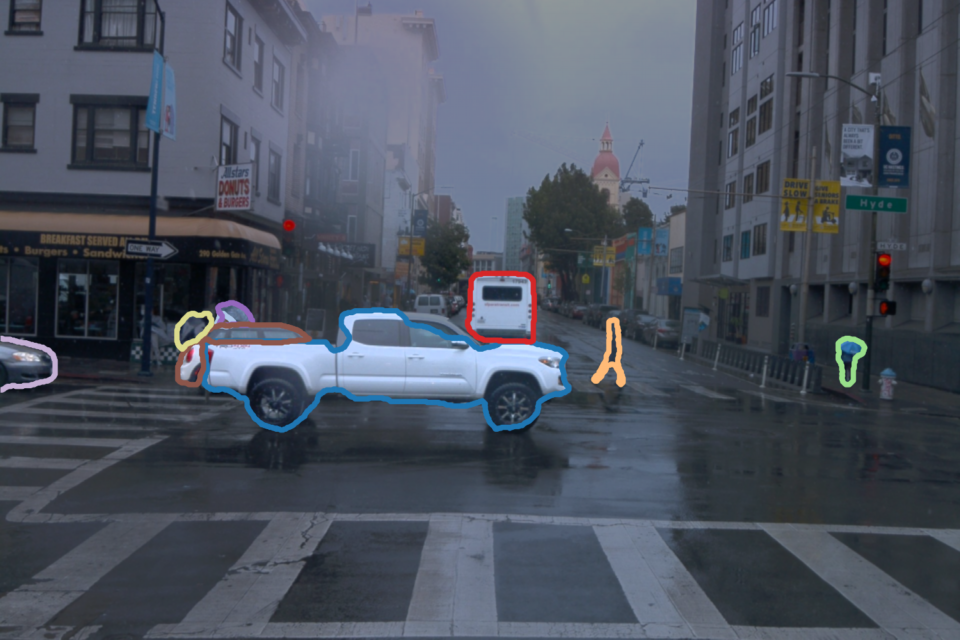}& %
         \gridImage{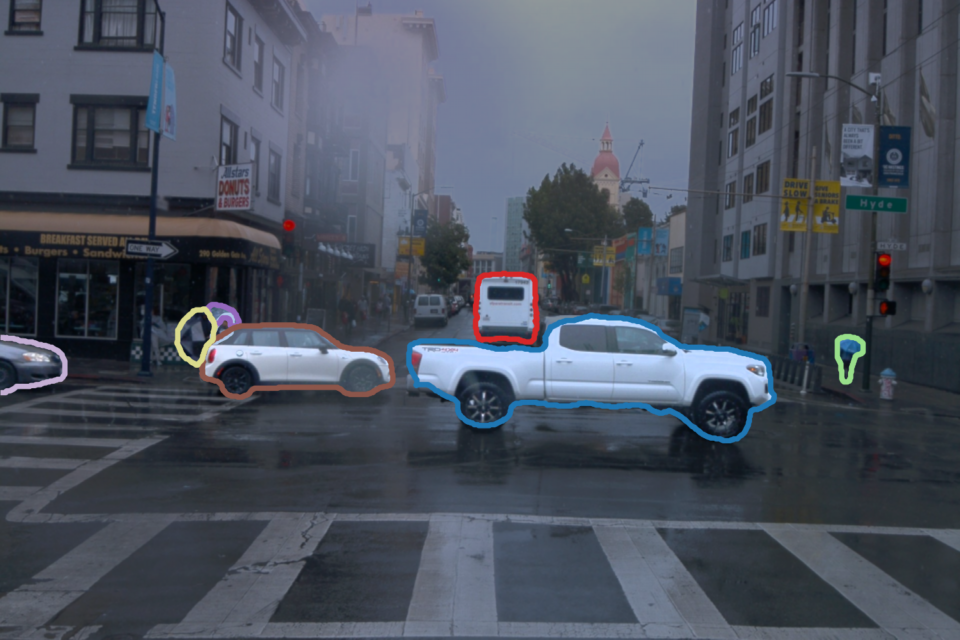}&  %
         \gridImage{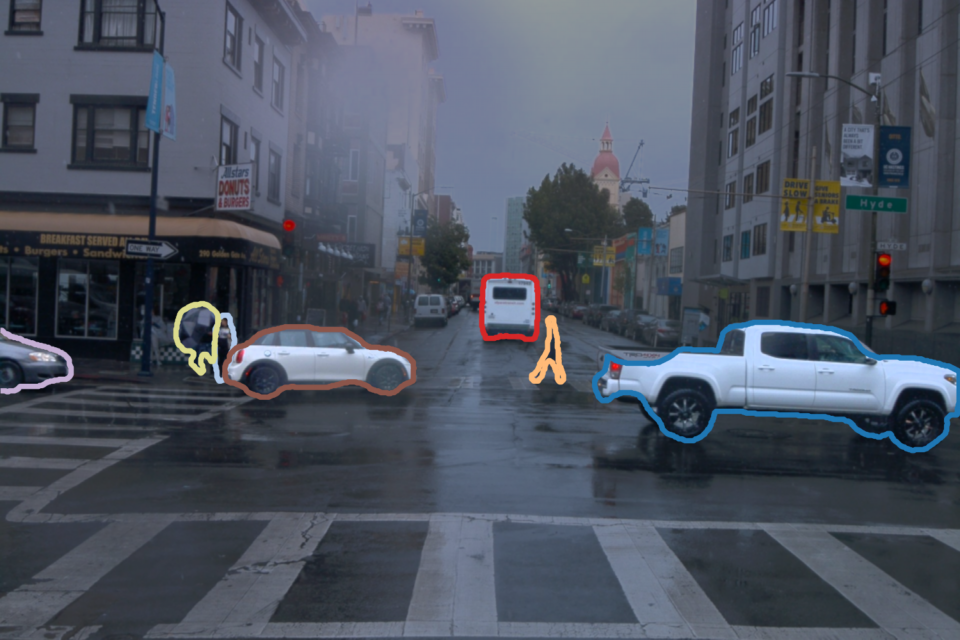}\\
         \gridImage{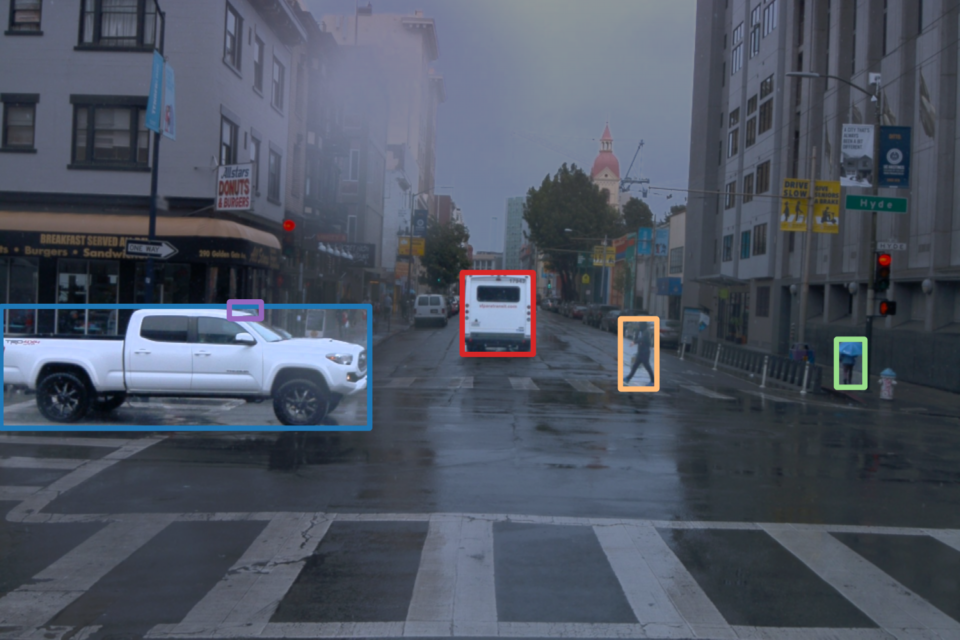}& %
         \gridImage{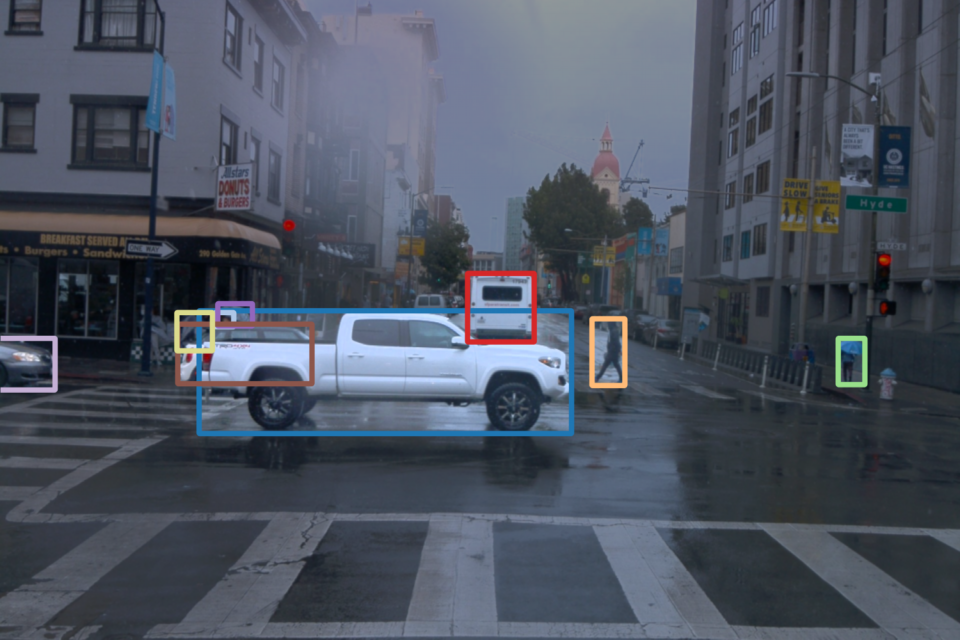}& %
         \gridImage{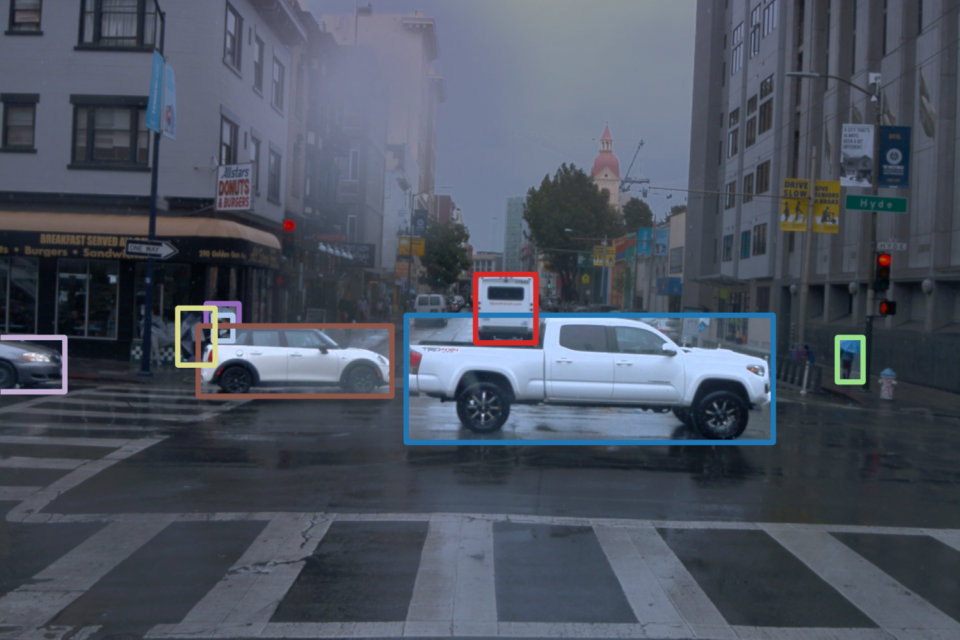}&  %
         \gridImage{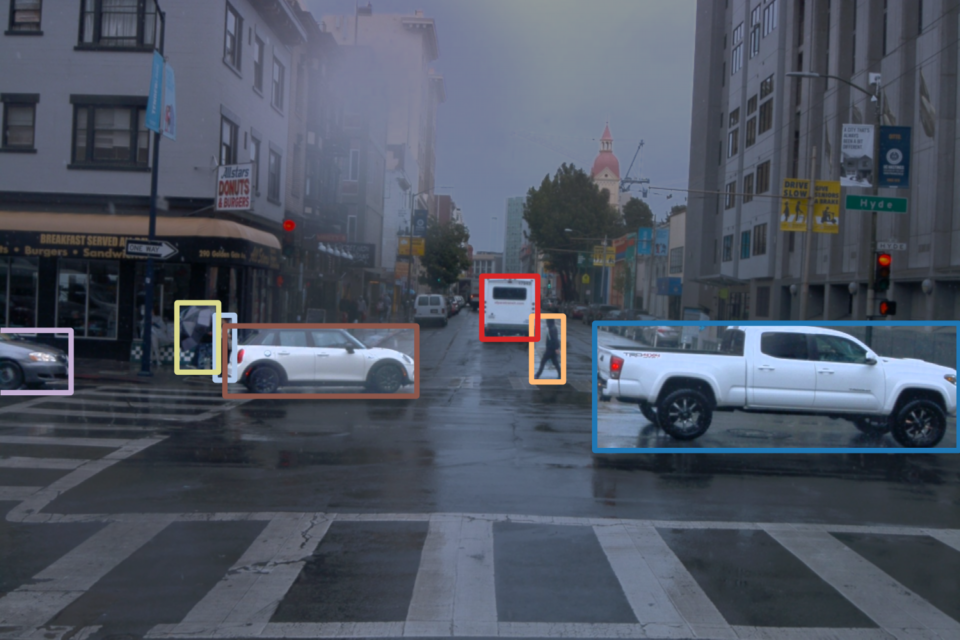}\\
    \end{tblr}
    \caption{Ground truth references and masks (top) and their corrupted versions using morphological operations (middle) as well as axis-aligned bounding box masks (bottom). We showcase four frames of the s-141 sequence (timestamps 40, 45, 50, 55), with 0.5 seconds spacing. The morphological masks show significant aberated and time-varying borders, while the imprecision of bounding boxes pose a challenge on adressing overlapping.}
    \label{fig:adverse_masks_gt}
\end{figure}

Second, we generated corrupted mask versions from our precise segmentations to simulate less ideal input and training conditions. We created two specific mask types for this analysis: the \textit{Morphological Masks} ("Morph. Masks") simulate imprecise segmentation by artificially corrupting the precise masks using morphological operations—specifically, smooth boundary erosion and dilation based on Perlin noise \cite{perlin1985image} maps. Alternatively, the \textit{Bounding Box Masks} ("Bounding. Masks") simulate the scenario where only axis-aligned bounding boxes are available, by using the bounding box of the original object mask. These two corrupted mask versions were then each trained both with and without the mask loss, resulting in the final set of conditions detailed in \tabref{tab:ablations}. Examples of these corrupted masks are presented in \figref{fig:adverse_masks_gt}.

Based on the quantitative metrics in \tabref{tab:ablations}, we observe that the mask quality and the mask loss term do not significantly affect the overall reconstruction quality (PSNR / SSIM / LPIPS). This relative resilience is expected, as the highly over-parametrized, node-based representation of the \ac{nag} has sufficient capacity to fit the scene even with minor initialization errors. However, we explicitly introduce the mask loss to suppress noise in the foreground and increase opacity in low-contrast segmentation areas (e.g., a grey car on a grey road).

\begin{figure}[tbh]
    \centering
    \ifdefined\imageWidth
    \else
        \newlength{\imageWidth}
    \fi
    \setlength{\imageWidth}{0.33\textwidth}
    \ifdefined\imagerowheight
    \else
        \newlength{\imagerowheight}
    \fi
    \settoheight{\imagerowheight}{\includegraphics[width=\imageWidth]{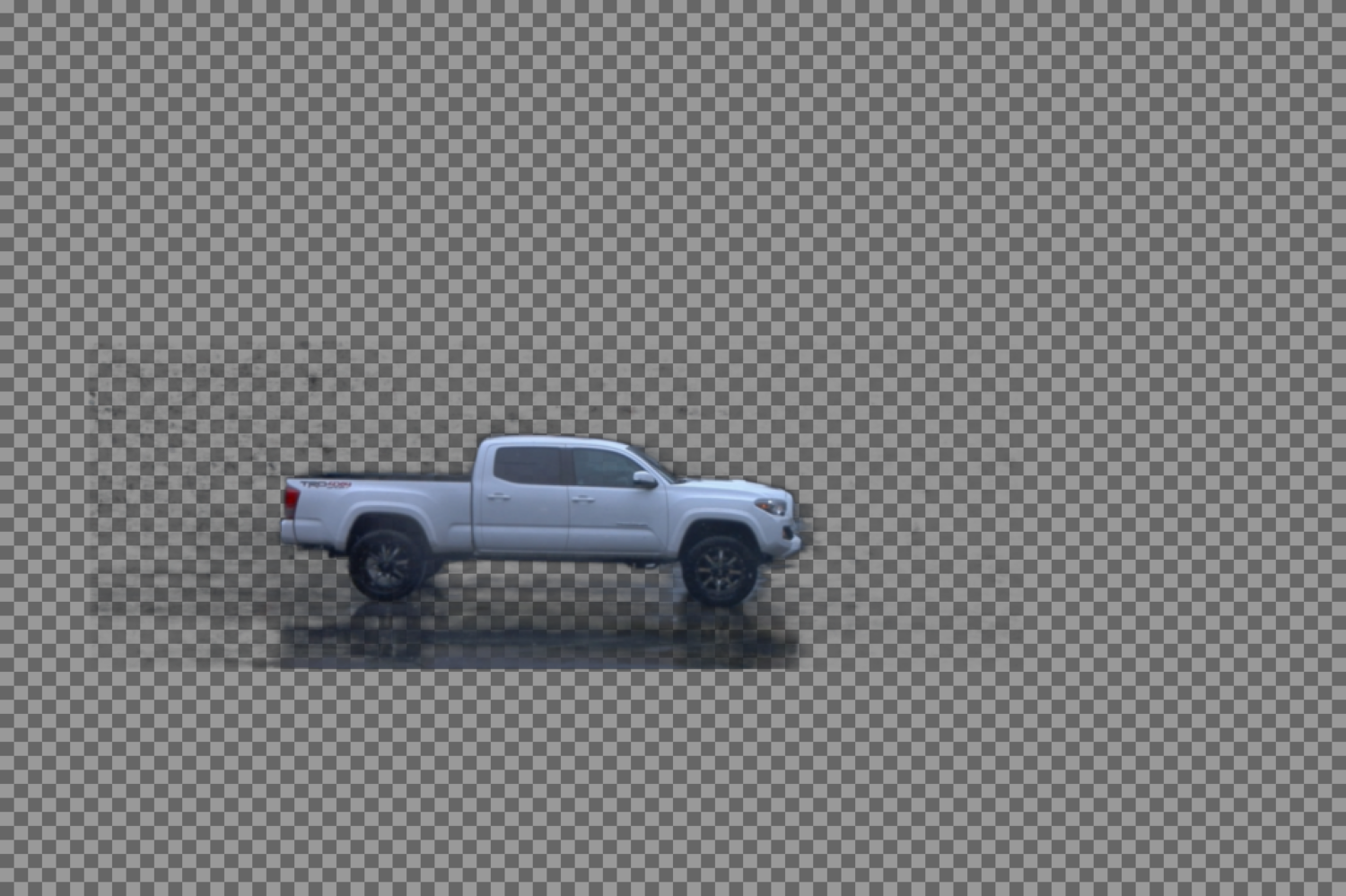}} 
    \ifdefined\gridImage
        \renewcommand{\gridImage}[1]{%
            \includegraphics[width=\imageWidth]{#1}%
        }
    \else
        \newcommand{\gridImage}[1]{%
            \includegraphics[width=\imageWidth]{#1}%
        }
    \fi
    \begin{tabular}{ccc}
        \parbox{\dimexpr0.33\textwidth-2\tabcolsep\relax}{\centering Precise} &%
        \parbox{\dimexpr0.33\textwidth-2\tabcolsep\relax}{\centering Morph.} &%
        \parbox{\dimexpr0.33\textwidth-2\tabcolsep\relax}{\centering Bounding Box}\\
    \end{tabular}
    \begin{tblr}{colspec={c c c},colsep=0mm, rowsep=-1.3mm, row{1-2}={ht=\imagerowheight}, cells={halign=c, valign=m}}
         \gridImage{images/ablation/masks/00/W141_002_043_large}& %
         \gridImage{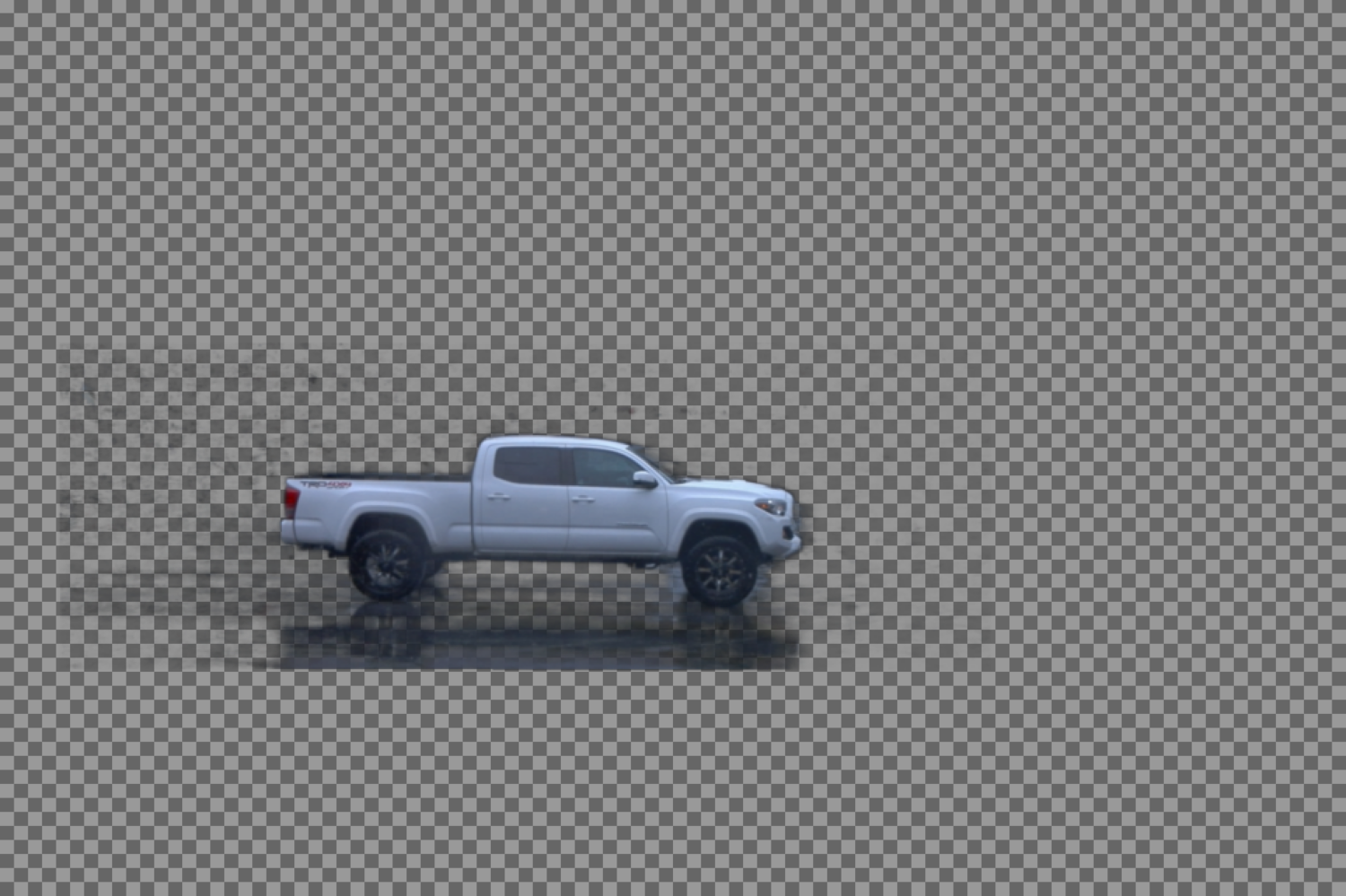}& %
         \gridImage{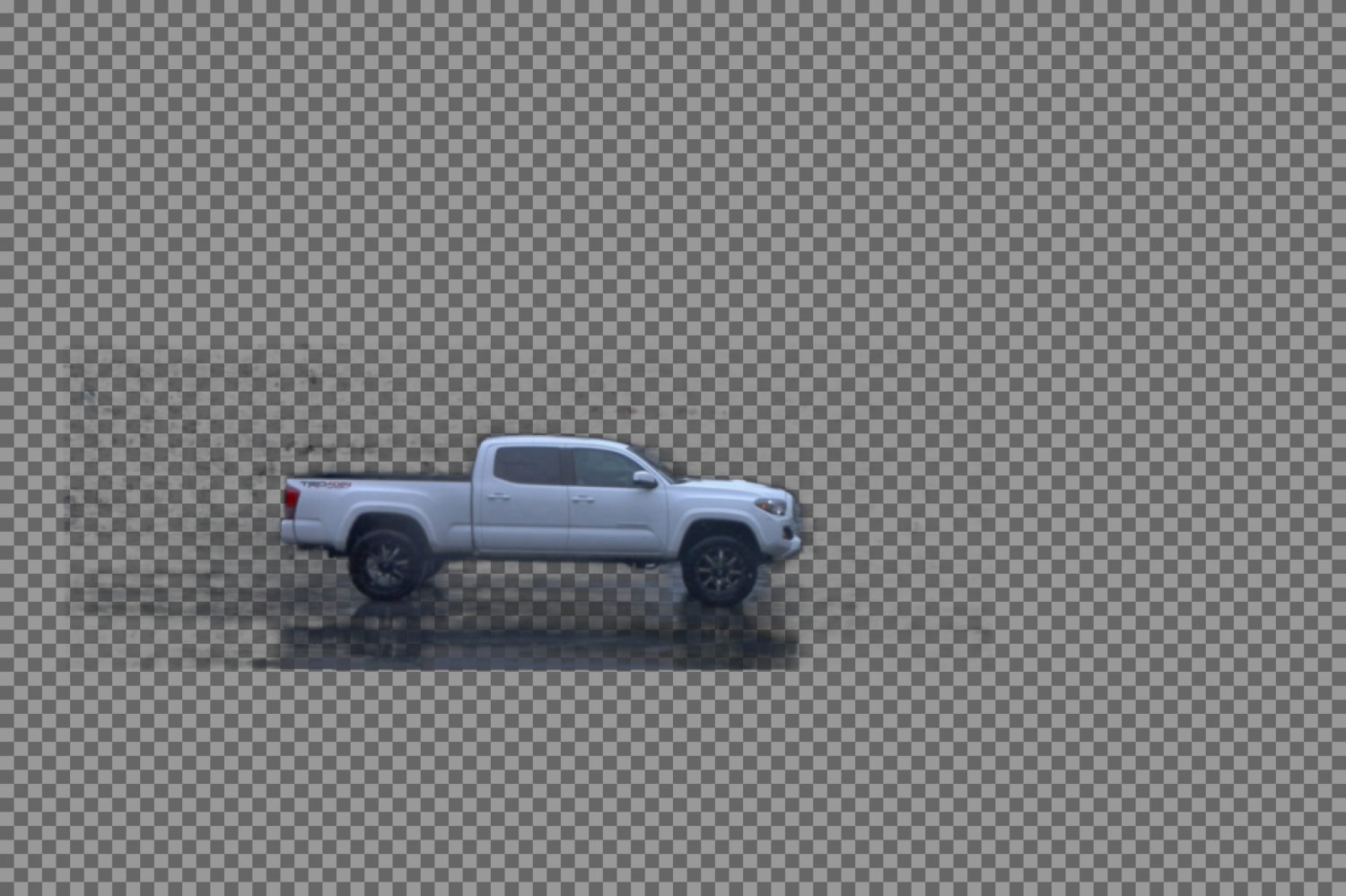}\\
         \gridImage{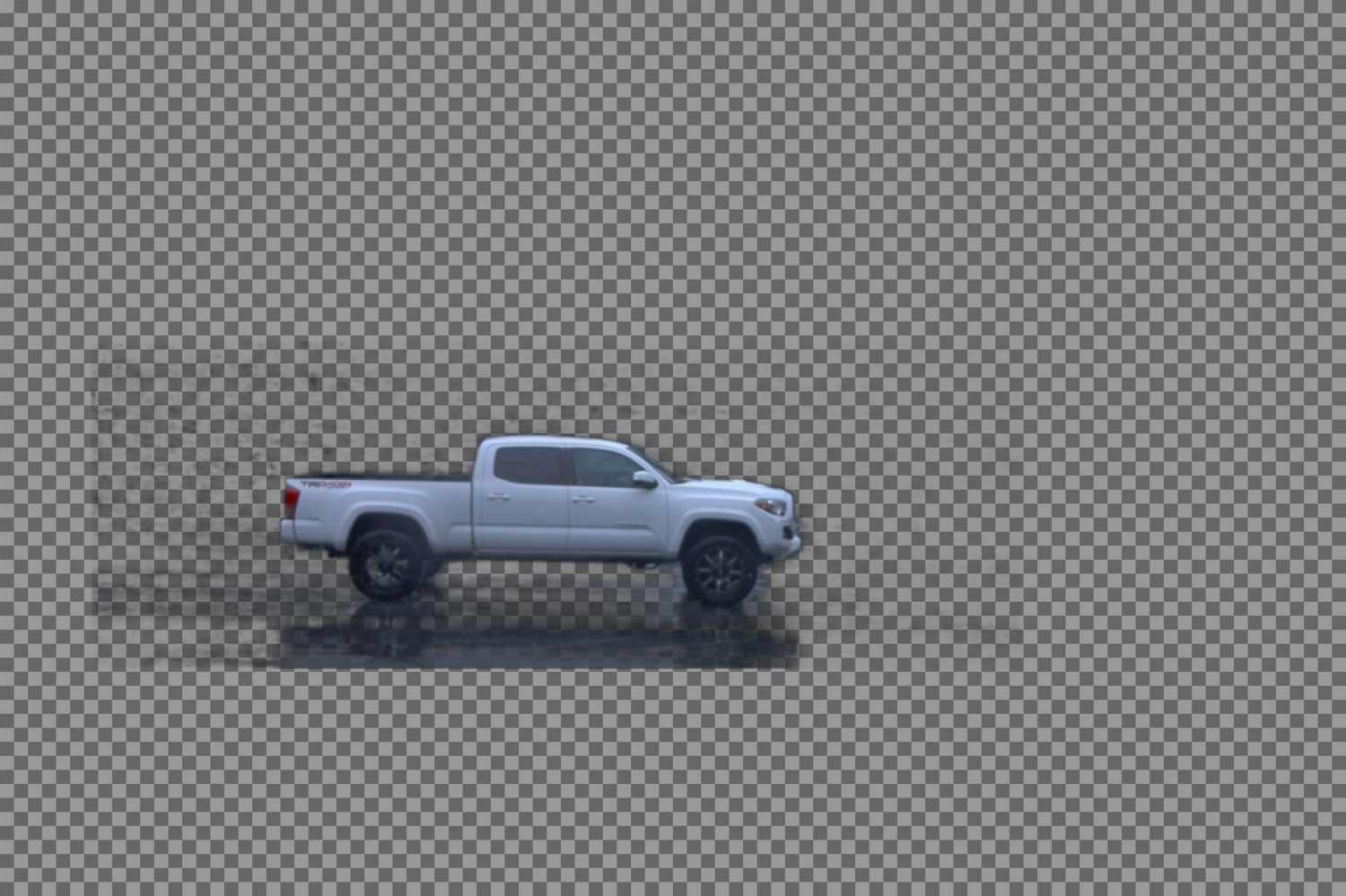}& %
         \gridImage{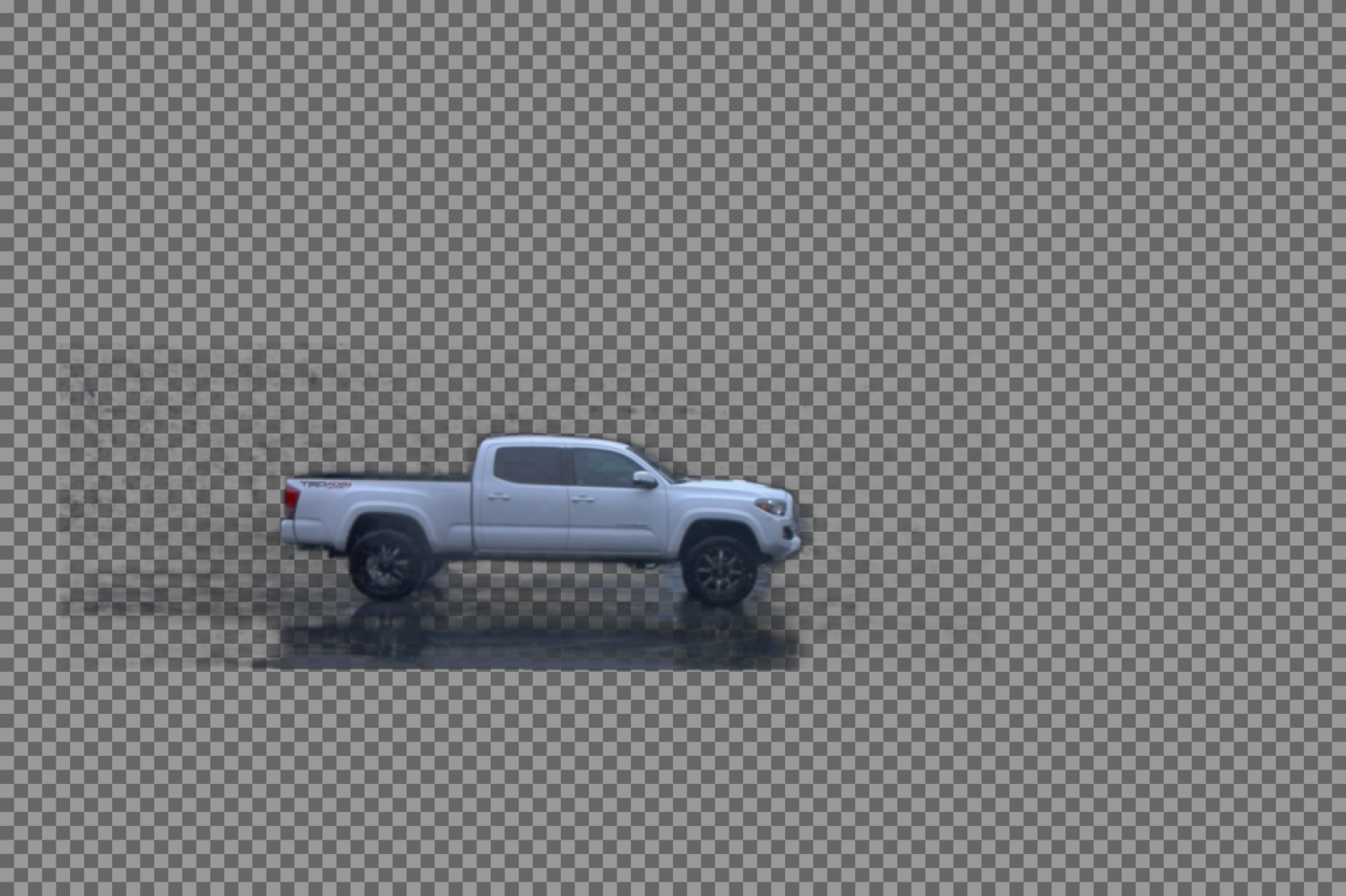}& %
         \gridImage{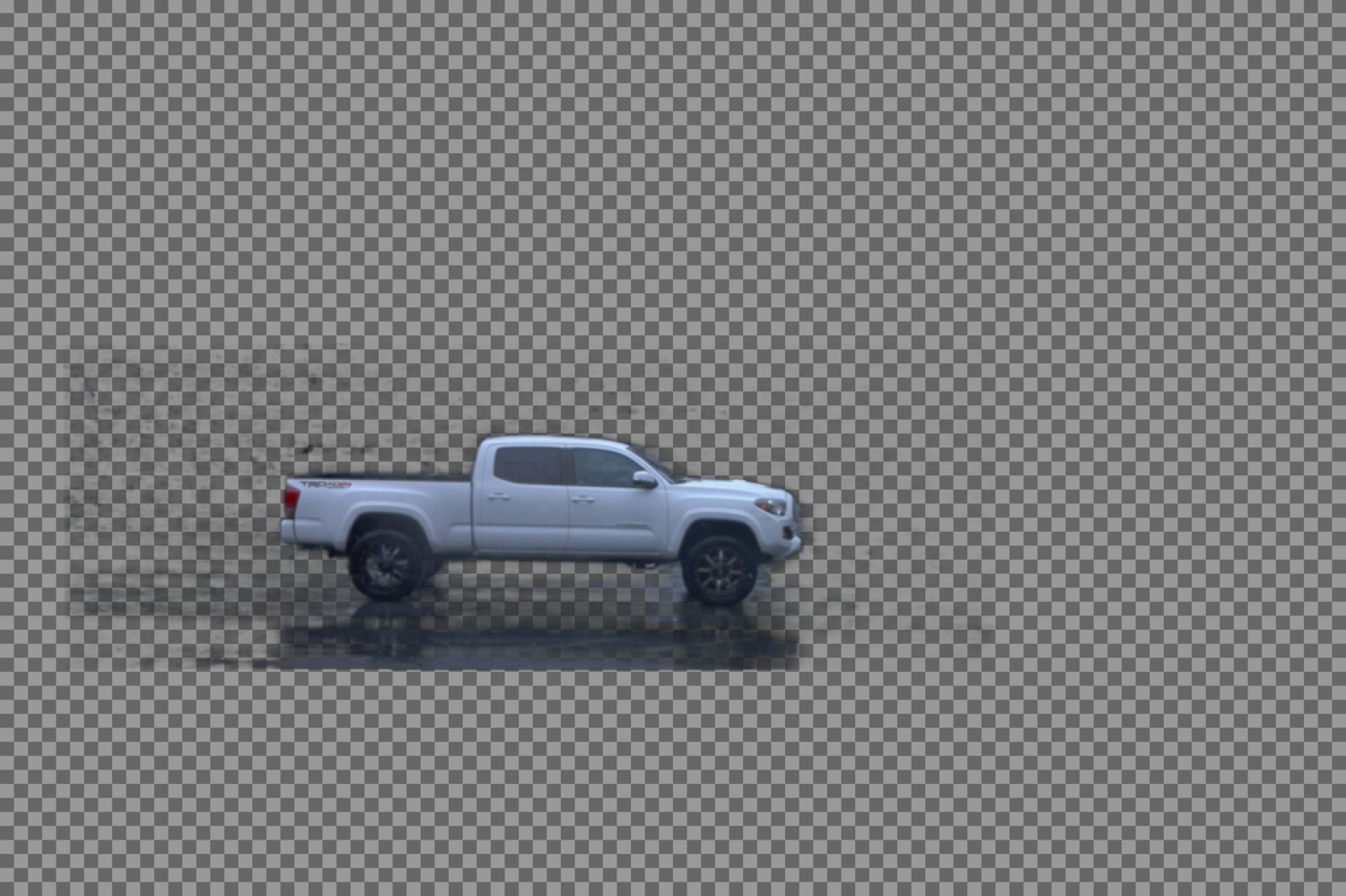}\\
    \end{tblr}
    \begin{tabular}{ccc}
        \parbox{\dimexpr0.33\textwidth-2\tabcolsep\relax}{\centering Precise w/o Mask Loss} &%
        \parbox{\dimexpr0.33\textwidth-2\tabcolsep\relax}{\centering Morph. w/o Mask Loss} &%
        \parbox{\dimexpr0.33\textwidth-2\tabcolsep\relax}{\centering Bounding Box w/o Mask Loss}\\
    \end{tabular}
    \caption{Object decomposition of the white truck from scene s-141 (timestamp 45; ref. \figref{fig:adverse_masks_gt}). The top row shows results trained with the mask loss, while the bottom row excludes it. Even under aberrated masks (Morph. and Bounding Box), the decomposition remains highly precise for contrast-rich and independently moving objects, showing only minor increases in background noise fitting.}
    \label{fig:adverse_masks_obj_0}
\end{figure}

\begin{figure}[tbh]
    \centering
    \ifdefined\imageWidth
    \else
        \newlength{\imageWidth}
    \fi
    \setlength{\imageWidth}{0.33\textwidth}
    \ifdefined\imagerowheight
    \else
        \newlength{\imagerowheight}
    \fi
    \settoheight{\imagerowheight}{\includegraphics[width=\imageWidth]{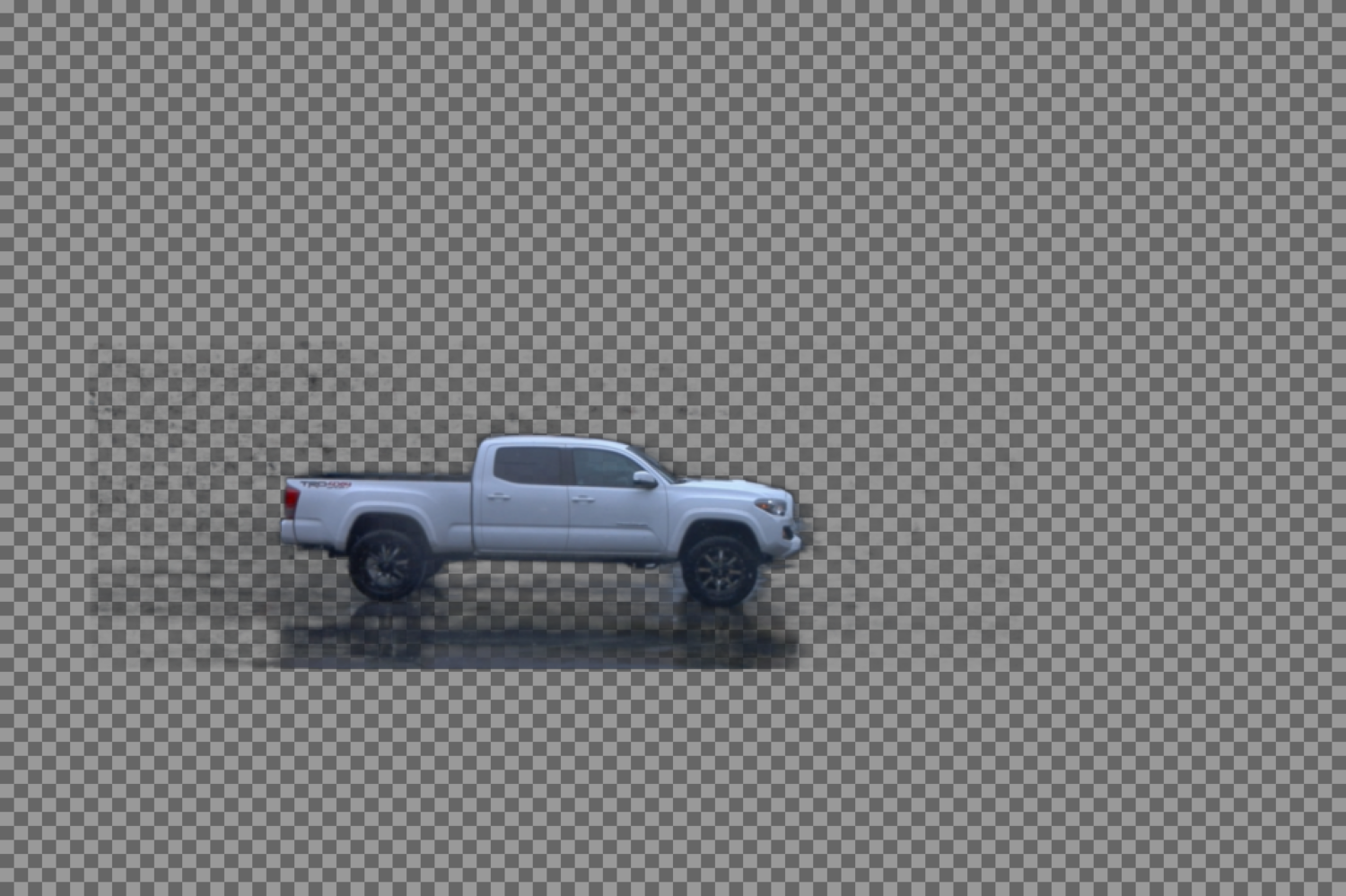}} 
    \ifdefined\gridImage
        \renewcommand{\gridImage}[1]{%
            \includegraphics[width=\imageWidth]{#1}%
        }
    \else
        \newcommand{\gridImage}[1]{%
            \includegraphics[width=\imageWidth]{#1}%
        }
    \fi
    \begin{tabular}{ccc}
        \parbox{\dimexpr0.33\textwidth-2\tabcolsep\relax}{\centering Precise} &%
        \parbox{\dimexpr0.33\textwidth-2\tabcolsep\relax}{\centering Morph.} &%
        \parbox{\dimexpr0.33\textwidth-2\tabcolsep\relax}{\centering Bounding Box}\\
    \end{tabular}
    \begin{tblr}{colspec={c c c},colsep=0mm, rowsep=-1.3mm, row{1-2}={ht=\imagerowheight}, cells={halign=c, valign=m}}
         \gridImage{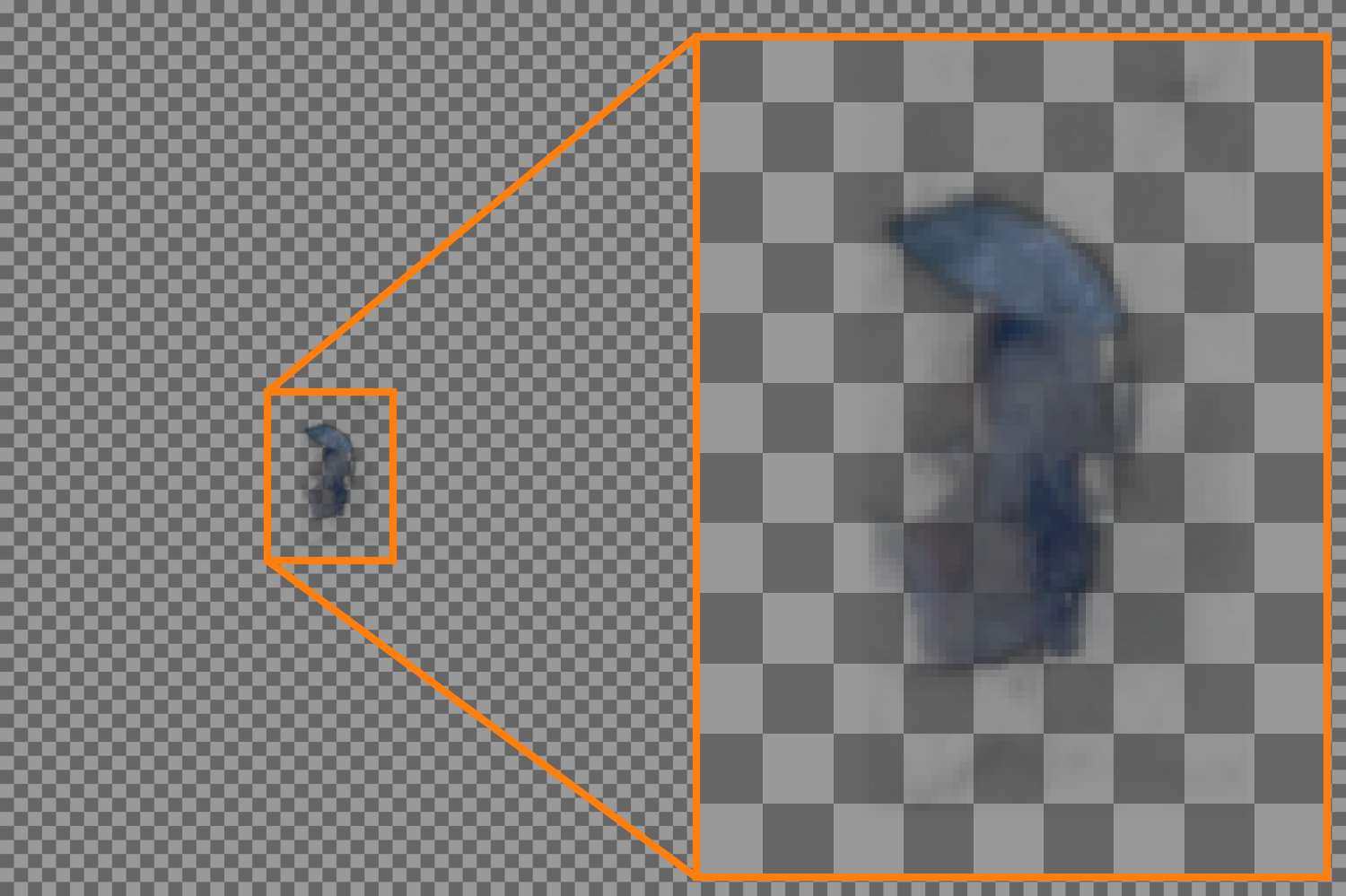}& %
         \gridImage{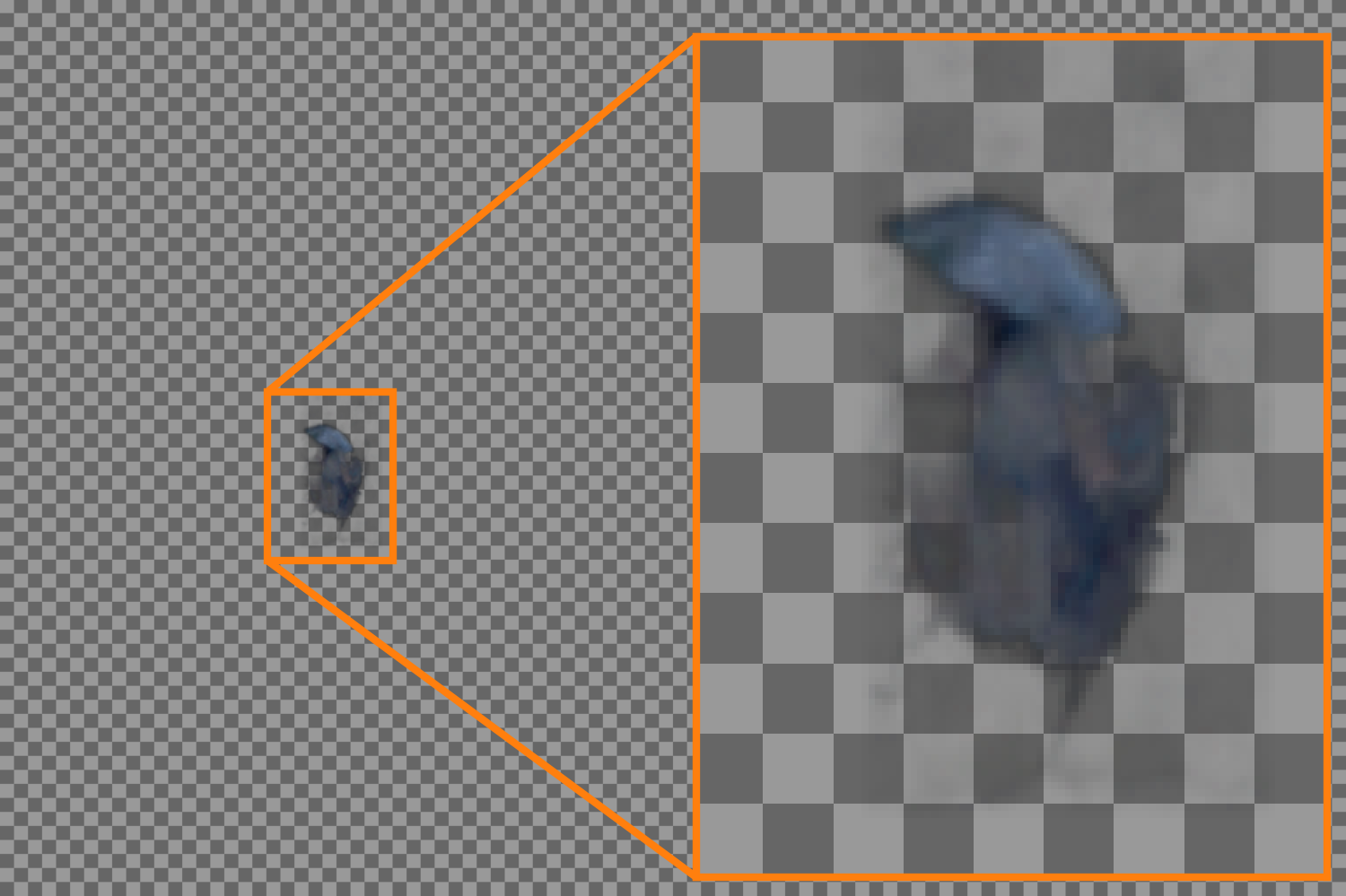}& %
         \gridImage{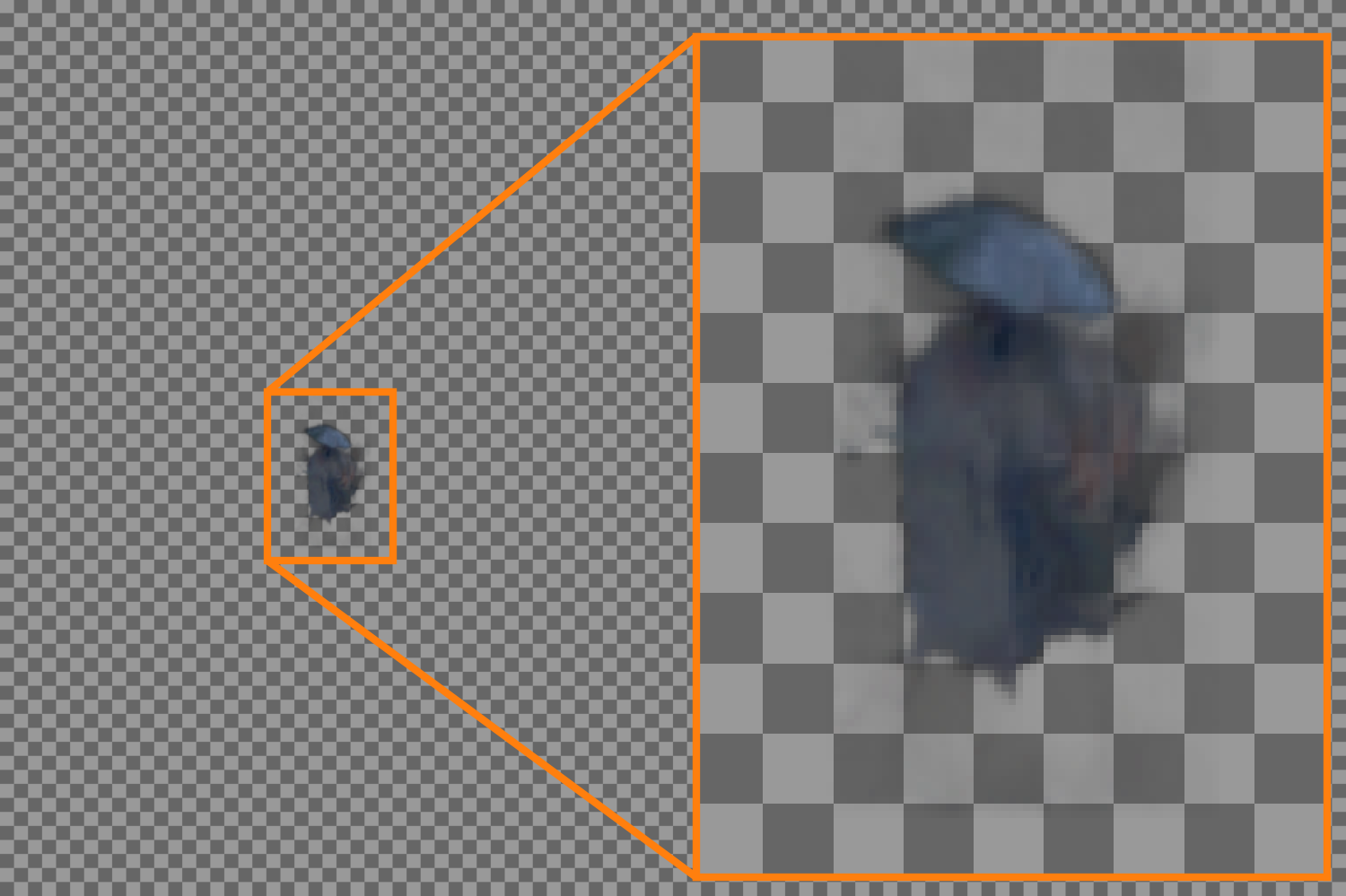}\\
         \gridImage{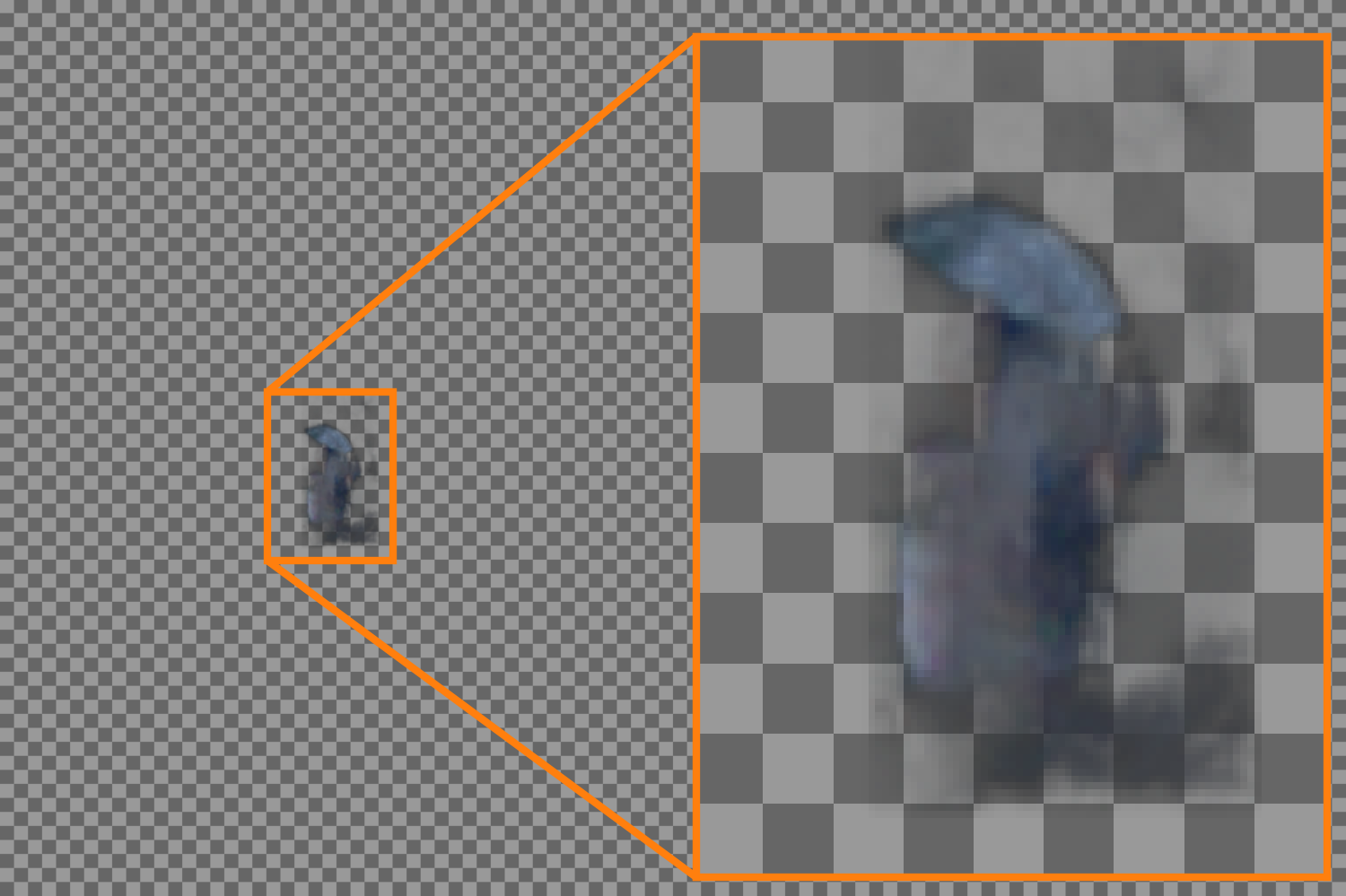}& %
         \gridImage{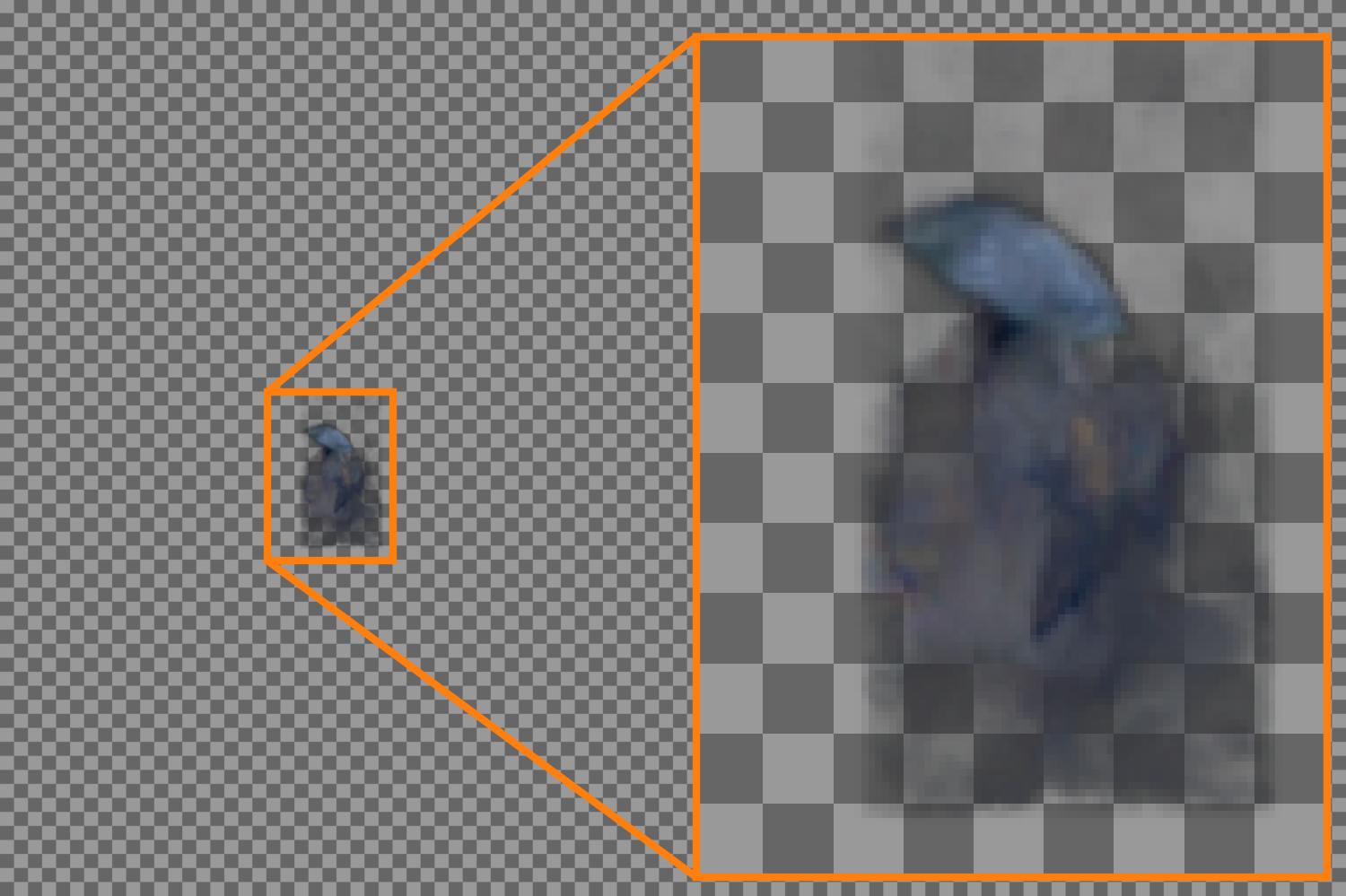}& %
         \gridImage{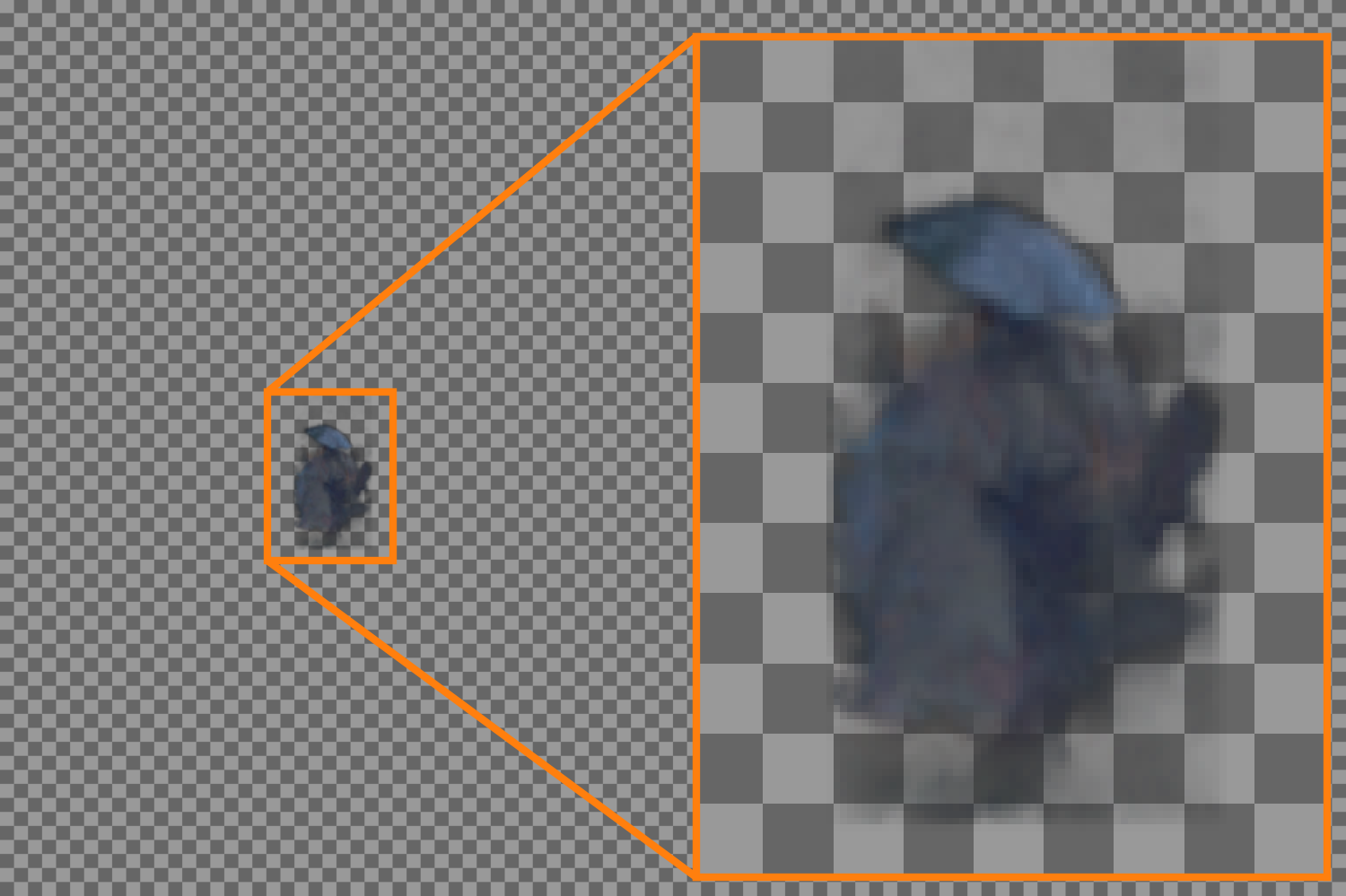}\\
    \end{tblr}
    \begin{tabular}{ccc}
        \parbox{\dimexpr0.33\textwidth-2\tabcolsep\relax}{\centering Precise w/o Mask Loss} &%
        \parbox{\dimexpr0.33\textwidth-2\tabcolsep\relax}{\centering Morph. w/o Mask Loss} &%
        \parbox{\dimexpr0.33\textwidth-2\tabcolsep\relax}{\centering Bounding Box w/o Mask Loss}\\
    \end{tabular}
    \caption{Object decomposition of an occluded person in scene s-141 (following \figref{fig:adverse_masks_obj_0}). While the original version maintains a consistent silhouette despite heavy occlusion, the use of aberrated masks (Morph. and Bounding Box) tends to produce a less consistent and visually disturbed representation in these challenging occluded regions.}
    \label{fig:adverse_masks_obj_6}
\end{figure}

Since the masks are used for initializing and refining individual atlas nodes to correctly factorize the scene, performance differences become apparent when qualitatively studying the decomposed objects. We present two decomposed objects from the s-141 scene in \figref{fig:adverse_masks_obj_0} and \ref{fig:adverse_masks_obj_6}.
For highly contrastive objects with clear, independent motion (like the white truck in \figref{fig:adverse_masks_obj_0}), the decomposition quality is minimally impacted by mask quality or loss. At most, a slight increase in opacity around the object boundaries fitting to background noise can be observed.
Conversely, for challenging, occluded objects (such as the person in \figref{fig:adverse_masks_obj_6}), the effect is significant: the representation severely degrades when using aberrated masks. The original precise masks were necessary to maintain a reasonable representation of the person, while disabling the mask loss further exacerbated the fitting to noise.

Furthermore, poor mask quality may indirectly affect texture editability, as fitting exterior noise or background content into the atlas can push information into the view-dependence field, potentially hindering subsequent texture modification. Nevertheless, the qualitative examples in \figref{fig:adverse_masks_obj_0} suggest that even using bounding box masks or segmentation models with coarser outputs could be sufficient to yield a reasonable scene decomposition when interest lies primarily in dominant foreground objects with clear motion patterns. While an in-depth discussion on our method's mask-quality sensitivity would require dedicated experiments on synthetic data, our exemplary study indicates certain usability even in the absence of precise segmentations, relying only on bounding boxes.

In summary, our ablation studies provide valuable insights into the contribution of individual components of our model, highlighting the importance of model size, flow estimation, view-dependent modeling, and translation learning for achieving high-quality reconstructions.

\FloatBarrier

\subsection{Additional Visual Results}
\label{sec:additional_visual_results}
This supplementary section provides extended visual results that further illustrate the capabilities of our proposed \ac{nag} representation. We present additional examples showcasing the editing potential of NAGs, including object insertion, retiming, and shifting. Furthermore, we offer supplementary visual examples from the Davis Dataset \cite{Perazzi2016}, including reconstructions and their corresponding scene decompositions, providing deeper insights into our model's performance and representation.

\begin{figure}[tb]
    \centering
    \setlength{\imageWidth}{0.25\textwidth}
    \settoheight{\imagerowheight}{\includegraphics[width=\imageWidth]{images/qual_results/W975_000_032_GT_z_8}} 
    \newcommand{\gridImage}[1]{%
        \includegraphics[width=\imageWidth]{#1}%
    }
    \newcommand{\rowLabel}[1]{%
        \raisebox{0.5\imagerowheight}{\vcenter{\hbox{\rotatebox{90}{#1}}}}%
    }
    \begin{tblr}{colspec={c c c c},colsep=0mm, rowsep=-1.2mm, row{1-3}={ht=\imagerowheight}, cells={halign=c, valign=m}}
         \gridImage{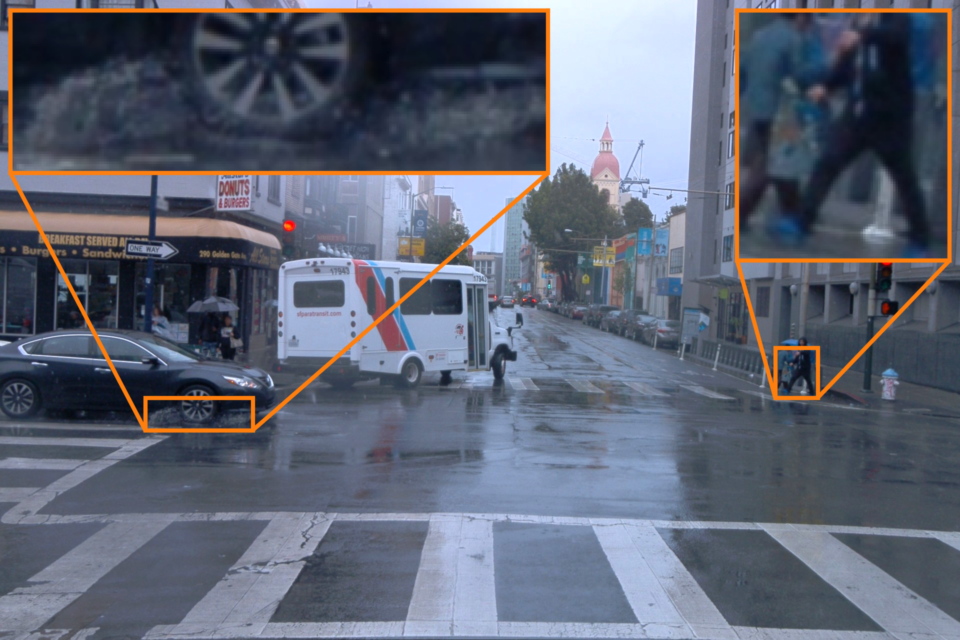} &%
         \gridImage{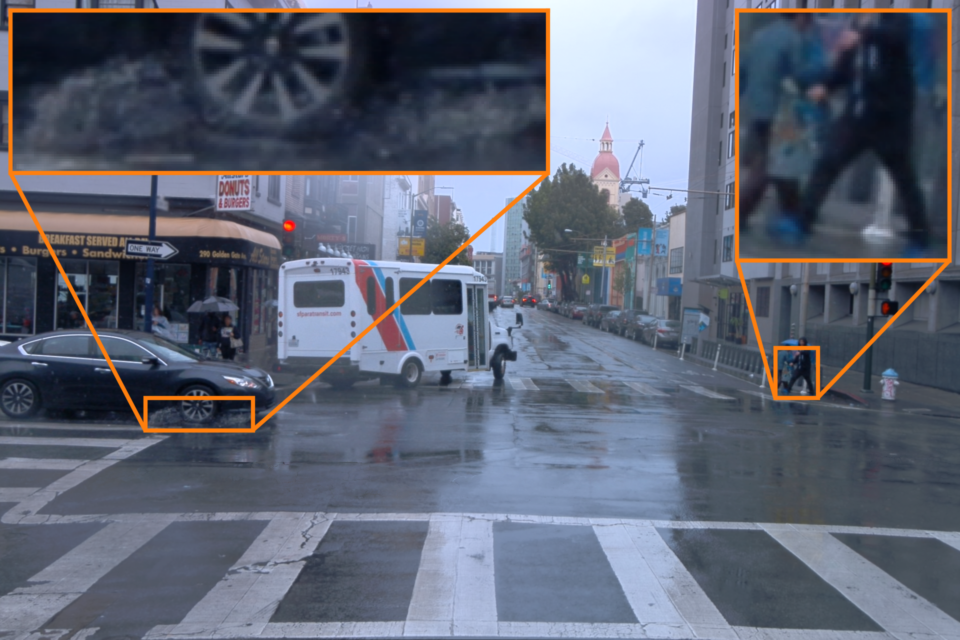} &%
         \gridImage{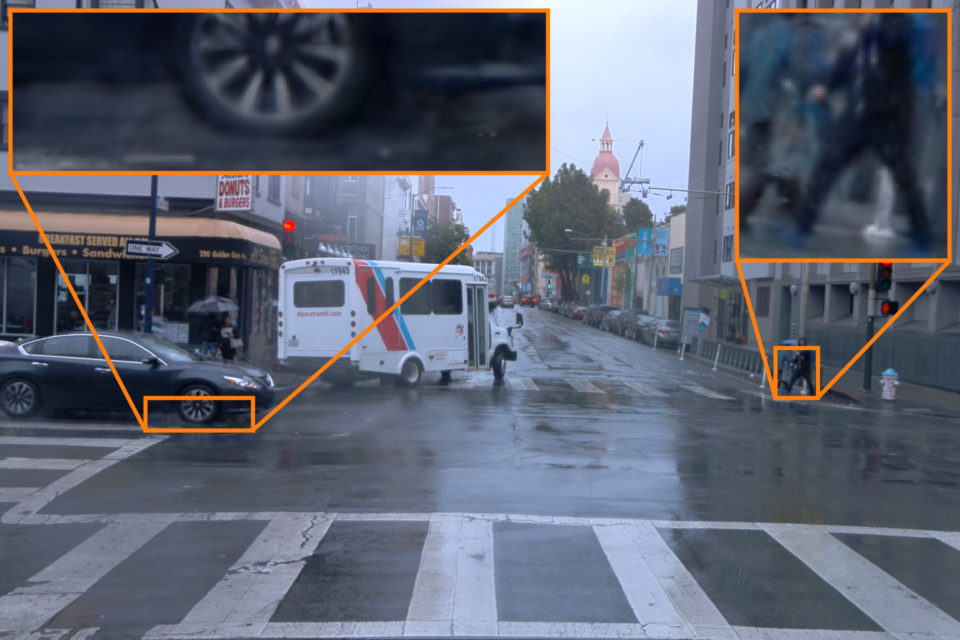}&  %
         \gridImage{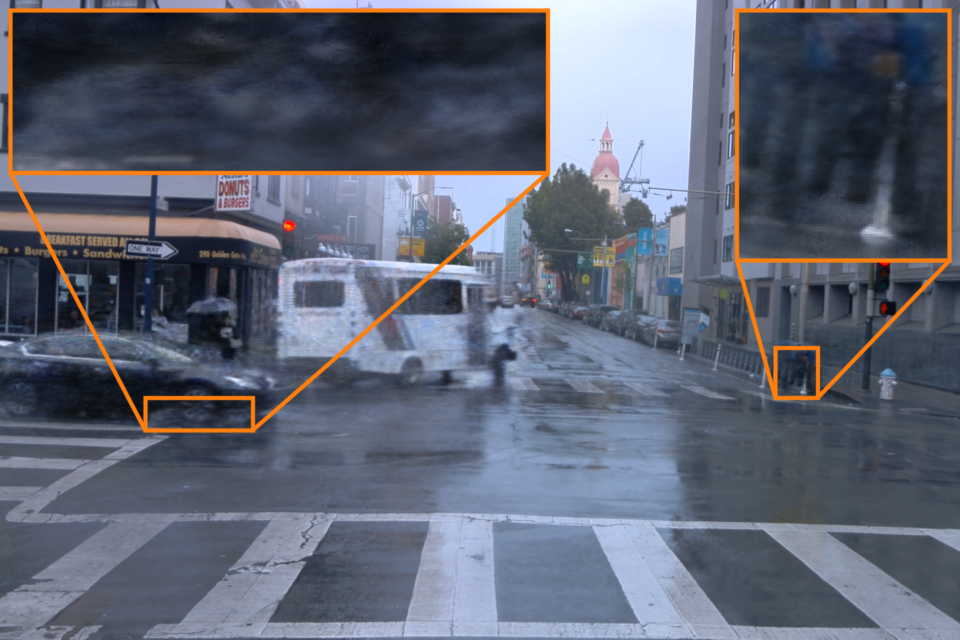}
         \\
         \gridImage{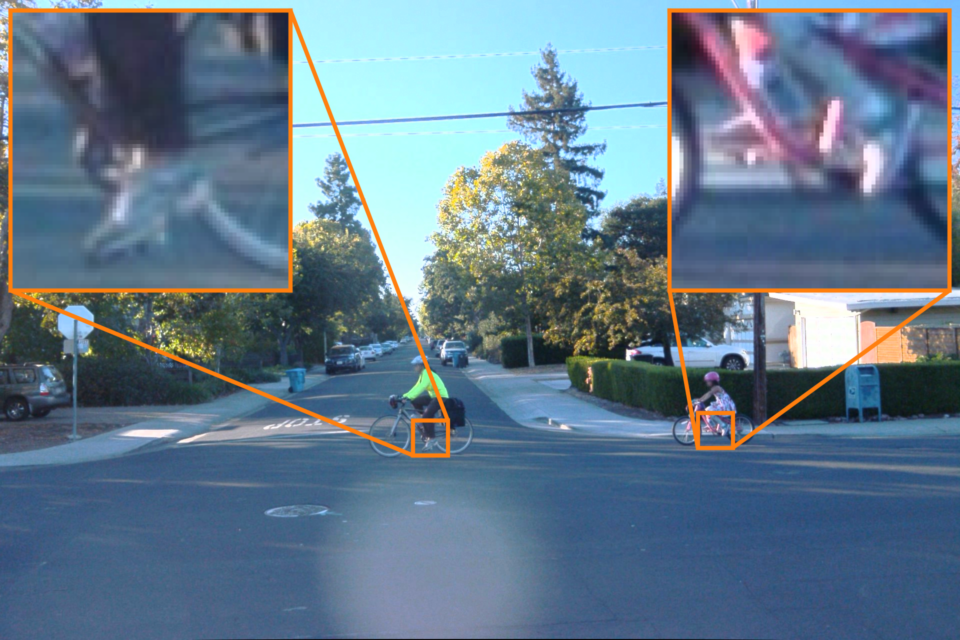}
         & %
         \gridImage{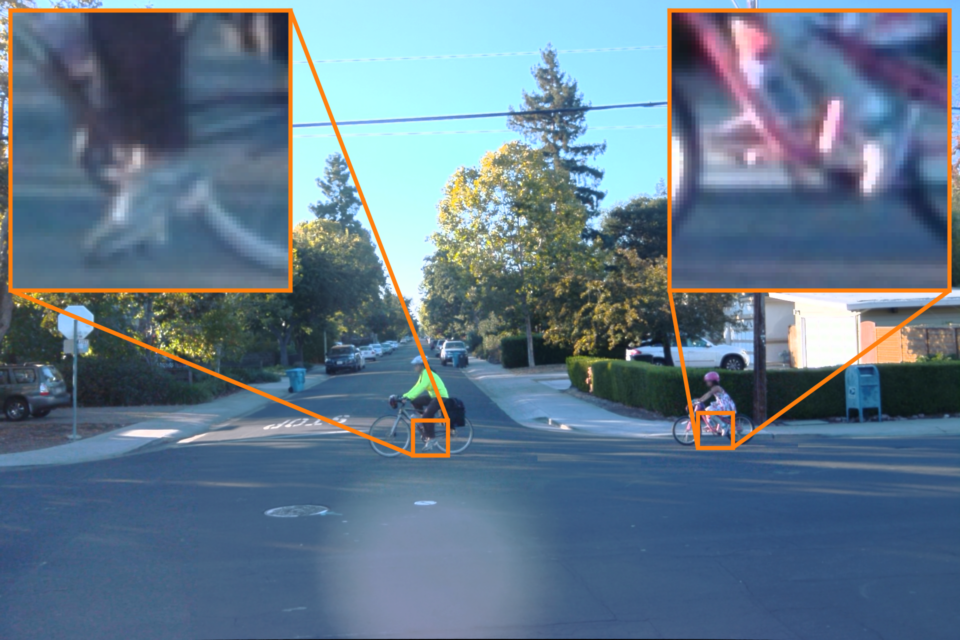}
         & %
         \gridImage{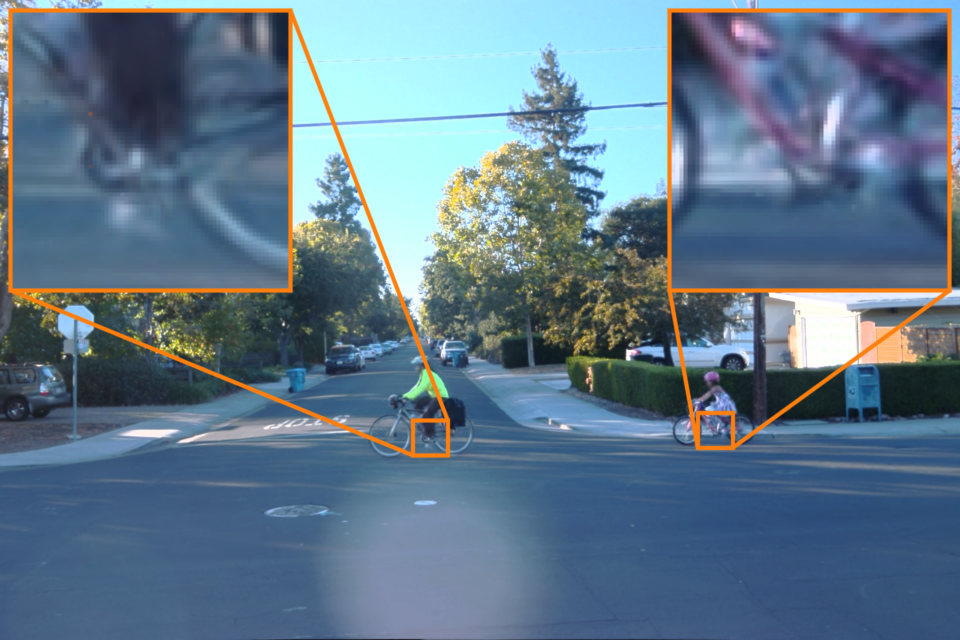}
         &  %
         \gridImage{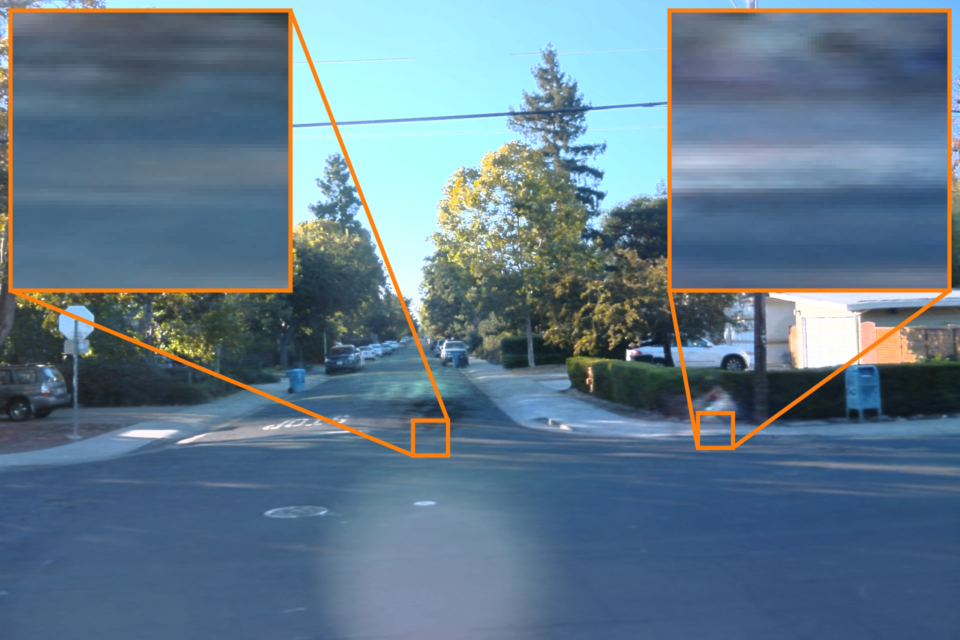}
         \\
         \gridImage{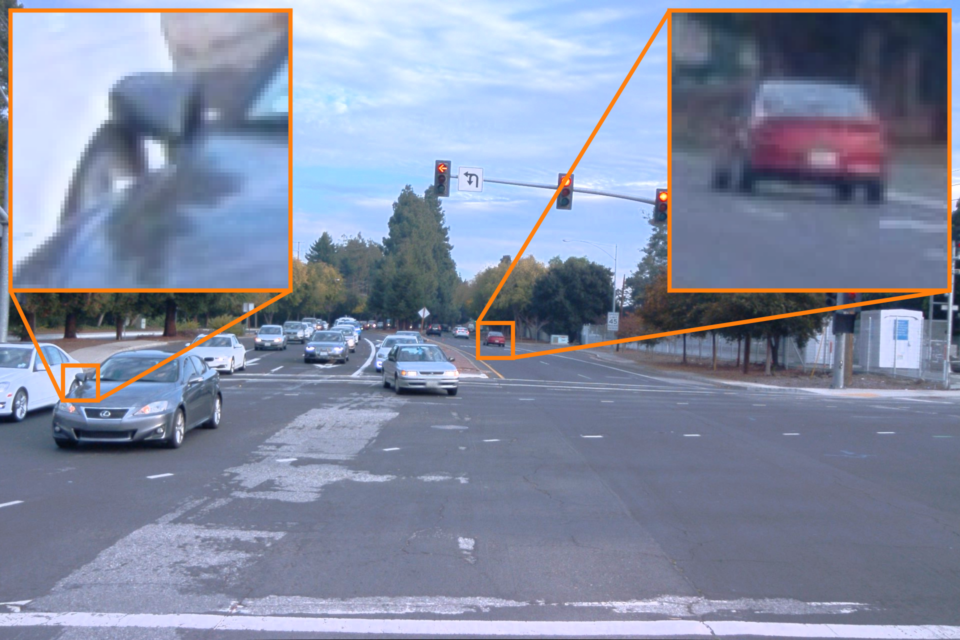}
         & %
         \gridImage{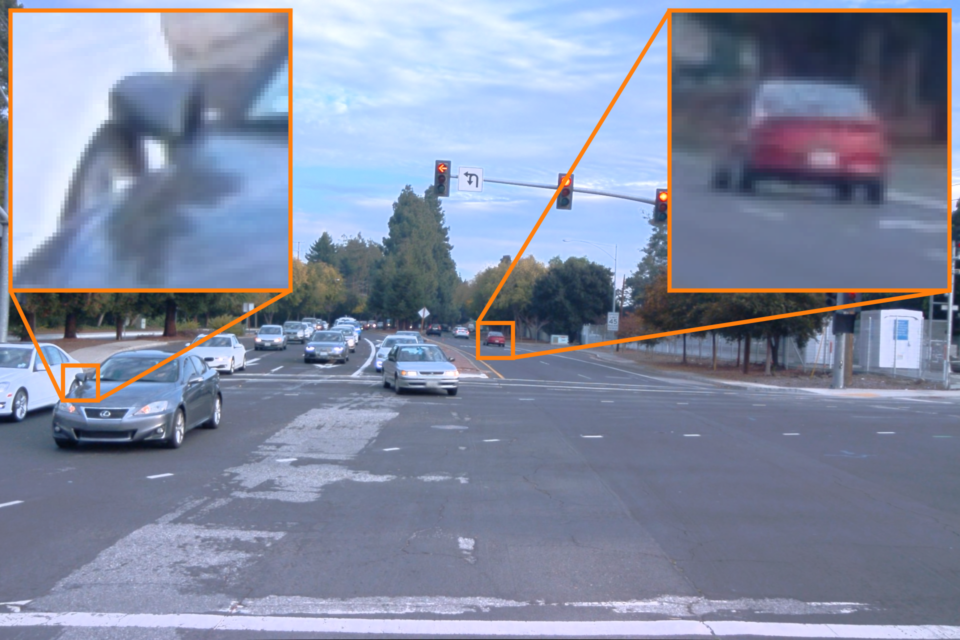}
         & %
         \gridImage{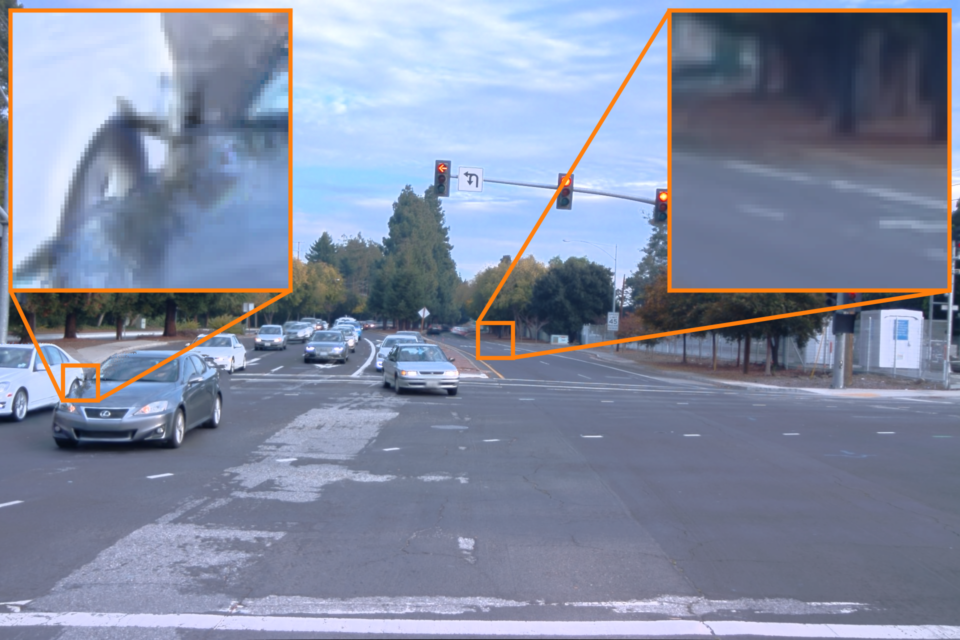}
         &  %
         \gridImage{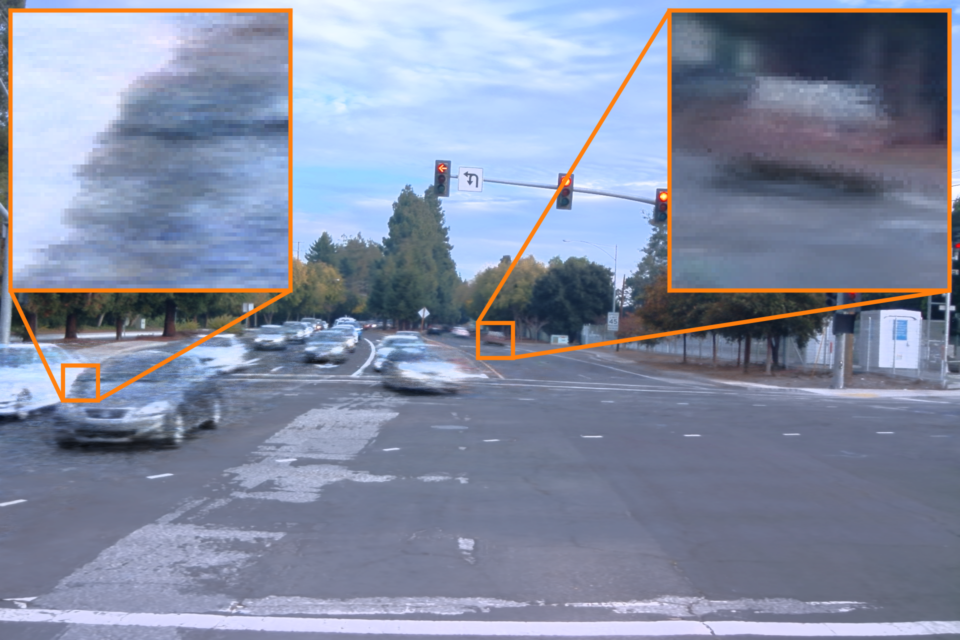}
         \\
    \end{tblr}
    \begin{tabular}{cccc}
        \centering \parbox{\dimexpr0.25\textwidth-2\tabcolsep\relax}{\centering Ground Truth} &
        \centering \parbox{\dimexpr0.25\textwidth-2\tabcolsep\relax}{\centering Ours} &
        \centering \parbox{\dimexpr0.25\textwidth-2\tabcolsep\relax}{\centering OmniRe \cite{chen2025omnire}} &
        \centering \parbox{\dimexpr0.25\textwidth-2\tabcolsep\relax}{\centering EmerNeRF \cite{yang2024emernerf}}
    \end{tabular}
    \caption{Extended visual results on the Waymo Dataset \cite{sun2020scalability}, showcasing reconstructed sequences s-141, s-975, and s-125. Our model demonstrates the ability to capture fine details, such as water droplets and the non-rigid motion of cyclists' feet, while exhibiting fewer artifacts compared to baseline methods. 
    }
    \label{fig:comparison_vis_quality_other}
\end{figure}
In \figref{fig:comparison_vis_quality_other} we demonstrate our reconstructions on 3 more scenes, showcasing its handling of complex and fine details like water droplets, the feets of cyclists which we are, despite their challenging motion, capable to represent accurately. Further we highlighted our improved handling of distant objects emphasizing our visual performance increases w.r.t our baselines.

\begin{figure}[tb]
    \setlength{\imageWidth}{0.25\textwidth}
    \settoheight{\imagerowheight}{\includegraphics[width=\imageWidth]{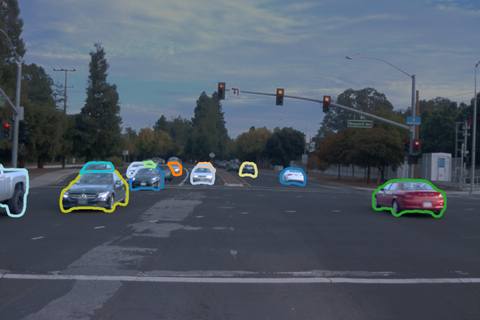}} 
    \newcommand{\gridImage}[1]{%
        \includegraphics[width=\imageWidth]{#1}%
    }
    \begin{tblr}{colspec={c c c c},colsep=0mm, rowsep=-1.2mm, row{1-2}={ht=\imagerowheight}, cells={halign=c, valign=m}}
         \gridImage{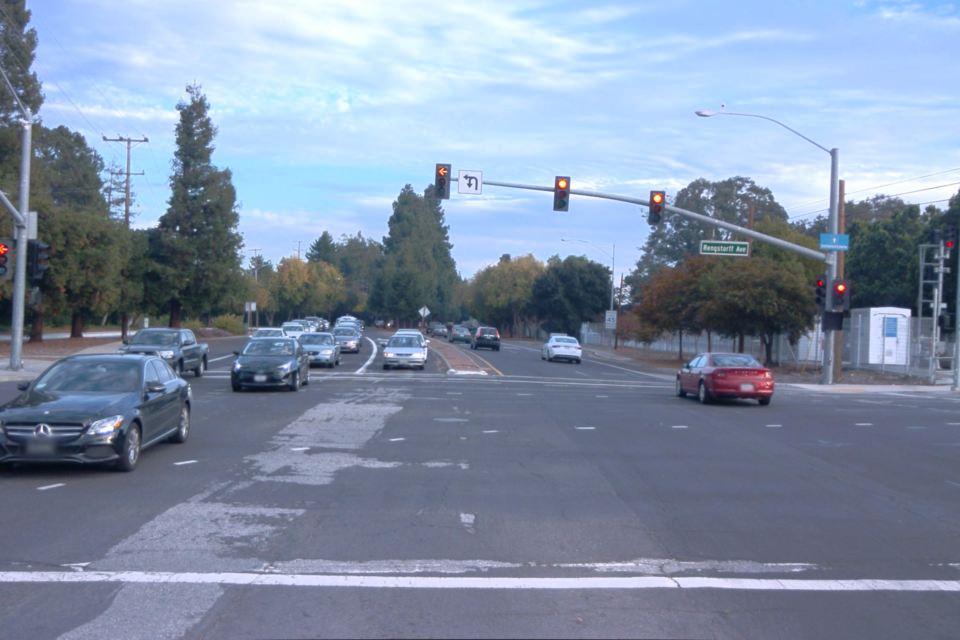}& %
         \gridImage{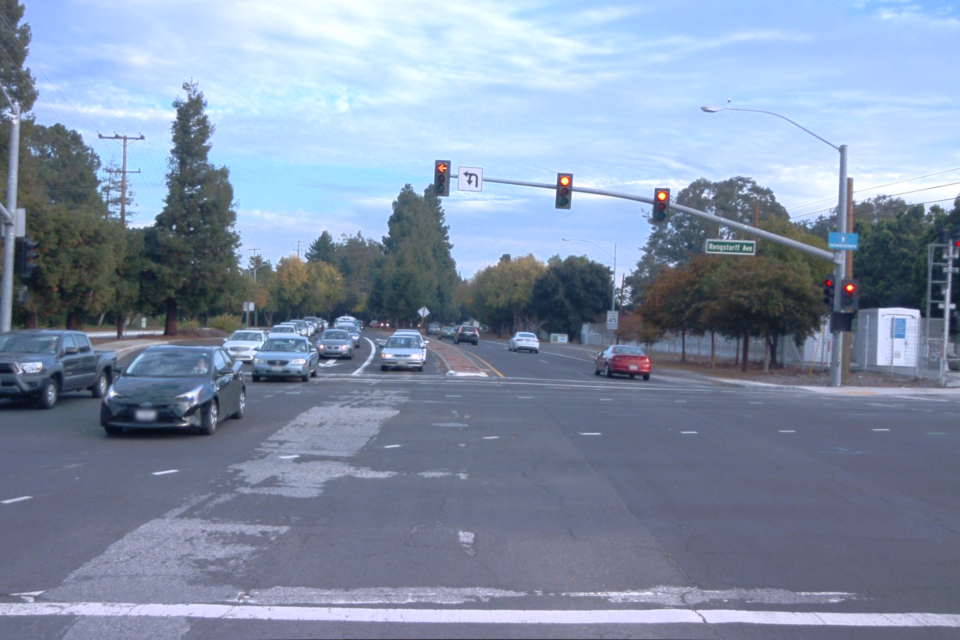}& %
         \gridImage{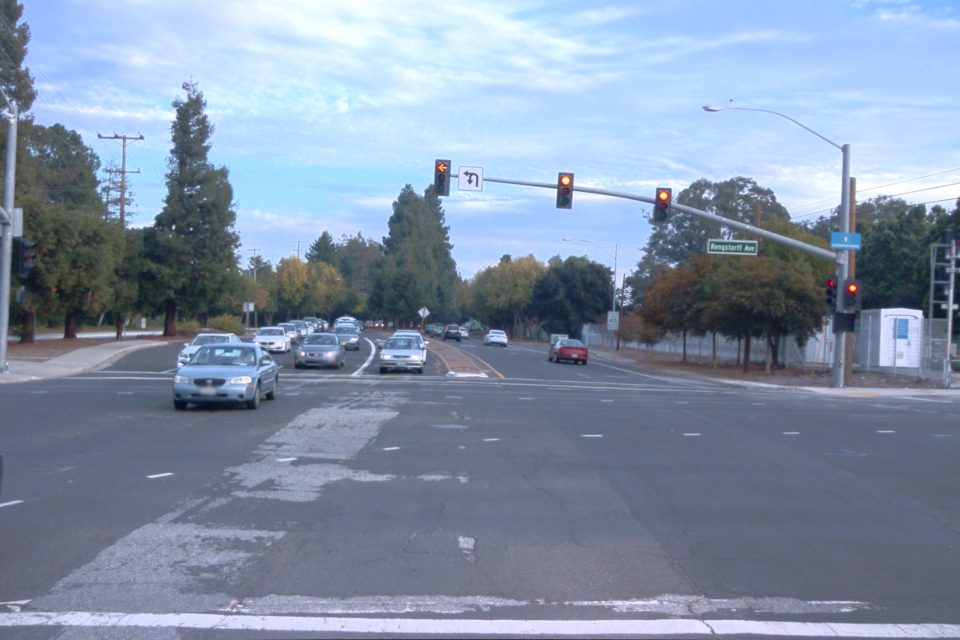}&
         \gridImage{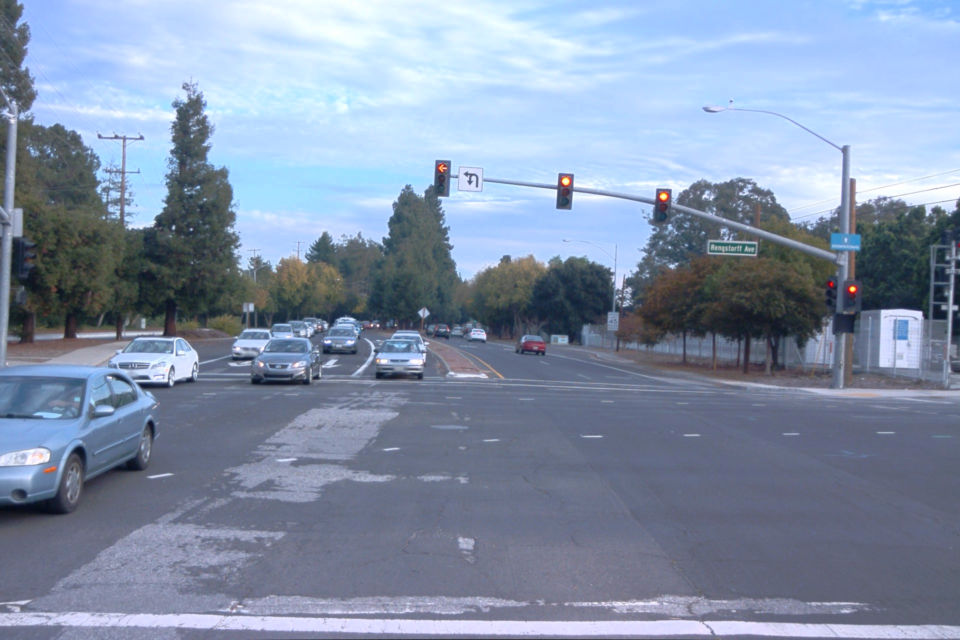}\\
        \gridImage{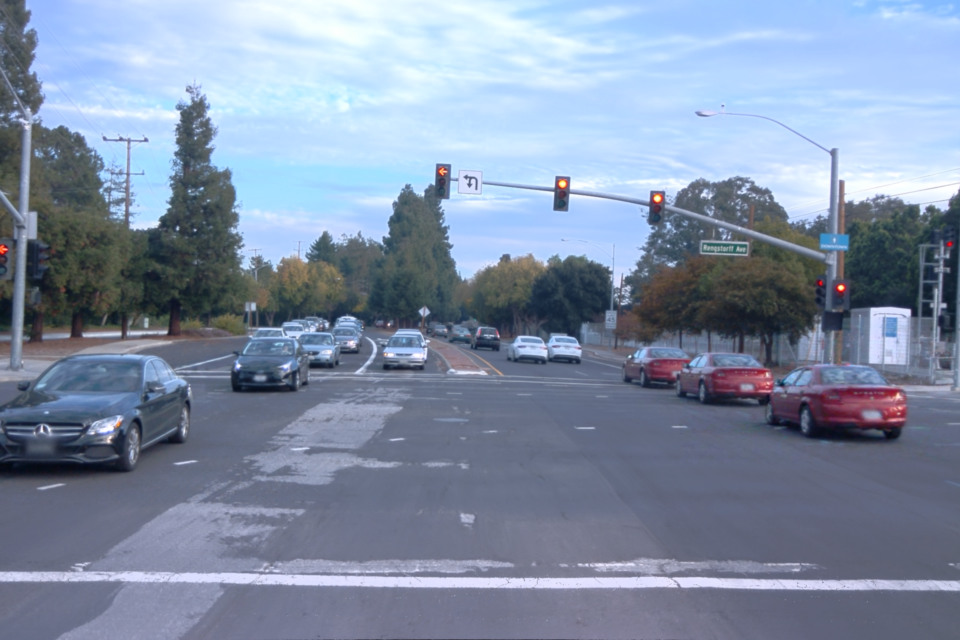}& %
         \gridImage{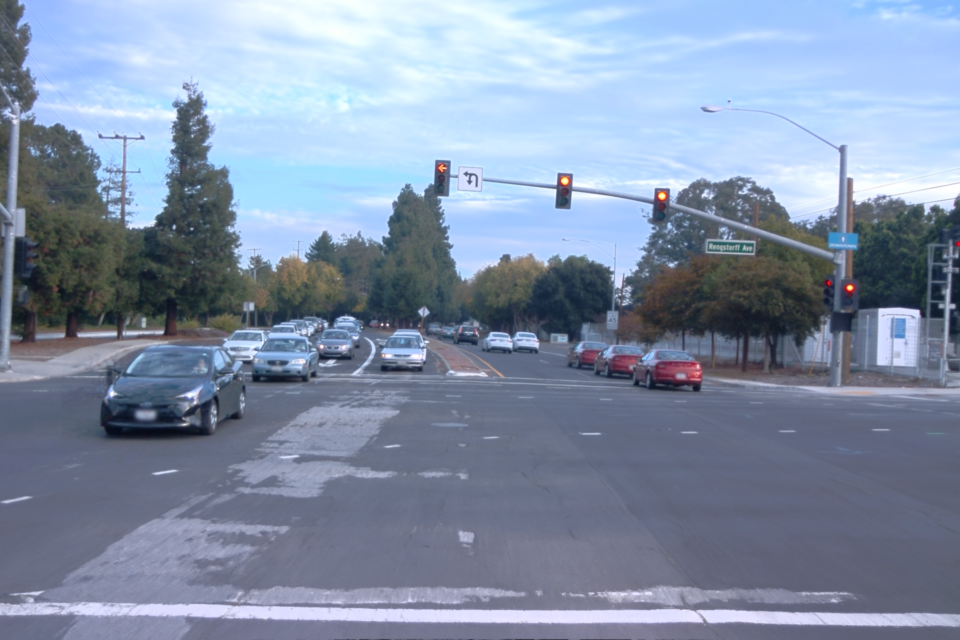}& %
         \gridImage{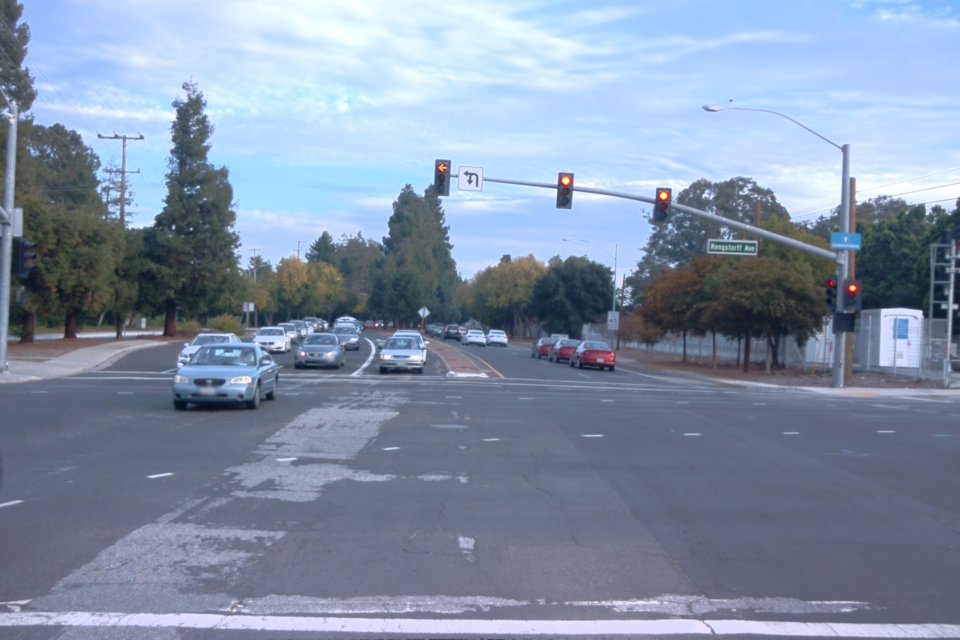}&
         \gridImage{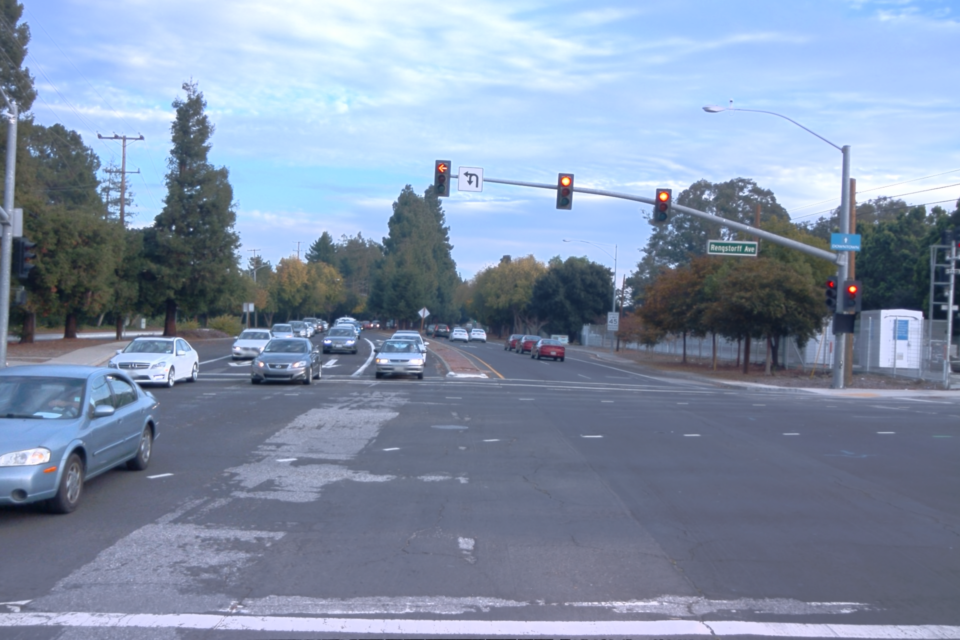}\\
    \end{tblr}
    \caption{Illustration of editing operations on the s-125 scene. We showcase: the removal of an existing object (left truck), copying and spatially shifting the white car, and the temporal manipulation of the red car through duplication and a ±5 timestamp shift.}
    \label{fig:waymo_editing_second}
\end{figure}

Further, we illustrate in \figref{fig:waymo_editing_second} the comprehensive editing capabilities of our method on the Waymo s-125 scene. We demonstrate three distinct types of manipulations: object removal (specifically, the truck on the left), object duplication / adding and precise spatial shifting (exemplified by the white car copied and moved by 2 units to the left and 0.5 units towards the camera), and temporal manipulation (achieved by duplicating the red car and shifting its presence by ±5 timestamps). These examples collectively highlight our method's versatility in handling edits that remain consistent with our flow- and view-dependent model, allowing for precise control over scene composition and dynamics.

\begin{figure}[tb]
    \centering
    \setlength{\imageWidth}{0.25\textwidth}
    \settoheight{\imagerowheight}{\includegraphics[width=\imageWidth]{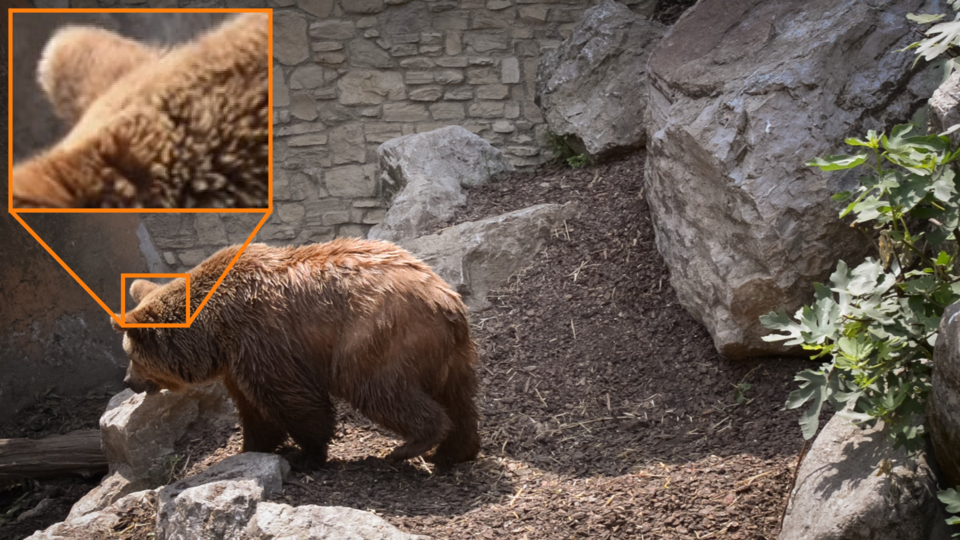}} 
    \newcommand{\gridImage}[1]{%
        \includegraphics[width=\imageWidth]{#1}%
    }
    \newcommand{\rowLabel}[1]{%
        \raisebox{0.5\imagerowheight}{\vcenter{\hbox{\rotatebox{90}{#1}}}}%
    }
    \begin{tblr}{colspec={c c c c},colsep=0mm, rowsep=-1.2mm, row{1-4}={ht=\imagerowheight}, cells={halign=c, valign=m}}
         \gridImage{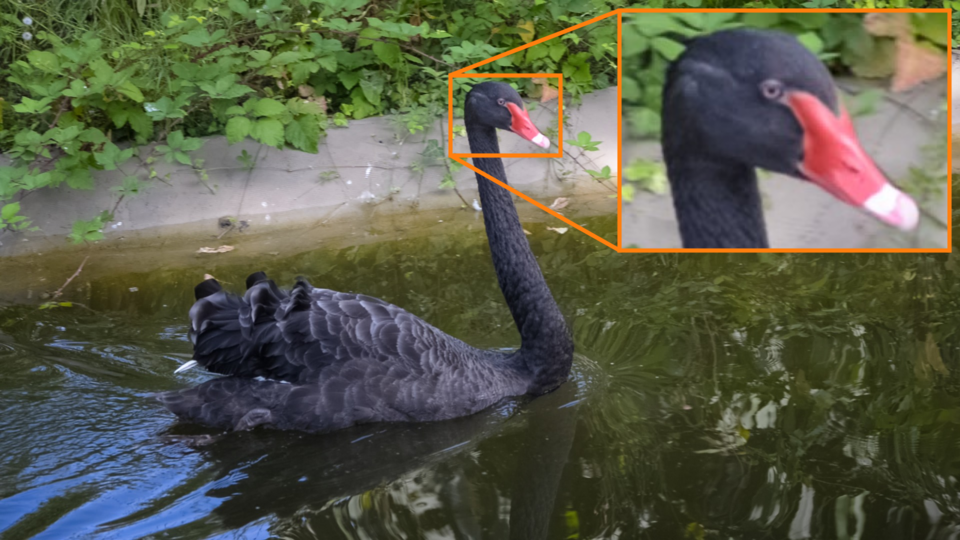} & %
         \gridImage{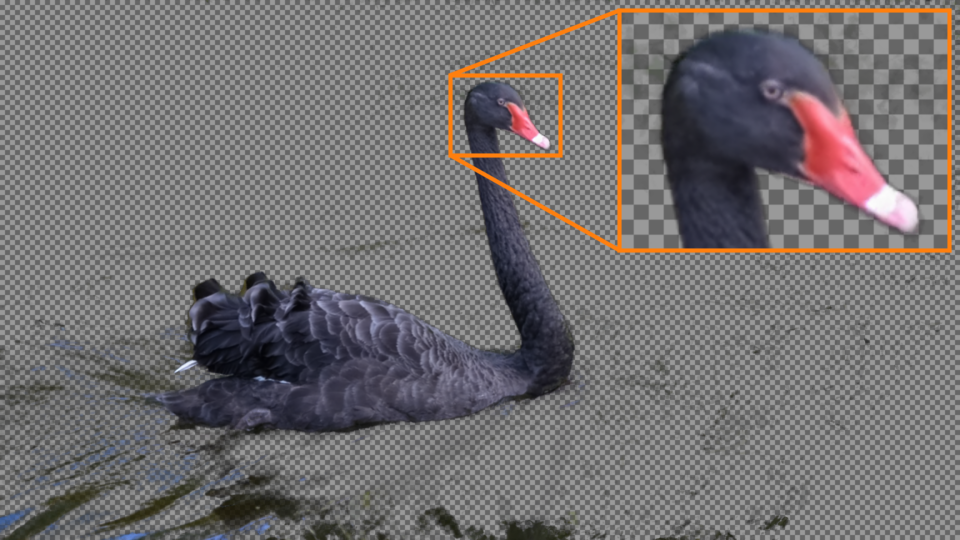} & %
         \gridImage{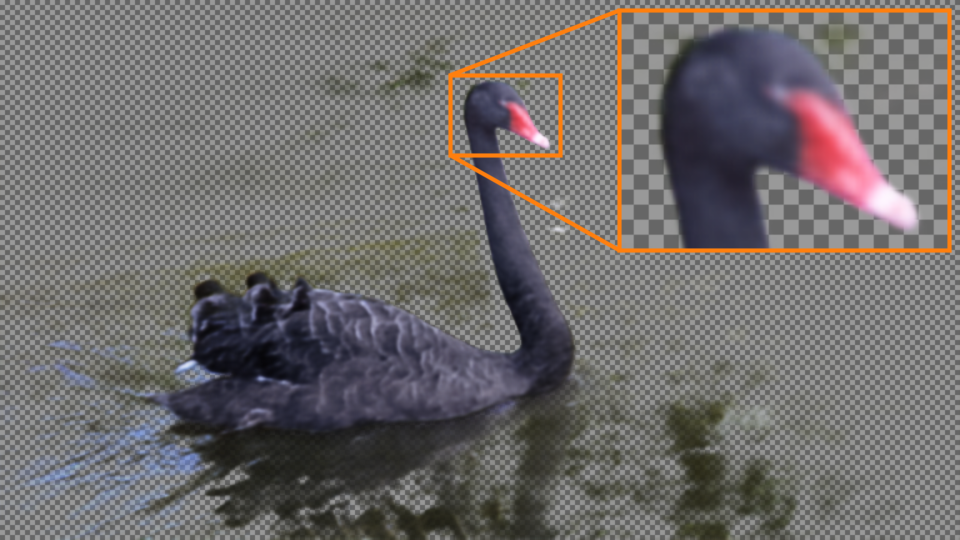} & %
         \gridImage{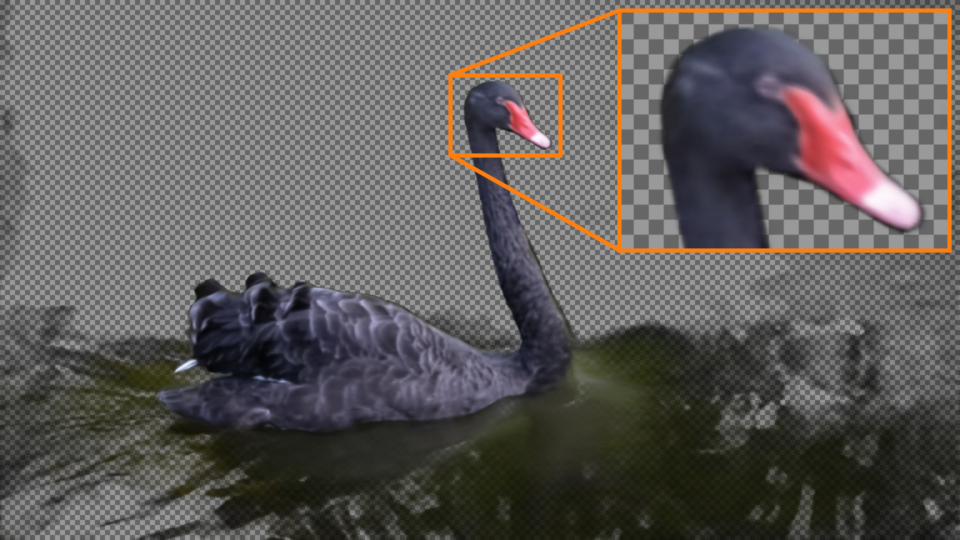} \\
         & %
         \gridImage{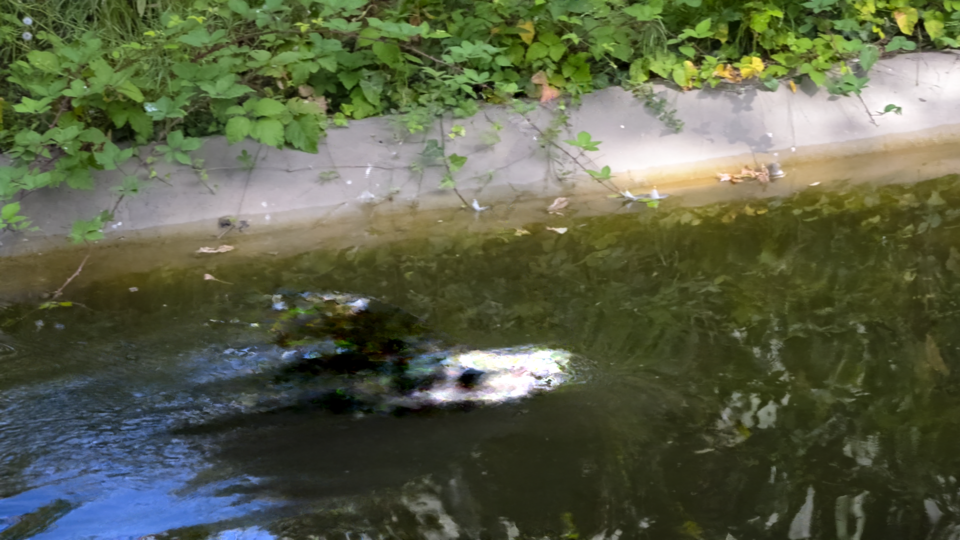} & %
         \gridImage{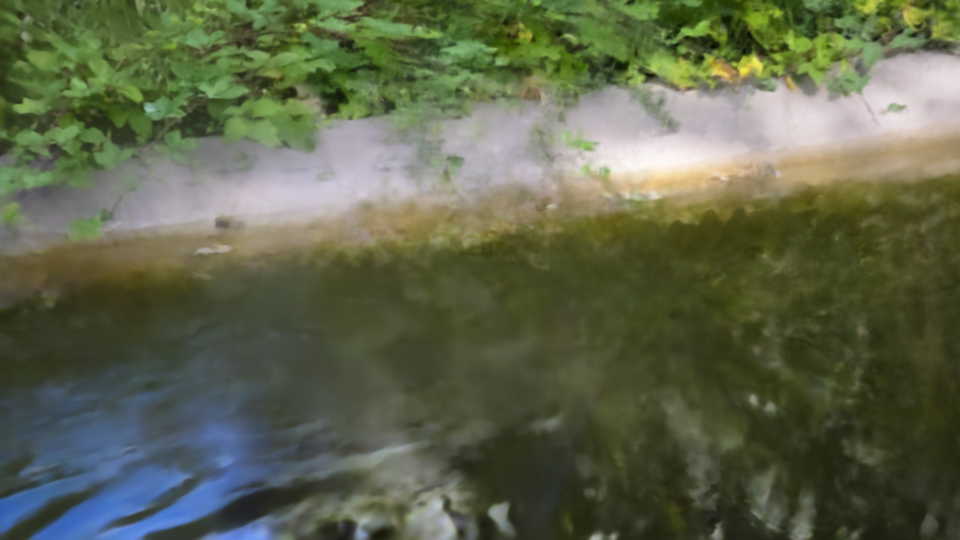} &  %
         \gridImage{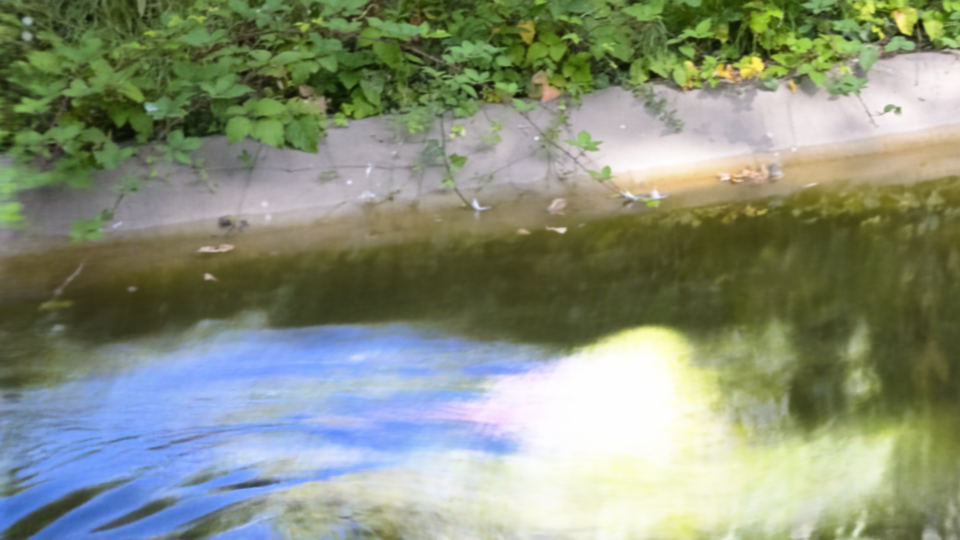} \\
         \gridImage{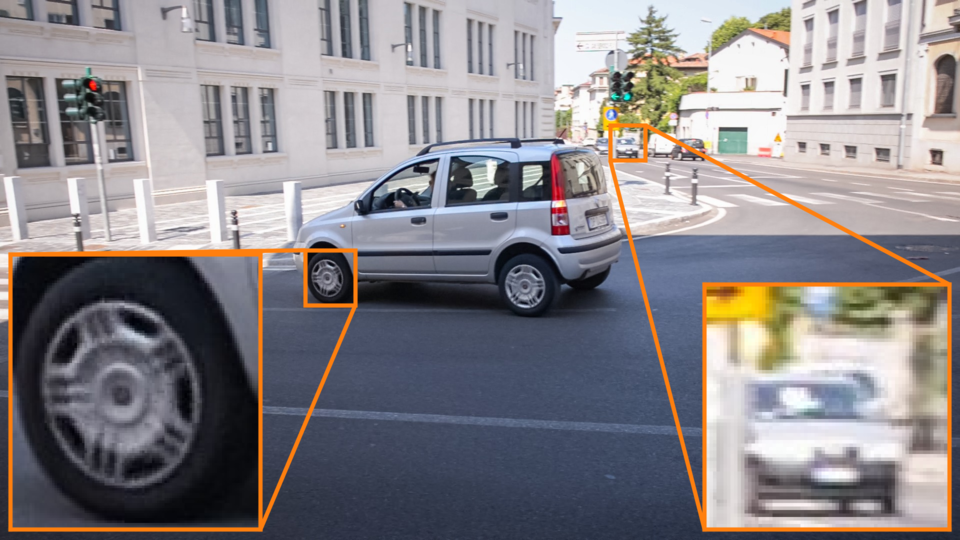} & %
         \gridImage{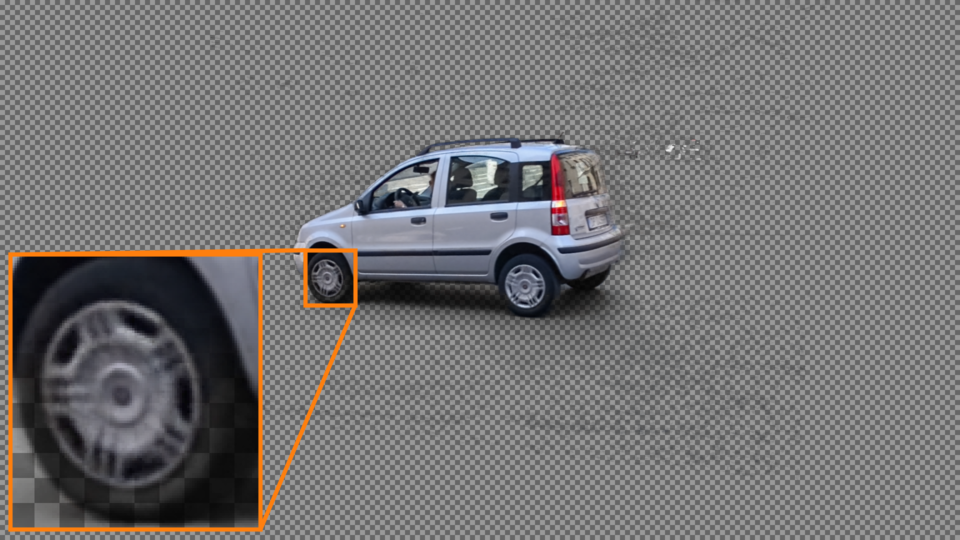} & %
         \gridImage{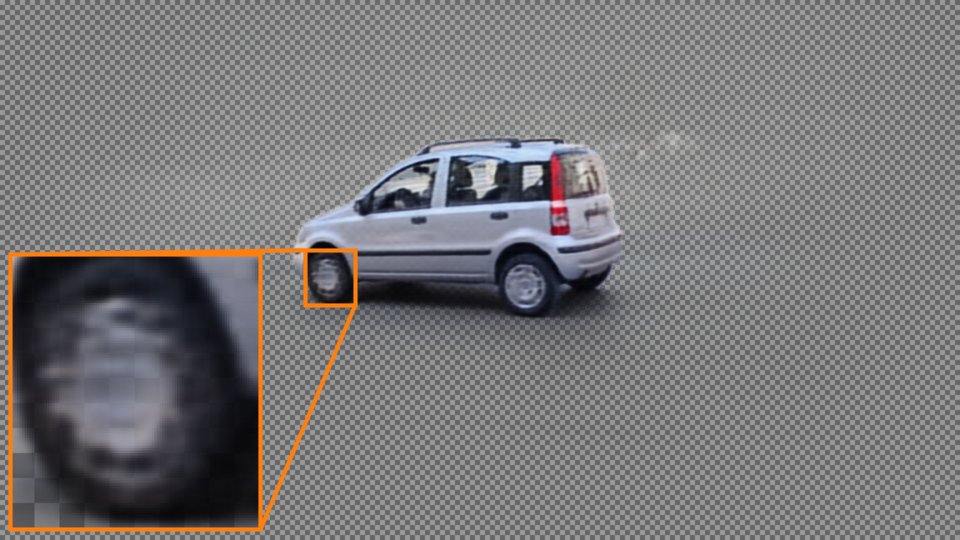} &  %
         \gridImage{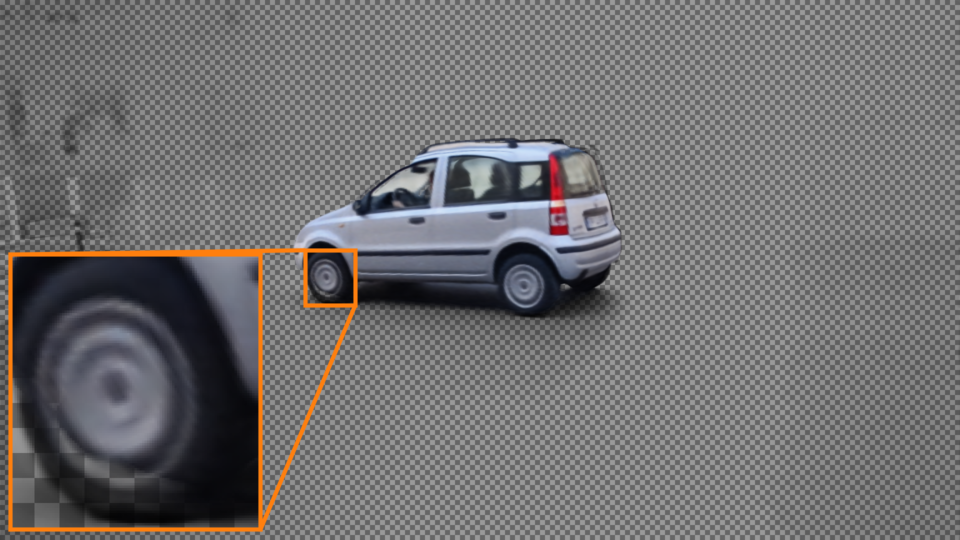} \\
         & %
         \gridImage{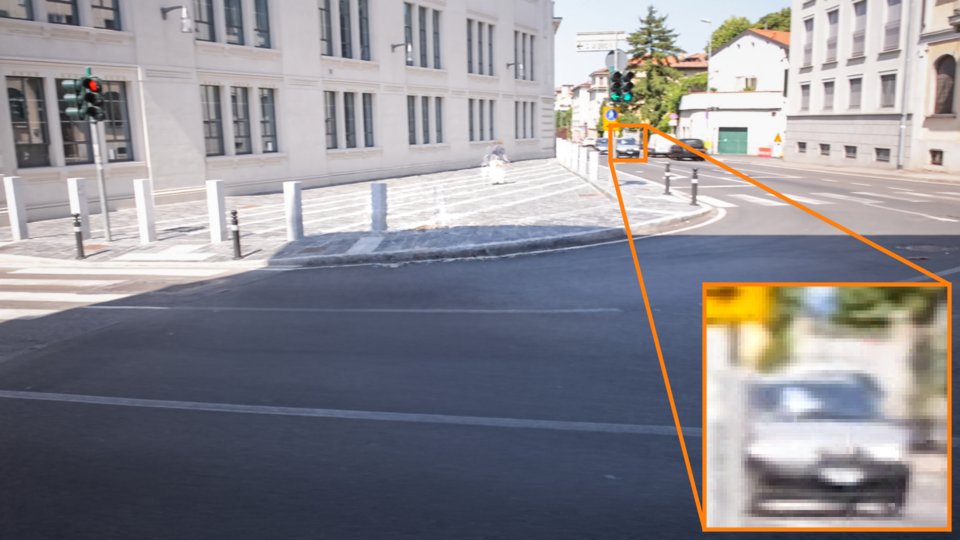} & %
         \gridImage{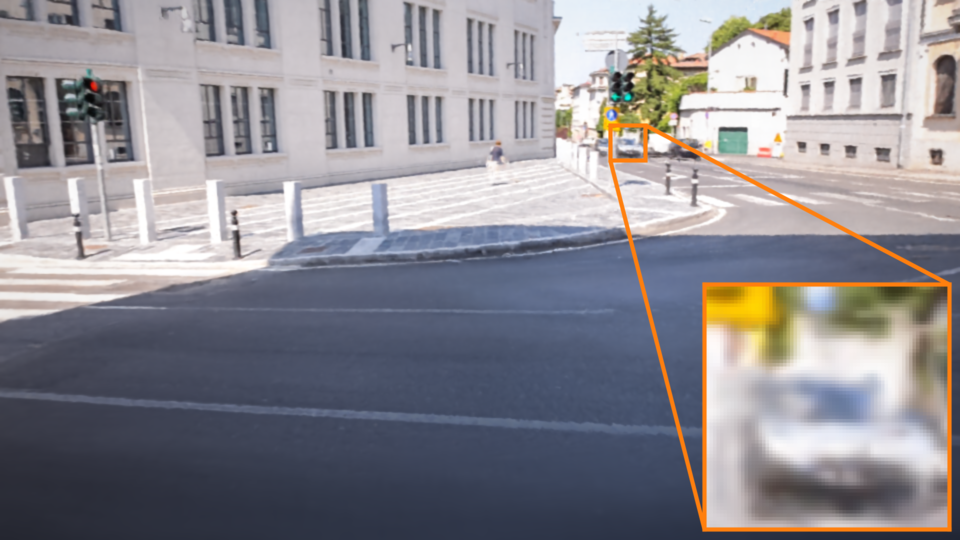} &  %
         \gridImage{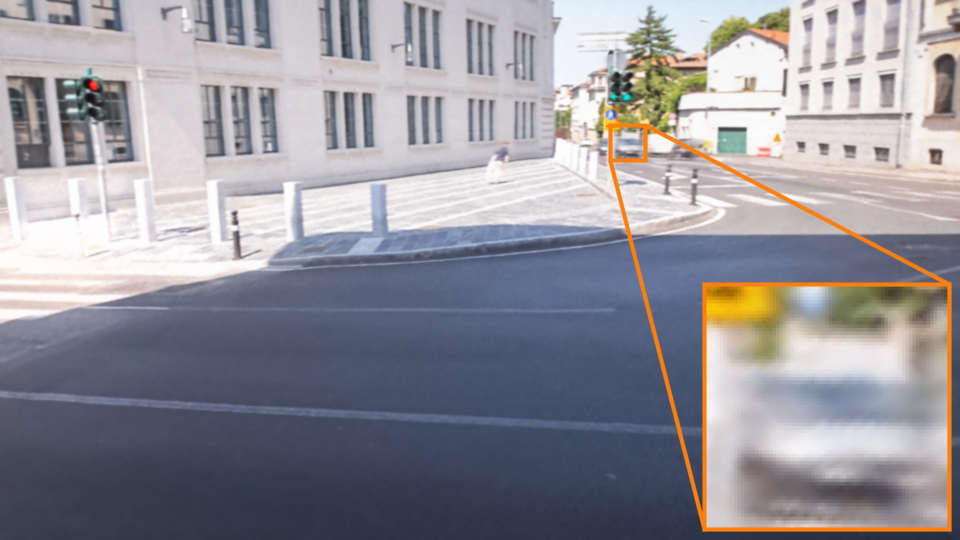} \\
    \end{tblr}
    \vspace{-1mm} 
    \begin{tabular}{cccc}
        \centering \parbox{\dimexpr0.25\textwidth-2\tabcolsep\relax}{\centering Ground Truth} &
        \centering \parbox{\dimexpr0.25\textwidth-2\tabcolsep\relax}{\centering Ours} &
        \centering \parbox{\dimexpr0.25\textwidth-2\tabcolsep\relax}{\centering ORF \cite{lin2023omnimatterf}} &
        \centering \parbox{\dimexpr0.25\textwidth-2\tabcolsep\relax}{\centering LNA \cite{kasten2021layered}}
    \end{tabular}
    \caption{Visual comparison and decomposition of the blackswan and car-shadow sequence within the DAVIS dataset. The insets stating difficult regions underlining the capabilities of our model in accurately representing highly textured regions (swan head), time-variant content (spinning wheels) and distant background objects.}
    \label{fig:davis_quality}
\end{figure}

\begin{figure}
    \centering
    \setlength{\imageWidth}{0.25\textwidth}
    \settoheight{\imagerowheight}{\includegraphics[width=\imageWidth]{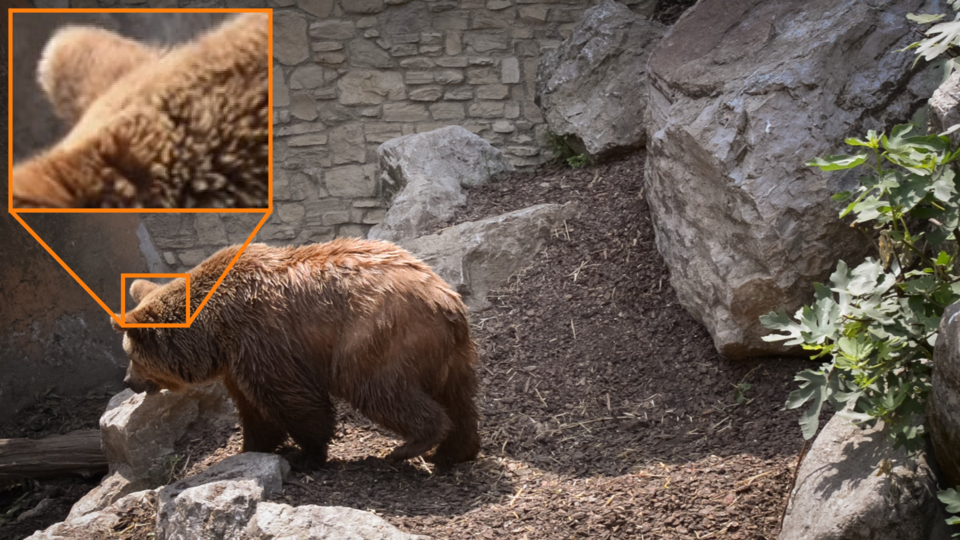}} 
    \newcommand{\gridImage}[1]{%
        \includegraphics[width=\imageWidth]{#1}%
    }
    \begin{tblr}{colspec={c c c c},colsep=0mm, rowsep=-1.2mm, row{1-3}={ht=\imagerowheight}, cells={halign=c, valign=m}}
         \gridImage{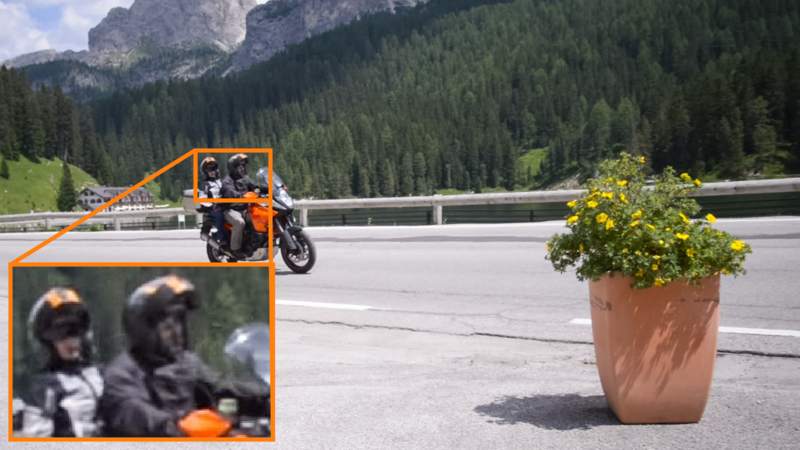}
         & %
         \gridImage{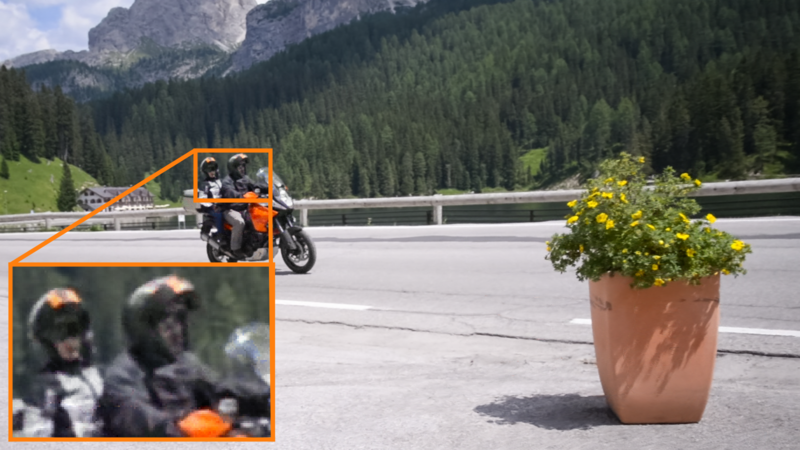}
         & %
         \gridImage{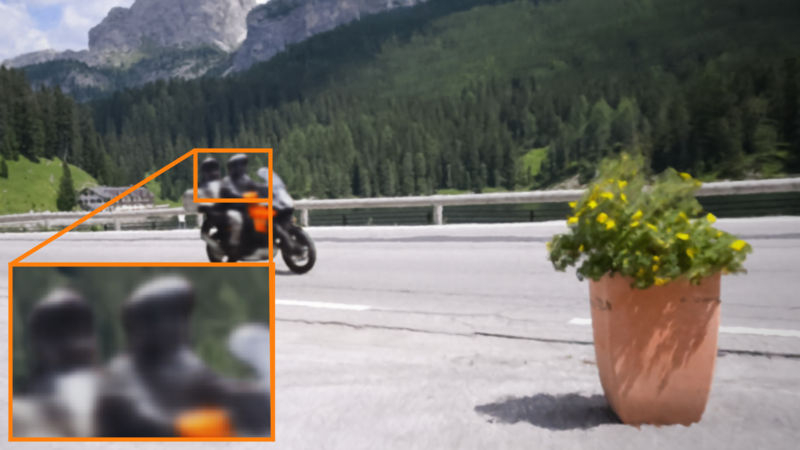}
         &  %
         \gridImage{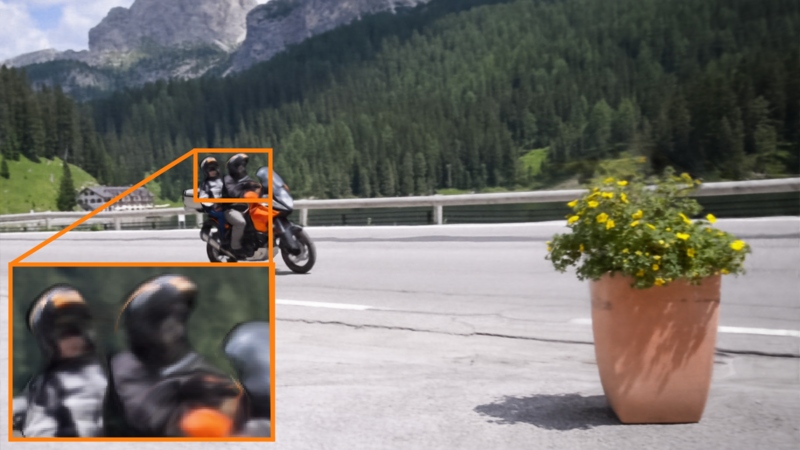}
         \\
         \gridImage{images/qual_results_davis/bear_010_GT_z_4_0}
         & %
         \gridImage{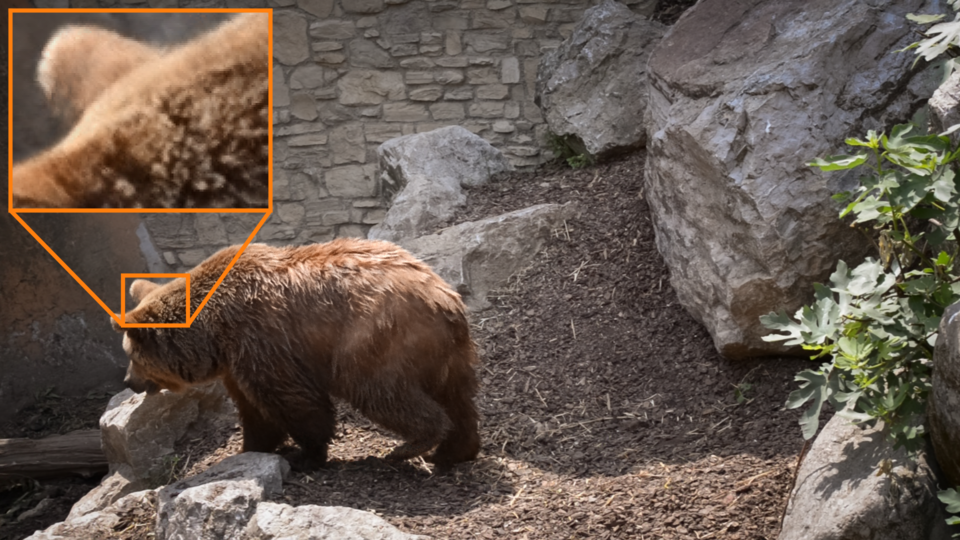}
         & %
         \gridImage{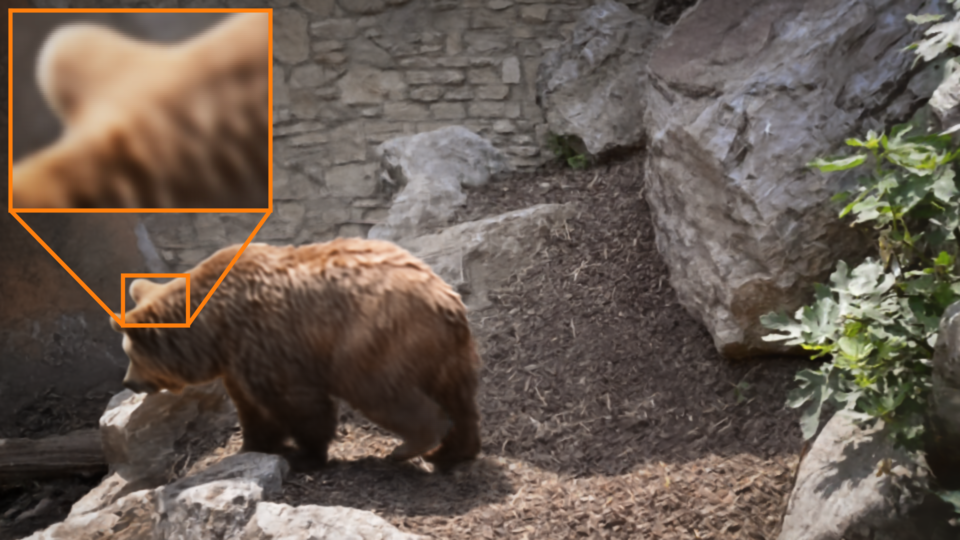}
         &  %
         \gridImage{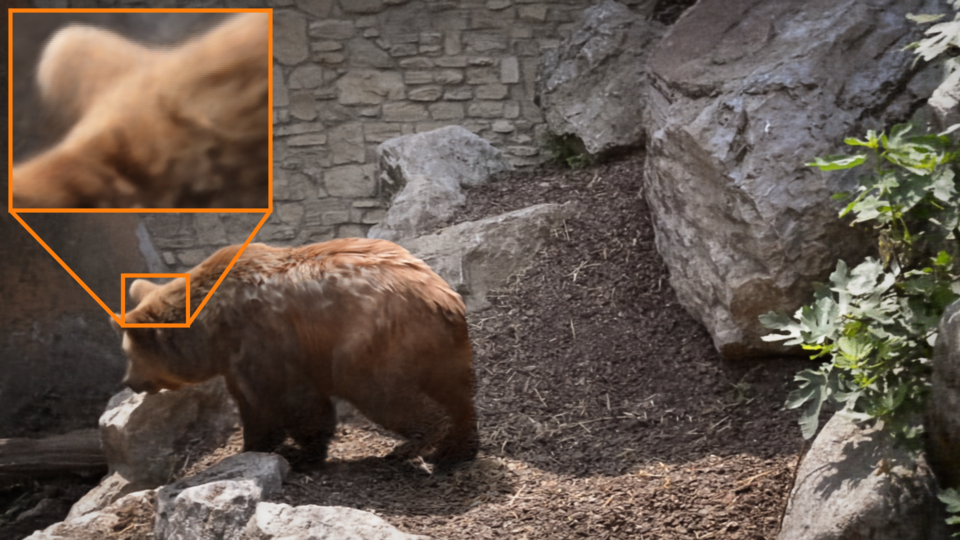}
         \\
         \gridImage{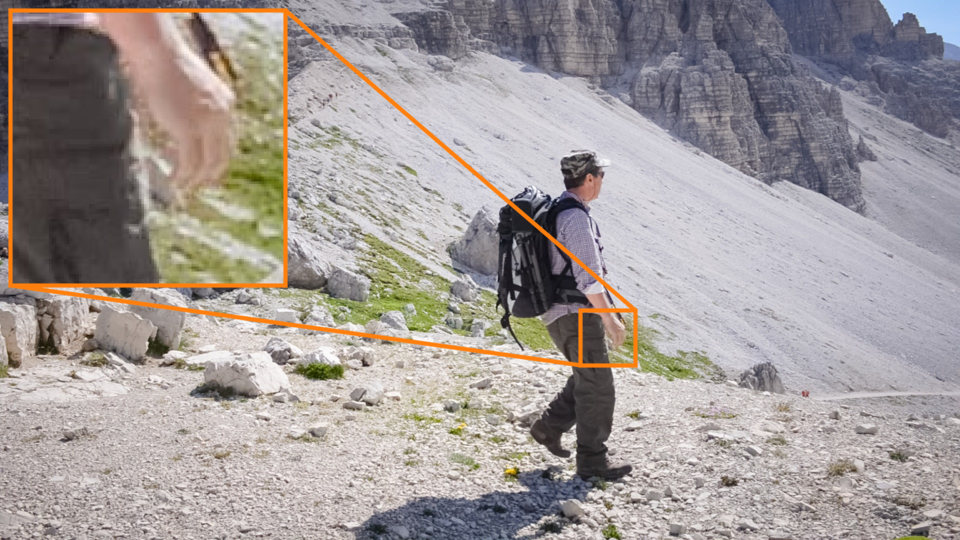}
         & %
         \gridImage{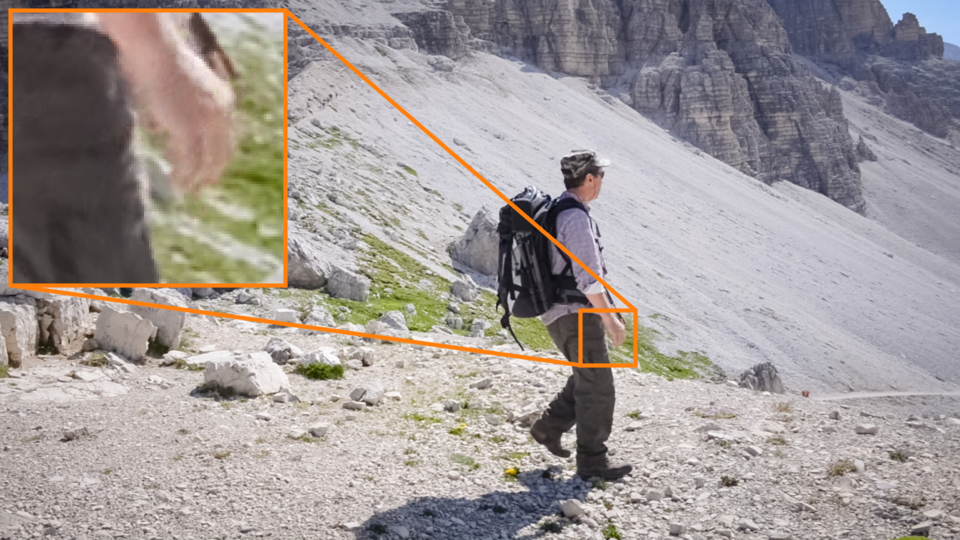}
         & %
         \gridImage{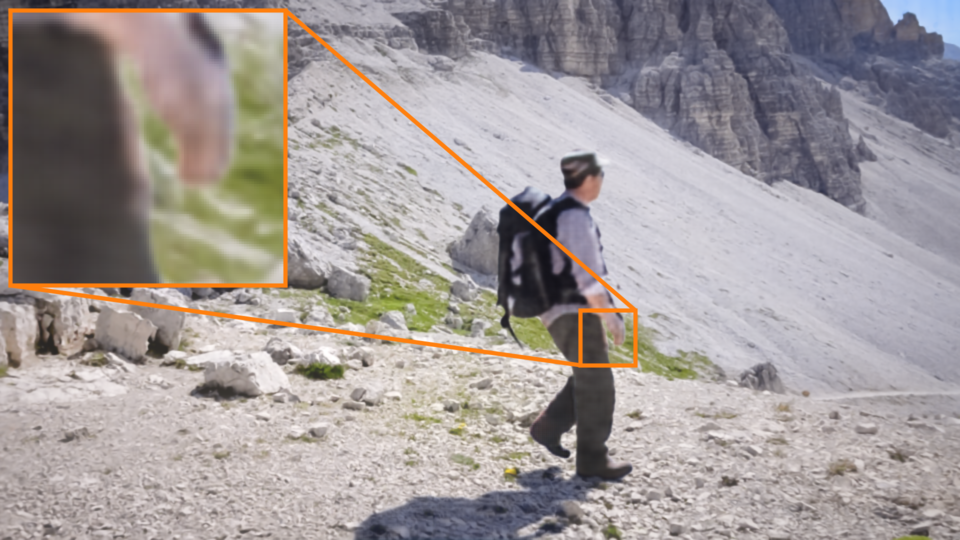}
         &  %
         \gridImage{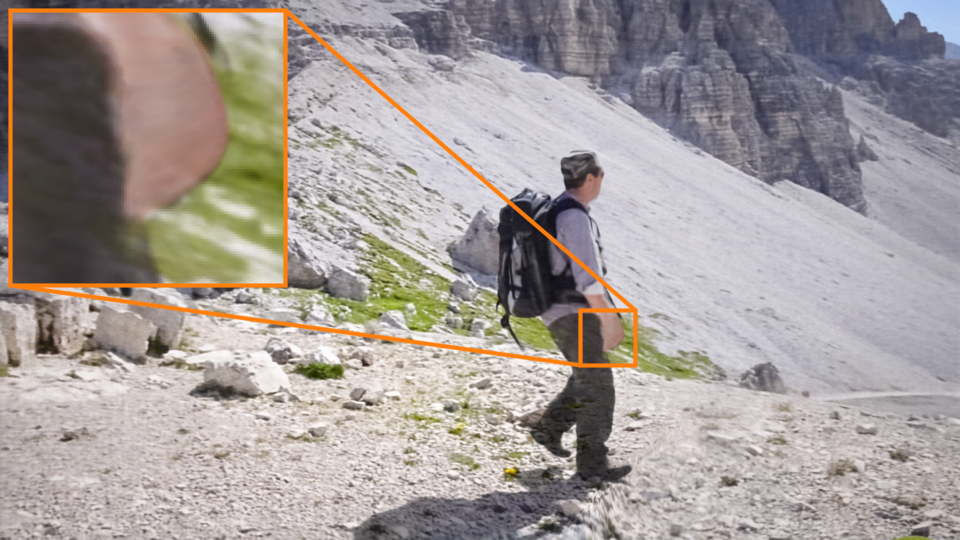}
    \end{tblr}
     \begin{tabular}{cccc}
        \centering \parbox{\dimexpr0.25\textwidth-2\tabcolsep\relax}{\centering Ground Truth} &
        \centering \parbox{\dimexpr0.25\textwidth-2\tabcolsep\relax}{\centering Ours} &
        \centering \parbox{\dimexpr0.25\textwidth-2\tabcolsep\relax}{\centering ORF \cite{lin2023omnimatterf}} &
        \centering \parbox{\dimexpr0.25\textwidth-2\tabcolsep\relax}{\centering LNA \cite{kasten2021layered}}
    \end{tabular}
    \caption{Additional visual examples of the DAVIS \cite{Perazzi2016} sequences motorbike, bear and hike, showcasing our models quality in representing fine and complex details on rigid and non-rigid foreground actors.}
    \label{fig:davis_quality_additional}
\end{figure}

Figure \ref{fig:davis_quality} showcases the visual effectiveness of our method on the DAVIS Dataset \cite{Perazzi2016}. Our view-dependent model components lead to significant improvements in visual quality compared to baselines, evident in increased sharpness and the detailed rendering of fine structures such as tire spokes. Notably, even with its view-dependent nature, our method produces reasonable background estimates in occluded regions, as illustrated by the car-shadow example. Figure \ref{fig:davis_quality_additional} presents results on three additional challenging DAVIS sequences where our method outperforms baselines: 1) motorbike: capturing the fast-moving foreground with fidelity; 2) bear: accurately modeling non-rigid motion and intricate fur texture; and 3) hike: handling actor movement against a complex, high-depth background.

\begin{figure}[tb]
    \setlength{\imageWidth}{0.25\textwidth}
    \settoheight{\imagerowheight}{\includegraphics[width=\imageWidth]{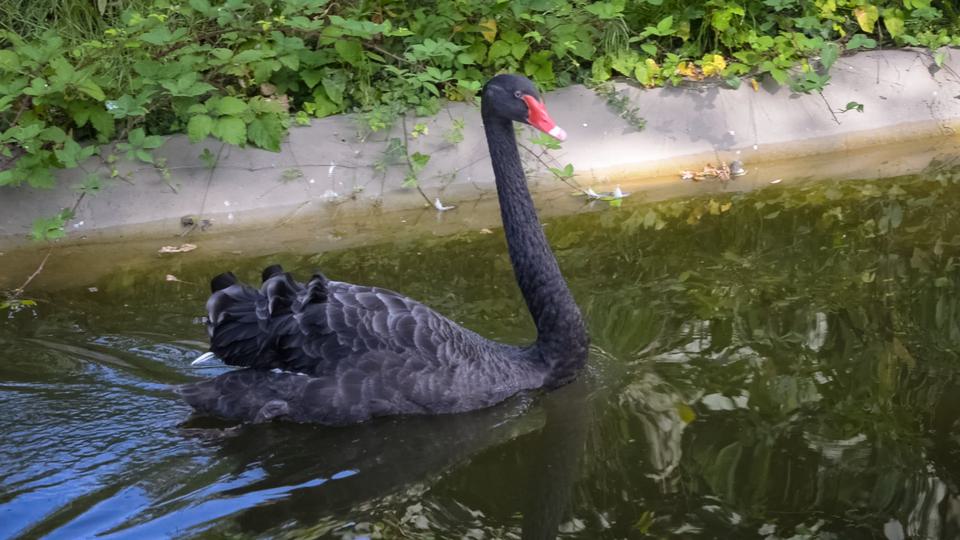}} 
    \newcommand{\gridImage}[1]{%
        \includegraphics[width=\imageWidth]{#1}%
    }
    \begin{tblr}{colspec={c c c c},colsep=0mm, rowsep=-1.2mm, row{1-3}={ht=\imagerowheight}, cells={halign=c, valign=m}}
         \gridImage{images/edit_results/blackswan_by2/gt/00005}& %
         \gridImage{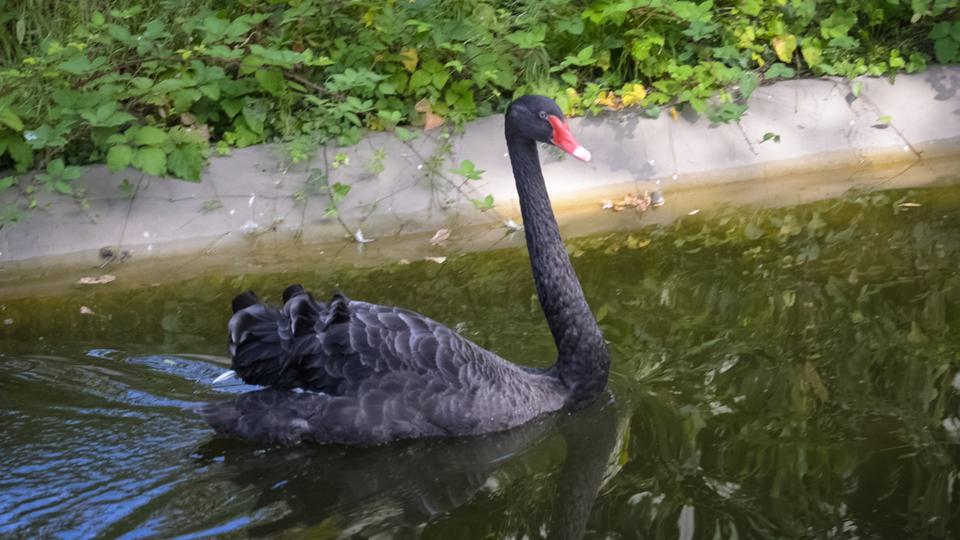}& %
         \gridImage{}& %
         \gridImage{} \\
         \gridImage{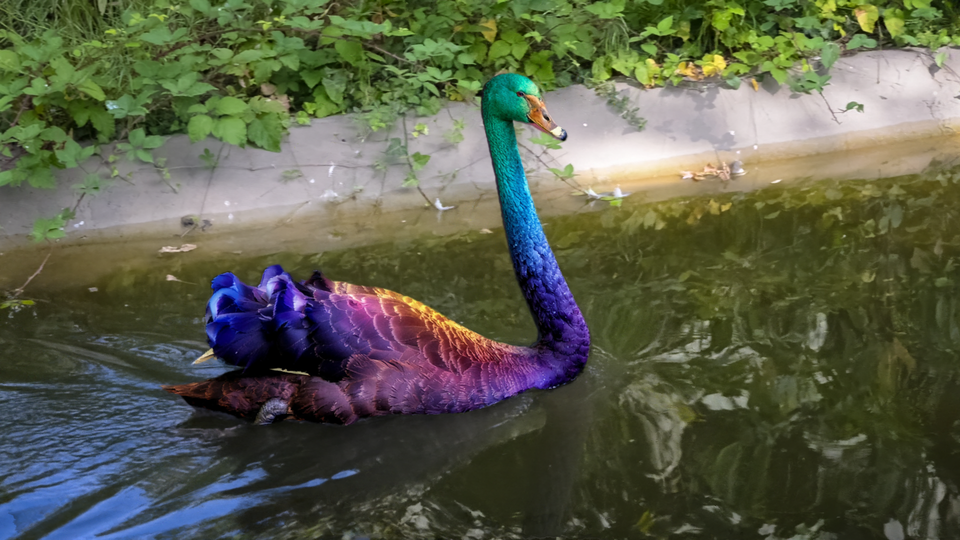}& %
         \gridImage{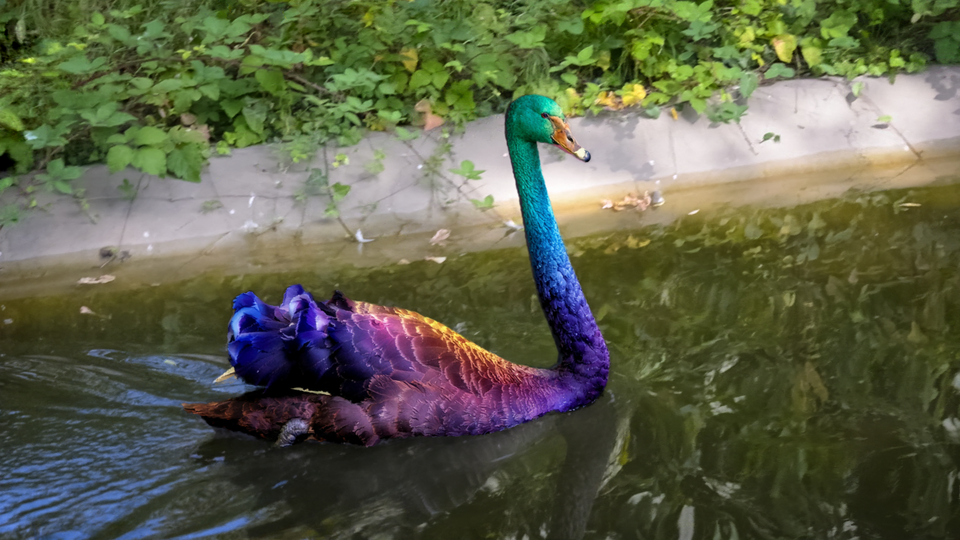}& %
         \gridImage{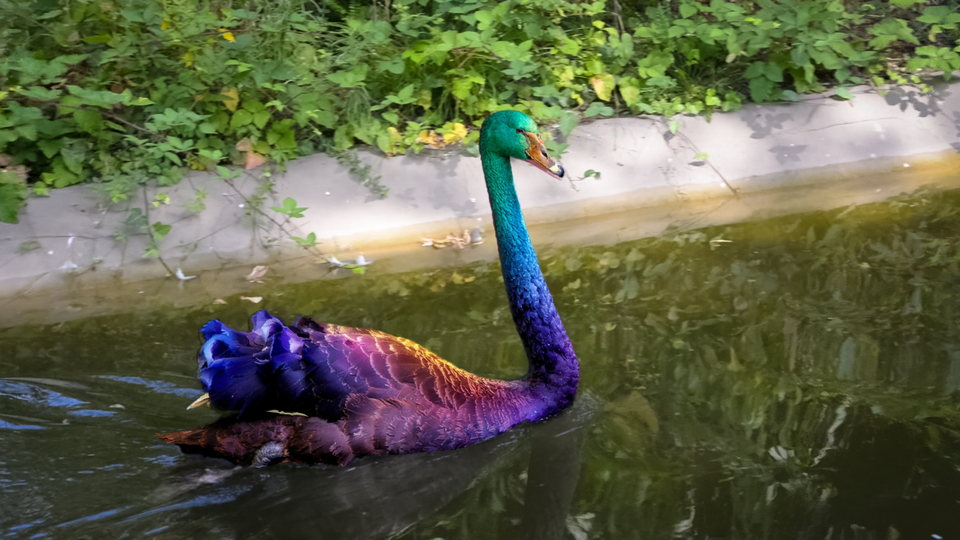}& %
         \gridImage{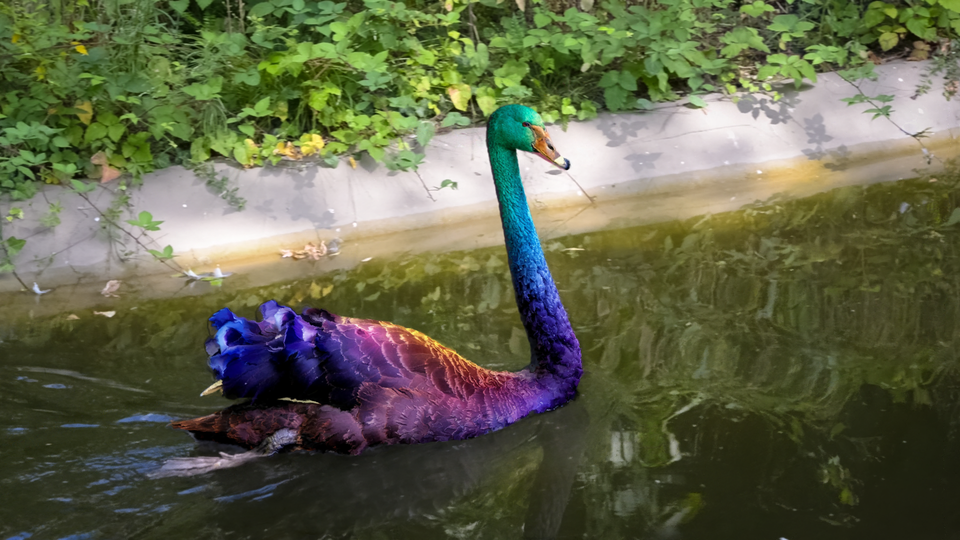} \\
         \gridImage{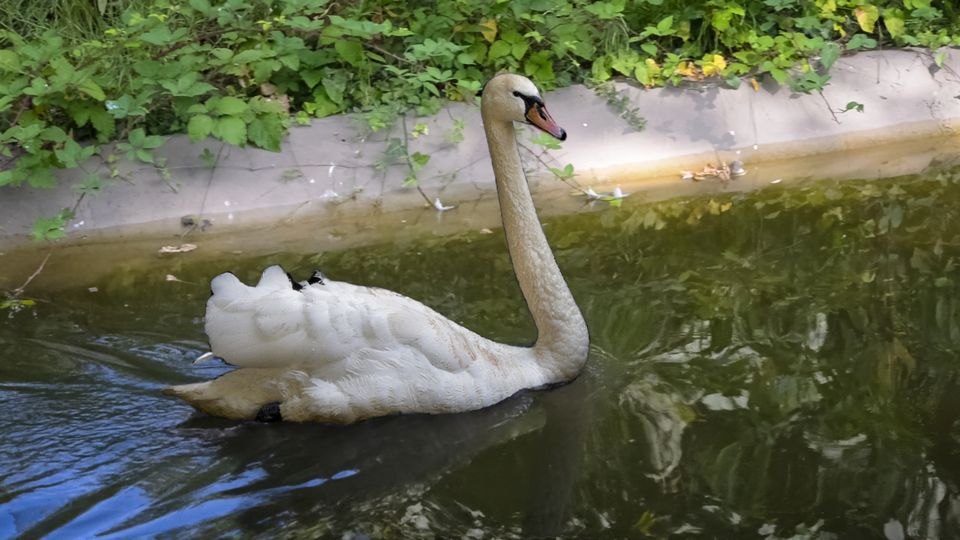}& %
         \gridImage{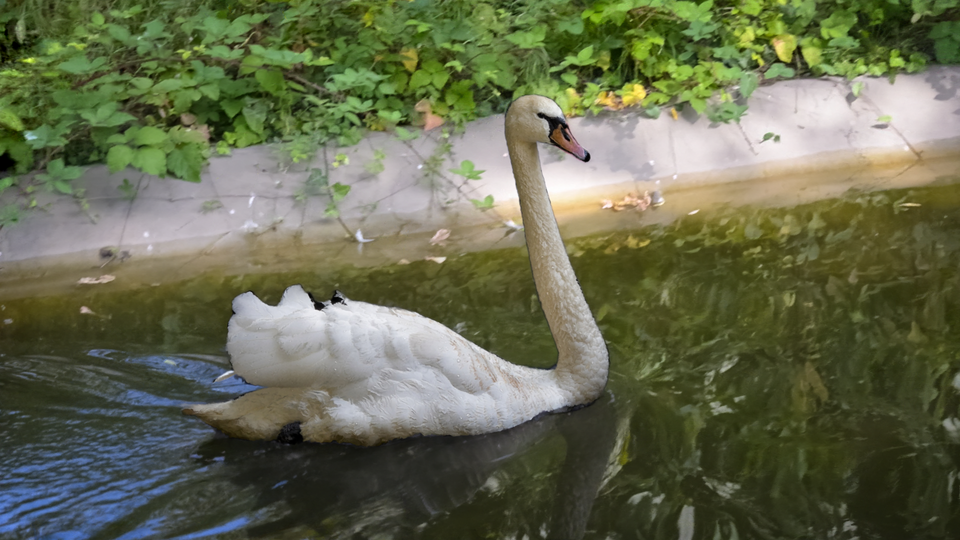}& %
         \gridImage{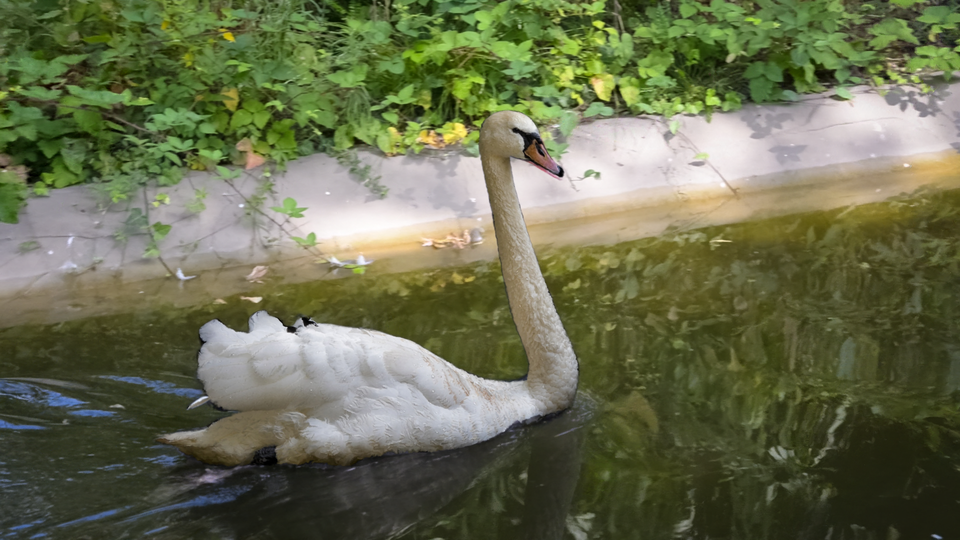}& %
         \gridImage{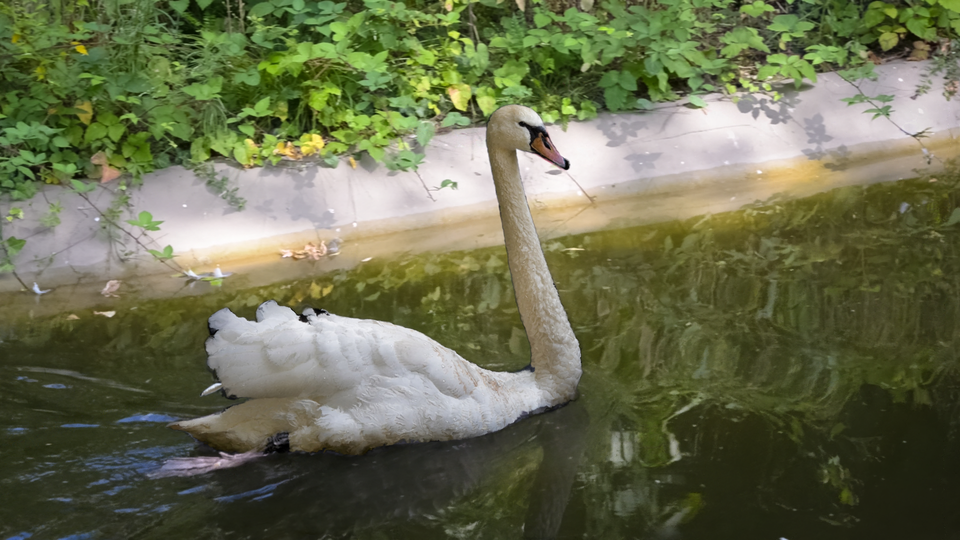} \\
    \end{tblr}
    \caption{Advanced texture editing applied to the blackswan sequence (DAVIS dataset \cite{Perazzi2016}). The top row presents the ground truth. The lower rows display edits where an off-the-shelf image generation model was leveraged to create rainbow and white swan texture variants. Using these generated textures, our method effectively propagates these localized changes consistently across all video frames, demonstrating robust temporal coherence.}
    \label{fig:davis_editing_swan}
\end{figure}

\begin{figure}[tb]
    \setlength{\imageWidth}{0.25\textwidth}
    \settoheight{\imagerowheight}{\includegraphics[width=\imageWidth]{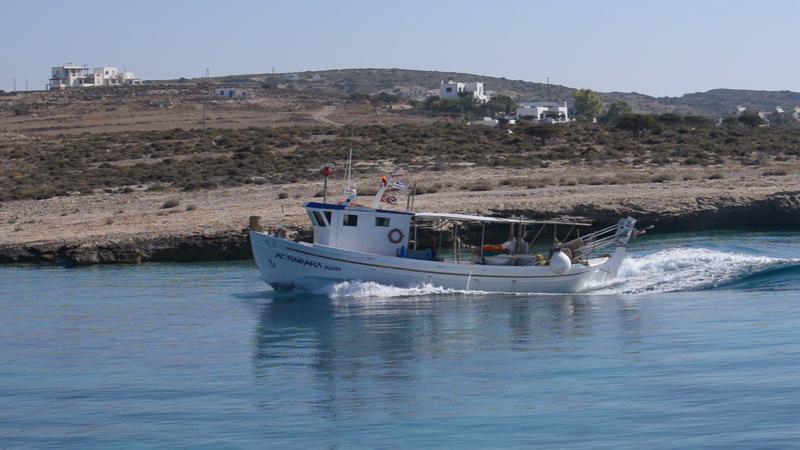}} 
    \newcommand{\gridImage}[1]{%
        \includegraphics[width=\imageWidth]{#1}%
    }
    \begin{tblr}{colspec={c c c c},colsep=0mm, rowsep=-1.2mm, row{1-3}={ht=\imagerowheight}, cells={halign=c, valign=m}}
         \gridImage{images/edit_results/boat_by2/gt/00000}& %
         \gridImage{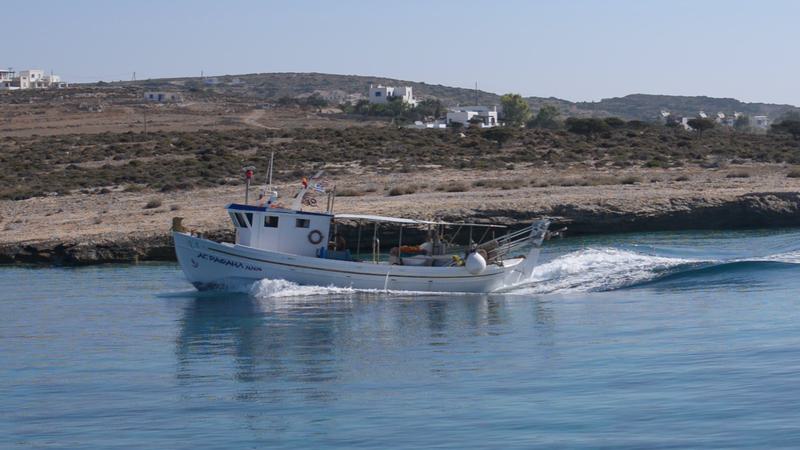}& %
         \gridImage{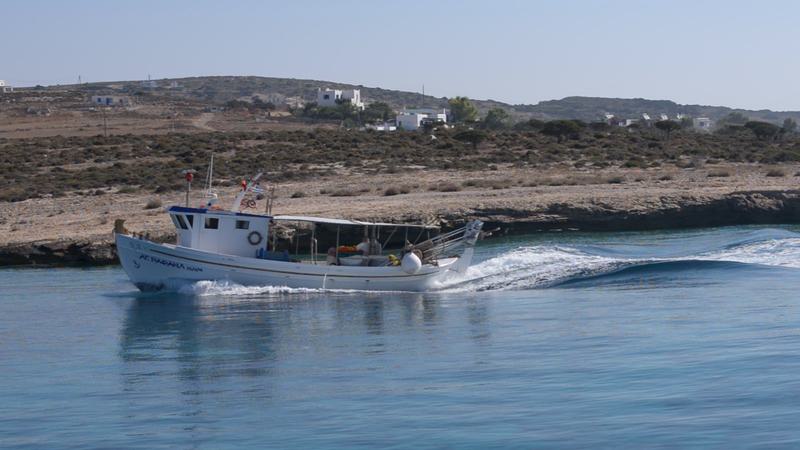}& %
         \gridImage{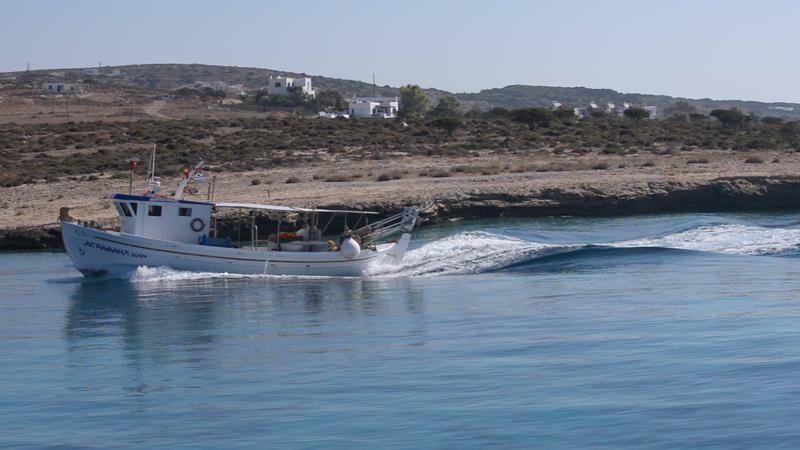} \\
         \gridImage{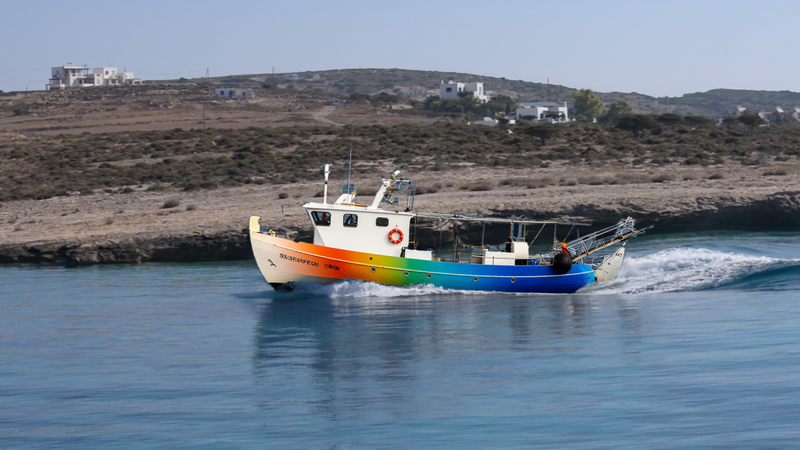}& %
         \gridImage{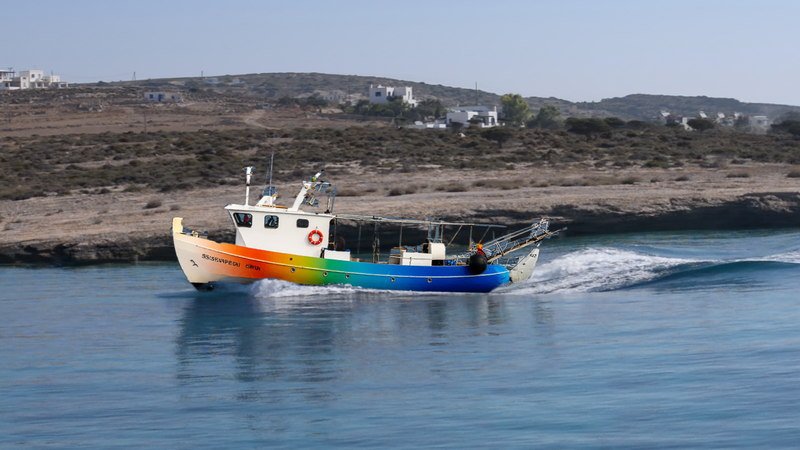}& %
         \gridImage{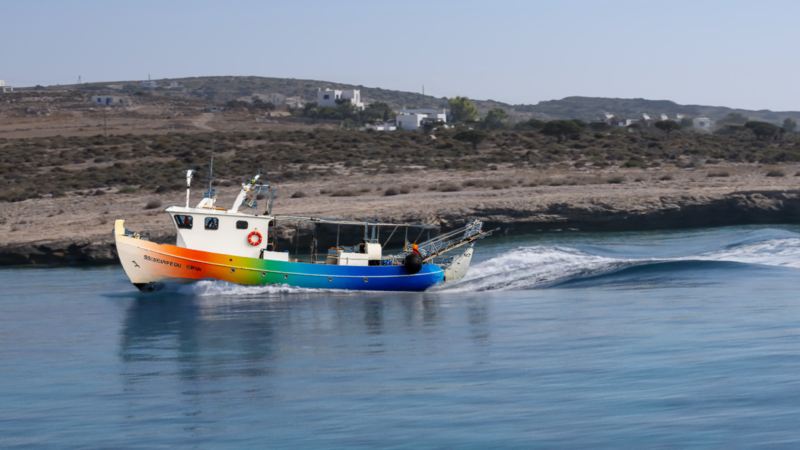}& %
         \gridImage{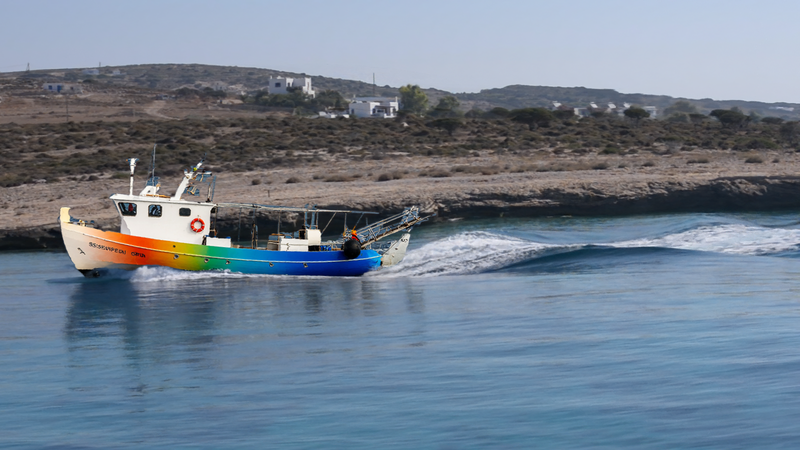} \\
         \gridImage{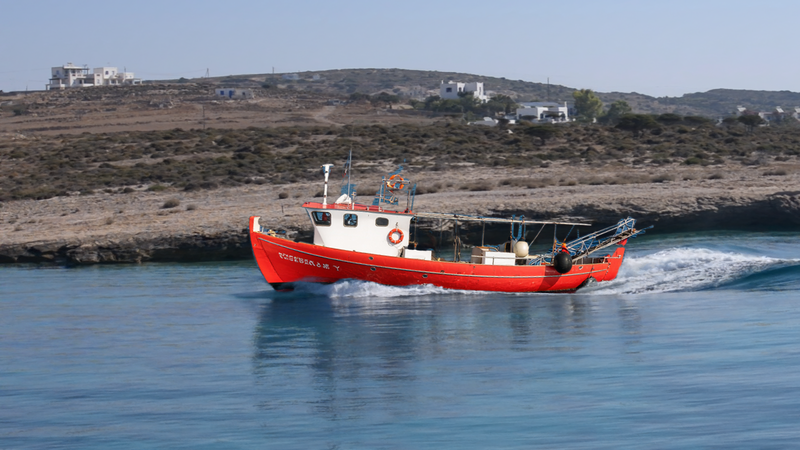}& %
         \gridImage{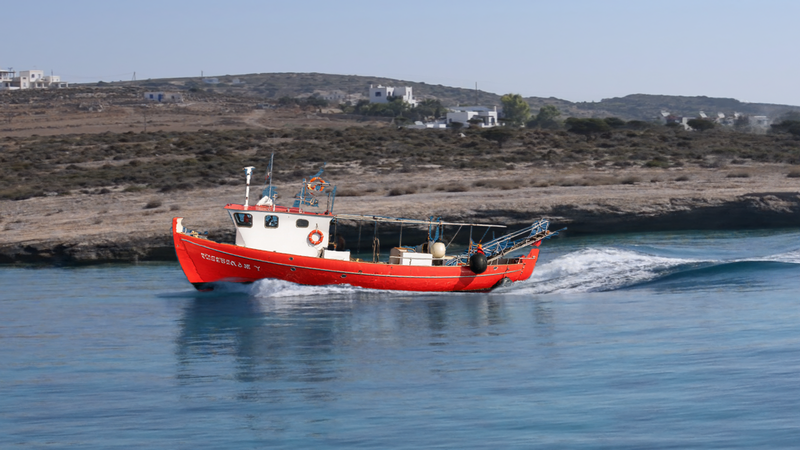}& %
         \gridImage{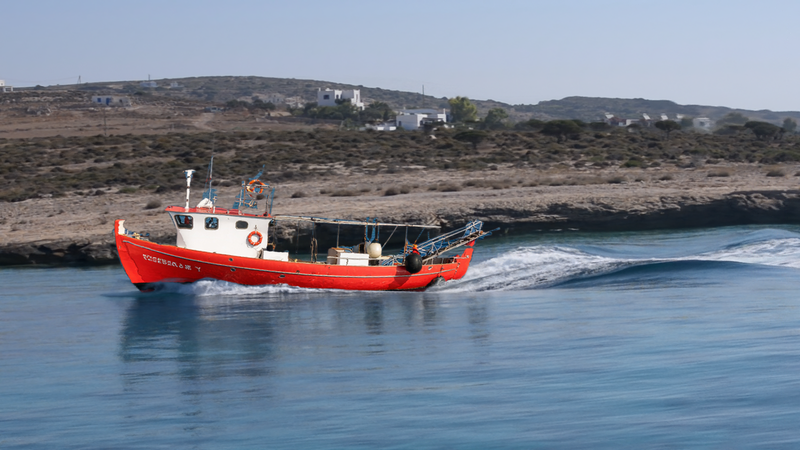}& %
         \gridImage{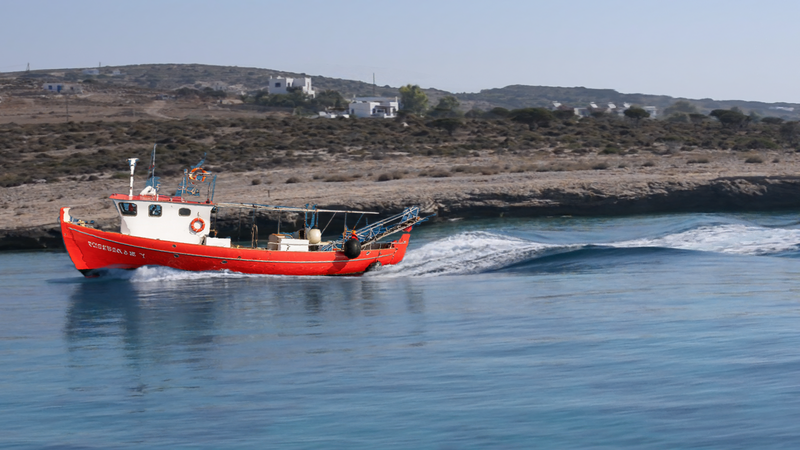} \\
    \end{tblr}
    \caption{Texture edits using DAVIS \cite{Perazzi2016} boat sequence. Similar to \figref{fig:davis_editing_swan}, we state the boat sequence, retexturing it with a rainbow and a red texture. The top row shows the ground truth, while the lower ones are the respective edits.}
    \label{fig:davis_editing_boat}
\end{figure}

Lastly, we state two more texture edits of DAVIS \cite{Perazzi2016} sequences in \figref{fig:davis_editing_swan} and \figref{fig:davis_editing_boat} where we utilized an off-the-shelf image generation model to create new textures for the decomposed foreground object, and applied these consistently along the video, yielding accurate edits even when changing the complete texture of these mostly rigid moving objects.

\subsection{Additional Gaussian Splatting Baselines}
\label{sec:additional_gaussian_splatting_baselines}

While \ac{ore} and \ac{erf} are recent 3DGS and NeRF baselines for object-specific and agnostic scene reconstruction, we broaden our analysis to include further dynamic 3D Gaussian Splatting methods.

To ensure comprehensive coverage against the state-of-the-art methods in dynamic 3DGS, we specifically evaluated three additional methods: Street Gaussians \ac{streetgs} \cite{yan2024street} (object-specific), \ac{pvg} \cite{chen2023periodic} (object-agnostic) and \ac{deformgs} \cite{yang2024deformable}, which uses a global, non-separable gaussian representation, on our selected Waymo sequences. We utilize the benchmark suite from \cite{chen2025omnire} for evaluation, with overall results presented in \tabref{tab:std_extra_results_waymo_main}. Our method, NAG, outperforms all these additional baselines in terms of PSNR (+4.08 dB) and SSIM (+0.014), while remaining competitive in LPIPS (+0.005). Furthermore, the object-specific results in \tabref{tab:std_extra_results_waymo_object} confirm NAG's superior performance in preserving structural details and enhancing dynamic object representation.

We note, that PVG yields a 0.85 dB higher PSNR score than \ac{ore}. Yet, given \ac{pvg} is an object-agnostic representation, while \ac{ore} is object-specific, with support for object-based editing, more closely matching our methods capabilities, we stick to \ac{ore} as our main GS comparison within our manuscript.

\begin{table*}[t]
    \centering
    \caption{Quantitative Evaluation on Dynamic Driving Sequences of the Waymo~\cite{sun2020scalability} Open Driving Dataset. The temporal consistency is measured by the inter-frame standard deviation ($\pm$ STD), which is calculated over sub-segments and mean-aggregated per sequence. Best results are in bold. PVG refers to Periodic Vibration Gaussian \cite{chen2023periodic}, DGS to Deformable 3D Gaussians \cite{ye2022deformable}, and SGS to Street Gaussians\cite{yan2024street}. Our method compares favorably against these additional object-agnostic (PVG, DGS) and object-specific (SGS) baselines, including high consistency in PSNR, SSIM and LPIPS.
    }
    \label{tab:std_extra_results_waymo_main}
    \setlength{\tabcolsep}{5pt}
    \resizebox{\textwidth}{!}{%
    \begin{tabular}{l S[table-format=2.2] S[table-format=2.2] S[table-format=2.2] S[table-format=2.2] S[table-format=1.3] S[table-format=1.3] S[table-format=1.3] S[table-format=1.3] S[table-format=1.3] S[table-format=1.3] S[table-format=1.3] S[table-format=1.3] }

         \toprule
          \multirow{2}{*}{Seq.} & \multicolumn{4}{c}{PSNR $\uparrow$} & \multicolumn{4}{c}{SSIM $\uparrow$} & \multicolumn{4}{c}{LPIPS $\downarrow$} \\
         & {Ours} & {PVG} & {DGS} & {SGS} & {Ours} & {PVG} & {DGS} & {SGS} & {Ours} & {PVG} & {DGS} & {SGS} \\
         \midrule
       \multirow{2}{*}{s-975} & \textbf{40.21} & 34.00 & 36.67 & 36.19 & \textbf{0.976} & 0.958 & 0.961 & 0.962 & 0.058 & 0.064 & 0.059 & \textbf{0.054} \\
      & \pm1.11 & \pm1.88 & \pm1.96 & \pm2.24 & \pm0.004 & \pm0.006 & \pm0.007 & \pm0.005 & \pm0.012 & \pm0.008 & \pm0.018 & \pm0.010\\
      
      \multirow{2}{*}{s-203} & \textbf{43.15} & 38.71 & 35.86 & 33.08 & \textbf{0.978} & 0.966 & 0.960 & 0.958 & 0.070 & \textbf{0.052} & 0.063 & 0.060 \\
      & \pm0.39 & \pm0.98 & \pm1.03 & \pm1.81 & \pm0.001 & \pm0.001 & \pm0.002 & \pm0.003 & \pm0.004 & \pm0.002 & \pm0.004 & \pm0.004\\
      
      \multirow{2}{*}{s-125} & \textbf{43.32} & 38.78 & 35.82 & 38.47 & \textbf{0.980} & 0.964 & 0.958 & 0.964 & 0.057 & 0.046 & 0.053 & \textbf{0.040} \\
      & \pm0.49 & \pm0.61 & \pm1.16 & \pm0.72 & \pm0.003 & \pm0.001 & \pm0.004 & \pm0.002 & \pm0.007 & \pm0.002 & \pm0.005 & \pm0.002\\
      
      \multirow{2}{*}{s-141} & \textbf{42.55} & 38.31 & 33.97 & 33.56 & \textbf{0.978} & 0.963 & 0.949 & 0.954 & 0.057 & \textbf{0.054} & 0.080 & 0.065 \\
      & \pm1.60 & \pm0.54 & \pm1.06 & \pm1.21 & \pm0.003 & \pm0.002 & \pm0.005 & \pm0.003 & \pm0.005 & \pm0.004 & \pm0.010 & \pm0.005\\
      
      \multirow{2}{*}{s-952} & \textbf{41.89} & 39.55 & 35.62 & 34.78 & \textbf{0.976} & 0.968 & 0.963 & 0.961 & 0.058 & \textbf{0.041} & 0.047 & 0.047 \\
      & \pm0.59 & \pm0.79 & \pm1.59 & \pm1.38 & \pm0.003 & \pm0.002 & \pm0.004 & \pm0.003 & \pm0.006 & \pm0.003 & \pm0.008 & \pm0.004\\
      
      \multirow{2}{*}{s-324} & \textbf{40.85} & 36.97 & 34.01 & 30.96 & \textbf{0.977} & 0.961 & 0.953 & 0.942 & \textbf{0.038} & \textbf{0.038} & 0.049 & 0.059 \\
      & \pm1.31 & \pm0.65 & \pm1.25 & \pm2.10 & \pm0.002 & \pm0.002 & \pm0.004 & \pm0.010 & \pm0.004 & \pm0.003 & \pm0.006 & \pm0.009\\
      
      \multirow{2}{*}{s-344} & \textbf{41.84} & 38.47 & 32.75 & 29.90 & \textbf{0.983} & 0.968 & 0.957 & 0.953 & 0.031 & \textbf{0.030} & 0.046 & 0.050 \\
      & \pm0.52 & \pm0.72 & \pm1.32 & \pm1.01 & \pm0.001 & \pm0.002 & \pm0.004 & \pm0.004 & \pm0.002 & \pm0.002 & \pm0.005 & \pm0.006\\
      
      \midrule
      \multirow{2}{*}{Mean} & \textbf{41.85} & 37.77 & 34.83 & 33.70 & \textbf{0.978} & 0.964 & 0.957 & 0.956 & 0.051 & \textbf{0.046} & 0.056 & 0.054 \\
      & \pm0.91 & \pm0.87 & \pm1.36 & \pm1.49 & \pm0.002 & \pm0.003 & \pm0.004 & \pm0.005 & \pm0.006 & \pm0.003 & \pm0.008 & \pm0.006\\

      \bottomrule
    \end{tabular}%
    }
\end{table*}

\begin{table*}[tbh]
    \centering
    \caption{Quantitative Evaluation of Human and Vehicle Rendering on Waymo~\cite{sun2020scalability} Driving Sequences for additional 3DGS methods. The temporal consistency is computed as in \tabref{tab:std_extra_results_waymo_main} - by the inter-frame standard deviation ($\pm$ STD), calculated over sub-segments and mean-aggregated per sequence.}
    \label{tab:std_extra_results_waymo_object}
    \setlength{\tabcolsep}{5pt}
    \resizebox{\textwidth}{!}{%
    \begin{tabular}{l 
    S[table-format=2.2] S[table-format=2.2] S[table-format=2.2] S[table-format=2.2]
    S[table-format=1.3] S[table-format=1.3] S[table-format=1.3] S[table-format=1.3]
    S[table-format=2.2] S[table-format=2.2] S[table-format=2.2] S[table-format=2.2]
    S[table-format=1.3] S[table-format=1.3] S[table-format=1.3] S[table-format=1.3]}
          \toprule
        \multirow{2}{*}{Seq.} & \multicolumn{4}{c}{Vehicle PSNR $\uparrow$} & \multicolumn{4}{c}{Vehicle SSIM $\uparrow$} & \multicolumn{4}{c}{Human PSNR $\uparrow$} & \multicolumn{4}{c}{Human SSIM $\uparrow$} \\
        & {Ours} & {PVG} & {DGS} & {SGS} & {Ours} & {PVG} & {DGS} & {SGS} & {Ours} & {PVG} & {DGS} & {SGS} & {Ours} & {PVG} & {DGS} & {SGS} \\
        \midrule
          \multirow{2}{*}{s-975} & \textbf{46.79} & 33.11 & 27.22 & 32.75 & \textbf{0.991} & 0.912 & 0.824 & 0.930 & \textbf{45.37} & 33.06 & 26.09 & 23.01 & \textbf{0.989} & 0.903 & 0.788 & 0.744 \\
          & \pm1.21 & \pm1.82 & \pm2.55 & \pm2.97 & \pm0.001 & \pm0.025 & \pm0.064 & \pm0.031 & \pm1.58 & \pm1.90 & \pm2.64 & \pm2.93 & \pm0.002 & \pm0.026 & \pm0.049 & \pm0.048 \\
          
          \multirow{2}{*}{s-203} & \textbf{41.90} & 34.46 & 28.39 & 30.19 & \textbf{0.986} & 0.946 & 0.848 & 0.897 & \textbf{45.40} & 35.79 & 27.28 & 16.22 & \textbf{0.986} & 0.947 & 0.870 & 0.665 \\
          & \pm1.89 & \pm2.05 & \pm2.94 & \pm2.96 & \pm0.005 & \pm0.019 & \pm0.063 & \pm0.050 & \pm1.65 & \pm1.50 & \pm2.88 & \pm2.57 & \pm0.004 & \pm0.006 & \pm0.031 & \pm0.067 \\
          
          \multirow{2}{*}{s-125} & \textbf{41.00} & 33.61 & 25.97 & 28.82 & \textbf{0.989} & 0.955 & 0.815 & 0.875 & N/A & N/A & N/A & N/A & N/A & N/A & N/A & N/A \\
          & \pm1.90 & \pm1.28 & \pm1.95 & \pm2.39 & \pm0.004 & \pm0.011 & \pm0.053 & \pm0.054 & N/A & N/A & N/A & N/A & N/A & N/A & N/A & N/A \\
          
          \multirow{2}{*}{s-141} & \textbf{43.21} & 33.76 & 26.15 & 32.07 & \textbf{0.981} & 0.936 & 0.801 & 0.916 & \textbf{44.22} & 36.13 & 29.78 & 25.08 & \textbf{0.986} & 0.940 & 0.828 & 0.725 \\
          & \pm1.44 & \pm1.84 & \pm2.29 & \pm2.34 & \pm0.007 & \pm0.015 & \pm0.050 & \pm0.032 & \pm1.61 & \pm1.67 & \pm3.31 & \pm2.91 & \pm0.005 & \pm0.016 & \pm0.070 & \pm0.098 \\
          
          \multirow{2}{*}{s-952} & \textbf{40.94} & 32.88 & 27.27 & 28.74 & \textbf{0.986} & 0.942 & 0.848 & 0.894 & \textbf{40.45} & 34.90 & 26.02 & 23.29 & \textbf{0.968} & 0.928 & 0.706 & 0.629 \\
          & \pm1.46 & \pm1.97 & \pm3.70 & \pm3.88 & \pm0.004 & \pm0.016 & \pm0.063 & \pm0.055 & \pm2.82 & \pm1.77 & \pm3.29 & \pm2.77 & \pm0.021 & \pm0.018 & \pm0.087 & \pm0.092 \\
          
          \multirow{2}{*}{s-324} & \textbf{41.71} & 34.03 & 29.20 & 29.29 & \textbf{0.986} & 0.948 & 0.869 & 0.880 & \textbf{44.12} & 34.16 & 26.20 & 22.56 & \textbf{0.988} & 0.918 & 0.727 & 0.626 \\
          & \pm1.56 & \pm2.19 & \pm3.29 & \pm3.73 & \pm0.004 & \pm0.016 & \pm0.057 & \pm0.062 & \pm1.95 & \pm2.21 & \pm2.69 & \pm2.40 & \pm0.005 & \pm0.027 & \pm0.076 & \pm0.098 \\
          
          \multirow{2}{*}{s-344} & \textbf{43.97} & 34.59 & 28.59 & 28.81 & \textbf{0.985} & 0.948 & 0.823 & 0.860 & \textbf{40.99} & 33.99 & 21.37 & 17.04 & \textbf{0.975} & 0.933 & 0.640 & 0.509 \\
          & \pm1.69 & \pm1.66 & \pm2.77 & \pm3.35 & \pm0.007 & \pm0.010 & \pm0.031 & \pm0.032 & \pm2.97 & \pm1.60 & \pm2.60 & \pm2.35 & \pm0.016 & \pm0.014 & \pm0.086 & \pm0.095 \\
          
          \midrule
          \multirow{2}{*}{Mean} & \textbf{42.88} & 33.74 & 27.54 & 30.10 & \textbf{0.986} & 0.940 & 0.832 & 0.893 & \textbf{42.94} & 34.63 & 25.93 & 21.82 & \textbf{0.981} & 0.927 & 0.740 & 0.639 \\
          & \pm1.56 & \pm1.83 & \pm2.81 & \pm3.13 & \pm0.005 & \pm0.016 & \pm0.054 & \pm0.045 & \pm2.21 & \pm1.81 & \pm2.92 & \pm2.64 & \pm0.010 & \pm0.019 & \pm0.074 & \pm0.088 \\
          \bottomrule
    \end{tabular}%
    }
\end{table*}

\subsection{Training Time}
\label{sec:training_time}

The training duration for the Neural Atlas Graph ranges from 2 to 6 hours per scene on an NVIDIA L40 GPU. To accurately assess this performance, we contextualize these training times within the domain of dynamic neural scene representation and we further detail the sources of this computational cost. State-of-the-art methods, such as  \ac{ore} \cite{chen2025omnire}, report comparable training durations (e.g., approximately one hour for a single scene at a lower 960×640 resolution). Crucially, the proposed method consistently trains at full resolution (1920×1280 for Waymo and up to 1920×1080 for DAVIS), which substantially increases the computational load compared to baselines often optimized for lower resolutions. Furthermore, highly optimized methods, such as those leveraging Gaussian Splatting, have benefited from dedicated native CUDA implementations and extensive optimization efforts over recent years \cite{kerbl_3d_2023, ye2025gsplat}.

 \tabref{tab:training_time} provides a direct comparison of observed training times across all baselines at their full respective resolutions. For Waymo, we used a subsequence from segment 975 with a length of 68 frames, and for DAVIS, the popular \textit{bear} sequence with 74 frames.

\begin{table}[h]
    \centering
    \caption{Observed training times at full resolution for a representative sequence within the datasets.}
    \label{tab:training_time}
    \begin{tabular}{llc}
    \toprule
    \textbf{Dataset} & \textbf{Method} & \textbf{Time (min.)} \\
    \midrule
    \multirow{3}{*}{Waymo} & Ours & 140 \\
     & ORe \cite{chen2025omnire} & 127 \\
     & ERF \cite{yang2024emernerf} & 42 \\
    \midrule
    \multirow{3}{*}{Davis} & Ours & 198 \\
     & ORF \cite{lin2023omnimatterf} & 279 \\
     & LNA \cite{kasten2021layered} & 444 \\
    \bottomrule
    \end{tabular}
\end{table}

Our training time is significantly shorter than the other atlas-based methods, \ac{lna} and \ac{orf}, on the DAVIS dataset. On Waymo, our time is comparable to \ac{ore} \cite{chen2025omnire} but slower, which is attributable to the latter's highly optimized Gaussian Splatting implementations. \ac{erf} \cite{yang2024emernerf} is significantly faster as it is not a scene graph method and lacks a dedicated model per object, reducing architectural overhead. It should also be noted that ERF generally produced less accurate results in our quantitative experiments.

We emphasize that three key components are essential for achieving the high-quality and editable \ac{nag} representation, inherently contribute to the overall training duration:

\begin{enumerate}
    \item \textbf{Extensive Ray-Casting}: Learning dynamic video at high resolution necessitates extensive ray-casting. For instance, our full training involves $2.8 \times 10^9$ rays over 80 epochs. Training time naturally reduces for lower resolutions or shorter videos (e.g., to under 20 minutes for highly reduced settings).
    \item \textbf{Multi-Stage Optimization}: A three-phase optimization strategy is employed to ensure stable convergence, accurate decomposition, and enhanced texture editability, yet may add an overhead in execution time.
    \item \textbf{Per-Object Networks}: The training time scales with scene complexity due to the presence of dedicated, independent neural networks for each object and the background, increasing the total parameter count and computation per step.
\end{enumerate}

These architectural choices are a necessary investment to deliver the decomposition and editing capabilities that are the focus of this work. Full training details and ablation studies are available in the Experimental Details (\secref{sec:dataset_details}) of this supplementary material. We see potential areas for future optimization in all these components, such as leveraging more efficient ray-casting implementations, employing object size-driven network architectures, or improving initialization strategies to significantly reduce training time.

\section{Discussion on Scene Representation Methods}

\acresetall 

As recent literature has introduced a wide range dynamic scene representation methods tailored to different domains, in this section we summarize their core ideas and priors to situate Neural Atlas Graphs (NAGs) within the broader research context. We separate these broadly into 2D-, 2.5D, and 3D model categories based on their representation domain. Furthermore, we classify them based on whether they are object-specific (separating objects based on modality/user input) or object-agnostic (differentiating only static and dynamic scenbe content). These classifications only serve as broad guidelines as their boundaries can be fluid. 

\paragraph{2D - Scene Models}

This family of models focuses primarily on 2D decomposition, rooted in classic video layer separation and sprite-based techniques. The key idea here is to decompose an abritrary input video into a (static) background and one or more dynamic layers.

\ac{lna} \cite{kasten2021layered} achieve this through a time-consistent \textit{atlas} or \textit{canvas} representation. LNA learns a 2D-to-2D warp that maps scene points onto this unwrapped atlas. This design results in a highly compact video representation that facilitates direct appearance editing on the 2D atlas -- similar to editing a regular image -- and propagating those changes consistently across all time steps. LNA is highly self-contained, requiring only the input video, masks for designated foreground layers, and pre-computed optical flow. Given that masks are used to separate objects, LNA can be treated as object-specific, relying on a separate masking pipeline. Subsequent works like Deformable Sprites \cite{ye2022deformable} take a similar approach, but learn a sprite-based deformation and rely on motion cues rather than explicit masks to separate individual layers.

\paragraph{2.5D - Mixed Scene Models}

This category is comprised of models which integrate elements of 3D reasoning, such as shared reference frames or volumetric models, yet remain grounded in a predominantly layered or planar framework.

As an example, Omnimatte \cite{Lu_2021_CVPR} focuses on separating objects along with their associated effects (shadows, dust, reflections). Omnimatte uses inputs similar to LNA but computes frame-by-frame homographies to yield a common reference system, which places it in the 2.5D domain due to its reliance on 3D planar motion assumptions while not explicitly making use of a 3D mesh representation. The layers are typically represented as U-Nets \cite{ronneberger2015u}, allowing for high fidelity, but they lack the direct, intuitive editing control of LNA’s atlas structure.

\ac{orf} \cite{lin2023omnimatterf}, a direct successor of Omnimatte, addresses the limited expressivity of a 2D background layer by replacing it with a volumetric 3D NeRF. This requires the model to have a camera pose initialization, which can be derived from Structure-from-Motion (SfM) techniques such as COLMAP \cite{schoenberger2016sfm, schoenberger2016mvs} or NeRF-based approaches \cite{liu2023robust}. Similar to Omnimatte, ORF relies on optical flow, masks, homographies, and a (monucular) depth estimation -- remaining object-specific.

A downside shared across these 2D/2.5D models is their reliance on a correct -- often pre-defined -- ordering of the foreground layers. As they compose image layers via alpha matting, an improper order can lead to transparency issues or incorrect information being picked up from objects in the background -- e.g., holes in the foreground content. This becomes a large problem under varying occlusions or in busy scenes where a consistent layer ordering is hard to define as objects move in front and behind one another. 

\paragraph{3D - Scene Models}

This family of models explicitly represents geometry and color content in 3D space, offering a more expressive and potentially more geometrically consistent representation of real-world scenes than 2D or 2.5D methods.

\ac{nsg} \cite{ost2021neural} addresses positional editability by explicitly factorizing the scene into discrete semantic components (e.g., background, individual vehicles, pedestrians), with each object represented as a node in the scene graph. These nodes are modeled as individual radiance fields, linked by edges representing spatial relationships and 3D pose transformations. Following on NSGs, StreetGS \cite{yan2024street} and further \ac{ore} \cite{chen2025omnire} use dynamic 3D Gaussian splatting techniques to represent objects, yielding faster render performance while maintaining positional editing capabilities.
However, editing texture or appearance in these volumetric, neural field, or Gaussian representations is non-trivial, as changes must be propagated in a visually consistent way throughout the continuous 3D content.

The core strength of these object-specific scene graph methods is their high degree of 3D editability; objects can be translated, rotated, and removed via simple graph manipulation. Conversely, these methods are rarely self-contained, requiring extensive semantic and motion priors derived from external structured inputs such as 3D bounding boxes, object trajectories, or LIDAR depth data. Their performance is thus heavily tied to the availability and quality of these priors, commonly found in specialized autonomous driving datasets \cite{sun2020scalability, Geiger2012CVPR}.

Beyond these object-specific approaches, object-agnostic methods have also emerged. These approaches, such as \ac{erf} \cite{yang2024emernerf} and Gaussian Splatting approaches like \ac{pvg} \cite{chen2023periodic}, separate only static from dynamic content. While these methods have a lower reliance on data priors, they are also more difficult to directly edit as their representations can lack semantic meaning without nodes tied to discrete objects.

\paragraph{Neural Atlas Graphs}

We position our Neural Atlas Graphs (NAGs) within the 2.5D domain, drawing inspiration from 3D-based NSGs for scene decomposition while relying on lower dimensional component representations for easier appearance editing. These design choice highlight a core trade-off of all of these methods: 3D expressivity is versus ease of texture editing.

This hybrid structure enables three key capabilities: \textit{Direct Appearance Editing} through manipulation of the 2D atlas canvas (similar to LNA); \textit{Positional Editing} by modifying the learned spline-based rigid motion trajectories for object planes (similar to NSGs); and \textit{Implicit Occlusion Handling} by explicitly modeling object movement and interaction along learned 3D trajectories, directly addressing the layer ordering issue.

In summary, NAGs occupy a unique point in the scene representation spectrum. NAGs adopt 3D structural priors to gain positional editing and robust occlusion handling, while retaining the ease of texture editing of a 2D atlas representations. While this approach requires sacrificing self-containment for structured priors (bounding boxes, trajectories, etc.), we showcase within our Outdoor Videos that even in domains of high self-containment (less geometric prior), NAGs construct a reasonable representation with high fidelity and demonstrate a clear advantage over 3D representations in low-ego motion and low-parallax scenes.

\end{document}